\documentclass[rmp,twocolumn,eqsecnum]{revtex4-1}
\usepackage{amsmath,amssymb,amsfonts}
\usepackage{datetime}
\usepackage{dcolumn}
\usepackage{bm}
\usepackage[]{graphicx}\graphicspath{{graph/},{NewGraphics/}}
\usepackage{epstopdf}
\usepackage{float}
\usepackage{color}
\usepackage{hyperref}
\hypersetup{
  colorlinks,
  citecolor=magenta,
  linkcolor=blue}

\makeatletter
\def\aliasbib#1#2{\expandafter\xdef
\csname b@#1\endcsname{\csname b@#2\endcsname}}
\makeatother

\newcommand{\be}{\begin{equation}}
\newcommand{\ee}{\end{equation}}
\newcommand{\bea}{\begin{eqnarray}}
\newcommand{\eea}{\end{eqnarray}}

\newcommand{\bc}{\begin{center}}
\newcommand{\ec}{\end{center}}

\newcommand{\pa}{\partial}

\renewcommand{\vec}[1]{\mathbf{#1}}

\renewcommand{\max}{{\mathrm{max}}}

\renewcommand{\Re}{{\rm Re}}

\def\lambdabar{\protect\@lambdabar}
\def\@lambdabar{%
\relax \bgroup
\def\@tempa{\hbox{\raise.73\ht0
\hbox to0pt{\kern.25\wd0\vrule width.5\wd0
height.1pt depth.1pt\hss}\box0}}%
\mathchoice{\setbox0\hbox{$\displaystyle\lambda$}\@tempa}%
{\setbox0\hbox{$\textstyle\lambda$}\@tempa}%
{\setbox0\hbox{$\scriptstyle\lambda$}\@tempa}%
{\setbox0\hbox{$\scriptscriptstyle\lambda$}\@tempa}%
\egroup }

\renewcommand{\vec}[1]{\textnormal{\boldmath$#1$}}

\begin{document}

\title{Beam by design: laser manipulation of electrons in modern accelerators}
\author{Erik Hemsing}
\email{Electronic address: ehemsing@slac.stanford.edu}
\author{Gennady Stupakov}
\email{Electronic address: stupakov@slac.stanford.edu}
\author{Dao Xiang}
\email{Electronic address: dxiang@slac.stanford.edu}
\affiliation{SLAC National Accelerator Laboratory, Menlo Park, CA 94025, USA}
\author{Alexander Zholents}
\email{Electronic address: azholents@aps.anl.gov}
\affiliation{Advanced Photon Source, Argonne National Laboratory Argonne, IL 60439, USA}
\begin{abstract}

Accelerator-based light sources such as storage rings and free-electron lasers use relativistic electron beams to produce intense radiation over a wide spectral range for fundamental research in physics, chemistry, materials science, biology and medicine. More than a dozen such sources operate worldwide, and new sources are being built to deliver radiation that meets with the ever increasing sophistication and depth of new research. Even so, conventional accelerator techniques often cannot keep pace with new demands and, thus, new approaches continue to emerge. In this article, we review a variety of recently developed and promising techniques that rely on lasers to manipulate and rearrange the electron distribution in order to tailor the properties of the radiation. Basic theories of electron-laser interactions, techniques to create micro- and nano-structures in electron beams, and techniques to produce radiation with customizable waveforms are reviewed. We overview laser-based techniques for the generation of fully coherent x-rays, mode-locked x-ray pulse trains, light with orbital angular momentum, and attosecond or even zeptosecond long coherent pulses in free-electron lasers. Several methods to generate femtosecond pulses in storage rings are also discussed.  Additionally, we describe various schemes designed to enhance the performance of light sources through precision beam preparation including beam conditioning, laser heating, emittance exchange, and various laser-based diagnostics. Together these techniques represent a new emerging concept of ``beam by design'' in modern accelerators, which is the primary focus of this article

\end{abstract}
\maketitle

\tableofcontents
\section{Introduction}

Large scientific user facilities driven by relativistic electron beams produced in particle accelerators, such as synchrotron light sources and free-electron lasers (FELs), have played a key role in the development of numerous scientific fields. By sending relativistic electron beams through bending magnets and undulators (a series of dipole magnets with alternating fields), these massive machines produce intense radiation from millimeter to x-ray wavelengths that serves an extraordinarily wide array of applications (see, for example \cite{Schneider}).

Since the discovery of synchrotron radiation in 1947 \cite{SR}, accelerator based light sources have evolved from the first generation to the fourth generation, each having a significant enhancement in radiation brightness. Currently, most of the light sources in operation are third generation light sources, i.e. dedicated electron storage rings built since the 1990's that produce radiation in a wide range of photon energies from a few eV to hundreds of keV~\cite{Zhao:RAST}. X-ray FELs, considered as fourth generation light sources, have emerged only recently and are expected to open a new era of x-ray science \cite{FLASH, LCLS_2010, SACLA, FERMI}.

The last two decades have witnessed particularly rapid growth both in the number of scientific users and in the diversity of new science enabled by third and fourth generation light sources. This growth is driven by the significant enhancements in the capabilities of these facilities, as well as the quality and quantity of light they can produce and deliver to experiments. Such enhancements are primarily from advancements in accelerator science and technology, which in many cases includes the use of lasers to manipulate the relativistic electron distributions to precisely tailor the properties of the emitted x-rays.

For instance, the duration of stored electron bunches at equilibrium in synchrotrons is typically on the order of a few tens of picoseconds. This sets the minimum x-ray pulse duration that they can produce, which inhibits investigations of ultrafast structural dynamics that occur on time scales on the order of 100 fs (an atomic vibrational period) or shorter.  Using a femtosecond laser pulse to manipulate the electron energy distribution in a short slice of the electron bunch, however, generates femtosecond time structures in the bunch that emit femtosecond long x-ray pulses \cite{laser-slicing96}. This so-called ``laser slicing'' method has been used in several synchrotrons \cite{laser-slicingLBNL, laser-slicingSLS, laser-slicingBESSY} and immediately opened up many new opportunities for capturing ultrafast dynamics at third generation light sources. The applications include femtosecond x-ray crystallography \cite{LSapplication1}, femtosecond time-resolved x-ray diffraction \cite{LSapplication2, LSapplication3}, and femtosecond time-resolved x-ray absorption spectroscopy \cite{LSapplication4, LSapplication5, LSapplication6}, just to name a few.

Laser-manipulation techniques are particularly well-suited for modern x-ray FELs, which are capable of producing ultrashort and ultra-intense x-ray pulses with peak brightness ten orders of magnitude higher than synchrotrons. There is significant work to be done, however, in order to meet ever-increasing demands for improved brightness, coherence, and control of the x-ray pulses in FELs over the broad range of scientific needs. Laser-based manipulation of the electron beam allows one to rearrange particle distribution in the phase space to meet the requirements of specific applications. With lasers, one can create micro-structures and imprint micro-correlations in relativistic electron beams with extremely high precision in both the temporal and spatial domain, thereby offering numerous options to tailor the properties of the emitted x-rays.

Some prime examples of laser-beam manipulations are laser seeding techniques, which can be used to produce fully coherent x-rays naturally synchronized with the external lasers \cite{HGHG1, EEHG1, XiangStupakov2009}; laser heaters, which are used to suppress the microbunching instability to produce higher radiation power \cite{huang_laser_heater}; periodically modulated beams for the generation of mode-locked x-rays \cite{PhysRevLett.100.203901, Kur, ML3, ML4}; generating localized current peaks and energy-chirps with few-cycle lasers for producing attosecond x-ray pulses \cite{attosecond1, attosecond2}; rearranging an electron beam into an x-ray scale helix for generation of x-rays with orbital angular momentum (OAM) which may be used to probe matter in new ways \cite{OAM1, HemsingOAMFEL}; and producing sequential (or simultaneous) x-ray pulses with varying wavelengths for time-resolved x-ray spectroscopy \cite{FERMI2color}.

This review provides an overview of the physics, challenges and promises relevant to the latest developments in advanced beam manipulation with lasers: the emerging concept of ``beam by design'' in accelerator physics. Section~\ref{sec:2} discusses the theory of electron-laser interactions, including the basics of lasers and relativistic electron beams, and the physics of their interactions in vacuum and in undulators. Section~\ref{sec:3} describes the general concept of using modulator-chicane modules to manipulate the beam phase space for creating fine structures ranging from THz to x-ray wavelengths, as well as for generation of light with OAM. Section~\ref{sec:4} discusses advanced beam manipulation for third generation light sources, focusing on generating femtosecond x-ray pulses. The wide array of different beam manipulation techniques and their applications in fourth generation light sources are discussed in Section~\ref{sec:5}. In Section~\ref{sec:6} several methods are reviewed that allow to improve the electron beam quality for FEL applications, as well some laser based diagnostic techniques to measure electron bunch properties.   A summary is presented in Section~\ref{sec:7}.

%
\section{Basics}\label{sec:2}
%
In this section, we give an overview of the basics of electron beams, laser beams, and their interaction in vacuum and in an undulator. Throughout this paper, we limit our discussions to relativistic electron beams produced in particle accelerators for which the relativistic factor $\gamma={\cal E}/mc^2$ is much larger than one and the normalized electron velocity $\beta=v/c=\sqrt{1-\gamma^{-2}}$ is close to unity. Here $\cal E$ is the electron beam energy, $m$ is the electron mass at rest, and $c$ is the speed of light. We also restrict our attention to purely classical systems where quantum recoil effects can be neglected, as well as systems in the absence of collective effects such as scattering and space-charge.

%
\subsection{Basics of electron beams}
%

\begin{figure}[htb]
\centering
\includegraphics[width=0.4\textwidth, trim=0mm 0mm 0mm 0mm, clip]{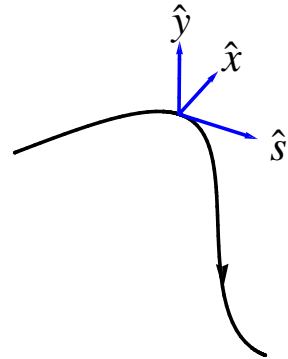}
\caption{Local coordinate system $x$, $y$, $s$ associated with the beam trajectory shown by black curve. The direction of beam propagation is shown by arrow.}
\label{fig:coordinates}
\end{figure}

Beams contain large number of electrons, $N_e$, moving together in space along the trajectory governed by external magnetic and electric fields. The motion of electrons with the reference energy ${\cal E}_0$ in a focusing channel define this trajectory (called the reference trajectory or orbit) and all other electrons move in close proximity with small oscillations around it.  The local coordinate system with unit vectors $(\hat{\vec x}, \hat{\vec y})$ is used to describe these oscillations with $\hat{\vec x}$ being in the plane of orbit curvature and being orthogonal to the velocity vector of the reference electrons and $\hat{\vec y}$ being orthogonal to $\hat{\vec x}$ and the tangent to the reference orbit, see Fig.~\ref{fig:coordinates}. 
The distance along the reference trajectory  is measured by $S$ and the relative position of electrons on this trajectory with respect to the beam center is measured by $s$. 
Thus, an electron location in the six-dimensional (6-D) phase space co-moving with the electron beam is characterized by the vector $\vec{X}=(x, x', y, y', s, \eta)^T$, where $x$ and $y$ are the transverse positions along $(\hat{\vec x}, \hat{\vec y})$ directions respectively, $x'=dx/dS$ and $y'=dy/dS$ are the transverse angles, and $\eta$ is the relative energy deviation, $\eta = \Delta {\cal E}/{\cal E}_{0}$, and the superscript $T$ denotes transposition. Note that the angular divergence of a relativistic beam is typically small, and $x'$ and $y'$ can be associated with the angles (in $x$ and $y$ directions, respectively) between the velocity of the particle and the reference orbit.

It is further convenient to characterize the electron beam distribution in the 6-D phase space by the second order moments, $\langle x^2\rangle$, $\langle xx'\rangle$, $\langle x'^2\rangle$, etc., where the brackets denote averaging over all the particles. Accordingly, the beam matrix is defined as,
\begin{equation} \label{bm}
\Sigma=
\left[
\begin{array}{cccc}
 \langle x^2\rangle & \langle xx'\rangle  & \cdots & \langle x\eta\rangle \\
\langle x'x\rangle & \langle x'^2\rangle  & \cdots &  \langle x'\eta\rangle\\
 \vdots &  &  &  \vdots  \\
 \langle \eta x\rangle  & \cdots & \langle \eta s\rangle &  \langle \eta^2\rangle  \\
\end{array}
\right].
\end{equation}
The absolute value of the determinant of this matrix defines a 6-D volume $V$ of the phase space ellipsoid occupied by the particles,  $V=(|\mathrm{det}\,\Sigma|)^{1/2}$.

The performance of a beam-driven facility usually depends critically on the specific partitioning of the phase space volume in each subspace $(x,x')$, $(y,y')$, $(s,\eta)$. Therefore, the root mean squared (rms) emittance is defined for each subspace individually, with that in $x-x'$ plane given as,
    \begin{equation} \label{emittance}
    \varepsilon_{x}
    = \sqrt{\langle x^2\rangle \langle x'^2\rangle-\langle xx'\rangle^2}
    \,,
    \end{equation}
assuming the first order moments $\langle x\rangle$ and $\langle x'\rangle$ are zero. In a similar way, the rms emittance in $y-y'$ and $s-\eta$ planes are defined as $\varepsilon_{y}$ and $\varepsilon_{s}$. It can be shown that if the beam is not coupled across the $x$, $y$ and $s$ planes then $V=\varepsilon_{x} \varepsilon_{y} \varepsilon_{s}$. With acceleration, however, $\varepsilon_{x}$ is not conserved and the normalized emittance $\varepsilon_{nx}=\beta\gamma\varepsilon_{x}$, that is conserved during acceleration, is used instead~\cite{SYLee}  (and similarly for $y$ and $s$ components). A widely used figure of merit for the electron beam quality is the beam brightness defined as
   \begin{equation} \label{emittance}
    B_e
    = \frac{N_e}{(2\pi)^3 \varepsilon_{nx} \varepsilon_{ny} \varepsilon_{ns}}
    .
    \end{equation}

In linear approximation, when the electron beam moves from $S=S_0$ to $S=S_1$, the phase space coordinates of each electron are transformed from the initial state to a final state as
    \begin{equation} \label{bd}
    \vec{X}_1=R\vec{X}_0
    \,,
        \end{equation}
where $R$ is a $6\times6$ transfer matrix. In a Hamiltonian system the transfer matrix is symplectic~\cite{SYLee}, i.e.,
   \begin{equation} \label{symplectic1}
    RJ_6R^T=R^TJ_6R=J_6
    \,,
        \end{equation}
where
   \begin{equation} \label{symplectic2}
    J_6=\left[
\begin{array}{ccc}
 J & 0 & 0 \\
0 &  J& 0\\
0 & 0 &  J \\
\end{array}
\right]
    \,,
\end{equation}
with $J$ being the unit symplectic matrix in 2-D phase space,
   \begin{equation} \label{symplectic3}
    J=\left[
\begin{array}{cc}
 0 & 1 \\
-1 &  0\\
\end{array}
\right].
\end{equation}

Corresponding to Eq.~(\ref{bd}),  the beam matrix $\Sigma_1$ at $S=S_1$ is connected with the initial beam matrix $\Sigma_0$ at $S=S_0$ as,
    \begin{equation} \label{bm2}
    \Sigma_1=R\Sigma_0R^T.
	\end{equation}

Note, it follows from Eq.~(\ref{symplectic1}) that $\mathrm{det}\,R=1$. It is then straightforward to see from Eq.~(\ref{bm2}) that $V=(|\mathrm{det}\,\Sigma|)^{1/2}$ is invariant under an arbitrary symplectic transformation. Furthermore, it can be shown that the trace of the product $\Sigma J_6\Sigma J_6$ is also an invariant. These two invariants lead to a so-called emittance exchange theorem which states that the emittances of the subspaces cannot be partially transferred from one plane to another if the beam is uncoupled before and after the transformation \cite{Courant}.  Advanced beam manipulation techniques have to obey these basic rules when rearranging the beam distributions in the phase space.

Consider now an uncoupled electron beam going through a focus along a straight line $z$ in free space. The coordinate $S$ in this case is equal to $z$. The rms beam size and divergence at the waist is given by $\sigma_x=\sqrt{\varepsilon_x \beta_{x0}}$ and $\sigma_{x'}=\sqrt{\varepsilon_x /\beta_{x0}}$, where $\beta_{x0}$ is the horizontal Twiss function at the waist \cite{Sands}.  Similar definitions are applicable for the $y$ plane, but the location of the $y$-waist in $z$ may not necessarily coincide with the location of the $x$-waist. The electron beam size varies along the distance $z$ measured from the waist according to $\sigma_x(z)=\sigma_x(0)\sqrt{1+(z/\beta_{x0} )^{2} }$ and it grows by a factor of $\sqrt{2}$ at a distance $z=\beta_{x0}$.
%
\subsection{Basics of laser beams}\label{Section:I.A}
\begin{figure}[!htb]
    \bc
    \includegraphics[draft=false, width=0.4\textwidth]{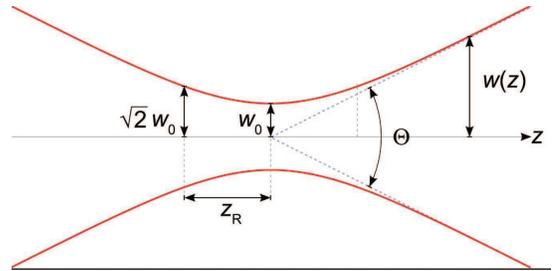}
    \caption{Gaussian laser beam width ${w}({z})$ as a function of the axial distance~$z$.
    \label{Fig.II.A.1}}
    \ec
    \end{figure}

%
Laser beams, particularly those in the form of Gaussian beams, are considered throughout of this paper \cite{siegman}:
	\begin{equation} \label{II.A.1}
	I\left(r,z\right)
	=
	I_{0}
	\left(\frac{w_{0} }{w(z)} \right)^{2}
	e^{-2{r^{2} }/{w(z)^{2} } }
	.
	\end{equation}
Here $I(r,z)$ is the intensity (power per unit area) of the beam, $w(z)$ is the beam radius defined as the distance from the beam axis where intensity drops to $1/e^{2} $ of the intensity on-axis, $I_{0} =I(0,0)$ is the on-axis intensity at the waist, and $w_{0} $ is the beam radius at the waist.  Analogous to an electron beam, the beam radius varies along the axial distance \textit{z} measured from the waist according to $w(z)=w_{0} \sqrt{1+(z/z_{R} )^{2} } $ and the length over which it diverges by a factor of $\sqrt{2} $ determines the Rayleigh length $z_{R} =\pi w_{0}^{2} /\lambda _{L} $, where $\lambda _{L} $ is the wavelength of the laser field, and $\omega_L=2\pi c/\lambda _{L}$ is the laser pulse carrier frequency. The laser beam diverges in the far-field region (i.e., for $z$ values much larger than $z_{R} $) with the angle of divergence $\theta =\lambda _{L} /\pi w_{0} $. All of these variables are shown in Fig.~(\ref{Fig.II.A.1}). The product $w_0 \theta/4$ defines the area in the phase space, typically called the rms light emittance $\varepsilon_L=\lambda_L/4\pi$.
Thus, the Rayleigh length $z_R$ for the light beam plays the same role as $ \beta_{x0}$  for the electron beam.

Gaussian beams are solutions to the so-called paraxial wave equation, which describes waves for which the longitudinal variation is negligible over the scale of the wavelength $\lambda_{L}$. In this limit, the waves are described by Gaussian transverse distributions, for which $\theta \ll 1$ is satisfied, and thus $w_{0} \gg \lambda _{L}$ also holds. Within this approximation, the lowest order solution for a linearly polarized electric field is
	\begin{align} \label{II.A.2}
	E(r,z,t)
      &=
	\frac{E_{0} }{\sqrt{1+(z/z_{R})^{2}}}
	e^{-{r^{2} }/{w(z)^{2} } }
    \\&
	\times
	\sin \left(
	k_{L}z-\omega_{L}t
	+
	\psi _{G}
	+
	k_{L}\frac{r^{2} }{2R(z)}
	+
	\psi_{0}\right)
	\nonumber
	,
	\end{align}
where $E_0$ is the amplitude of the field, $k_{L}=2\pi /\lambda _{L} $ is the wave number, $R(z)=z[1+\left(z_{R} /z\right)^{2}]$ is the radius of curvature of the wavefronts, $\psi _{G} =-\arctan (z/z_{R})$ is the Gouy phase shift and $\psi_{0}$ is an arbitrary phase. The quantity $E$ in this equation (and in the subsequent equations of this section) is any transverse component of the electric field in the beam. The magnetic field has the same functional form as $E$ and is perpendicular both to the electric field and the direction of propagation.
The  laser beam intensity equals the time-averaged Poynting vector and is given by~\eqref{II.A.1} with $I_{0} = (c/8\pi)E_{0}^{2}$.

While these Gaussian laser beams are the simplest and most common, the paraxial wave equation allows also solutions in terms of higher order transverse modes. There are, in fact, a whole family of TEM modes that describe free-space fields with more complicated amplitude and phase structures that have applications in electron beam manipulation techniques. In cartesian coordinates ($x$ and $y$), Hermite-Gaussian modes describe a common TEM$_{nm}$ basis, where the indices $n$ and $m$ determine the field shape in the horizontal and vertical directions, respectively \cite{siegman}. In terms of Hermite polynomials $H_n$, these modes are mutually orthogonal and are,
\begin{align}\label{II.A.2HG}
&E(x,y,z,t)
=
E_{0}\frac{w_0}{w(z)}\exp\left[-\frac{x^2+y^2}{w(z)^2}\right]
\\
&\times
H_n\left(\frac{x\sqrt{2}}{w(z)}\right)
H_m\left(\frac{y\sqrt{2}}{w(z)}\right)
\nonumber\\
&\times
\sin\left[
k_{L}z - \omega_{L}t
+
\frac{k_{L}(x^2+y^2)}{2R(z)}
+
(n+m+1)\psi _{G}\right]
\nonumber.
\end{align}
An explicit expression for the field amplitude of the TEM$_{10}$ mode, which is used in several electron beam manipulation schemes, is:
	\begin{align} \label{II.A.3}
	&E(x,y,z,t)
	=
	\frac{E_{0} }{1+(z/z_{R} )^{2}}
	\frac{2\sqrt{2}\, x}{w_0}
       e^{-({x^2+y^2 })/{w(z)^{2} } }
      \nonumber \\&
      \times
       \sin \left(
       k_{L}z - \omega_{L}t+\psi _{G}^{(1)} +
       k_{L}\frac{x^2+y^2 }{2R(z)}+\psi_{0} \right)
       .
	\end{align}
Both the TEM$_{00}$  and TEM$_{10}$ field modes in \eqref{II.A.2} and \eqref{II.A.3} are examined explicitly in Section \ref{Section:II.C}.

In a cylindrically symmetric system, Laguerre-Gaussian (LG) modes describe paraxial waves in cylindrical coordinates $(r,\phi,z)$,
	\begin{align}\label{II.A.2LG}
	&E(r,\phi,z,t)
	=
	E_{0}\frac{w_0}{w(z)}
	{\exp}\left[-\frac{r^2}{w(z)^2}\right]
	\\
	&\times\left(\frac{r\sqrt{2}}{w(z)}\right)^{|l|}L_p^{|l|}\left(\frac{2r^2}{w(z)^2}\right)
	\nonumber\\
	&\times\sin
	\left[
	k_{L}z - \omega_{L}t
	+
	l\phi
	+
	\frac{k_{L}r^2}{2R(z)}
	+
	(2p+l+1)\psi _{G}\right]
	.
	\nonumber
	\end{align}
The radial mode number is given by the integer $p$, and the functions $L_p^l$ are associated Laguerre polynomials. These modes are of specific interest due to their azimuthal phase structure, given by their dependence on $\phi$ and on the azimuthal mode number $l$. This helical phase structure of the $|l|>0$ modes leads to a non-zero, discrete component of the linear photon momentum spiraling about the propagation axis and thus a well-defined orbital angular momentum (OAM) component carried by the field \cite{Allen}. With this feature, these modes can interact in novel ways with matter, and are examined in the context of advanced accelerator-based light sources in more detail in Section \ref{sec:3-5}.

Expressions~\eqref{II.A.1}-\eqref{II.A.2LG} refer to long laser beams when the variation of the field amplitude with time can be neglected. For a short laser pulse with a Gaussian profile, one usually adds an additional factor $\exp[-(z-ct)^{2}/4c^{2}\sigma_{t}^{2}]$ on the right-hand side of the field expressions to account for finite duration of the pulse, with $\sigma_{t}$ the rms duration of the pulse intensity. Such a pulse is referred to as being \emph{transform limited}, in that the pulse duration corresponds to a minimum spectral bandwidth of $\sigma_\omega=1/2\sigma_{t}$.  Given the inverse relationship, narrow band pulses $\sigma_\omega/\omega_L \ll 1$ are long in time, whereas short temporal pulses consisting of only a few cycles extend over a broad spectral bandwidth.

\begin{figure}[!htb]
    \bc
    \includegraphics[draft=false, width=0.4\textwidth]{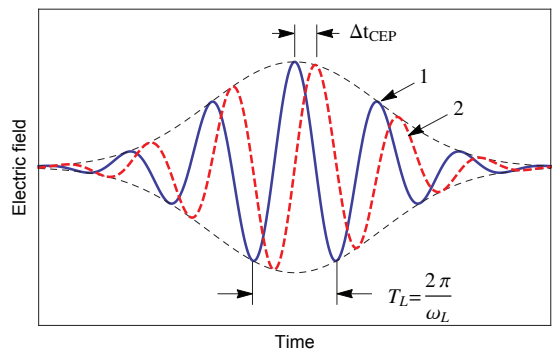}
    \caption{Laser pulses containing two oscillation cycles
     within the full width at half maximum of their intensity profile,
      with cosine (1) and sine (2) shaped waveforms.
    \label{Fig.II.A.2}}
    \ec
    \end{figure}

Finally, we note that $\psi_{0}$ determines the timing of the field oscillations with respect to the laser pulse peak $\Delta t_{\rm CEP}=\psi_{0}/\omega_L$ and is usually referred to as the carrier-envelope phase (CEP, see, e.g., \cite{Jones}). In the case of long pulses, the CEP typically plays no role in the laser dynamics, and is usually neglected. But when the laser pulse is comprised of only a few oscillation cycles of the electromagnetic field, as shown in Fig.~(\ref{Fig.II.A.2}), the CEP defines the temporal evolution of the field. As we will see, this can be important to the structure of the electron beam modulations generated by the laser, and thus on the structure of the emitted x-rays. A technique of CEP stabilization for IR lasers was developed to ensure reproducibility of $\Delta t_{\rm CEP}$ from pulse-to-pulse with tens of attosecond precision~\cite{Borchers}.

%
\subsection{Interaction of a point charge with an electromagnetic field in free space}\label{sec:2-2}
%

A fundamental question to laser-beam interactions is whether an electromagnetic field can effectively interact with a moving point charge in vacuum far from material boundaries. As it turns out, the answer to this question is no if a particle is moving with a constant velocity along a straight line (that is, its motion is not affected either by the electromagnetic field\footnote{Neglecting the deviation of the orbit from a straight line due to the Lorentz force of the electromagnetic field, we eliminate the Thomson scattering effect from the analysis. This appears in the second order of the perturbation theory and is proportional to the square of the magnitude of the field.} or by any other external field in the system). This statement is often referred to as the ``General Acceleration Theorem''~\cite{Palmer:1987hi,Palmer:1994tj}. The formal proof of this theorem is given below.

We calculate the energy gain $\Delta{\cal E}$ of a point charge $e$ passing through electromagnetic field $\vec{E}(\vec{r},t)$ assuming that it moves with a constant velocity $\vec{v}$,
   \begin{align}\label{eq:I-C.1}
    \Delta{\cal E}
    =
    e
    \int_{-\infty}^\infty
    \vec{v}\cdot
    \vec{E}(\vec r(t),t)
    dt
    ,
    \end{align}
where $\vec r(t)$ is the particle's trajectory. For motion with constant velocity, the trajectory is $\vec r(t) = \vec{r}_0 + \vec{v}t$, with $\vec r_{0}$  the particle's position at $t=0$. The electromagnetic field in vacuum can be represented as a superposition of plane electromagnetic waves that propagate with the speed of light:
   \begin{align}\label{eq:I-C.2}
    \vec{E}(\vec{r},t)
    =
    \int
    d^3 k
    \vec{\tilde E}(\vec{k})
    e^{i\vec{k}\cdot\vec{r}
    -i\omega t}
    \,,   
    \end{align}
with $\omega = ck$. Subtituting~\eqref{eq:I-C.2} to~\eqref{eq:I-C.1} we obtain
    \begin{align}\label{eq:I-C.3}
    \Delta{\cal E}
    &=
    e\vec{v}\cdot
    \int_{-\infty}^{\infty} dt
    \int
    d^3 k
    \vec{\tilde E}(\vec{k})
    e^{i\vec{k}\cdot(\vec{r}_0 + \vec{v}t)
    -i\omega t}
    \nonumber\\
    &=
    2\pi e
    \int
    d^3 k
    \,
    \vec{v}\cdot
    \vec{\tilde E}(\vec{k})
    e^{i\vec{k}\cdot\vec{r}_0 }
    \delta(\omega -\vec{k}\cdot\vec{v})
    \,.
    \end{align}
The argument in the delta function in the last integral is never
equal to zero, because
   \begin{align}\label{eq:I-C.4}
    \omega -\vec{k}\cdot\vec{v}
    =
    ck(1-\beta \cos\alpha)
    >0
    \,,
    \end{align}
where $\alpha$ is the angle between $\vec k$ and $\vec v$. Hence the integral~\eqref{eq:I-C.3} vanishes and $\Delta{\cal E}=0$.

If follows from the General Acceleration Theorem that to accelerate (or decelerate) charges with electromagnetic fields, one has to break at least one of the conditions assumed in the above proof. There are several ways in which this can be achieved. One way is to introduce close material boundary conditions that would allow for evanescent fields in the system and invalidate the assumption of free-space plane waves in ~\eqref{eq:I-C.2}. This is the mechanism behind radio frequency (rf) acceleration in rf cavities and structures, as well as plasma and dielectric wakefield acceleration~\cite{plasma_acceleration,PhysRevLett.54.693, Shersby, Gai}. Another approach is to bend the particle's trajectory and violate the assumption of the straight orbit in Eq.~\eqref{eq:I-C.1}. This can be done efficiently by introducing an external magnetic field in the interaction region. This concept forms the foundation of the laser modulator using magnetic undulators considered both throughout this review and in detail in Section~\ref{Section:II.C}.

%
\subsection{Acceleration by electromagnetic fields and energy balance}\label{sec:I-B-2}
%

In the language of quantum mechanics, acceleration by electromagnetic fields in free space corresponds to the absorption of a photon by a freely moving charged particle. However, such an absorption is forbidden by conservation of energy and momentum unless the particle also radiates a photon in the process of the interaction. This is the well-known Compton scattering effect in quantum electrodynamics.
In a similar fashion, the apparent absence of sustained energy exchange between the electromagnetic field and a charge in the classical model of the previous section can be attributed to the absence of radiation effects if a particle is assumed to move with a constant velocity. This example indicates that radiation and acceleration under the influence of the external field are intimately related phenomena. Indeed, as we show in this section through  analysis of the energy balance equation,  the interference of the accelerating field with the particle's radiation field provides an independent method for calculation of the rate of acceleration~\cite{huang04sz,xie03}. 

\begin{figure}[htb]
\centering
\includegraphics[width=0.3\textwidth, trim=0mm 0mm 0mm 0mm, clip]{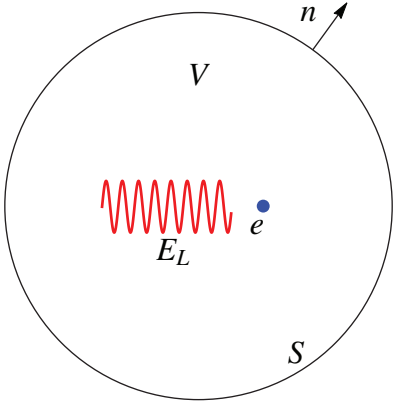}
\caption{Illustration of interaction of a point charge with electromagnetic field.}
\label{fig:poynting}
\end{figure}

Some general properties of the radiation-acceleration connection can be analyzed without specifying the mechanism of the interaction between the particle and the fields. Consider a moving point charge $e$ interacting with an external laser field $\vec E_{L}$.  We assume that the interaction occurs in a volume $V$ enclosed by surface $S$ of large radius $R$ (which we later assume approaches infinity, $R\to \infty$) as shown Fig.~\ref{fig:poynting}.
Initially, at $t \rightarrow -\infty$, the charge and the laser pulse are located outside of the surface $S$. At some time they cross the boundary and interact inside the volume. After the interaction, at $t \rightarrow \infty$, they leave the volume.

 The energy balance equation for the total electromagnetic field in $V$ reads (see, e.g., \cite{landau_lifshitz_ctf,jackson}):
   \begin{align}\label{eq:I-C.5}
    \frac{\partial }{\partial t}
    \int_V dV
    \frac{E^2 + B^2}{8\pi}
    +
    \int_V  dV \vec{j}\cdot \vec{E}
    =
    -\int_S \vec{S}\cdot \hat{\vec n}\, dS
    \,,
   \end{align}
where the indices $V$ and $S$ under the integral signs indicate that the integration goes over the volume $V$ and  the surface $S$, respectively, $\hat{n}$ is a unit vector in the outward direction normal to the surface, $\vec{j}$ is the current density, $\vec{S}$ is the Poynting vector, $\vec{S} = (c/4\pi)\vec{E}\times\vec{B}$, and $\vec{B}$ is the magnetic field. Integrating this equation over time, from $t = -\infty$ to $t=\infty$, and taking into account that at $t = \pm \infty$ there is no electromagnetic field inside $V$, we obtain
   \begin{align}\label{eq:I-C.6}
    \int_{-\infty}^\infty dt
    \int_V dV \vec{j}\cdot \vec{E}
    =
    -\int_{-\infty}^\infty dt
    \int_S \vec{S}\cdot \hat{\vec n}\, dS
    \,.
   \end{align}
The current density $\vec{j}$ associated with the moving point charge is $\vec{j} = e \vec{v}\delta(\vec{r}(t))$.  Hence the integral on the left hand side of~\eqref{eq:I-C.6} reduces to
   \begin{align}
    \label{eq:I-C.6a}
    \Delta{\cal E}
    =e\int_{-\infty}^\infty \vec{v}\cdot \vec{E}\,dt\,
   \end{align}
taken along the trajectory $\vec{r}(t)$. It is equal to the energy gain (or loss, if negative) $\Delta{\cal E}$ of the particle due to the interaction with the field. This, in turn, is equal to the negative of the time-integrated electromagnetic field power through the surface $S$,
   \begin{align}
    \label{eq:I-C.7}
    \Delta{\cal E}
    =
    -\int_{-\infty}^\infty dt
    \int_S \vec{S}\cdot \hat{\vec n}\, dS
    \,.
   \end{align}
Note that this formula is exact and is valid for arbitrary curvilinear motion of the particle.

It is convenient to take the Fourier transformation of the fields
    \be\label{fourier}
    \left\{
    \begin{array}
    {r}
    \vec{ \tilde E}(\omega)\\
    \vec{ \tilde B}(\omega)
    \end{array}
    \right\}
    =
    \frac{1}{2\pi}
    \int_{-\infty}^\infty
    dt
    e^{i\omega t}
    \left\{
    \begin{array}
    {r}
    \vec{ E}(t)\\
    \vec{ B}(t)
    \end{array}
    \right\}
    \,.
    \ee
Using Parseval's theorem we rewrite~\eqref{eq:I-C.7} in the following form
    \begin{align}\label{eq:I-C.8}
    \Delta{\cal E}
    &=
    -\frac{c}{2}\int_{-\infty}^\infty d\omega
    \int_S
    \Re
    [\vec{\tilde E}(\omega)
    \times
    \vec{ \tilde B^*}(\omega)]
    \cdot \hat{\vec n}\, dS
    ,
    \end{align}
where the asterisk denotes complex conjugation. The field in Eq.~(\ref{eq:I-C.8}) is a superposition of the laser field, $\vec{\tilde E}_{L}$, and the particle field. The latter can be split into the Coulomb field of the particle, which formally appears in~\eqref{eq:I-C.8} due to the particle's crossing of the surface $S$, and the radiation field $\vec{\tilde E}_r$. One can show that the Coulomb field does not contribute to the integral~\eqref{eq:I-C.8}~\cite{xie03}, so the fields in~\eqref{eq:I-C.8} can be considered as a superposition of only $\vec{\tilde E}_{L}$ and $\vec{\tilde E}_r$. For these fields we can use the relations $\vec{\tilde B} = \hat{\vec n}\times \vec{\tilde E}$ and $\hat{\vec n}\cdot \vec{\tilde E} = 0$ which are valid in the far zone, as we assume $R\to \infty$. This simplifies~\eqref{eq:I-C.8} to
    \begin{align}\label{eq:I-C.8-1}
    \Delta{\cal E}
    &=
    -\frac{c}{2}\int_{-\infty}^\infty d\omega
    \int_S
    \Re
    \left(
    \vec{\tilde E}(\omega)
    \cdot
    \vec{ \tilde E^*}(\omega)
    \right)
    \, dS
    .
    \end{align}

The final step in the derivation is to substitute $\vec{\tilde E} = \vec{\tilde E}_{L} + \vec{\tilde E}_r$ and note that, if $\vec{\tilde E}_r$ is not included, we retrieve the result $\Delta{\cal E}=0$ from the previous section. Otherwise we obtain
    \begin{align}\label{eq_for_energy1}
    \Delta{\cal E}
    =
    -{c}
    \int_{-\infty}^\infty d\omega
    \int_S
    \Re
    (
    \vec{\tilde E}_L
    \cdot
    {\vec{\tilde E}_\mathrm{r}^*}
    )
    \, dS
    \end{align}
which is responsible for the acceleration (or deceleration) due to the interaction with the laser field, and is proportional to $e\vec{\tilde E}_{L}$ from the linear dependence of $\vec{\tilde E}_r$ on $e$, as expected. Therefore, we see that a particle can be accelerated (or decelerated) only if it radiates. Note that the purely radiative term $\vec{\tilde E}_r\cdot \vec{\tilde E}_\mathrm{r}^*$ scales as $e^2$ and describes the energy loss of the particle due to the radiation. In applications of interest for this paper it is typically small and is excluded from Eq.~(\ref{eq_for_energy1}).

Eq.~\eqref{eq_for_energy1} identifies the connection between the energy change of a particle in an external field with the interference of its radiation with that field. An practical application of this relationship for optimization of the electron-laser interaction will be given in Section~\ref{Section:II.C}.
%
\subsection{Electron-laser interaction in an undulator}\label{Section:II.C}
%

\begin{figure}[!htb]
    \bc
    \includegraphics[draft=false, width=0.4\textwidth]{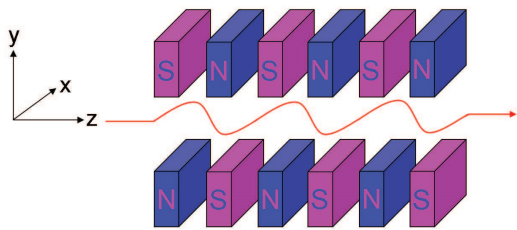}
    \caption{A schematic of a planar undulator and a sine-like trajectory of the electron.
    \label{Fig.II.C.1}}
    \ec
    \end{figure}

As discussed in the previous section, the interaction of relativistic particles with a laser beam is extremely inefficient in free space. The interaction becomes much more pronounced, however, if the laser beam interacts with the electron inside an undulator. Undulators, as depicted in Fig.~\ref{Fig.II.C.1}, are composed of a periodic series of dipole magnets that make the electrons wiggle back and forth transversely. They are commonly used to generate radiation from the beam in accelerator-based light sources. When used in tandem with a laser, they can also be used to create modulations in the electron beam distribution on the scale of the laser wavelength. In this section we present a basic theoretical analysis of the laser-electron interaction in the undulator, obtain the optimal conditions for the interaction, and provide additional insight into the mechanism of energy exchange by considering interference between the laser light and the electron spontaneous radiation fields.

Consider the interaction of an electron with a laser beam in a planar undulator whose magnetic field in the plane of symmetry is
\begin{equation}
{{B}_{y}}(z)={{B}_{0}}\cos ({{k}_{u}}z),
\end{equation}
with $k_{u} = 2\pi/\lambda_{u}$, where $\lambda_{u}$ is the undulator period. We use a Cartesian coordinate system $(x,y,z)$ as shown in Fig.~\ref{Fig.II.C.1}, with the center located in the middle of the undulator and $z$ the direction of propagation.

Taking $z$ as the independent variable, we assume a highly relativistic electron $\gamma\gg1$ whose motion is due only to the planar undulator fields. The Lorentz force on the longitudinally moving electron from the vertical magnetic field generates sinusoidal motion in the $x$ direction. This, in turn, affects motion in the longitudinal direction, as described by the following equations:
	\begin{align} \label{eq:II-C.3}
 	{{\beta }_{x}}(z)
 	&=
 	-\frac{K}{\gamma }\sin \left( {{k}_{u}}z \right),
 	\nonumber \\
 	{{\beta }_{z}}(z)
 	&=
 	\sqrt{1-\frac{1}{{{\gamma }^{2}}}-\beta _{x}^{2}}
 	\nonumber \\ &
 	\approx
 	1-\frac{1}{2{{\gamma }^{2}}}
 	\left( 1+\frac{K^{2}}{2} \right)
 	+
 	\frac{{{K}^{2}}}{4{{\gamma }^{2}}}\cos \left( 2{{k}_{u}}z \right)
	\end{align}
where ${{\beta }_{x,z}}={{v}_{x,z}}/c$ is the normalized velocity and
    \begin{align}\label{und_parameter}
    K=\frac{eB_{0}}{{k}_{u}mc^2}
    \end{align}
is the undulator parameter\footnote{Note that strong undulators with $K\gg 1$ are often called wigglers.}. The longitudinal oscillations trace out a figure eight in the co-moving beam frame, and lead to higher harmonic interactions. The normalized longitudinal velocity, averaged over one undulator period, is ${{\bar{\beta }}_{z}}=1-\left( 1+{{K}^{2}}/2 \right)/2\gamma^{2}$. Thus, an electron slips back a distance
    \begin{align}\label{und_wavelength}
    \lambda_{r} = \lambda_{u}\frac {1+K^2/2} {2\gamma^{2}}
    \end{align}
after traveling one undulator period with respect to the distance
propagated by the light it generates. Due to this periodic slippage, light emitted at this wavelength interferes constructively each undulator period. $\lambda_r$ is therefore
called the undulator's resonant wavelength~\cite{McNeilNature}. Thus, sustained
energy transfer can occur between the co-propagating electromagnetic
field with wavelength $\lambda_L\simeq\lambda_r$ and the electrons. This
is the principle behind the laser modulator, where electrons can be
accelerated (or decelerated) by a co-propagating laser.

To couple to the horizontal electron motion, we consider a simple linearly polarized laser field $E_{x}$. From Eq. (\ref{eq:I-C.6a}), the wiggling electron changes its energy due the work of the laser electric field, as governed by
	\begin{align} \label{eq:II-C.1}
	\frac{d\gamma}{dz}
    =
    \frac{e}{mc^2v_z}{E}_{x} v_{x}
    \approx
    \frac{e}{m{{c}^{2}}}{E}_{x} {\beta }_{x}
    ,
	\end{align}
where $dt=dz/v_z\approx dz/c$ is assumed, valid for a highly relativistic electron. To give a feel for the variety of different regimes in which laser modulators are typically used, a couple of different forms for the electric field are considered in the following sections.

%
\subsubsection{Energy modulation by a plane electromagnetic wave}
%

Let us first consider the simplest case of a plane wave of the form,
	\begin{align} \label{eq:II-C.2a}
	E_{x}(z,t)=E_0\sin \left[ k_{L}(z-ct)\right]
	.
	\end{align}
This model assumes that the laser Rayleigh length $z_R$ is much longer than the undulator length $L_u=N_u\lambda_u$ where $N_u$ is the number of periods, that the rms width of the laser pulse intensity envelope $c\sigma_t$ is much longer than the total slippage length in the undulator $N_u\lambda_r$, and that the laser waist is large compared to the horizontal and vertical electron bunch rms sizes in the undulator, ${{w}_{0}}\gg{{\sigma }_{x,y}}$. This is a reasonable approximation for many beam manipulations with the beam energy $\lesssim$ 1 GeV and picosecond or longer optical laser pulses. Using Eq.s (\ref{eq:II-C.3} -
\ref{eq:II-C.2a}), one obtains
	\begin{align} \label{eq:II-C.1a}
	\frac{d\gamma}{dz}
    =-\frac{eKE_0}{\gamma m c^2}\sin \left( {{k}_{u}}z \right)\sin \left[ k_{L}(z-ct)\right]\, .
	\end{align}

It is convenient to rewrite $t$ in terms of the independent variable $z$,
\begin{equation}
\begin{aligned}
ct(z)&=\int\frac{dz}{\beta_{z}(z)}\\
&\approx-s/{{\bar{\beta }}_{z}}+z+\frac{z}{2{{\gamma }^{2}}}
 	\left( 1+\frac{K^{2}}{2} \right)-\frac{{{K}^{2}}}{8{{k}_{u}}{{\gamma }^{2}}}\sin
	\left(2{{k}_{u}}z\right),
	\end{aligned}
	\label{eq:II-C.4-2a}
\end{equation}
where we have used the co-moving coordinate
\begin{equation}
s=z-c{{\bar{\beta }}_{z}}t \, ,
\end{equation}
which is the longitudinal position of the electron in an electron bunch, with positive $s$ corresponding to the bunch head.
It is further convenient to define the resonant electron energy, also called an FEL resonant energy,
$\gamma_{r}^{2}=k_{L}\left(1+K^{2}/2 \right)/2k_{u}$, and assume a small energy deviation $(\gamma -\gamma_{r})/\gamma_{r}\ll1$ for the electron energy. Therefore, we find from~\eqref{eq:II-C.4-2a} that
	\begin{equation} \label{II.C.6)}
	k_{L}(z-ct)
	=
	k_{L}s/{{\bar{\beta }}_{z}}
	-
	{{k}_{u}}z\frac{\gamma _{r}^{2}}{\gamma^{2}}
	+
	\frac{\xi }{2}\sin \left( 2{{k}_{u}}z \right),
	\end{equation}
where $\xi ={{K}^{2}}/(2+{{K}^{2}})$. Inserting this into Eqs (\ref{eq:II-C.1a}) and averaging over the undulator period $\lambda_u$, the energy change of the electron at resonance is given by
	\begin{align} \label{eq:II-C.1b}
	\frac{d\gamma}{dz}
    =-\frac{eKE_0{\cal J} }{2\gamma m c^2}\cos(k_Ls/{{\bar{\beta }}_{z}})\, .
	\end{align}
where ${\cal J} = J_0\left( \xi/2 \right)- J_1\left( \xi/2 \right)$. In a laser modulator, the energy modulation is typically small compared to the average beam energy such that, to lowest order, the particles' change in position $ds/dz$ can be neglected. This can be integrated directly over the undulator length, and written in terms of the laser beam power $P_{L}=\left( cE_{0}^{2}/8\pi  \right)\left( \pi w_{0}^2/2 \right)$. The energy change at the end of the undulator is then given by,
\begin{align} \label{eq:II-C.13}
	\Delta \gamma (s)
    =\sqrt{\frac{P_L}{P_0}}\frac{2K L_u \cal J}{\gamma w_0} \cos \left( k_{L}s \right)
	\end{align}
where we have assumed ${{\bar{\beta }}_{z}}\approx 1$ and we see that the modulation period along $s$ is the same as that of the laser.
Here we have defined $P_{0}=I_{A}mc^{2}/e\approx8.7~$GW, and $I_{A}=mc^3/e\approx17~$kA is the Alf\'{e}n current.

This plane wave  approximation provides an accurate description of the sinusoidal modulation imprinted onto the electron beam energy distribution in many cases. In some devices however, such as those with much larger electron beam energies, the modulators tend to be longer so that diffraction and slippage effects become important. These are examined in the next section.

%
\subsubsection{Energy modulation by a finite duration laser pulse}\label{Section:II.C-3}
%

In this section we consider a Gaussian laser pulse with an arbitrary pulse width limited only by a condition $c\sigma_{t} \gg 1/k_{L}$. We also include the effects of diffraction on the laser over long distances, but maintain the prior assumption that the waist is large compared to the beam, ${{w}_{0}}\gg{{\sigma }_{x,y}}$. In this case we can neglect the radial dependence in Eq. \eqref{II.A.2} and use the following simplified equation for the field:
	\begin{align} \label{eq:II-C.2}
	E_{x}(z,t)
	&=
	\frac{{{E}_{0}}}{\sqrt{1+{{\left( z/{{z}_{R}} \right)}^{2}}}}
    \sin \left[ k_{L}(z-ct)+\psi  \right]
    \nonumber\\
    &\times
    {{e}^{-{{(z-ct)}^{2}}/4\left(c\sigma _{t}\right)^{2}}}
    ,
	\end{align}
with $\psi =-\arctan (z/z_{R})+{{\psi }_{0}}$, where ${{\psi }_{0}}$ is an arbitrary phase of the laser wave. Combining this field with \eqref{eq:II-C.3} and \eqref{eq:II-C.1} we write:
	\begin{align} \label{eq:II-C.5}
	\frac{d\gamma }{dz}
	&=
	-\frac{e{E_{0}}K}{m{c^{2}}
	\gamma
	\sqrt{1+{{\left( z/z_{R} \right)}^{2}}}}
	\sin \left( k_{u}z \right)
	\nonumber\\&
	\times
	\sin \left( k_{L}(z-ct)+\psi  \right)
	{e^{-{(z-ct)^{2}}/4\left(c\sigma _{t}\right)^{2}}}.
	\end{align}

We choose the center of the undulator at $z=0$ and introduce the dimensionless variable $\hat{z}=z/L_u$ so that the undulator occupies the region $- \frac{1}{2}<\hat z < \frac{1}{2}$. Note that according to our choice of the coordinate system the center of the laser pulse arrives in the middle of the undulator at $t=0$. From~\eqref{eq:II-C.5} one obtains:
	\begin{align} \label{eq:II-C.8}
	\frac{d\gamma }{d\hat{z}}
	&=
	\frac{e{{E}_{0}}K{{L}_{u}}{\cal J}}{2m{{c}^{2}}\gamma }
	\frac{\cos\hat\psi}
	{\sqrt{1+{{\left(q\hat{z}\right)}^{2}}}}
	{{e}^{-{{(\hat{z}/{\tau}-s/c{\sigma}_{t})}^{2}/4}}}
	\end{align}
where the definition $\hat\psi=2\pi\nu\hat{z}-\arctan\left(q\hat{z}\right)+{\psi}_{0}+k_{L}s$ is used for brevity and we have also defined
    \begin{align}\label{eq:II-C.9}
    \nu
    =
    \frac{2{N}_{u} (\gamma-\gamma_{r})}{{\gamma }_{r}}
    ,\qquad
    q
    =
    \frac{L_{u}}{z_{R}}
    ,\qquad
    \tau
    =
    \frac{c{\sigma}_{t}}
    {{N}_{u}\lambda_{L}}
    .
    \end{align}
Eq.~\eqref{eq:II-C.8} is valid inside the undulator and ${d\gamma }/{d\hat{z}} =0$ outside of it.

By introducing the pulse energy ${A_{L}}={P_{L}}\sqrt{2\pi }\sigma_{t}$, an electron's energy change is finally obtained by integrating \eqref{eq:II-C.8} \cite{Zhol3},
\begin{align} \label{eq:II-C.11}
	\Delta \gamma (q,\nu ,\tau,s)
	&=
	\frac{2}{m{{c}^{2}}}
	{\cal J}
	\sqrt{{{A}_{L}}\alpha\hbar{{\omega}_{L}} \xi }
	f(q,\nu,\tau,s)
	\nonumber\\
	&\times\cos(k_{L}s+{{\psi}_{0}})
	\end{align}
where
	\begin{align} \label{eq:II-C.12}
	f(q,\nu ,\tau,s)
	&=
	\sqrt{\frac{2q}{\sqrt{2\pi} \tau}}
	\int\limits_{-1/2}^{1/2}
	\frac{\cos\left(2\pi\nu\hat{z}-\arctan\left(q\hat{z}\right)\right)}
	{\sqrt{1+{{\left(q\hat{z}\right)}^{2}}}}
	\nonumber\\
	&\times
	e^{-{{(\hat{z}/\tau-s/c{\sigma}_{t})}^{2}/4}}
	d\hat{z}
	\end{align}
and $\alpha=e^2/\hbar c \approx 1/137$ is the fine structure constant and $\hbar$ is Planck's constant.
	
\begin{figure}[!htb]
    \centering
    \includegraphics[draft=false, width=0.2\textwidth]{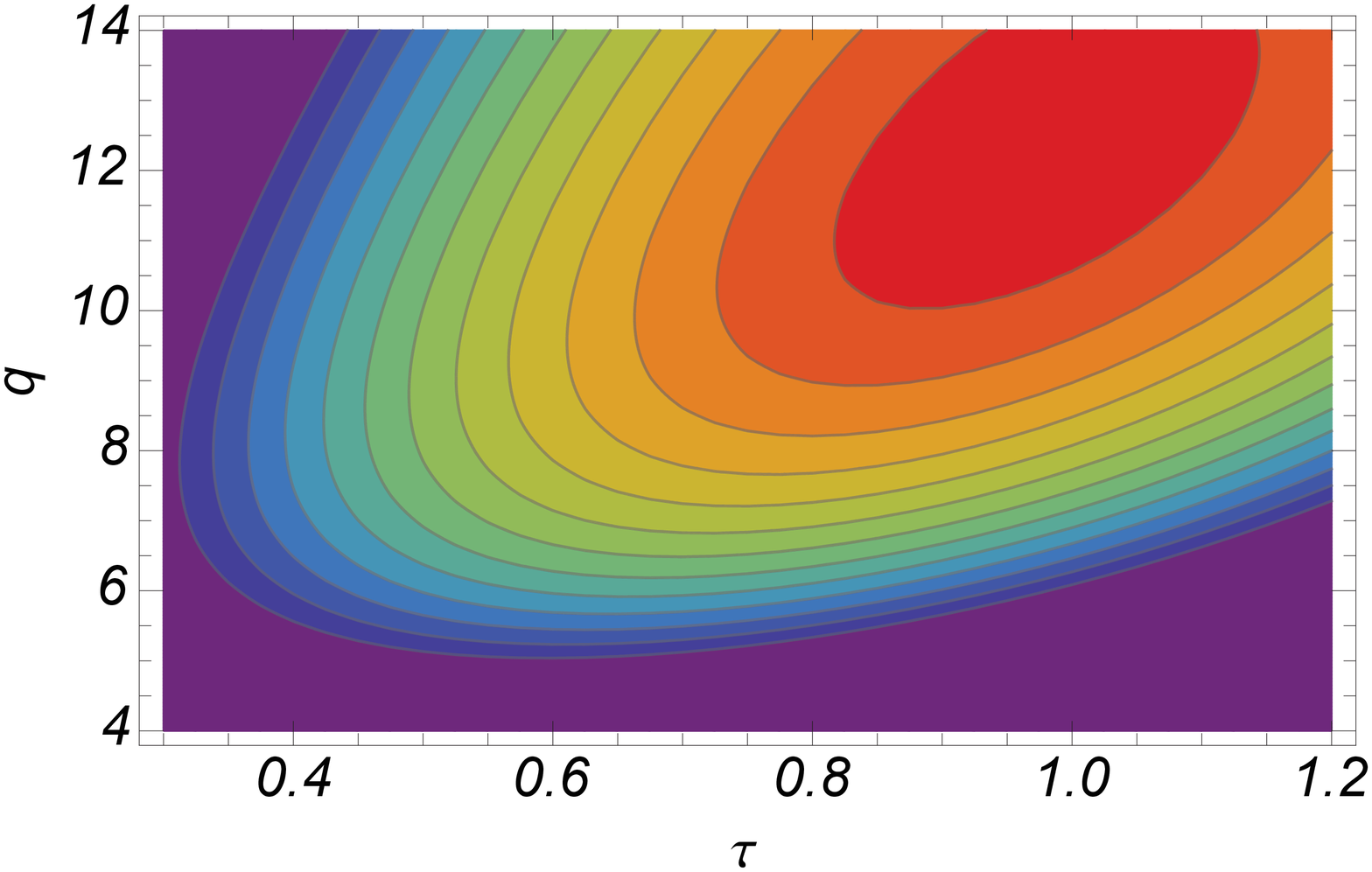}
    \includegraphics[draft=false, width=0.2\textwidth]{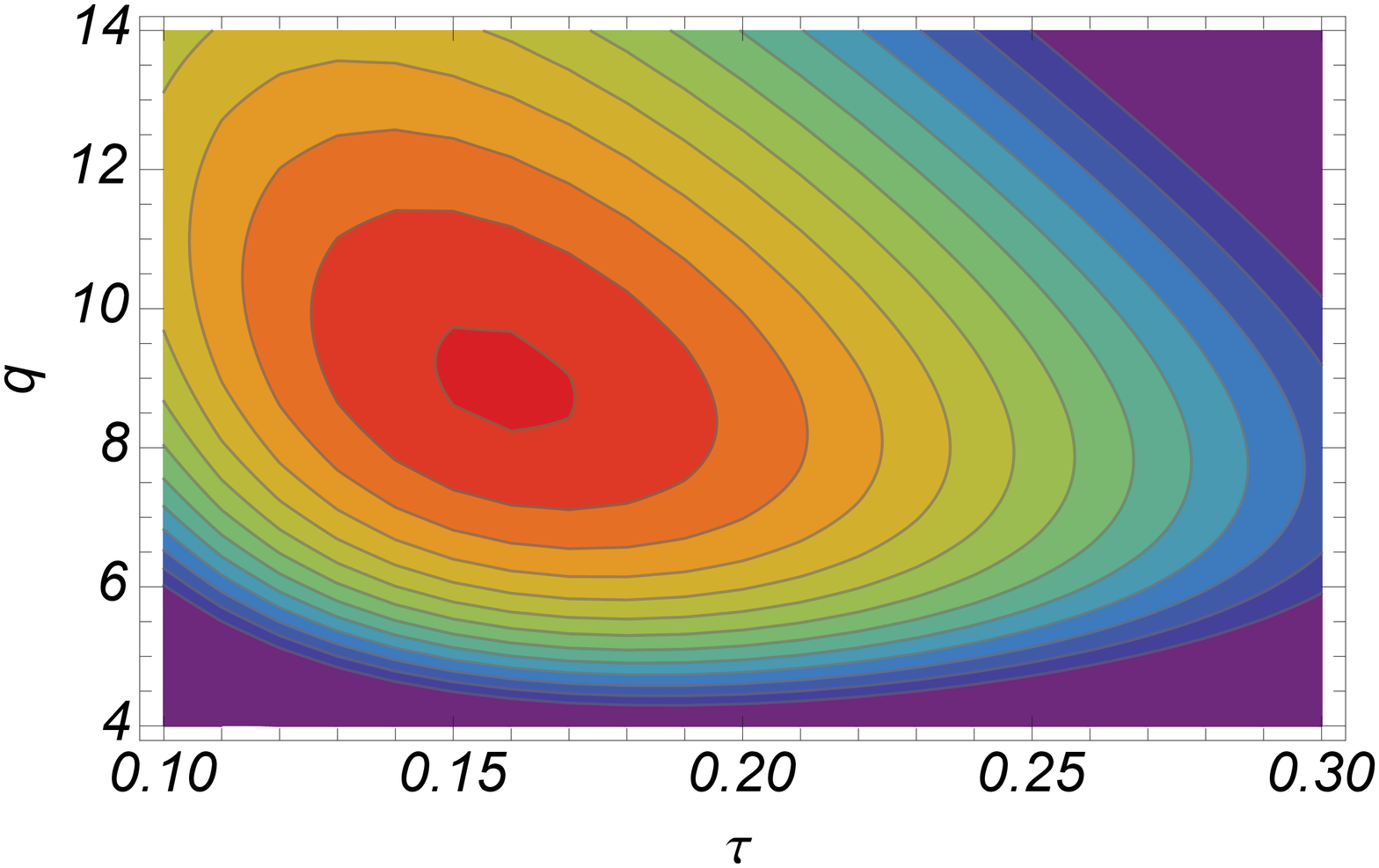}
    \includegraphics[draft=false, width=0.036\textwidth]{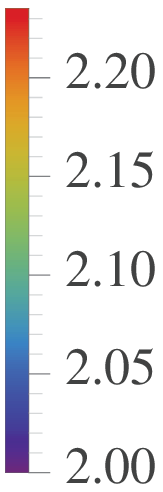}
    \caption{Contour plots of function $f$ using $\tau=0.12, s=0$ (left panel) and $\nu=0.7, s=0 $ (right panel) showing the range of function between 2.0 and 2.3 with 15 equidistant grade step.
    \label{Fig.II.C.2}}
    \end{figure}

Analysis of this equation shows that the function $f$ reaches its maximum $f_{\max}=2.24$ for the following values of its arguments: $q=11.85$, $\nu=0.95$, $\tau = 0.12$. It has a broad maximum around these values which is illustrated by Fig.~\ref{Fig.II.C.2}.
One can see that in practice the $\tau$ can be made considerably larger (i.e., longer pulse length) and $q$ smaller (longer Rayleigh length) to avoid tight laser beam focusing without a large decrease in $f$. The physical interpretation of these optimizations is discussed in the next section.

We note that it is also possible to generalize Eq.~\eqref{eq:II-C.11} to the case with an arbitrary ratio between ${{w}_{0}}$  and ${{\sigma }_{x,y}}$ by adding a factor $\exp(-(x^2+y^2)/{w_{0}^{2}(1+{{(q\hat{z})}^{2}})})$ under the integral in formula \eqref{eq:II-C.12}. This is accurate in the limit that the transverse dependence of the laser phase in the area occupied by the electron beam is negligible.

It is worth pointing out that electrons also interact with the laser light in the undulator when the laser frequency matches the undulator harmonic frequency, {\it i.e.} when $\lambda_L\simeq \lambda_{r}/n$, where $n$ is the odd harmonic number $n=1,3,5, ...$. In this case Eq.~\eqref{eq:II-C.13} and \eqref{eq:II-C.11} should be used with the substitutions ${\cal J} \rightarrow {\cal J}\left(n\right) = (-1)^{(n-1)/2}\left(J_{(n-1)/2}\left( n\xi/2 \right) - J_{(n+1)/2}\left( n\xi/2 \right) \right)$ and $N_u \rightarrow nN_u$ followed by substitutions $\nu \rightarrow n\nu$  and $\tau \rightarrow \tau/n$ \cite{Colson1, Schmuser}. When $K \gg 1$, the coupling of the light to electrons only weakly depends on $n$ since $|{\cal J}\left(n\right)| \simeq 0.68 n^{-2/3}$.

%
\subsubsection{A connection between laser-beam coupling and spontaneous radiation}
%

While Eq.~\eqref{eq:II-C.11} gives a formal solution of the beam energy modulation problem, it is instructive to analyze the result and to understand the optimal values of the various parameters. This can be done by recalling the results of Section~\ref{sec:I-B-2} where it was shown that the energy gain of a particle is due to the interference of the laser field in the far zone with the radiation field of the particle. It follows from Eq.~\eqref{eq_for_energy1} that obtaining the maximum energy modulation amplitude for a given laser pulse energy requires achieving the best possible overlap of the laser field with the field of electron spontaneous emission in the far field region. The spectra must also overlap. These conditions are achieved by selecting the undulator radiation frequency $\omega_{r}$ to match with the carrier frequency of the laser pulse ${{\omega }_{L}}$, the laser pulse bandwidth to match the bandwidth of the spontaneous emission, and also the laser field rms size and divergence in the far field region to match with the size and divergence of the field of spontaneous emission. The latter is regulated by adjusting the focusing of the laser light into the undulator.

Note that the optimal values of $q$ and $\tau$ obtained in the previous section correspond to the optimization analysis outlined above from the point of view of field interference. Indeed, the laser pulse relative bandwidth is $\Delta \omega_{L}/\omega_{L} \sim (\omega_{L}\sigma_{t})^{-1} = (2\pi \tau N_{u})^{-1}$, and for the optimal value $\tau = 0.12$ it is on the order of the relative bandwidth $1/N_{u}$ of the resonant frequency of the undulator radiation. 

Using the maximum value of $f_{\max}=2.24$, we note that the product $f_{\max}^{2}\xi {\cal J}^2 \alpha \hbar {{\omega }_{L}}$ is approximately equal to the energy $A_{S}$ of spontaneous emission of the electron in the fundamental mode. The optimized Eq.~\eqref{eq:II-C.11} can then be written as
	\begin{equation} \label{II.C.11}
	\Delta {{\gamma }_{\max }}
    \approx
    \frac{2}{m{{c}^{2}}}
    \varkappa
    \sqrt{{{A}_{L}}{{A}_{S}}}
	\end{equation}
where $\varkappa$ is a numerical parameter of order of one.

The phase velocity of the
laser field at the focus is greater than velocity of light in vacuum (this effect is known as a Guoy phase shift). Therefore,
in order to maintain optimal interaction with the electron over the entire undulator length, the laser frequency $\omega_L$ must be
red-shifted relative to the frequency of the electron spontaneous emission in the undulator $\omega_r=2\pi c/\lambda_r$. This explains why the maximum amplitude of energy modulation is obtained by detuning $\nu\approx 0.95$ from the undulator resonance.

The phenomena of the spectral shift  can be equivalently explained by the angular-frequency correlation
in the electron undulator emission $\omega_r\left(\theta\right)=\omega_r\left(0\right)\left[1+{\gamma^2\theta^2}/({1+K^2/2})\right]^{-1}$, where $\theta$ is the observation angle. Because of the angular dependence, the maximum spectral intensity of the electron undulator emission integrated over the solid angle is red-shifted relative to $\omega_r\left(0\right)$. Therefore, for better overlap of the laser and spontaneous radiation fields in the far field region, the laser frequency should be red-shifted correspondingly.

%
\subsubsection{Angular modulation}\label{Section:II.C.AM}
%

In certain scenarios, higher-order laser modes provide an additional level of control over the electron distribution. In Sections~\ref{Section:IV.D}, ~\ref{V.H} and~\ref{Sec:MeasuringUltraShortBeams}, for example, TEM$_{10}$ laser modes are examined to impart an optical-scale angular kick to the electrons. Repeating the same analysis as above, but now using the TEM$_{10}$ mode given by \eqref{II.A.3}, one can show that the interaction with this field changes both the electron energy and the electron transverse momentum \cite{zholents2008}. The energy modulation in the beam is now calculated using the field
	\begin{align} \label{eq:II-C.14}
	{{E}_{x}}
	&=
	\frac{{{E}_{0}}}{1+{{(z/{{z}_{R}})}^{2}}}\frac{2\sqrt{2}x}{{{w}_{0}}}
	\sin (k_{L}(z-ct)+{{\psi }^{(1)}})
	\nonumber\\
	&\times
	{e^{-{(z-ct)^{2}}/4\left(c\sigma _{t}\right)^{2} }}
	,
	\end{align}
where  ${{\psi }^{(1)}}=\psi _{G}^{(1)}+{{\psi }_{0}}$. Repeating the calculations of the previous section one finds
    \begin{align}\label{eq:II-C.15}
  	\frac{\Delta \gamma}{\gamma }
	 (q,\nu ,\tau,s)
	&=
	\frac{2 K}{\gamma^2}
	\sqrt{\frac{P_L}{P_0} } {\cal J}k_{L} x_0
	\nonumber \\&
	\times
	 f_1(q,\nu,\tau,s)
	\cos(ks+\psi_0)
    \end{align}
where $x_0$ is the electron horizontal offset and
	\begin{align}\label{eq:II-C.18}
	{{f}_{1}}(q,\nu ,\tau,s)
	&=
	q
	\int\limits_{-1/2}^{1/2}
	\frac{\cos(2\pi\nu\hat{z}-2\arctan(q\hat{z}))}
	{1+{{\left(q\hat{z}\right)}^{2}}}
	\nonumber\\&\times
	{{e}^{-{{(\hat{z}/\tau-s/c{\sigma}_{\tau})}^{2}/4}}}
	d\hat{z}
	.
	\end{align}
Note that now, electrons on-axis are unchanged in energy. One can then use the Panofsky-Wenzel theorem \cite{panofsky56w} which relates the transverse variation of $\Delta \gamma$ with the longitudinal variation of the angular kick $\Delta {x}'$,
	\begin{align}\label{eq:II-C.16}
	\frac{\partial \Delta {x}'}{\partial s}
	=
	\frac{\partial }{\partial {{x}_{0}}}
	\left( \frac{\Delta \gamma }{\gamma } \right)
	.
	\end{align}
The result is
	\begin{align} \label{eq:II-C.17}
	\Delta{x}'(q,\nu ,\tau,s)
	&=
	\frac{2 K}{\gamma^{2}}
	\sqrt{\frac{P_L}{P_0} }{\cal J}
	\nonumber \\&
	\times
	f_1(q,\nu,\tau,s)\sin(ks+{{\psi}_{0}})
	\end{align}
	
As in the previous section, Eq. \eqref{eq:II-C.17} can be further simplified in the long laser pulse regime, or generalized to the case with an arbitrary ratio between ${{w}_{0}}$  and ${{\sigma }_{x,y}}$ by inserting $\exp[-{(x_{0}^{2}+{y}_0^{2})}/{w_{0}^{2}(1+{{(q\hat{z})}^{2}})}]$ under the integral in Eq.~\eqref{eq:II-C.18}.

%
\subsection{Incoherent and coherent radiation from beams}\label{incoh_coh_radiation}
%

As was emphasized in the Introduction, radiation from relativistic beams is widely used as a powerful scientific tool in various areas of research. The radiation properties critically depend on how electrons are distributed in the beam. If electrons are positioned randomly throughout the bunch, short wavelength radiation (that is the radiation with the wavelength shorter than the bunch length) is \emph{incoherent}, and its intensity is proportional to the number of particles in the beam. Much more intense \emph{coherent} radiation can be generated if the particle positions in the beam are correlated, and if the correlation length is comparable with the wavelength of the radiation. The intensity of the coherent radiation scales as the number of particles squared and can greatly exceed the incoherent radiation. A remarkable example of beams that radiate coherently is a free-electron laser, where the correlations between the particles' positions are achieved through an instability developed in the long FEL undulator.

To better understand the relation between the two types of radiation, we assume that each particle of the beam, passing through a radiator, emits electromagnetic field whose time Fourier transform is $\tilde E(\omega)$. Within a numerical factor, the spectral intensity $I(\omega)$ of radiation from one particle is equal to $|\tilde E(\omega)|^{2}$. The radiation field of the beam with $N_e$ particles is
	\begin{align}\label{eq:coh-field}
	\tilde E_{\mathrm{b}}(\omega)
	=
	\sum_{j=1}^{N_e}\tilde E(\omega)e^{i\omega t_{j}}
	,
	\end{align}
where $t_{j}$ is the arrival time to the radiator of particle $j$, and the factor $e^{i\omega t_{j}}$ takes into account the phase shifts between radiation fields of different particles in the beam. The intensity of the beam radiation $I_{\mathrm{b}}$ is the absolute value of the total field squared,
    \begin{align}\label{eq:coh-rad-1}
    I_{\mathrm{b}}(\omega)
    =
    I(\omega)
    \bigg|
    \sum_{j=1}^{N_e}e^{i\omega t_{j}}
    \bigg|^{2}
    =
    I(\omega)
    \left(
    N_e
    +
    \sum_{j\ne m}
    e^{i\omega (t_{j}-t_{m})}
    \right)
    .
    \end{align}
The first term on the right hand side describes incoherent radiation, $I_{\mathrm{b}}^{(\mathrm{incoh})}(\omega)=N_eI(\omega)$. The second one involves correlations between positions of pairs of particles. To clarify the scaling of this term we assume that arrival times can be adequately described by a distribution function $f(t)$ such that the probability of the arrival time $t_{j}$ (for $j=1,2,\ldots N_e$) to be equal to $t$ within an interval $dt$ is given by $f(t)dt$. Then the second term in~\eqref{eq:coh-rad-1}, which is the intensity of the coherent radiation, can be written as
    \begin{align}\label{eq:coh-rad-2}
    I_{\mathrm{b}}^{(\mathrm{coh})}(\omega)
    &=
    N_e(N_e-1)
    I(\omega)
    \int dtdt'
    f(t)f(t')
    e^{i\omega(t-t')}
    \nonumber\\&
    \approx
    N_e^{2}
    I(\omega)
    |\hat f(\omega)|^{2}
    ,
    \end{align}
where $\hat f(\omega) =    \int_{-\infty}^{\infty} dt f(t)e^{i\omega t}$, and we used $N_e\gg 1$. We see that, indeed, the coherent radiation scales as $N_e^{2}$.

In many cases, for short wavelength radiation, the coherent term can be neglected. Indeed assuming a Gaussian distribution, $f(t) \propto e^{-c^2t^{2}/2\sigma_{s}^{2}}$, with $\sigma_{s}$ the rms bunch length, it is easy to find that $|\hat f(\omega)|^{2} = e^{-\omega^{2}\sigma_{s}^{2}/c^2}$,
and for the frequencies larger than $c/\sigma_{s}$, $|\hat f(\omega)|^{2}$ becomes exponentially small. This is typical for x-ray radiation in storage rings.

Laser manipulations with relativistic beams, as we will see in subsequent sections, allow one to introduce a density modulation with the wavelength $\lambda \ll \sigma_{s}$. To understand radiation properties of such a beam we consider a simple model where the sinusoidal modulation with the wavenumber $k_{0} = \omega_{0}/c \gg 1/cT$ is imposed on a beam with a flat profile. The resulting distribution function of the beam is
    \begin{align}\label{eq:coh-rad-3}
    f(t)
    =
    \frac{1}{T}
    \left[
    1
    +
    2b
    \sin(\omega_{0}t)
    \right]
    ,
    \,\,\,
    \mathrm{for}
    \,\,\,
    -\frac{1}{2T}
    <t<
    \frac{1}{2T}
    ,
    \end{align}
and $f=0$ otherwise. In this equation $T$ is the beam duration and the half-amplitude of the relative modulation $b$ is traditionally called the \emph{bunching factor} of the modulated beam. Concentrating on the contribution to $|\hat f(\omega)|^{2}$ due to the modulation only, one finds that the dominant term at $\omega > 0$ is
    \begin{align}\label{eq:coh-rad-4}
    |\hat f(\omega)|^{2}
    =
    \frac{4b^{2}}{T^{2}}
    \frac{\sin^{2}[(\omega-\omega_{0})T/2]}{(\omega-\omega_{0})^{2}}
    \to
    2\pi
    \frac{ b^{2}}{T}
    \delta(\omega-\omega_{0})
    ,
    \end{align}
where in the last expression we formally took the limit $T\to\infty$.

Using~\eqref{eq:coh-rad-4} we can now estimate when the coherent radiation of a modulated beam exceeds the incoherent one. Substituting~\eqref{eq:coh-rad-4} into~\eqref{eq:coh-rad-2} and integrating over the frequency we find
    \begin{align}
    \int d\omega
    I_{\mathrm{b}}^{(\mathrm{coh})}(\omega)
    =
    2\pi
    \frac{N_e^{2} b^{2}}{T}
    I(\omega_{0})
    ,
    \end{align}
while $\int d\omega I_{\mathrm{b}}^{(\mathrm{incoh})}(\omega)$ can be evaluated as $N_eI(\omega_{0})\Delta \omega$, where $\Delta \omega$ is the characteristic width of the radiation spectrum. Comparing the two we see that the coherent radiation is dominant if
    \begin{align}\label{eq:coh-rad-6}
    b
    >
    b_\mathrm{sn}
    \equiv
    \left(
    \frac{T}{2\pi N_e}
    \Delta \omega
    \right)^{1/2}
    =
    \left(
    \frac{1}{N_0\lambda_{0}}
    \frac{\Delta \omega}{\omega_{0}}
    \right)^{1/2}
    ,
    \end{align}
where $N_0 = N_e/cT$ is the number of particles per unit length, and $\lambda_{0}=2\pi/c\omega_{0}$ is the radiation wavelength. The quantity $b_\mathrm{sn}$ can be interpreted as the bunching factor associated with the shot noise in the beam. When the beam bunching greatly exceeds $b_\mathrm{sn}$ its radiation intensity is strongly enhanced. Moreover, as indicated by~\eqref{eq:coh-rad-4} its spectrum can become narrower than the original spectrum if $T^{-1} \ll \Delta \omega$. This property of radiation of modulated beams forms the foundation for several  \emph{FEL seeding} methods~\cite{HGHG1,EEHG1} which allow narrowing of the FEL spectrum.
As an illustration of typical values of $b_\mathrm{sn}$ in FELs, we note that for a beam with peak current of 1 kA, wavelength $\lambda_0 = 1$ nm, and relative FEL bandwidth of $\Delta \omega/\omega_0 = 10^{-3}$ we find $b_\mathrm{sn} \approx 2\times 10^{-4}$.

%
\section{Optical manipulation of electron beams with lasers}\label{sec:3}
%
For most of the cases discussed in this paper, lasers are used to change the longitudinal distribution of electrons in the beam. In the longitudinal plane, beam manipulation typically requires a dispersive transport element. This is because relativistic electrons with $\gamma\gg1$ travel close to the speed of light (e.g., for 1 GeV electron, $1-v/c \approx 1.3\cdot 10^{-7}$). As a result, in modern beams with typical small energy spreads, the relative longitudinal velocities of electrons are so small that electrons do not change their relative positions when the beam travels along a straight line in a drift. With a dispersive element, however, one can force the electrons with different energies to follow different paths in order to rearrange them longitudinally. In this section we review a wide range of techniques that use lasers and dispersive elements to manipulate beam distributions.

%
\subsection{Beam manipulation using modulator-chicane modules}\label{sec:3-1}
%

\begin{figure}[!ht]
    \includegraphics[width=0.45\textwidth]{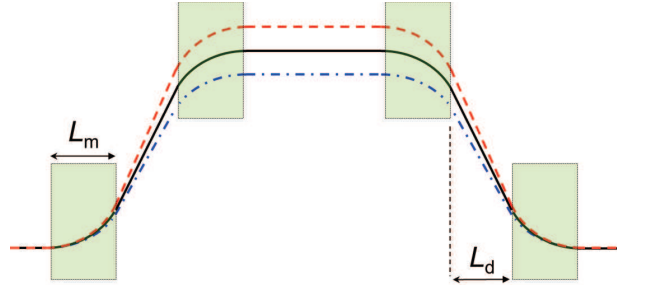}
    \caption{Schematic of orbits in a chicane for the reference particle (black solid line), for a particle with slightly higher energy (blue dashed dotted line) and for a particle with slightly lower energy (red dashed line). The 4 dipole magnets are illustrated with green blocks.
    \label{chicane}}
    \end{figure}

The most widely used longitudinally dispersive element is a chicane which typically consists of four dipole magnets, as shown in Fig.~\ref{chicane}. In a chicane, particles with lower energies are bent more and have longer path lengths, while particles with higher energies are bent less and have shorter path lengths. In a negatively chirped bunch where the bunch tail has higher energy than bunch head, for example, the tail particles catch up with the head particles in the chicane and as a result the bunch is compressed (assuming that the tail does not overtake the head). In contrast, a positively chirped bunch will be decompressed when it passes through a chicane.

One primary application of a chicane is to compress the beam to obtain high peak currents. The process of bunch compression, to first order, can be described as a linear transformation where the bunch length is reduced while the energy spread and peak current
are both increased, as expected from conservation of phase space and charge, respectively. Typically, the required negative chirp for compression is achieved by accelerating the beam off-crest in rf cavities.

For more advanced beam manipulations in the longitudinal plane, the rf cavity is replaced with a laser and an undulator. The sinusoidal modulation induced by the electron-laser interaction in an undulator described in \ref{Section:II.C} consists of both positive and negative chirp regions in electron beams longer than the laser wavelength. Therefore, after passing through a chicane with proper dispersive strength characterized by the transport matrix element $R_{56}=\partial s/\partial \eta$, a density modulation at the scale of the laser wavelength is generated. (See Figure \ref{HGHG} and the discussion in the next section.) A handy formula for the dispersion of the typical symmetric four-dipole chicane (all four magnets are identical) shown in Fig. \ref{chicane} is $R_{56}\simeq 2\alpha_0^2(L_d+\frac{2}{3}L_m)$, where $\alpha_0$ is the bend angle. Depending on the desired application, the undulators and chicanes are configured in different ways to produce precisely tailored beam distributions.

%
\subsubsection{Combination of one modulator and one chicane}\label{sec:3-1-1}
%

\begin{figure}[!ht]
    \includegraphics[width=0.4\textwidth, trim=0mm 21mm 0mm 5.5mm, clip]{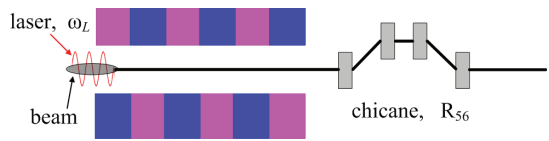}
    \caption{The beam energy is modulated in an
    undulator due to the interaction with a laser beam. The beam then passes
    through a dispersion section to form a density modulation.
    \label{fig1-1}}
    \end{figure}

We consider the standard setup shown in Fig.~\ref{fig1-1} which is used to imprint optical density modulations on relativistic beams.
This setup is typically used for the coherent harmonic generation (CHG) \cite{LURE, Kinkaid}  (see, also Section ~\ref{sec:4-D}) and for high-gain harmonic generation (HGHG) in FELs ~\cite{BenZvi1991181,HGHG1}.

We assume an initial Gaussian beam energy distribution with an average energy ${\cal E}_0$ and the rms energy spread $\sigma_{\cal E}$, and use the variable $ p = ({\cal E}-{\cal E}_0) /\sigma_{\cal E} $ for the dimensionless energy deviation of a particle. The initial longitudinal phase space distribution can then be written as $f_0(p)=N_0(2\pi )^{-1/2} e^{-p^{2} /2} $, where $N_0$ is the number of electrons per unit length of the beam. Here the bunch length is taken to be much larger than the wavelength of the modulation, and we neglect local variations of the beam current and assume a longitudinally uniform beam.

After passage through the undulator, the beam energy is modulated with the amplitude $\Delta {\cal E}$, so that the final dimensionless energy deviation $p'$ is related to the initial one $p$ by the equation
    \begin{align}\label{eq:II-A.0}
    p' = p + A \sin(k_L s),
    \end{align}
where $A = \Delta {\cal E} /\sigma_{\cal E}$, and $s$ is the longitudinal coordinate in the beam. The distribution function after the interaction with the laser becomes $ f_1(\zeta,p ) = N_0(2\pi)^{-1/2} \exp \left[- ( p - A \sin \zeta)^2/2 \right]$ where we now use the dimensionless variable $ \zeta = k_L s$. The beam then passes through the dispersion section with dispersive strength $R_{56}$, which converts the longitudinal coordinate $s$ into $s'$, $s' = s + R_{56} p \,\sigma_{\cal E}/{\cal E}_0$ (where $p$ now refers to the value at the entrance to the dispersion section). The distribution function is then,
    \begin{align}\label{eq:II-A.1}
    f_2(\zeta,p )
    =
    \frac{N_0}{\sqrt{2\pi}}
    \exp
    \left[
    -\frac{1}{2}
    \left(
    p
    -
    A
    \sin (\zeta-Bp )
    \right)^2
    \right]
    \,,
    \end{align}
where $ B = R_{56}k_L\sigma_{\cal E}/{\cal E}_0$ (for notational clarity, we dropped primes in the arguments of $f$).

Integration of $f$ over $p$ gives the 1-D beam density (number of particles per unit length) $N$ as a function of the coordinate $\zeta$,
    \begin{align}\label{eq:II-A.2}
    N(\zeta)
    =
    N_0
    \int_{-\infty}^\infty
    dp  f(\zeta,p )
    \,.
    \end{align}
Noting that this density is a periodic function of $\zeta$ one can
expand it into Fourier series
    \begin{align}\label{eq:II-A.3}
    \frac{N(\zeta)}{N_0}
    &=
    1
    +
    \sum_{n=1}^\infty
    2b_n \cos( n\zeta)
    \,,
    \end{align}
where the coefficient $b_n$ is the bunching factor for the harmonic $n$ (see Section~\ref{incoh_coh_radiation}). Calculations with the function (\ref{eq:II-A.1}) give an analytical expression for $b_n$ (see, e.g., \cite{HGHG1} and references therein) of the form,
    \begin{align}\label{eq:II-A.4}
    b_n
    =
    e^{-\frac{1}{2} B^2 n^2}
   J_n(-A B n)
   ,
   \end{align}
where $J_n$ is the Bessel function of order $n$.

Equation~\eqref{eq:II-A.4} indicates that by properly choosing the energy modulation amplitude and the chicane's dispersive strength, considerable bunching may be generated, not only at the laser wavelength but also at higher harmonics.

\begin{figure}[h]
    \includegraphics[width = 0.23\textwidth]{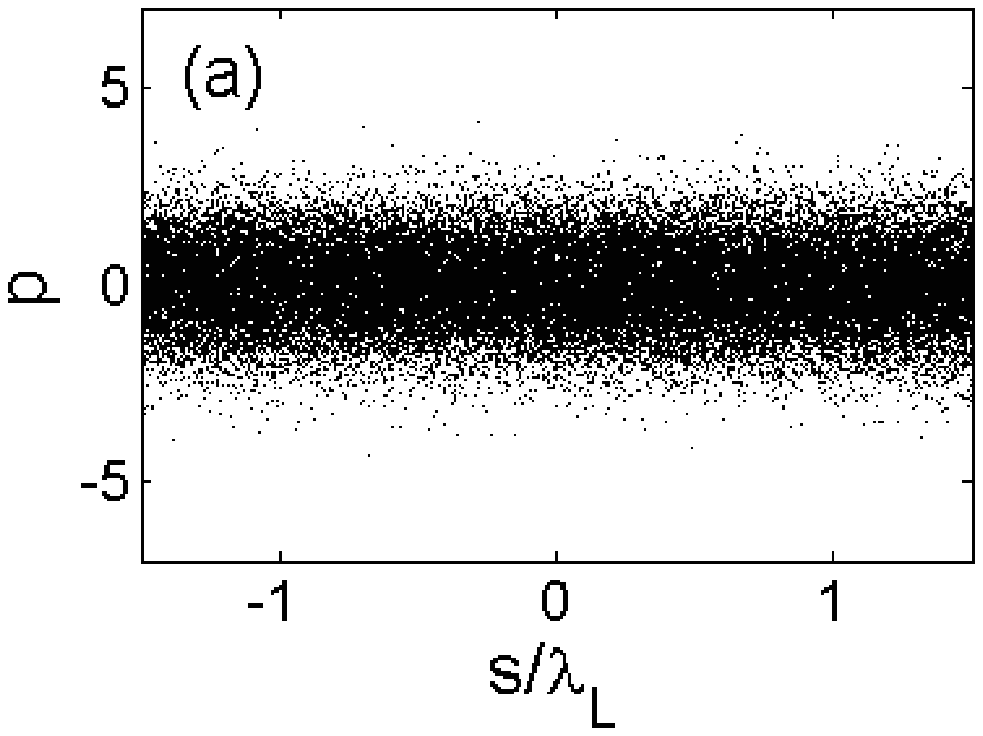}
    \includegraphics[width = 0.23\textwidth]{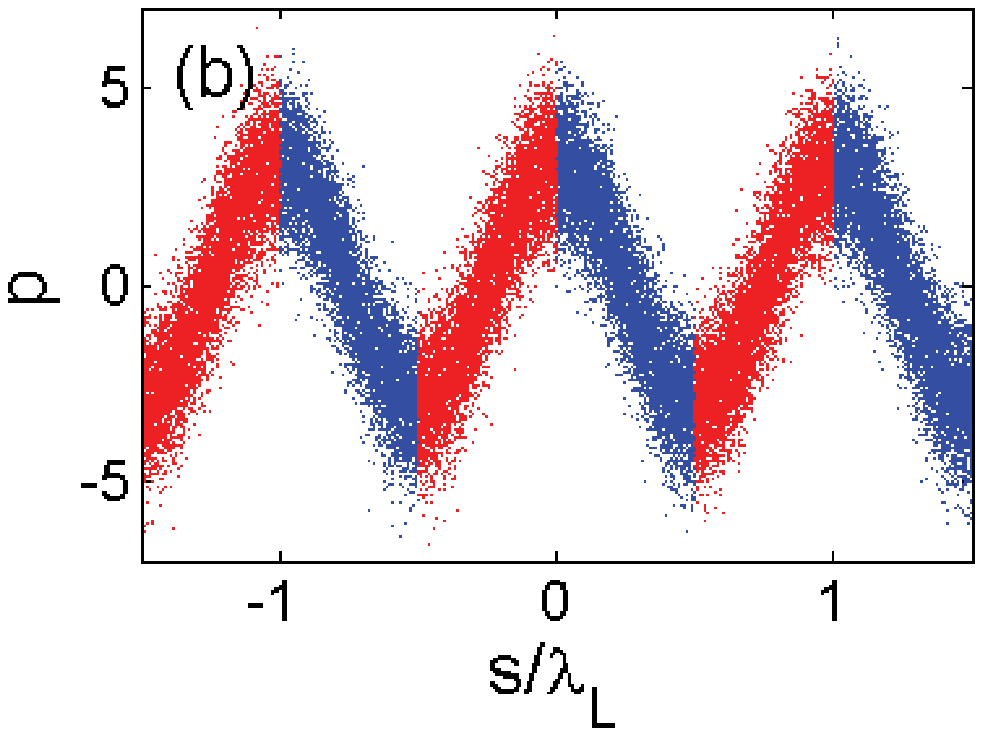}
    \includegraphics[width = 0.23\textwidth]{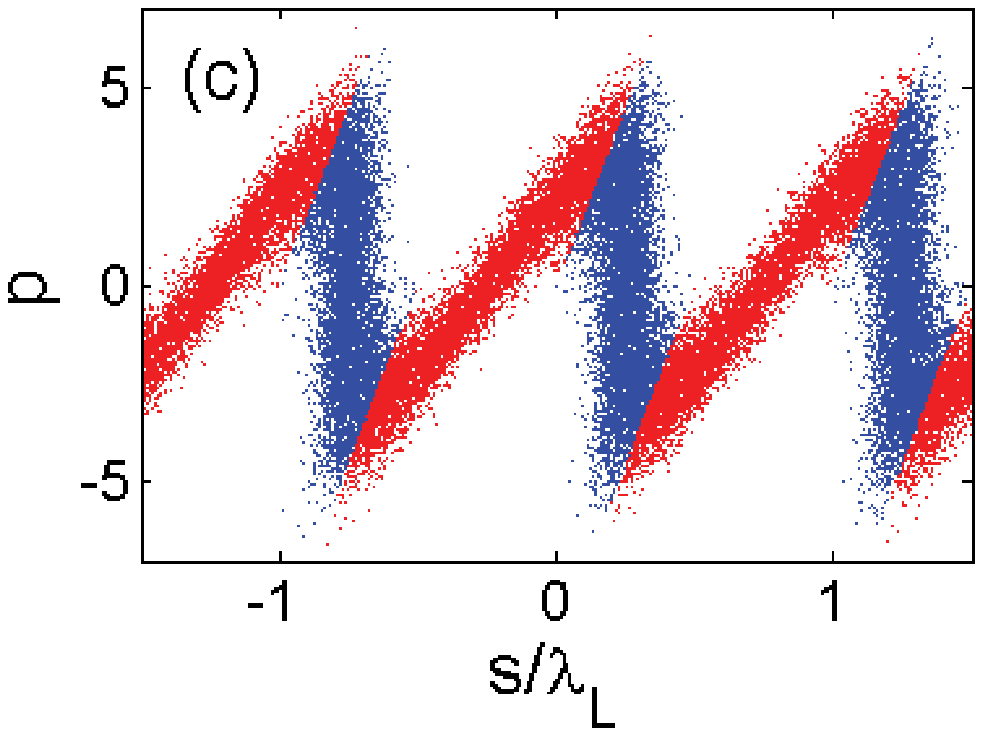}
    \includegraphics[width = 0.23\textwidth]{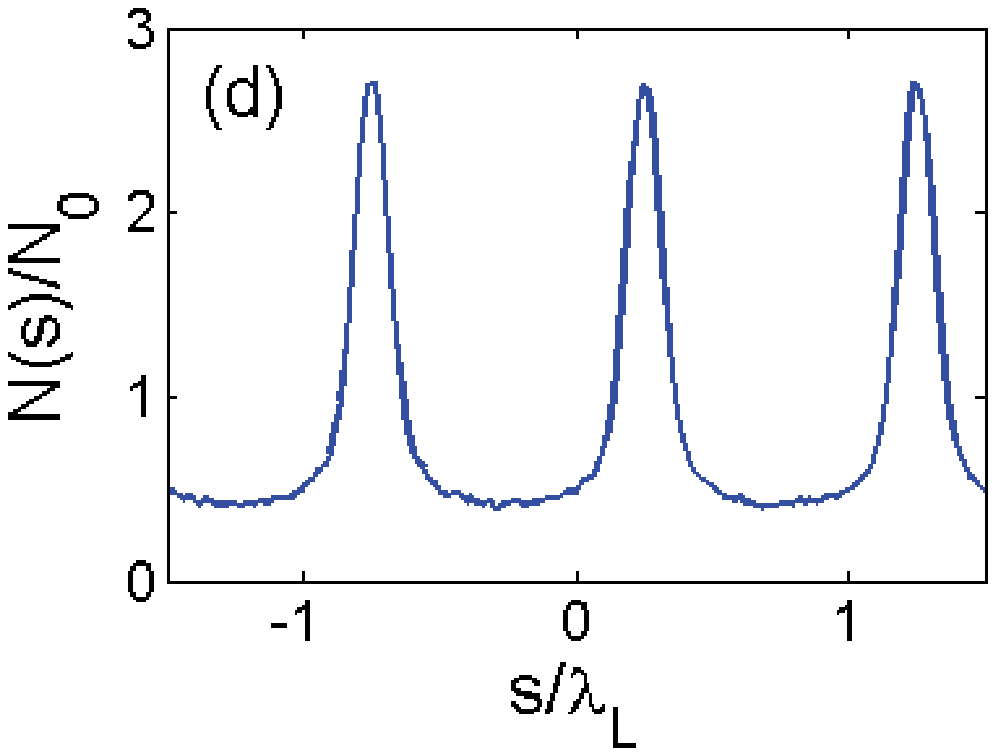}
    \caption{Evolution of the longitudinal phase space in the HGHG scheme with $A=3$. (a) before the modulator; (b) after the modulator; (c) after the chicane; (d) density distribution after the chicane.
    \label{HGHG}}
    \end{figure}

The phase space evolution in this scenario is illustrated in Fig.~\ref{HGHG}. The sinusoidal energy modulation imprinted on the beam by the laser is shown in Fig.~\ref{HGHG}b. After passing through a small chicane, half of the particles that have the negative energy chirp (blue dots in Fig.~\ref{HGHG}b and Fig.~\ref{HGHG}c) are compressed, while the other half with the positive energy chirp (red dots in Fig.~\ref{HGHG}b and Fig.~\ref{HGHG}c) are decompressed. As a result of this transformation, the energy modulation is converted into a density modulation as in Fig.~\ref{HGHG}d, where the beam density consists of many spikes equally separated by the laser wavelength. These sharp peaks contain frequency components at both the laser frequency and harmonic frequencies which can be radiated coherently when the electron beam is sent through a radiator.

In the limit of a large modulation amplitude, $A \gg 1$, the density (and hence current) spikes shown in Fig.~\ref{HGHG}d become much larger than the initial beam density (current) and the spike FWHM width $\Delta s$ much shorter than the laser period. Analysis of~\eqref{eq:II-A.1} shows that asymptotically for large values of $A$,
    \begin{align}\label{eq:esase_theory}
    \frac{N_\mathrm{max}}{N_0}
    \approx
    1.5A^{2/3}
    ,\qquad
    \Delta s
    \approx
    0.5
    \frac{\lambda_L}{A}
    .
    \end{align}
More accurate fitting formulas for moderate values of $A \sim 3-20$ were obtained in Ref.~\cite{Zhol4}, where it was proposed to use the enhanced peak current in the spikes for generation of a train of short pulses in an FEL. The concept, dubbed ESASE, is described in detail in Section~\ref{Section:IV.D}.

As described by Eq.~\eqref{eq:II-A.4}, the exponential suppression factor makes it difficult to obtain usable bunching factors for large values of $n$ unless the dispersion $B$ is also reduced to $B \sim n^{-1}$. Because the Bessel function is peaked when its argument is $\sim n$, this in turn requires an increase in the amplitude of the laser modulation to approximately $A \sim n$. Physically, this is because the longitudinal phase space area is conserved, and when the particles are locally compressed by $n$ times to produce harmonics up to the $\sim n$th order, the slice energy spread in this narrow region also increases by a factor of $n$. For a beam with vanishingly small energy spread, very high harmonics can be obtained since the maximal bunching factor scales as $b_n \sim n^{-1/3}$, as dictated by $J_n$.
In reality, however, it is typically desirable to keep the induced energy spread small. Therefore, in practice, generation of high harmonics is typically limited to values $n$ around 10.     
%
\subsubsection{Combination of two modulators and one chicane}
%

The significant growth in beam energy spread in one modulator and one chicane bunching technique can be mitigated
to some extent with two modulators.
    \begin{figure}[h]
    \includegraphics[width = 0.48\textwidth]{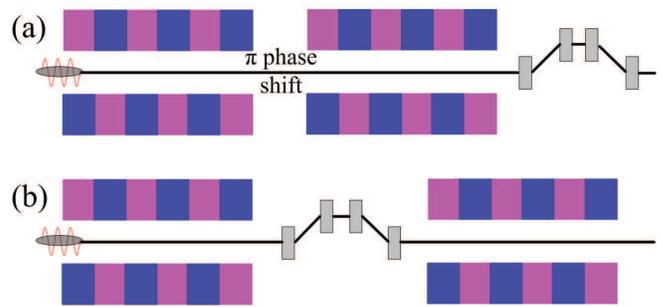}
    \caption{Schematic of two HGHG variants to reduce beam energy spread with two modulators and one chicane.
    \label{HGHGvariant}}
    \end{figure}

In the scheme considered in \cite{HGHGS1, HGHGS2} (Fig.~\ref{HGHGvariant}a), the modulator is subdivided into two undulator sections, and a phase shifter that delays the electron bunch by $\pi$ in laser phase is inserted in-between. The $\pi$ phase shift can be achieved with a small chicane with $R_{56}=\lambda_L$, or with one additional undulator period resonant at $1.5\lambda_L$. A high power laser (typically on the order of 10 GW) is first used to generate a large energy modulation in the first undulator section such that when the electron bunch goes through the second undulator section, its energy modulation is gradually converted into density modulation only from the dispersive strength $R_{56}=2N_u\lambda_L$ of the second undulator. Therefore, in the second undulator section the same laser partially reverses the modulation imprinted in the first undulator section, reducing the induced energy spread. Finally, the electron bunch is sent through a weak chicane that is used to maximize the bunching at high harmonics. Overall, the reduced energy spread achieved in this scheme allows bunching at approximately twice the harmonic number of that practically obtained in the conventional scheme with only one modulator \cite{HGHGS2}.

\begin{figure}[h]
    \hspace*{1.7mm}
    \includegraphics[width = 0.55\textwidth]{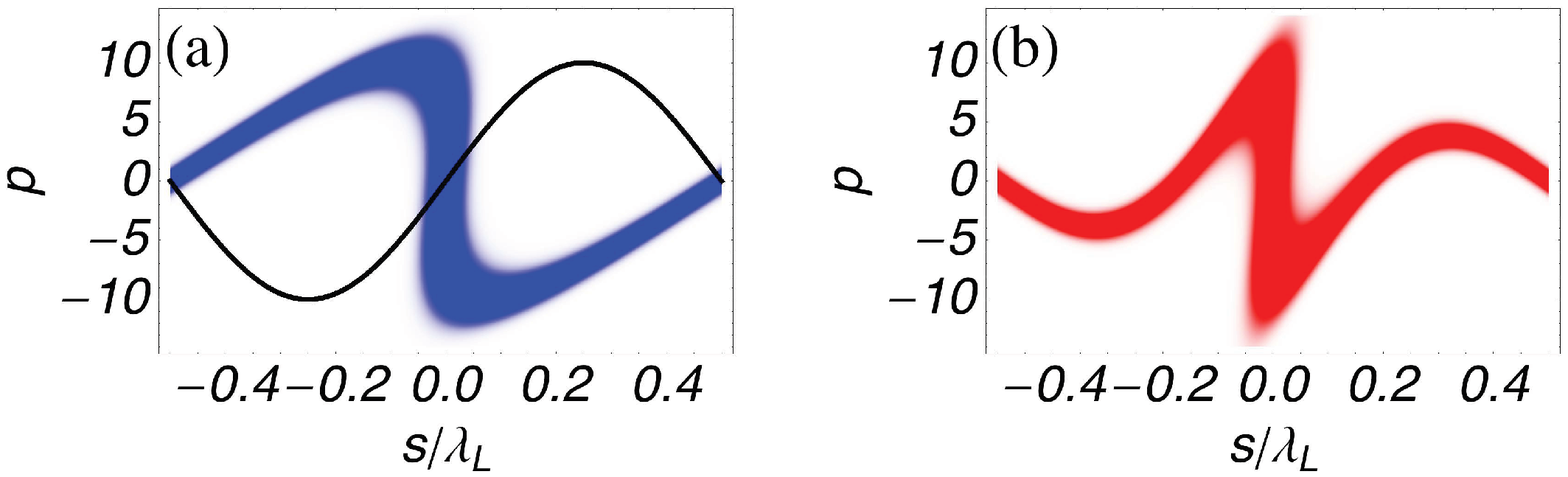}
    \includegraphics[width = 0.48\textwidth]{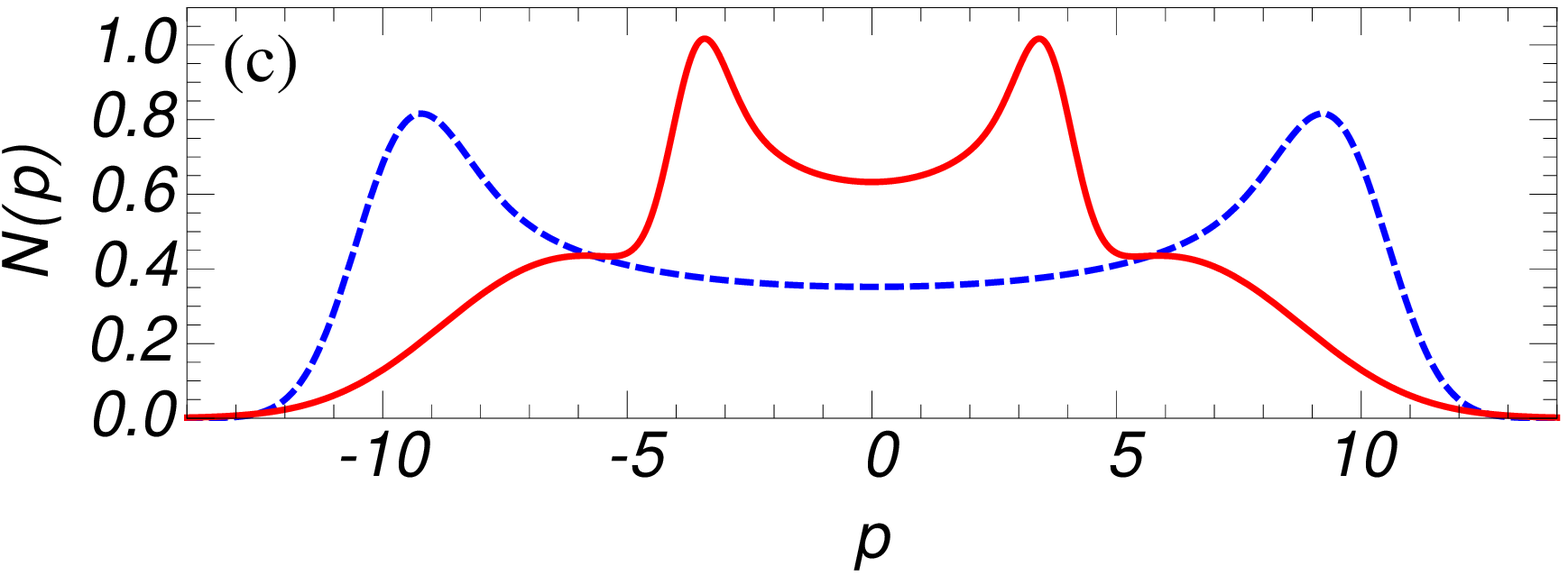}
    \caption{
    (a) Phase space after passing through the chicane and the corresponding
energy modulation (black curve) in the second modulator;
(b) phase space after the reverse modulation in the second
modulator; (c) energy distribution before (blue dashed curve) and
after (red curve) the reverse modulation.
    \label{HGHGS}}
    \end{figure}

An alternate way to reduce the energy spread is to use a second modulator after the beam goes through the chicane \cite{HGHGS3} (Fig.~\ref{HGHGvariant}b). This is illustrated in Fig.~\ref{HGHGS} where for simplicity we assume particles' longitudinal positions do not change in the modulators. In this example, the energy modulation is ten times larger than beam slice energy spread. As can be seen in Fig.~\ref{HGHGS}a, after passing through a small chicane, half of the particles stand up to provide about 15\% bunching at the 10th harmonic \cite{HGHGS3}. A laser shifted by $\pi$ is then used in the second modulator section to cancel part of the modulation (red line in Fig.~\ref{HGHGS}a). This partially removes the induced correlated energy spread for the un-bunched particles and leads to the phase space distribution shown in Fig.~\ref{HGHGS}b. The particles' projected energy distribution before and after the second laser modulation is shown in Fig.~\ref{HGHGS}c where one can clearly see that the beam energy spread is reduced (by about 30\% in this example). 

It should be pointed out that for both schemes, the beam longitudinal phase spaces deviate from sinusoidal distributions when the beam interacts with the laser in the second modulator. Therefore, cancelation of the energy modulation in these schemes is not complete. This imperfection is illustrated in Fig.~\ref{HGHGS}a where one can see that the bunched particles are at the zero-crossing of the second laser and their energy spread cannot be reduced. For HGHG schemes, however, this reduction of the energy spread in the unbunched portion of the beam can lead to better overall FEL performance as these electrons are recollected to a narrowed region of the gain bandwidth.

%
\subsubsection{Combination of two modulators and two chicanes}\label{sec:3-1-3}
%

\begin{figure}[!ht]
    \centering
    \includegraphics[width=0.45\textwidth, trim=0mm 0mm 0mm 0mm, clip]{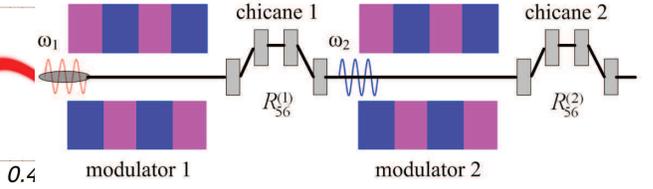}
    \caption{Schematic of the EEHG.
    \label{fig1}}
    \end{figure}

Consider now two modulators and two dispersion sections shown in Fig.~\ref{fig1}. This scheme was proposed in~\cite{EEHG1,XiangStupakov2009} under the name of Echo-Enabled Harmonic Generation (EEHG). Compared to HGHG and its variants, EEHG can produce a much higher harmonic with a relatively small energy modulation. In this scheme, a laser pulse with frequency $\omega_{1}$ is used to modulate the beam energy with amplitude $\Delta {\cal E}_1$ in the first undulator
tuned at that frequency. After passing through the first dispersion section with $R_{56}^{(1)}$, the beam energy is then modulated in the second undulator (modulator 2) with amplitude $\Delta {\cal E}_2$ tuned to the frequency $\omega_2$ of the second laser beam ($\omega_2$ can be equal to $\omega_1$). The beam then passes through the second dispersion section with $R_{56}^{(2)}$ to produce a density modulation at the frequency $n \omega_1+ m \omega_2$, where $n$ and $m$ are integers.

The mathematical formulation of EEHG process is similar to the derivation outlined in Section~\ref{sec:3-1-1}. The final distribution function at the exit from the second dispersion section can be easily found by consecutively applying two more transformations to (\ref{eq:II-A.0}) and (\ref{eq:II-A.1}). The resulting final distribution function $f_f$ is:
    \begin{align}\label{eq:II-A.6}
    f_f(\zeta,p )
    &=
    \frac{N_0}{\sqrt{2\pi}}
    \exp
    \Bigl[
    -\frac{1}{2}
    \bigl(
    p
    -
    A_2
    \sin ( K\zeta-K B_2p +\psi )
    \nonumber\\
    &-
    A_1
    \sin (
    \zeta-(B_1 + B_2)p
    \nonumber\\
    &+
    A_2 B_1
    \sin ( K\zeta - K B_2p +\psi )
    )
    \bigr)^2
    \Bigr]
    \,,
    \end{align}
where $ \zeta = k_1 s$, $ K = {k_2}/{k_1}$ with $k_1 = \omega_1/c$ and $k_2 = \omega_2/c$, $A_1 = \Delta {\cal E}_1 /\sigma_{\cal E}$, $A_2 = \Delta {\cal E}_2 /\sigma_{\cal E}$, $ B_1 = R_{56}^{(1)}k_1\sigma_{\cal E}/{\cal E}_0$, $ B_2 = R_{56}^{(2)} k_1 \sigma_{\cal E}/{\cal E}_0 $, and $\psi$ is the phase difference of the two lasers.

Integration of this formula over $p$ again gives the beam density $N$ as a function of $\zeta$, $N(\zeta) = \int_{-\infty}^\infty dp f_f(\zeta,p)$. Analysis shows~\cite{XiangStupakov2009} that at the exit from the system the beam turns out to be modulated at a combination of multiple wavenumbers of both lasers,
    \begin{align}\label{eq:II-A.7}
    \frac{N(s)}{N_0}
    &=
    \sum_{n,m=-\infty}^\infty
    2b_{n,m} \cos[(nk_{1} + mk_{2})s + \psi_{n,m}]
    \,,
    \end{align}
where $\psi_{n,m}$ is the modulation phase. Assuming $A_1B_1\gg1$, The bunching factors $b_{n,m}$ are found to be independent of the relative phase of the two lasers and are given by
    \begin{align}\label{eq:II-A.9}
    b_{n,m}
    &=
    e^{-\frac{1}{2} (n B_1+(K m+n)
    B_2){}^2}
    J_m(-(K m+n) A_2 B_2)
    \nonumber\\
    &\times
    J_n(-A_1 (n B_1+(K m+n)
    B_2))
    ,
    \end{align}
where $J_{n}$ is the Bessel function of order $n$.

\begin{figure}[htb]
	\centering
	\includegraphics[width=0.45\textwidth, trim=0mm 0mm 0mm 0mm, clip]{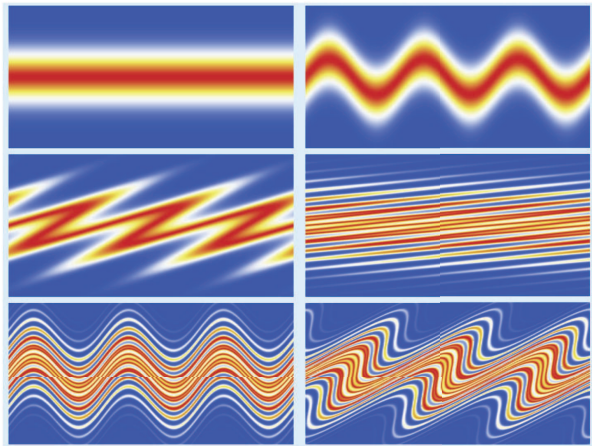}
	\caption{Evolution of the longitudinal phase space of the beam through an EEHG system in a 1-D model. Top left: initial phase space, top right: phase space after the first modulator, middle left: phase space in the center of the first chicane, middle right: phase space after the first chicane, bottom left: phase space after the second modulation, bottom right: phase space at the exit after the second chicane. The vertical axis is $p$ and the horizontal axis is $s/\lambda_{L}$ (both lasers are assumed of the same frequency). Shown are three laser periods.}
	\label{fig:echo_phase_plot}
	\end{figure}

The evolution of the phase space through the system is graphically illustrated in Fig.~\ref{fig:echo_phase_plot}. A crucial feature of the EEHG technique is that the first energy modulation is macroscopically smeared after beam passes through the first chicane with strong dispersion (middle right panel in Fig.~\ref{fig:echo_phase_plot}). This introduces complicated fine-scale structures (separated energy bands) into the phase space of the beam. After a second modulation, a density modulation then reappears after beam passes through a second chicane (bottom right panel in Fig.~\ref{fig:echo_phase_plot}), like an echo.

\begin{figure}[htb]
	\centering
	\includegraphics[width=0.4\textwidth, trim=0mm 0mm 0mm 0mm, clip]{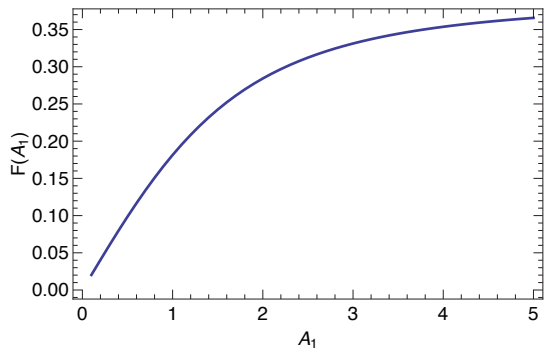}
	\caption{
	Function $F(A_{1})$ from Eq.~\eqref{eq:II-A.11}.
	}
	\label{fig:max_echo_1}
	\end{figure}

It is of practical interest to maximize the bunching factor at a high laser harmonic by varying the modulation amplitudes $A_{1}$ and $A_{2}$ and the strength of the dispersive elements $B_{1}$ and $B_{2}$. It was shown in Ref.~\cite{XiangStupakov2009} that the maximum is achieved when $n = -1$ and $m>0$, and for large values of $m$ the maximized value of $b_{-1,m}$ is given by
    \begin{align}\label{eq:II-A.11}
    b_{-1,m}
    \approx
    \frac{F(A_{1})}{m^{1/3}}
    ,
    \end{align}
where function $F(A_{1})$ is shown in Fig.~\ref{fig:max_echo_1}.
Asymptotically, for $A_{1} \gg 1$, function $F$ approaches the value of $0.39$. Remarkably, and in contrast with~\eqref{eq:II-A.4}, Eq.~\eqref{eq:II-A.11} does not show the exponential suppression of high harmonics; rather, in an optimized setup, the bunching factor slowly decays as $m^{-1/3}$ even for modest values of $A_1$ and $A_2$. This is the main advantage of the EEHG scheme.

Note that the bunching factor at high harmonics in the EEHG, HGHG and CHG related schemes can be considerably increased if one can use a synthesized laser waveform that approximates a sawtooth profile~\cite{FEL11stupakov_1,FEL11ratner_chao}. This can, in principle, be achieved by combining two or three laser harmonics with properly adjusted amplitudes and phases, or through manipulation of the beam phase space with a single harmonic as discussed in Section~\ref{sec:3-6}.

As one can see from Fig.~\ref{fig:echo_phase_plot}, the evolution of the beam phase space passes through a stage where it is ``shredded'' horizontally into narrow slices (the middle right panel in the figure, corresponding to the position after the first chicane). This means that the beam energy distribution is split into multiple narrow spikes with the width of each spike much smaller than the original energy spread of the beam. Analysis shows that the width of the spikes $\Delta\cal E$ is of the order of $\Delta {\cal E} \sim A_1\sigma_{\cal E}/m$, inversely proportional to the harmonic number $m$. Such a distribution function is sensitive to the energy diffusion processes in the system, such as quantum diffusion due to incoherent synchrotron radiation in magnetic fields, and intra-beam scattering. It seems likely that these effects set the upper limit on the maximally achievable harmonic multiplication factor in the EEHG scheme in practice~\cite{FEL11stupakov_2,FEL13stupakov_1}.

%
\subsection{Optical microbunching for laser acceleration}\label{StagedIFELs}
%

Laser manipulation schemes designed to generate coherent microbunching in the electron beam also have applications in advanced laser-based accelerators such as the inverse FEL (IFEL), inverse transition radiation accelerator \cite{PhysRevLett.95.134801}, dielectric laser accelerator \cite{PhysRevLett.111.134803, Peralta:2013vpa}, and acceleration through stimulated emission of radiation in an excited medium~\cite{PASERtheoryPRA1995}. Of particular interest is the IFEL where acceleration can be maintained over a large distance to yield GeV electrons. In contrast to FELs where energy is transferred from the electron beam to the radiated electromagnetic fields, in IFELs, a high-power input laser resonantly transfers energy to the transversely wiggling electrons, boosting them to high energies using potentially large (0.5-1 GeV/m) accelerating gradients.

The IFEL concept dates back over 40 years to Palmer, who suggested that relativistic electrons could be continuously accelerated by electromagnetic waves in a helical undulator \cite{palmer:3014}. Later, the authors of Ref.~\cite{PhysRevA.32.2813} presented a comprehensive analysis for both planar and helical undulators that included the effects of synchrotron radiation losses, as well as energy transfer enhancements obtained from undulator tapering. Based on these principles, several single stage IFEL experiments followed, with accelerating wavelengths in the microwave regime \cite{PhysRevLett.86.1765}, at 1.6 mm \cite{PhysRevA.46.3566}, and at 10.6 $\mu$m, both at the fundamental undulator resonance \cite{PhysRevLett.77.2690}, and including the second harmonic of a planar undulator \cite{PhysRevLett.94.154801}. The latter experiment at UCLA used a strongly tapered design to achieve a 70 MeV/m accelerating gradient for $5\%$ of the electrons with a 2$\times$10$^{14}$ W/cm$^2$ CO$_2$ laser. Recent experiments with a strongly tapered helical undulator show the energy doubling of a 50 MeV beam over a 60 cm undulator \cite{duris:440}. 

While the IFEL concept holds potential to reduce the size and cost of modern particle accelerators through enhanced accelerating gradients enabled by modern laser systems, a drawback common to laser-based single-stage IFELs is the large final electron beam energy spread (100$\%$) and poor capture efficiency. This occurs because the electron beam is typically much longer than the laser wavelength, so electrons are evenly distributed across both the accelerating and decelerating phases during the IFEL interaction. The result is a somewhat inefficient acceleration process and a large global energy spread, since only a fraction of the electrons are at the correct phase to be captured and boosted to the final peak energy. As noted in the original paper by Palmer, both the efficiency and the energy spread can be significantly improved by first microbunching the beam so that electrons are piled together within a small fraction of the accelerating phase. Such staged schemes rely on an initial modulator/dispersive section (pre-buncher) where a laser generates a constant sinusoidal energy modulation and a downstream dispersive section (either a drift or a chicane) then generates spatial density bunching at the period of the accelerating laser wavelength.
   \begin{figure}
   \centering
   \includegraphics[width=85mm]{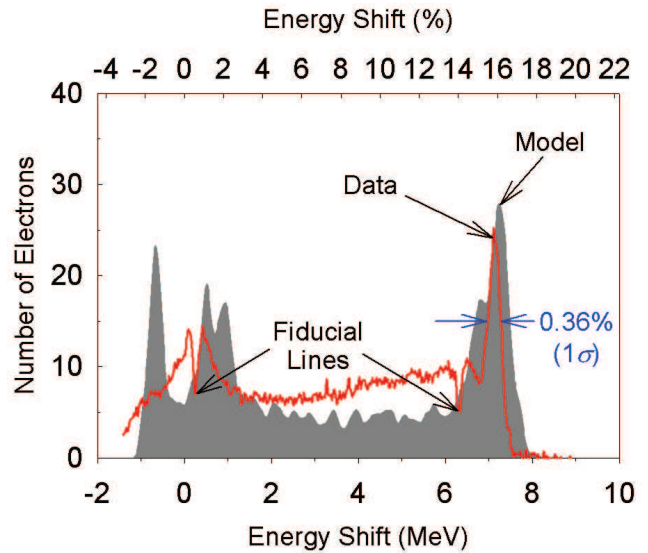}
   \caption{Two stage IFEL energy output spectrum (From \cite{KimuraSTELLAPRL2004}).}
   \label{KimuraIFELimage}
   \end{figure}

Several experiments based on the two-stage laser acceleration technique have been performed in recent years. In 2001, the STELLA experiment at Brookhaven National Laboratory used a high peak power ($>100$ MW) 10.6 $\mu$m CO$_2$ laser to first modulate and then boost electrons in a 45 MeV beam by $\sim$1-2 MeV over a 33 cm undulator \cite{KimuraSTELLAPRL2001}. While the final relative energy spread of the accelerated electrons was somewhat large (2$\%$), subsequent upgrades showed improved capture efficiency and reduced energy spread with the addition of a small chicane (to reduce the required first energy modulation) and gap tapered undulator \cite{KimuraSTELLAPRL2004}. Results showed $\sim$80$\%$ of the electrons were captured and accelerated, with $14\%$ boosted by 7 MeV with a $0.36\%$ relative energy spread (see Fig.~\ref{KimuraIFELimage}). More recently, two-stage accelerators in the 800 nm range have been demonstrated at the NLCTA at SLAC; one using inverse transition radiation to accelerate the microbunched beam \cite{PhysRevSTAB.11.101301}, and another using the IFEL interaction at the 3rd harmonic of a planar undulator \cite{PhysRevLett.110.244801}.

\begin{figure}
   \centering
   \includegraphics[width=85mm]{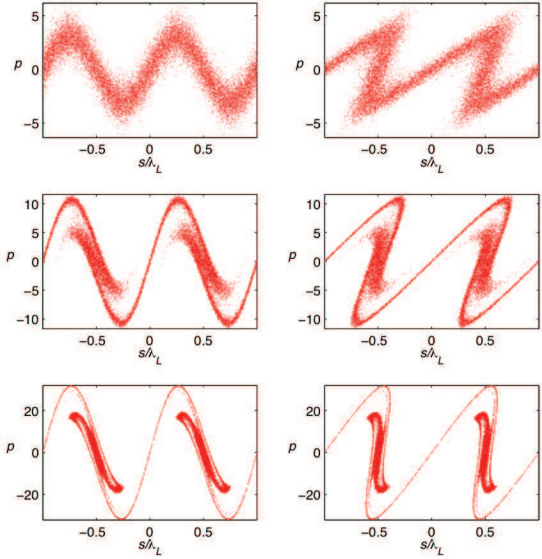}
   \caption{Phase space in staged adiabatic buncher with three modulator-chicane modules (modified from \cite{PhysRevSTAB.16.010706}). Phase space evolution in the first modulator-chicane module (top), in the second modulator (middle) and in the third module (bottom).}
   \label{adiabaticbuncher}
   \end{figure}

The staged approach to laser accelerator injectors has distinct advantages compared to a single stage, both in terms of the efficiency and the final beam parameters. For example, an energy modulation generated by multiple harmonic laser frequencies with the proper relative phases and amplitudes allows one to linearize the electron beam phase space at the optical wavelength prior to IFEL injection. This places more electrons into the accelerating phase and enhances the capture efficiency \cite{WangHermIFEL2002, PhysRevE.72.016501}. Such a scheme can be implemented in a single planar undulator to generate a sawtooth-type energy distribution, where the electron beam couples with both the odd and the even harmonics by injecting the laser at a small angle \cite{PhysRevSTAB.15.061301}. Alternatively, multiple staged modulations at the same laser frequency, mediated by dispersion, can adiabatically fold electrons up into the accelerating bucket to increase the bunching factor to near unity (Fig.~\ref{adiabaticbuncher})\cite{PhysRevSTAB.16.010706}. For instance, Fig.~\ref{adiabaticbuncher} shows the phase space evolution in a staged adiabatic buncher with three modulator-chicane modules in which the modulation amplitude is increasing and the dispersion strength is decreasing in each stage. In this scheme, each chicane is used to rotate the local phase space so as to deposit the maximum number of particles into the phase region $\pi/2 <k_L s < 3\pi/4$, leading to a bunching factor of about 92\% at the laser frequency.

%
\subsection{Creating long-scale structures for narrow-band Terahertz radiation}
%

Beams with density modulations at sub-mm wavelengths may be used to resonantly excite wake fields for advanced accelerators \cite{Muggli, LHC}, and for the generation of narrow-band THz radiation \cite{Bielawski, Shen}). Analysis shows that in addition to creating fine structures in the beam density distribution for the generation of fully coherent radiation at shorter wavelengths, the EEHG technique may also be used to produce long-scale density modulations (with modulation periods much longer than the laser wavelengths). For instance, it can be seen from Eq.~\eqref{eq:II-A.7} that with $n=1$ and $m=-1$ (or vice versa), a density modulation at the difference frequency $|k_1-k_2|$ of the two lasers can be generated with a double modulator-chicane system. In particular, if the wavelengths of the two lasers are close to each other, the difference frequency will be much lower than the laser frequency. Therefore, one may generate long-scale periodic structures in the electron beam through short-wavelength laser modulations.

\begin{figure}[h]
    \includegraphics[width = 0.23\textwidth]{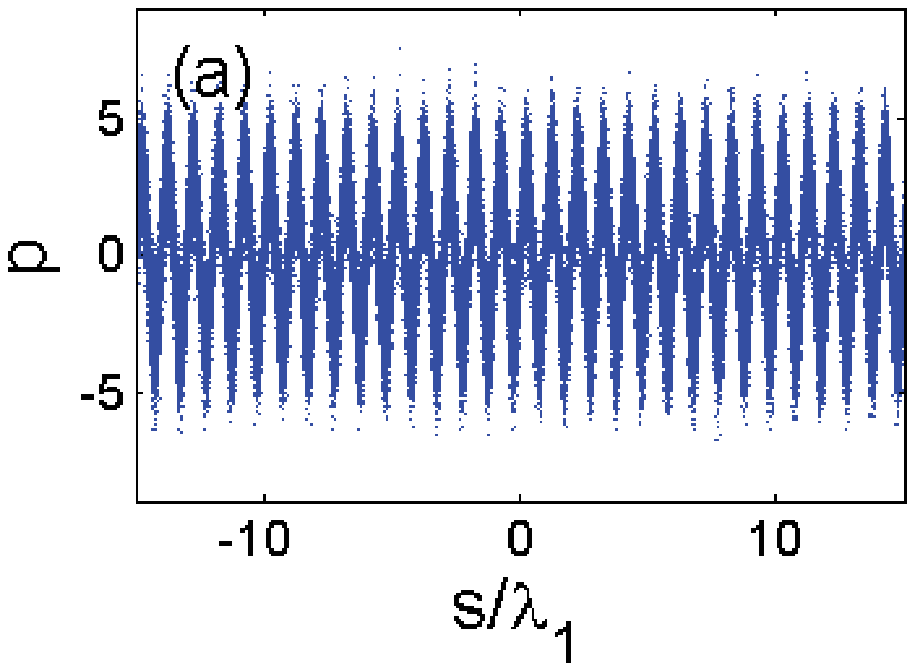}
    \includegraphics[width = 0.23\textwidth]{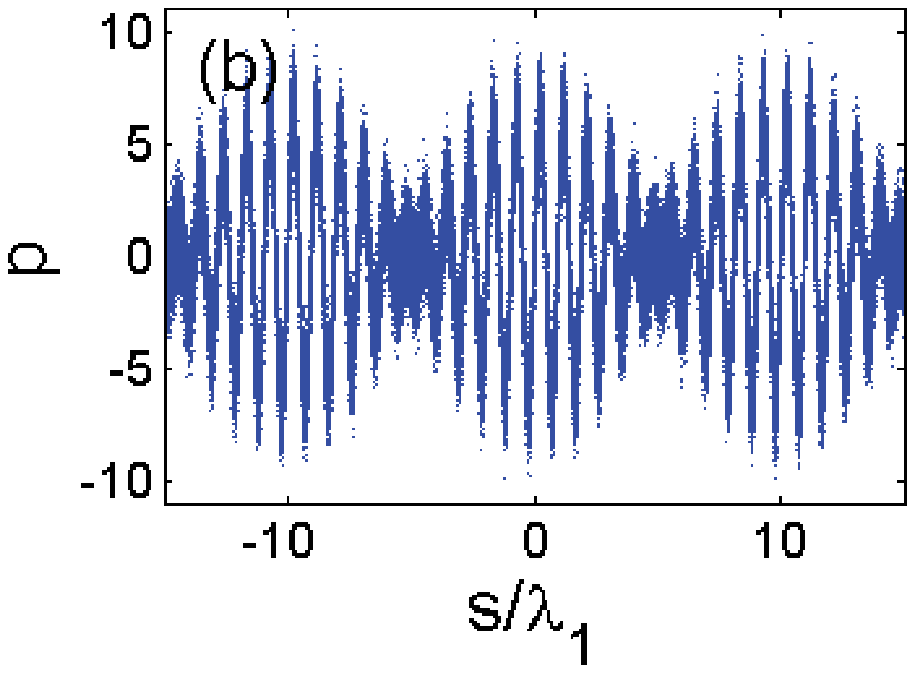}
    \includegraphics[width = 0.23\textwidth]{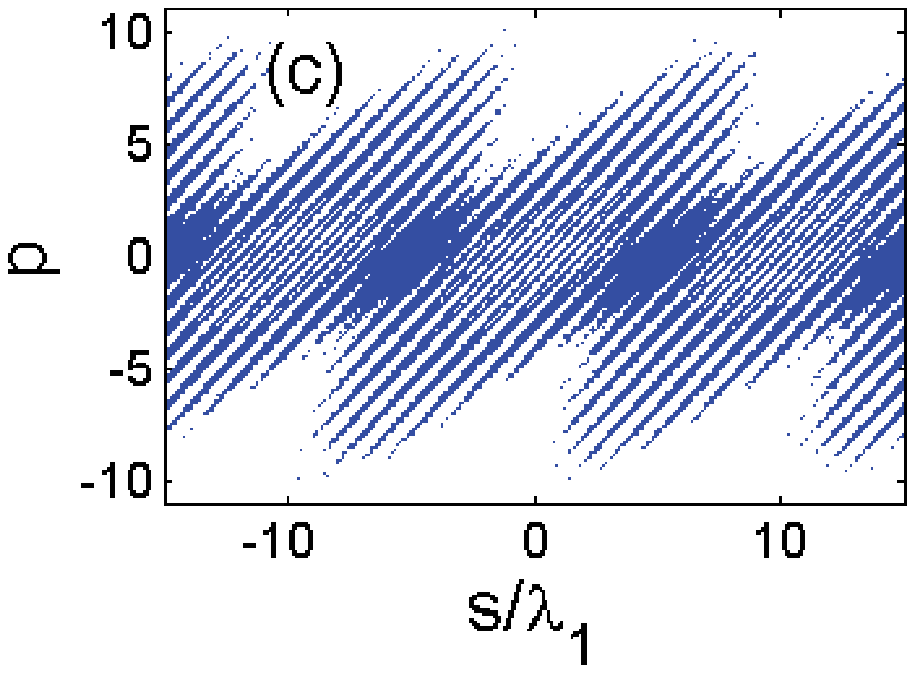}
    \includegraphics[width = 0.23\textwidth]{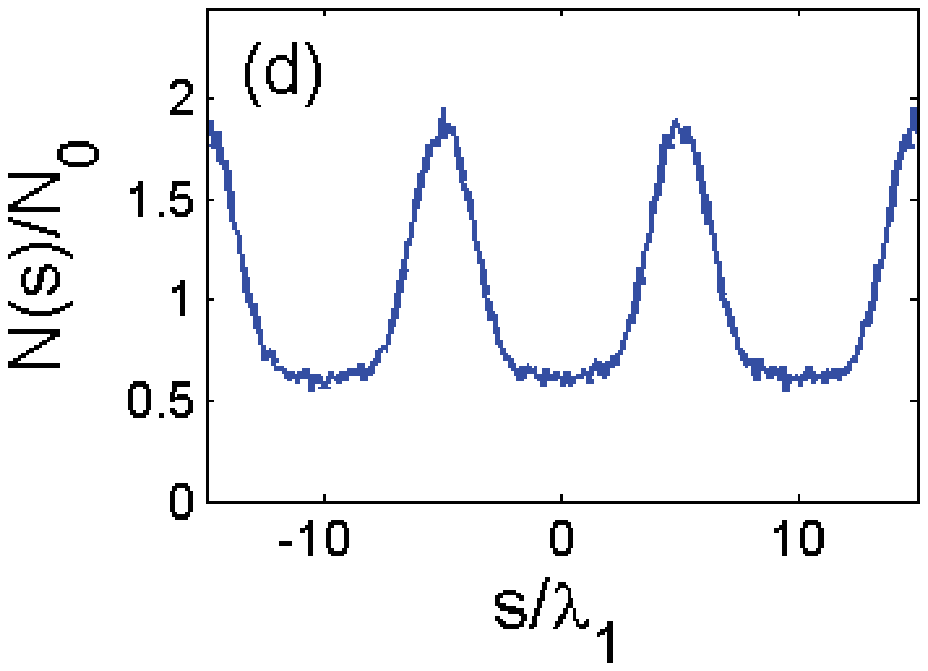}
    \caption{Evolution of the longitudinal phase space in difference frequency generation with two modulators and one chicane. (a) after the first modulator; (b) after the second modulator; (c) after the  chicane; (d) final current distribution.
    \label{figTHz}}
    \end{figure}

In contrast to the EEHG scheme, the chicane between the two modulators is not necessary for this application \cite{XiangStupakov2009} although it can provide an additional tuning knob. To illustrate the physics, the beam longitudinal phase space evolution in a simpler configuration with two modulators and one chicane is shown in Fig.~\ref{figTHz}. As before, in the first undulator a laser with wavelength $\lambda_1$ is used to generate energy modulation in the beam phase space (Fig.~\ref{figTHz}a). After interaction with the second laser with wavelength $\lambda_2=0.9\lambda_1$, the beam phase space consists of a slow modulation at the difference frequency superimposed on the initial sinusoidal modulation (Fig.~\ref{figTHz}b). After passing through a chicane, the energy modulation at the difference frequency (with wavelength at $\lambda_1\lambda_2/(|\lambda_1-\lambda_2|)$) is converted into a density modulation (Fig.~\ref{figTHz}c). The resulting beam current has a modulation period of about 10$\lambda_1$ (Fig.~\ref{figTHz}d). In this scheme, the relativistic electron beam is used as the nonlinear medium to down-convert the frequency of two optical lasers to the THz range.

This difference-frequency generation scheme does not necessarily require the wavelengths of the two lasers to be close to each other. Actually it applies to the general cases when $\lambda_1/n$ is close to $\lambda_2/m$. For instance, a density modulation around 10 THz can be generated by two lasers with wavelengths $\lambda_1=780~$nm and $\lambda_2=800~$nm (corresponding to $n=1, m=-1$), $\lambda_1=780~$nm and $\lambda_2=1600~$nm (corresponding to $n=1, m=-2$), or $\lambda_1=780~$nm and $\lambda_2=2400~$nm (corresponding to $n=1, m=-3$), etc. The scenario with $n=1$ and $m=-2$ is particularly useful for providing tunable THz radiation in a wide frequency range with commercially available IR laser systems. For example, an optical parametric amplifier (OPA) pumped by Ti:Sapphire laser at $\sim800$ nm can easily provide a tunable signal beam from  $1.1~\mu$m to $3~\mu$m \cite{OPA}. When combined with a Ti:Sapphire laser at $\sim800$ nm, the scenario with $n=1$ and $m=-2$ allows one to generate density modulations covering the whole THz gap.

While only one chicane is needed to generate THz density modulations in principle, inclusion of a chicane between the two modulators offers more flexibility in that it allows the generation of THz structures in an energy-chirped beam with two lasers of the same wavelength. This scenario can be understood as a four-step process. First, the laser interacts with the beam in the first modulator and generates an energy modulation at the wave number $k_1$. Second, the modulation frequency is compressed (or decompressed) to $C_1k_1$ from the combination of the global energy chirp and momentum compaction of the first chicane, where $C_1$ is the compression factor. Third, the energy modulation at $C_1k_1$ is superimposed on top of the energy modulation at $k_1$ from the second laser in the second modulator. Last, the difference frequency of the energy modulation at $(C_1-1)k_1$ is compressed (or decompressed) again with compression factor $C_2$, and is further converted into density modulation at $(C_1-1)k_1C_2$ after passing through the second chicane. This technique was demonstrated at SLAC's NLCTA where a density modulation around 10 THz was generated by down-converting the frequencies of an 800 nm laser and a 1550 nm laser \cite{THzNLCTA}. Once the density modulation is formed, it is then straightforward to send the beam through a bending magnet or a metallic foil to generate coherent narrow-band THz radiation. One of the many advantages of this technique is the flexibility it offers to tune the central frequency of the modulation, which can be achieved through tuning of the laser wavelengths, beam energy chirp, and/or chicane momentum compaction.

%
\subsection{Creating 3-D fine structures for the generation of light with orbital angular momentum}\label{sec:3-5}
%

While the primary aim of most laser-based electron beam manipulation techniques is tailoring of the longitudinal phase space, there are also schemes targeted at generating highly-correlated, complex distributions in the larger phase space. Because the light emitted by the electron beam will have a phase structure determined by the microbunching distribution, precision manipulation of the 3-D coherent electron distribution allows emission of exotic light beams that can expand the repertoire of modern light sources.

Of specific recent interest are light beams that carry orbital angular momentum (OAM). These ``optical vortex'' modes have an annular-shaped intensity profile and carry discrete values $l\hbar$ of OAM per photon as a result of a $e^{il\phi}$ dependence, where $\phi$ is the azimuthal coordinate and $l$ is an integer. OAM beams were first investigated in Ref.~\cite{Allen}, and have become the subject of intense interest in numerous contexts \cite{AllenBook}. The multitude of emerging applications enabled by optical OAM light suggests exciting new research opportunities in the hard x-ray regime where well-defined OAM provides an additional degree of freedom to probe the deep structure and behavior of matter \cite{PadgettNature, MolinaNature}.

\begin{figure}
\includegraphics[width = 0.5\textwidth]{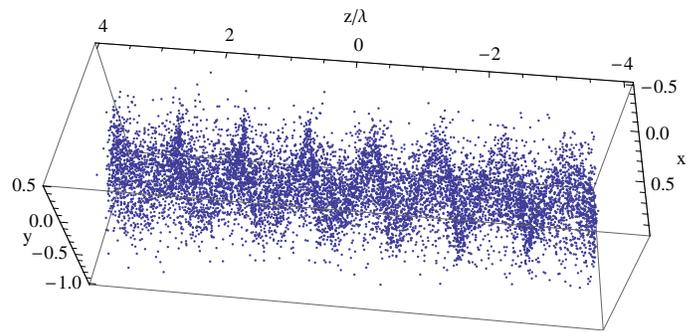}
\caption{Helically bunched beam for the emission of coherent OAM light. (From Ref.~\cite{HemsingFEL09})}
\label{helicalbeam}
\end{figure}

The harmonics of radiation produced in a helical undulator naturally carry approximately $\pm(n-1)\hbar$ units of OAM per photon, where $n$ is the harmonic number \cite{Sasaki,PhysRevLett.111.034801}. Thus, one can select a specific OAM value with a simple monochromator. Alternatively, coherent OAM light can be emitted at the fundamental frequency in an undulator (either planar or helical) by an electron beam that is helically microbunched, such that the electrons are arranged in a twist about the beam axis, with a helical period that matches the resonant wavelength of the undulator radiation. An illustration of such a beam is shown in Figure~\ref{helicalbeam}. Such a fine scale 3-D structure can be generated by the interaction of the beam with a laser in an undulator in two different ways, both described briefly in this section. A straightforward method is to modulate the beam with OAM laser modes at the fundamental frequency. In this case the transverse spatial field dependence maps directly to the energy kick experienced by the particle. A second method, which may be useful at short wavelengths, relies on coupling to harmonics of helical undulators. In this case, a laser mode with a simple structure (such as a TEM$_{00}$ mode) can be used and the correlated helical structure emerges naturally as the electrons interact with different regions of the laser field profile.

%
\subsubsection{Helical energy modulation}
%
To illustrate the concept, consider the rate of energy change equation in \eqref{eq:I-C.6a}
\be
\frac{d\gamma(\mathbf{x}_\perp,z)}{dz}=\frac{e}{mc^2}\vec{E}(\mathbf{x}_\perp,z)\cdot\frac{d\mathbf{x}_\perp}{dz}
\label{secOAMenergyexchange}
\ee
where here the energy change is shown to depend explicitly on the electron's transverse position $\mathbf{x}_\perp(z)$ in the laser field,
\be
\mathbf{x}_\perp(z)=\mathbf{x_0}+\mathbf{\tilde{x}}(z),
\ee
where $\mathbf{x_0}=(r,\phi)$ the secular offset and $\mathbf{\tilde{x}}$ the motion due to the undulator fields.

Let us assume that the electric field polarization is matched to the direction of electron motion, and that the laser Rayleigh length $z_{R}$ is much longer than the undulator length $L_{u}$ such that we neglect variation of the electric field amplitude versus $z$ in the undulator.
Expanding the field about $\mathbf{x_0}$ gives two terms that illustrate the two different ways to produce the helical structure,
\be
E(\mathbf{x}_\perp)\simeq E(\mathbf{x_0})+\left(\mathbf{\tilde{x}}\cdot\nabla\right)E(\mathbf{x_0}).
\label{secOAMfieldexp}
\ee
The first term describes the interaction at the fundamental frequency where the spatial distribution of the higher order laser field is imprinted directly on the electron beam energy distribution.
Considering only this term for the moment, for simplicity we assume a Laguerre-Gaussian type laser seed mode and write the field as
    \begin{align}\label{eq:OEM-field}
E(\mathbf{x_0})=\tilde{E}(r)\cos\left[k_{L}(z-ct)+l\phi\right],
    \end{align}
where the function $\tilde{E}(r)$ can be found by comparing this equation with~\eqref{II.A.2LG}. Substituting~\eqref{eq:OEM-field} into the energy rate of change equation (\ref{secOAMenergyexchange}) and integrating over the undulator length, the energy deviation is,
    \begin{align}\label{secOAMfirstharm}
	\Delta\gamma(r,s,\phi)
	=-\frac{eK\tilde{E}(r)L_u}{\sqrt{2}\gamma mc^2}\sin(k_Ls+l\phi),
    \end{align}
where $r$ is the radial offset of the electron orbit relative to the axis of the laser beam. As before, the change in particle position is neglected to lowest order. Further, because a helical correlation exists between the $s$ and $\phi$ coordinates in the argument of sine function in ~\eqref{secOAMfirstharm}, the electron beam obtains a helical energy modulation, defined by the helices with constant values of $\Delta\gamma(r)$. Passing this beam through a chicane will then generate a helically density modulated beam, as discussed in the following section. Note that this laser interaction also produces an angular modulation of the electrons, but in all practical cases, this modulation is negligibly small.

At short wavelengths, however, an OAM light beam may not be available. In this case, a similar helical energy modulation is obtained through the interaction of a light beam without OAM (like a simple Gaussian laser) with the electrons in a helical undulator. The modulation is produced through coupling to the
$\left(\mathbf{\tilde{x}}\cdot\nabla\right)E(\mathbf{x_0})$ term in (\ref{secOAMfieldexp}) that is realized when the carrier frequency of the light beam is resonant with a higher-harmonic emission frequency in a helical undulator \cite{Sasaki,PhysRevLett.111.034801}, as discussed in \cite{HemsingOAMFEL}. Accordingly, this effect enables tailored manipulation of electrons at much shorter wavelengths because the input laser beam can be generated by an upstream x-ray FEL, for example, in an arrangement similar to the self-seeding technique reported in~\cite{2012NaPho...6..693A}. To illustrate, consider a simple light beam with a radially symmetric Gaussian mode without azimuthal dependence, ($l=0$). At the second harmonic resonance in the helical undulator, the change in energy according to (\ref{secOAMenergyexchange}) and (\ref{secOAMfieldexp}) is,
\be
\begin{aligned}
\Delta\gamma=\mp\frac{eK^2L_u}{2\sqrt{2}\gamma^2mc^2k_u}\frac{\pa}{\pa r}\tilde{E}(r)\cos(k_Ls\mp \phi)
\label{secOAMsecondharm}
\end{aligned}
\ee
where the upper (lower) sign refers to the right (left) handedness of the helical undulator field. Much like (\ref{secOAMfirstharm}), the energy modulation is precisely that of an OAM-type light beam with a helical phase and axial null (no modulation on axis).

%
\subsubsection{Creating 3-D density structures in the beam}
%

To convert the helical energy modulation into helical microbunching for OAM light emission, the electrons must move longitudinally so that they create a helical density distribution. At low energies, this longitudinal rearrangement occurs naturally through a drift section, or even during transport through the modulator. In Ref \cite{hemsing:091110}, evidence of helical microbunching was first observed in a 12.5 MeV electron beam modulated in a helical undulator at the second harmonic by a Gaussian 10.6 $\mu$m CO$_2$ laser. Transport through the modulator was sufficient to generate helical bunching, which was characterized indirectly from the coherent transition radiation (CTR) of the electrons.

For higher energy beams however, the longitudinal motion of the particles is too small to generate density bunching, so the helical distribution is generated by passing the beam through a longitudinally dispersive section characterized by the matrix element $R_{56}$. Calculations of the density modulation introduced in the beam follow closely those of Section~\ref{sec:3-1}, so we will only briefly outline them here using the same notation as in Section~\ref{sec:3-1}. For the initial distribution function of the beam we now assume $f_{0}(r,p)=(2\pi)^{-1/2} N_0(r)e^{-p^2/2}$, where $N_{0}(r)$ gives the radial profile of the beam density.

Eq.~\eqref{eq:II-A.0} for the energy modulation in either (\ref{secOAMfirstharm}) or (\ref{secOAMsecondharm}) is now replaced by
    \begin{align}
	p'=p+{A}(r)\sin(k_Ls+l\phi)
	\label{secOAMfirstmod}
    \end{align}
which, in addition to having the phase $\phi$ dependence, takes into account the radial dependence of the energy modulation. After passage through the chicane the beam density becomes modulated along $s$ as well as in the azimuthal direction,
    \begin{align}
    N(r,\phi,s)
    =
    N_{0}(r)
    \left[
    1
    +
    \sum_{n=1}^{\infty}
    2b_{n}(r)
    \cos(nk_{L}s+nl\phi)
    \right]
    ,
    \end{align}
where
    \begin{align}\label{eq:bunchingOAM}
    b_n(r)
    =
    e^{-\frac{1}{2} B^2 n^2}
    J_n[-A(r) B n]
   ,
   \end{align}
(see Eq.~\eqref{eq:II-A.3}). Note that all final azimuthal density modes excited in the beam have the same up-conversion factor $n$ as the frequency.
Higher-order OAM light can therefore be emitted from the beam at harmonic frequencies of the light used to generate the energy modulation which, through Eq. (\ref{secOAMsecondharm}), need not be OAM light by virtue of the harmonic interaction in the helical modulator.

\begin{figure}
	\includegraphics[width=85mm]{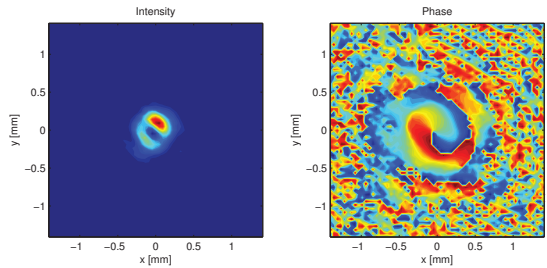}
	\caption{OAM light emitted by a 120 MeV electron beam that had been helically microbunched by a Gaussian mode laser at the second harmonic of a helical undulator. The transverse phase structure (right) was reconstructed from intensity images (left) of the far-field undulator emission. (From \cite{hemsing2013coherent}).}
	\label{OAMMeasurement}
	\end{figure}

This technique was recently demonstrated at high energy in \cite{hemsing2013coherent}. OAM light emitted from a 120 MeV beam at the first harmonic in a linearly polarized undulator was observed. Helical microbunching was produced as the beam passed through an $R_{56}=1.9$ mm chicane, after being modulated in a helical undulator tuned to a 1.6 $\mu$m wavelength at the fundamental resonance. There, the harmonic interaction, governed by Eq. (\ref{secOAMsecondharm}) with a 800 nm Gaussian laser profile, produced a helical energy modulation. The spiral density distribution emitted coherent OAM light (Fig. \ref{OAMMeasurement}) with a characteristic hollow intensity profile and $l$=1 helical phase, in agreement with expectations.

The helical microbunching principle can be expanded to multi-modulator/chicane techniques such as EEHG to further shorten the wavelength of the coherent emission of the OAM light, or also increase the OAM mode number, $l$. Dubbed Echo-v (for vortex) \cite{HemsingECHOOAM}, the beam is modulated in two undulators with the dimensionless energy modulation given by $p'=p+A_{1}(r)\sin(k_1s+l_{1}\phi)$ and $p'=p+A_{2}(r)\sin(k_2s+l_{2}\phi)$, respectively, with each modulation followed by a chicane. Calculation of the density modulation in this case repeats the EEHG analysis in Section~\ref{sec:3-1-3} and gives [see Eq.~\eqref{eq:II-A.7}]
    \begin{align}
    N(r,\phi,s)
    &=
    N_{0}(r)
    \sum_{n,m=-\infty}^\infty
    2b_{n,m}(r)
    \\
    &\times
    \cos[(nk_{1} + mk_{2})s + (nl_{1} + ml_{2})\phi+\psi_{n,m}]
    \nonumber
    ,
    \end{align}
with the coefficients $b_{n,m}(r)$ given by~\eqref{eq:II-A.9}. Note that the azimuthal dependence is now given by $l\phi$ with
	\begin{equation}
	l=nl_1+ml_2.
	\label{ECHOl}
	\end{equation}
Optimization of the bunching factor shows that large frequency up-conversion takes place for $m$ large, and $n$ small. Thus, depending on the individual values of $l_1$ and $l_2$ in each modualtor, Echo-v enables large up-conversion of the frequency and $l$ mode either simultaneously, or independently. By up-converting both together, (i.e., $l_1=0$, $l_2\ne0$) one can emit optical vortices with large $l$ at high-harmonics. Alternately, large frequency harmonics can be generated with little or no change in $l$, (i.e., $l_1\ne0$, $l_2=0$), so that the helical distribution generated at one frequency can be passed to a different frequency.

%
\subsection{Synthesis of radiation with an arbitrary waveform}\label{sec:3-6}
%

Laser modulators combined with dispersive chicanes provide fine scale control of the electron beam current distribution. While high-harmonics of the laser frequency can be generated in the electron beam by tuning the modulation amplitude and chicane dispersion (see Section~\ref{sec:3-1}), recently proposed schemes employing several modulator-chicane cascades also allow one to precisely tailor the harmonic content of the electron peak current distribution and radiation at the scale of the optical wavelength, in some cases using only a single laser \cite{PhysRevSTAB.16.010706}. As described further below, such systems can behave as optical analogues of conventional rf function generators, suggesting new ways to produce radiation with customized waveforms.

\begin{figure}
	 \includegraphics[width=90mm]{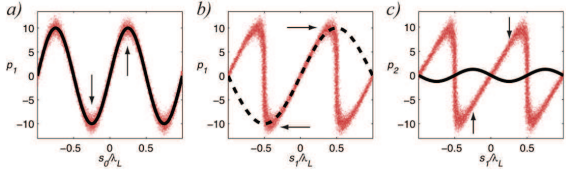}
	 \caption{Optical scale linearization of the longitudinal phase space performed by two laser modulators at the same frequency, mediated by dispersive chicane. (From \cite{PhysRevSTAB.16.010706}).}
	\label{SingleLaserChiLinear}
	\end{figure}

As an example manipulation procedure, in Fig.~\ref{SingleLaserChiLinear} we show how one can partially linearize a portion of a sinusoidal modulation. The energy modulation generated by an initial sinusoidal laser field (Fig.~\ref{SingleLaserChiLinear}a) creates a local chirp in the beam as a function of $s$ given by $h=dp/ds=A_1k_L\cos(k_Ls)$. After passage through the following dispersion section (Fig.~\ref{SingleLaserChiLinear}b), the regions where $h>0$ become stretched, while the the regions where $h<0$ become compressed. At a dispersion strength of $B\simeq 1/A_1$, the negatively chirped regions at $k_Ls=\pm\pi$ in each optical cycle are fully compressed, and half of the particles are localized in this narrowed region of phase. The other half are stretched over the decompressed regions extending in each optical cycle from $-\lambda_1/2$ to $\lambda_1/2$ and form a sinusoid with twice the wavelength of $\lambda_1$, as shown with a dashed line in Fig.~\ref{SingleLaserChiLinear}b.
Now the second laser with $k_2=k_L$ can interact on electrons in this region like the second harmonic frequency, and can therefore be used to partially linearize the modulation as shown in Fig.~\ref{SingleLaserChiLinear}c. In principle, this piecewise manipulation of specific phase space regions allows successive phase-stable modulations  precise control over the phase space distribution within a single wavelength, and can lead to enhancements in the achievable microbunching factors in harmonic generation and IFEL schemes.

Radiation waveform synthesis is another example of how cascaded laser manipulations may be used in the ``beam by design'' concept. In this case, the electron beam current distribution is arranged to have the precise spectral components of a desired waveform, which is then emitted by the electron beam. This works because the electric field spectrum produced in the radiator is proportional to the Fourier transform of the beam current distribution, which is just the bunching factor. Thus, in the typical scenario where the electron beam has transverse dimensions smaller than the transverse coherence size of the radiation field, the electric field spectrum is dominated by the longitudinal electron beam distribution and is given by
 Eq.~\eqref{eq:coh-field}. We can replace the summation over the particles in this equation by averaging over the longitudinal distribution function,
	\begin{align}
	\tilde E_b(\omega)
	&=
	N_e
	\int
	\tilde E(\omega)
	f(p,s)e^{-i\omega s/c}dpds
	\nonumber\\
	&=
	\frac{1}{ec}
	\int
	\tilde E(\omega)
	I(s)e^{-i\omega s/c}ds
	,
	\label{instantaneousfields}
	\end{align}
where $\tilde E(\omega)$ is the Fourier transform of the single particle emission field and $I(s) = ec N_e\int f(p,s)dp$ is the beam current. If the radiator emission spectrum is flat over the relevant microbunching spectrum then $\tilde E(\omega)$ can be treated as a constant. In this case, the emitted spatial electric field distribution is simply given by the beam current distribution, $E_b(t)\propto I(t)$.

\begin{figure}
    \includegraphics[width=85mm]{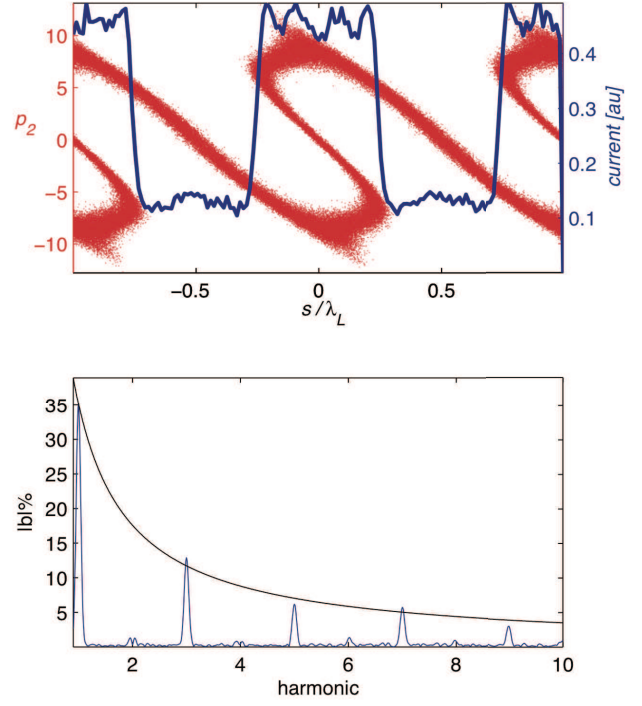}
    \caption{Generation of odd-harmonic bunching with amplitudes $b((2n-1)k_L)=b(k_L)/(2n-1)$ for the emission of square wavefrom fields with a double modulator-chicane system. Top: Modified phase space and current distribution with $A_1=10$, $A_2=1.388$, $B_1=0.295$, $B_2=-0.551$, $\phi_2=\phi_3=0$; Bottom: bunching factors for various harmonics.(From \cite{PhysRevSTAB.16.010706}).}
 	\label{RadiationSquarewave}
	\end{figure}

Consider the example of a square waveform, which is described by the sum odd harmonics $2n-1$ of the carrier frequency $k_L$ with amplitudes that scale as $1/(2n-1)$.  Shown in Fig.~\ref{RadiationSquarewave}, successive laser modulations at the frequency $k_L$, mediated by dispersion in two chicanes, generate a complicated electron beam phase space (red) that has a square wave current projection (blue) on the scale of the laser wavelength. The associated bunching spectrum displays a series of harmonic peaks whose magnitudes decrease in accord with the decreasing amplitudes $1/(2n-1)$ of the square waveform harmonics (decreasing line). As a result of this coherent structure, synthesized by modulations at a single laser frequency and dispersion, this beam with then radiate a square waveform in a broadband radiator \cite{PhysRevSTAB.16.010706}.

%
\section{Beam manipulation for synchrotron light sources}\label{sec:4}
\begin{figure}[!htb]
    \bc
    \includegraphics[draft=false, width=0.4\textwidth]{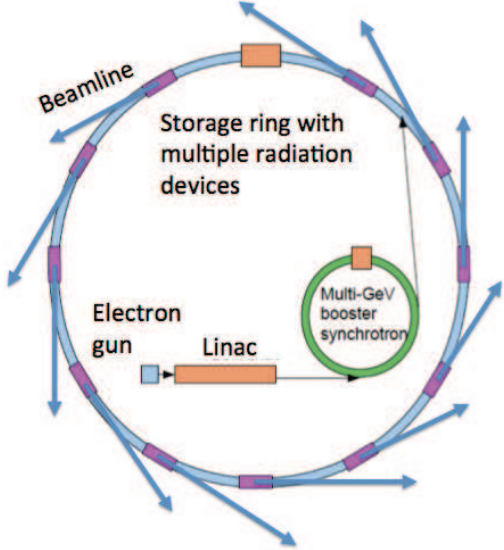}
    \caption{Typical layout of a storage ring light source.
    \label{Fig.III.A.1}}
    \ec
   \end{figure}

%
Storage ring synchrotron light sources produce high-brightness electromagnetic radiation in the spectral region extending from infrared to hard x-rays. A typical facility, illustrated in Fig.~\ref{Fig.III.A.1}, consists of an injector, booster synchrotron, transport lines between accelerators, a large storage ring with radiation-production devices, and numerous dedicated beamlines and experimental stations \cite{Winick}. The injector normally includes an electron linac equipped with an electron gun and a booster synchrotron. Most of the contemporary synchrotron light sources operate with a ``top-up'' injection \cite{Emery} to keep the electron beam current in the storage ring practically constant and, thus, use a full energy booster synchrotron. Electron storage ring technology has evolved considerably over the last 50 years, and in doing so, has led to highly reliable machines with many attractive features (discussed below) including simultaneous service to many users with diverse experimental programs.

A brief description of the historical development of synchrotrons is given in \cite{Zhao:RAST}. In this section, we give an overview of the properties of the electron beams and radiation produced in storage ring light sources, and show how laser-based beam manipulations may enhance their capabilities.

%
\subsection{Storage ring synchrotron light sources}\label{Section:III.A}
%

In a storage ring light source, an electron beam circulates around the ring repetitively for hours with remarkable orbit stability (on the order of a few microns in position and few microradians in angle \cite{Decker}), producing electromagnetic radiation via spontaneous emission in dipole magnets, wigglers, and undulators. Undulators provide the most concentrated radiation, with the rms radiation divergence given by $\sigma_{r'}=\sqrt{\lambda_r/\pi L_u}$, where $\lambda_r$ is the radiation wavelength and $L_u$ is the undulator length, and an rms transverse radiation size at the source of $\sigma_{r}= \frac{1}{4}\sqrt{\lambda_r L_u/\pi}$ \cite{attwood}. The on-axis spectrum is concentrated at the fundamental wavelength and its odd harmonics, each with bandwidth $\Delta\lambda_r/\lambda_r\approx1/nN_u$ at FWHM, where $n$ is the harmonic number. It should be noted that the undulator bandwidth is affected by the energy spread of electrons, but in most practical cases this is only visible at high undulator harmonics. The spectral photon flux $F$ measures the rate of photon flow within a particular bandwidth, and is usually expressed in the units of number of photons per unit time in 0.1\% of the relative spectral bandwidth. Typically $F \sim10^{13} -10^{15}$  (ph/s/0.1\% BW). The brightness of the radiation $B$ is defined as the photon flux per unit source area and per unit solid angle of the radiation:
	\begin{equation}\label{eq:III-A.2}
	B
	=
	\frac{F}{
	\left(2\pi\right)^2\sqrt{\sigma_x^2+\sigma_r^2}
	\sqrt{\sigma_{x'}^2+\sigma_{r'}^2}
	\sqrt{\sigma_y^2+\sigma_r^2}\sqrt{\sigma_{y'}^2+\sigma_{r'}^2}
	}
   	\end{equation}
where $\sigma_{x,y}$ are the horizontal and vertical electron beam sizes, and $\sigma_{x',y'}$ are horizontal and vertical divergence of the electron beam in the undulator. The brightness is traditionally expressed in the following units \cite{KJKim}:
	\begin{equation}\label{eq:III-A.3}
	\frac{{\rm number \, of \, photons}}
	{{\rm s \, \left(0.1\% \, bandwidth\right) \left(mm\right)^2\left(mrad\right)^2} }.
   	\end{equation}
The highest brightness is achieved when the horizontal and vertical beta functions in the undulator correspond to $\beta_x=\beta_y
\simeq L_u/4$. In this case
	\begin{equation}\label{eq:III-A.4}
	B
	=
	\frac{F}{(2\pi)^2
	\epsilon_r^2
	\left(1+{\epsilon_x}/{\epsilon_r}\right)\left(1+{\epsilon_y}/{\epsilon_r}\right)}
   	\end{equation}
where $\epsilon_r=\lambda_r/4\pi$ is the radiation emittance. In all contemporary storage ring light sources the horizontal electron beam emittance $\epsilon_x$ is much larger than the vertical emittance $\epsilon_y$ and also much larger than $\epsilon_r$, which is typically comparable to $\epsilon_y$. Thus, one finds from Eq.~\eqref{eq:III-A.4} that the undulator brightness is smaller than the brightness of a hypothetical diffraction-limited light source by a large factor ~$\epsilon_x/\epsilon_r \sim 10^2 - 10^3$, depending on the photon energy. Presently, efforts are ongoing to design a such source \cite{Borland2012, Cai2012, Leemann, Einfeld} that would have $B \sim10^{20} -10^{23}$  (ph/s/0.1\% BW/mm$^2$/mrad$^2$).

The origin of the electron beam horizontal emittance in the storage rings is due to electron orbit excitations from the random emission of  photons and synchrotron radiation damping~\cite{Wiedemann}. The same phenomenon also contributes to the electron beam energy spread $\sigma_{\cal E}$, which is related to the beam emittance by $\epsilon_x \simeq C ({\rho}/{\nu_x^3})({\sigma_{\cal E}}/{{\cal E}_0})^2$, where $C$ is the numerical coefficient that depends on the details of the storage ring lattice, $\rho$ is the bending radius, and $\nu_x$ is the horizontal betatron tune \cite{Sands}.
Designs of diffraction limited synchrotron light sources seek to take advantage of the strong $\epsilon_x$ dependence on $\nu_x$ to lower $\epsilon_x$.

The equilibrium energy spread $\sigma_{\cal E}$ can be calculated by considering the rms deviation of the number of photons emitted in one damping time from the mean number $n_q$. To get a qualitative feeling for the order of magnitude of the effect, we use the critical frequency of the synchrotron radiation in the bending magnet
$\omega_c=\frac{3}{2}c \gamma^3/\rho$ \cite{Wiedemann} and obtain a mean number of emitted photons $n_q \simeq {\cal E}_0/\hbar\omega_c$. The rms deviation from $n_q$ is equal to $\sqrt{n_q}$ (assuming a Poisson distribution) which gives an rms energy spread of $\sigma_{\cal E}\simeq\hbar\omega_c \sqrt{n_q}\simeq\sqrt{{\cal E}_0 \hbar\omega_c}$.

%
\subsection{Challenges in producing short x-ray pulses in synchrotrons}\label{Section:III.B}
%

The rms beam energy spread defined in \ref{Section:III.A} also affects the rms bunch length which, in the zero-current approximation, equals \cite{Sands}:
	\begin{align} \label{eq:III-B.1}
    \sigma_s
    =
    \bar{R}\frac{\alpha _c}{\nu_s}\frac{\sigma_{\cal E}}{{\cal E}_0},
	\end{align}
where $\bar{R}$ the average machine radius, $\alpha_c \simeq 1/\nu_x^2$ is the momentum compaction factor, and  $\nu_s$ is the synchrotron tune which is the number of longitudinal oscillations per revolution. Typically $\sigma_s/c$  is on the order of a few tens of ps. However, as the single-bunch electron beam current increases to a few mA, the bunch length also increases due to impact of the self-induced fields \cite{Pell, Bane} and the microwave instability \cite{Chao}. Therefore, all storage ring light sources operate with electron bunches that, unaltered, are too long for the investigation of fast processes at time scales below a few ps.

Several approaches to shorten the electron bunch have been tried, and one that takes advantage of a small (close to zero) $\alpha_c$ had been found to be the most successful \cite{Feik1, Feik2}. However, the synchrotron tune also decreases with $\alpha_c$ ($\nu _{s} \sim \sqrt{\alpha_c}$), and the less frequent change of particle positions inside the electron bunch leaves more time for instabilities to build up. As a result, short bunches on the order of 1 ps can only be obtained along with a dramatic reduction in the full electron bunch current \cite{Feik1, Limb}. This seems to be acceptable for generation of coherent synchrotron radiation in the THz part of the radiation spectrum  \cite{Wust} but not for spontaneous emission of photons in the x-ray regime.  Furthermore, accelerator-based x-ray sources also require transport lattices optimized to yield the smallest $\epsilon_x$, but low $\alpha_c$ storage ring lattice needs a negative dispersion function in a large number of their bending magnets, which is incompatible with a lowest-emittance lattice.

Implicit in these challenges is the assumption that the x-ray pulse have the same length as the electron bunch. However, much shorter x-ray pulses can be obtained if one can select the radiation emitted by electrons from a short section of the electron bunch and separate it from the radiation of all other electrons.

One way to achieve this is with a pair of rf deflecting cavities \cite{Sasha2TCAV}. Specifically, the first rf deflecting cavity imposes a time-dependent angular kick to the beam. Then the beam is sent through undulators or bending magnets to produce a radiation pulse in which transverse position or angle is also correlated with the time. The radiation pulse is then further shortened, either with an asymmetrically cut crystal that acts as a pulse compressor, or with an angular aperture such as a narrow slit positioned downstream. The second rf deflecting cavity then cancels the initial spatial chirp on the electron beam, minimizing the perturbations to the beam dynamics in the rest of the ring. Analysis shows that while this method can be readily used to provide x-ray pulses with a FWHM on the order of 1 ps \cite{BorlandAPSU}, it is challenging to push the x-ray pulse down to $\sim100~$fs. Such time scales can be achieved by laser manipulation, however, in an alternate method that uses a laser pulse to select an ultrashort slice from the electron bunch. This is discussed in detail in the following section.

%
\subsection{Laser-slicing for the generation of femtosecond x-rays in synchrotrons}\label{Section:III.C}
%

The ``slicing'' technique proposed in \cite{laser-slicing96} uses an ultra-short laser pulse to generate femtosecond x-ray pulses. At the time of writing, it had been implemented at the ALS \cite{laser-slicingLBNL}, BESSY \cite{laser-slicingBESSY}, and SLS {\cite{Schl} facilities. Fig.~\ref{Fig.III.C.1} shows a schematic of this technique. A laser pulse of moderate energy ($\sim$1 mJ) and pulse width $\sigma_t \sim100~$ fs modulates the energy of an ultrashort slice ($\sim 2c\sigma_t$) of a stored electron bunch as they co-propagate in an undulator (Fig.~\ref{Fig.III.C.1}a).  The energy-modulated electrons within this slice are spatially separated from the main bunch in a dispersive section of the storage ring  (Fig.~\ref{Fig.III.C.1}b)  and produce a femtosecond x-ray pulse  (Fig.~\ref{Fig.III.C.1}c) at a bend-magnet or other insertion-device (Fig.~\ref{Fig.III.C.1}b).
   \begin{figure}[!htb]
    \bc
    \includegraphics[draft=false, width=0.48\textwidth]{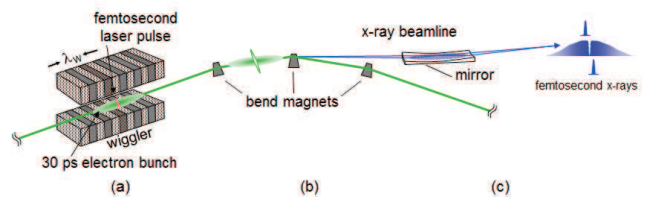}
    \caption{A schematic of the laser slicing method for generating femtosecond x-ray pulses (from~\cite{laser-slicingLBNL}). The three essential components are a) a modulator, b) a radiator and c) an x-ray beamline.
    \label{Fig.III.C.1}}
    \ec
    \end{figure}

	\begin{figure}[!htb]
    \bc
    \includegraphics[draft=false, width=0.45\textwidth]{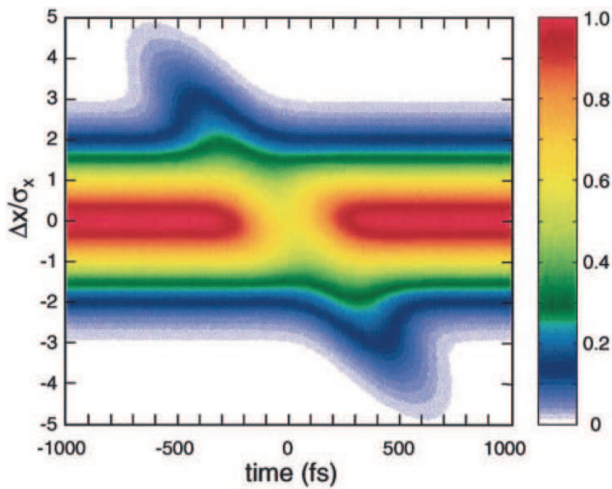}
    \caption{Calculated electron density distribution (as a function of time and horizontal displacement normalized to the rms electron beam size) after electron bunch propagation through one and one-half arc sectors at the ALS from the undulator to the bend magnet. Only a short section of the actual electron bunch is shown. Note that the path-length differences caused by time-of-flight properties of dispersive section give rise to the time-skew observed in the distribution, with electrons having $\Delta {\cal E}<0$ moving toward the bunch head and those with $\Delta {\cal E}>0$ toward the bunch tail (from~\cite{laser-slicingLBNL}).
    \label{Fig.III.C.2}}
    \ec
    \end{figure}

Figure~\ref{Fig.III.C.2} shows the calculated electron distribution of a laser-modulated bunch following propagation through a dispersive section.
X-ray optics positioned downstream of the radiator are used to image the radiation source point onto an aperture that selects only those x-rays originating from the transversely displaced electrons (see Fig. ~\ref{Fig.III.C.1}c). However, a long-pulse background from the electrons in the tails of the transverse distribution can also accompany this signal. Measurements of the electron-beam profile at the ALS and other electron storage rings indicate that it follows a Gaussian distribution approximately out to 5$\sigma_x$. Beyond this point, the population of electrons decreases with the distance from the beam core at a much slower rate \cite{Zhol2}. Another source of the long-pulse background is non-specular scattering from the x-ray optics, which mixes x-rays originating from different transverse coordinates. Thus, in order for the short-pulse signal to dominate over the background radiation, the slice electrons should be displaced by at least 5$\sigma_x$ or greater. This defines the minimal energy modulation amplitude and the required dispersion. It is worth pointing out that, because the electron beam has a smaller emittance in the vertical direction, the signal-to-background ratio can be improved if the beam is instead dispersed vertically \cite{Stei}.

The femtosecond x-rays from an undulator can also be isolated by introducing angular dispersion in the electron beam, as reported in \cite{laser-slicingBESSY}. Because no imaging optics were placed in front of the cutoff mask, an excellent signal-to-background ratio was obtained.
Another approach to obtain a short x-ray pulse from a laser-modulated beam is to use a high-resolution monochromator and take advantage of the fact that the highest or lowest energy electrons produce x-ray photons in the undulator that are shifted in energy. In this case, the long-pulse background will be determined by the combined spectral characteristics of the undulator and monochromator~\cite{Scho1}.

Note that energy modulation of an ultrashort slice leaves behind a hole in the main electron bunch (see Fig.~\ref{Fig.III.C.2}).  This will manifest as a dip  in the generated x-ray signal and, in principle, can be used for time-resolved spectroscopy in the same manner as an ultra-short pulse.

The time-of-flight properties of the dispersive section also cause temporal smearing of the sliced portion of the distribution due to the parameters $\sigma_{\cal E}$, $\sigma _{x}$, and $\sigma _{x'}$. Together these limit the duration of the ultrashort synchrotron x-ray pulse according to:
	\begin{align} \label{eq:III-C.1}
      {\sigma _{x-ray}^2}
      &=(2{\sigma _{t})^2} \\
    &+\frac{1}{c^2}\left[\left(R_{51}\sigma _{x}\right)^2
     + \left(R_{52}\sigma _{x'}\right)^2
     + \left(R_{56}\sigma_{\cal E}/{\cal E}_0\right)^2\right] \nonumber
	\end{align}
where $R_{51}$, $R_{52}$ and  $R_{56}$ are elements of the transport matrix from the modulator to the radiation source that contribute to the time-of-flight effects. The factor of 2 in front of $\sigma _{t}$ accounts for the slippage between the laser pulse and electron bunch as they traverse the modulator, assuming that $\sigma _{t}$ is appropriately chosen to yield the maximum amplitude of energy modulation.

The average flux, brightness, and spectral characteristics of the femtosecond x-ray pulses are determined from the nominal characteristics of the radiating bend magnet, undulator or wiggler source of x-rays, as well as the laser repetition rate. Increasing the latter provides the greatest opportunity to maximize the femtosecond x-ray flux. The practical limit is determined by the synchrotron radiation damping, which provides recovery of the natural electron bunch energy distribution between interactions. By arranging the timing such that the laser interacts sequentially with each bunch in the storage ring, the time interval between interactions is given by $N_B/f_L$, where $N_B$ is the number of bunches in the ring.  It turns out that, since the laser affects only a small fraction of the total bunch, an interaction interval corresponding to 30\% of the storage ring damping time (e.g., on the order of a few ms) is sufficient to allow recovery between laser interactions.

In addition to femtosecond x-rays, the time structure of the energy-modulated electron bunch with a hole in the central core (as seen in Fig.~\ref{Fig.III.C.2}) gives rise to coherent emission in the THz part of the spectrum.  This strong longitudinally and spatially coherent signal was used at ALS \cite{Byrd1}, BESSY \cite{Holl}, and SLS \cite{Schl} for an initial optimization of the laser-electron beam interaction, and for feedback control of spatial and timing drifts between the laser and electron beams in experiments with sub-ps x-ray pulses. It is also used for dedicated scientific experiments utilizing the broadband nature of the THz pulse. The same set-up also allows the production of a tunable narrow bandwidth THz pulse if one uses an intensity modulated laser \cite{Bielawski}. An elegant way of achieving this intensity modulation with the help of chirped pulse beating was proposed in \cite{Weling} and subsequently adopted by many groups to produce a density modulated electron beam \cite{Bielawski, Shen}.

On a final note, time-dependent phenomena in physics and chemistry are typically studied with pump-probe techniques in which fast dynamic processes are first initiated by a femtosecond laser or by a laser-driven ultrafast source and then probed, after a time delay, with an ultra-short x-ray pulse. Thus, the femtosecond laser that initiates the x-ray pulse in the slicing technique can also be utilized as a pump in this scheme, enabling a precise time delay between the pump and probe pulses.

%
\subsection{Coherent femtosecond radiation in synchrotrons}\label{sec:4-D}
%

In the laser-slicing method, the femtosecond x-ray pulse is generated by physically separating the modulated beam from the unmodulated part. As a result, the number of photons contained in the short pulse is reduced by a factor that approximately equals the ratio of the laser pulse length to the electron bunch length. Alternatively, one can use a laser to modulate a short section of the electron bunch and produce electron microbunching, thereby significantly increasing the intensity of the radiation of electrons in this section. The number of photons $N_{\text{ph}}$ radiated in a helical undulator at the fundamental frequency by a short section of an electron bunch containing $N_e$ pre-bunched electrons was derived in Ref.~\cite{ORS} for the case of a round beam, $\sigma_x = \sigma_y$.
In a synchrotron typically $\sigma_y \ll \sigma_x$ and the result of~\cite{ORS} cannot be used directly, however, it can be easily generalized for the case $\sigma_x \ne \sigma_y$. Analysis shows that in this more general case, $N_{ph}$ can be written as
	\begin{align}\label{eq:IV-D.1a}	
	{N_{ph}}
	&=
    \pi\alpha
    b^2
    \frac{K^{2}}{1+K^{2}}
    \frac{N_e^2N_u}{N_b}
    {\cal F}
    \left(
    \frac{\sigma_x}{d}
    ,
    \frac{\sigma_y}{d}
    \right)
    ,
    \end{align}
where $d=2\sigma_r=\frac{1}{2}\sqrt{\lambda_r L_u/\pi}$ is the transverse coherence length and $N_b$ is the number of the microbunches given by the length of the modulated section divided by the laser wavelength (a condition $N_{b} \gg N_{u}$ is assumed in~(\ref{eq:IV-D.1a})). A complicated expression for function $\cal F$ reduces to simple forms in limiting cases: for $\sigma_x,\sigma_y \ll d$, ${\cal F}=1$; for $\sigma_x,\sigma_y \gg d$, ${\cal F}=d^2/\pi\sigma_x \sigma_y$; for $\sigma_y \ll d \ll \sigma_x$, ${\cal F} = 0.71 d/\sigma_x$. An added benefit of the described approach is that the pre-bunched beam produces radiation with greatly improved temporal coherence compared to spontaneous radiation (relative bandwidth $\sim 1/N_{b}$ versus $1/N_{u}$), which makes this kind of radiation particularly suitable for spectroscopic applications.

Unlike the linac-based light sources where an electron beam with high peak current and small energy spread is used to drive exponential gain in a seeded single pass FEL, the beams in synchrotrons typically have relatively low peak currents and large energy spreads that inhibit amplification. Therefore, most of the methods to generate ultra-short coherent radiation in synchrotrons use the coherent harmonic generation (CHG) technique, which is similar to the HGHG technique in FELs for obtaining coherent emission except that there is no FEL gain. In the standard CHG scheme (see Fig.~\ref{figCHG}), a seed laser with wavelength $\lambda_L$ is first used to generate an energy modulation in the beam in a modulator. After passing through a dispersive element (e.g. a chicane), the energy modulation is converted to a density modulation (microbunching). Finally the density-modulated beam is used to generate coherent radiation at the wavelength $\lambda_L/n$ in the radiator, where $n$ is the harmonic number.
    \begin{figure}[h]
    \includegraphics[width = 0.48\textwidth]{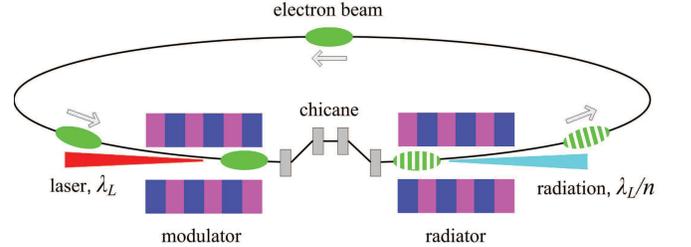}
    \caption{Schematic layout of CHG setup to generate ultra-short coherent radiation in synchrotrons. The electron beam is energy modulated by a laser with wavelength $\lambda_L$ in a modulator, microbunched by a chicane, and then used to generate intense coherent radiation with wavelength $\lambda_L/n$.
    \label{figCHG}}
    \end{figure}

The CHG technique was first demonstrated at LURE where coherent radiation at 355 nm was obtained with a 1.06 $\mu$m wavelength Nd:YAG laser (corresponding harmonic number is $n=3$) \cite{LURE}. By reducing the laser wavelength to 264 nm, intense ultra-short coherent radiation at 132 nm with peak brightness four orders of magnitude higher than the spontaneous radiation has been produced at Elettra \cite{Elettra}. Recently, coherent emission down to 89 nm has also been observed at UVSOR with a 800 nm Ti:Sapphire laser \cite{UVSOR1, UVSOR2}. So far this is the shortest wavelength achieved with CHG in synchrotrons. As discussed in Sec.\ref{sec:3-1-1}, generating the $n$-th harmonic of the seed laser typically requires the energy modulation amplitude $\Delta \cal E$ to be approximately $n$ times larger than the beam energy spread $\sigma_{\cal{E}}$. Because the required laser pulse energy scales with $(\Delta {\cal E})^2$ (see Eqs.~(\ref{eq:II-C.13}, \ref{eq:II-C.11})),  obtaining radiation at high harmonics can be problematic at high repetition rates. Furthermore, increasing the energy modulation beyond the energy acceptance of the synchrotron, which is typically on the order of (1-2)\% of the beam energy, will lead to significant beam loss. Accordingly, these constraints typically limit the harmonic number from the standard CHG scheme in synchrotrons to below $n=10$.

%
\subsubsection{Accessing shorter wavelengths}
%

To extend the coherent femtosecond radiation in synchrotrons to shorter wavelengths, it may be possible to adapt the EEHG technique to generate high harmonics with relatively small energy modulations, thus avoiding increasing the beam energy spread beyond the synchrotron energy acceptance. Because incoherent synchrotron radiation tends to wash out the fine structures in EEHG for high energy electron beams, EEHG is particularly suited to low or medium energy synchrotrons.

Such a scheme has been proposed at the synchrotron SOLEIL \cite{EEHGSOLEIL} at around 2.75 GeV. There, the first modulator in EEHG is located in one of the straight sections, additional dipoles are added, and the quadrupoles are properly set to reduce the $R_{56}$ and to zero the $R_{51}$ and $R_{52}$ transport terms. Finally, the second modulator, second small chicane, and radiator are together put in the following straight section. With this configuration, coherent femtosecond radiation may be generated at EUV and soft x-ray wavelengths, with peak power levels several orders of magnitude higher than the spontaneous radiation. A similar proposal is being pursued at the synchrotron DELTA \cite{DELTA} in which the bending dipoles are exchanged to make a long straight section (16 m) to house the two modulator-chicane modules, similar to the implementation in a linac FEL.

In addition to modulating the beam energy, it is also possible to modulate the beam divergence to generate high harmonics \cite{CHGdivergence}.  With a high power TEM$_{01}$ mode laser, the sinusoidal angular modulation discussed in Section \ref{Section:II.C} can be imprinted on the electron beam's vertical angles and later converted into a density modulation using a special magnetic chicane integrated with quadrupoles to provide a non-zero $R_{54}$ transfer matrix element. Analysis shows that in this case the achievable harmonic number is approximately $\Delta y'/\sigma_{y}'$, where $\Delta y'$ is the angular modulation amplitude and $\sigma_{y}'$ is the rms vertical divergence. For a storage ring with an ultimately small horizontal emittance operating in a low coupling mode with a geometric vertical emittance on the order of one picometer, this technique may be used to extend the harmonic number to beyond $n=50$ to provide coherent femtosecond soft x-rays.

%
 \subsubsection{Accessing higher peak power}
%

For the goal of obtaining a high peak power while using moderate harmonic numbers, one may adapt the chirped pulse amplification (CPA) technique \cite{CPA} which is well established in the laser community and operate CHG in CPA mode to make full use of the entire electron bunch. In this mode \cite{CPACHG1, CPACHG2}, an ultrashort laser with a relatively large bandwidth is first stretched to match the length of the electron beam. The frequency-chirped laser then interacts with the electron beam to generate frequency-chirped bunching at the harmonic frequency of the laser. The frequency-chirped high harmonic radiation generated in the radiator is then further compressed to provide an intense ultrashort coherent radiation.

According to the simulation results shown in Fig.~\ref{figCHGSINAP} this technique may increase the peak power in CHG by $2\sim3$ orders of magnitude and proof-of-principle experiments are being considered \cite{CPACHG1, CPACHG2}. In reality, however, the power enhancement also depends crucially both on the efficiency of the dispersive element (e.g. a grating) and the configuration of the compressor for shortening the pulse. A recent design of pulse compressor at 13.5 nm has shown an overall efficiency of about 20\% \cite{GratingEFF}, which suggests that CHG operating in CPA mode can in practice lead to a marked improvement in photon number.
    \begin{figure}[h]
    \includegraphics[width = 0.48\textwidth]{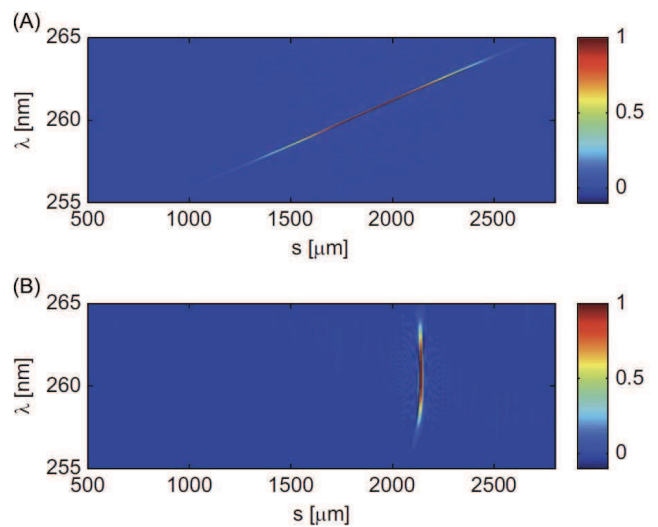}
    \caption{Wigner distributions of the radiation pulse before (A) and after (B) the compressor simulated for a proof-of-principle experiment to demonstrate CPA in CHG. The seed laser has a large frequency chirp with central wavelength at 786 nm. From \cite{CPACHG2}.
    \label{figCHGSINAP}}
    \end{figure}

%
\subsection{Steady state microbunching}\label{sec:SSMB}
%

Conventional FELs driven by linacs produce coherent radiation orders of magnitude brighter than incoherent sources, but have low duty cycles because each electron bunch is used only once. Fully filled storage rings, on the other hand, deliver continuous wave (CW) repetition rates but do not generally support high peak currents or sustained microbunching for the production of coherent light.

As a potential way to combine both features into a high-brightness, high-duty cycle light source, the steady state microbunching (SSMB) concept \cite{PhysRevLett.105.154801} proposes to generate sustained microbunching at a specific position in a ring in order to produce intense coherent radiation with MHz to CW repetition rates. Using either one or a pair of laser modulators, the SSMB idea works on the notion that in a storage ring, particles tend to collect around fixed points in phase space. A particle is at a fixed point if, after one turn it returns to its initial coordinates. For example, each accelerating bucket in the drive rf system has one fixed point per rf wavelength. Electrons forward of the fixed point are accelerated but slip backward due to dispersion, while those behind the stable point are decelerated but slip forward. The result is a train of electron bunches separated by the rf wavelength, which is on the order of a few tens of centimeters.

To generate microbunches at optical and short wavelengths, SSMB suggests the use of a laser modulator and the natural dispersion of the ring to stack electrons at stable points separated by the laser wavelength. In Ref. \cite{PhysRevLett.105.154801}, Ratner and Chao consider a particle at the zero crossing of the laser field with a specific energy deviation $\Delta \cal{E}$ from the stable fixed point energy. At this energy, the particle slips by $\Delta s=\lambda_L$ after each turn around the ring due to dispersion, and thus returns to another stable point at an identical zero crossing. Likewise, particles with energies $n\Delta \cal{E}$ for $n=0,~\pm1,~\pm2$ etc., also slip by $n\Delta s$ each turn, such that at each zero crossing the microbunches are stacked in energy. In principle, this well-aligned stacking occurs only in the modulator, as after a fraction of a turn $1/h$, the microbunches slip by $n\lambda_L/h$. If each microbunch is shorter than this slippage, the beam is microbunched at the wavelength $\lambda_L/h$ at this portion of the turn. At this position, one can insert a radiator tuned to emit resonantly at this wavelength, which is the $h^{th}$ harmonic of the laser wavelength.

Further variations include the addition of a second laser modulator at the same wavelength to improve the harmonic structure. This includes an EEHG-version of SSMB, which may lead to yet higher harmonics. A slightly different wavelength can also be used in the second modulator to produce lower frequency beat waves in the microbunching structure in the THz regime. While laser-based SSMB has tight tolerances on beam synchronization and time-of-flight transport elements, an alternate implementation of SSMB for high repetition rate coherent THz production has been suggested in \cite{Jiao2011} where the laser modulators are replaced with rf cavities.


%
\section{Beam manipulation for x-ray free-electron lasers}\label{sec:5}
%

Invented by John Madey in the 1970s \cite{Madey, Madey1, Madey2, Madey3}, free-electron lasers (FELs) now provide tunable, high power radiation 10 orders of magnitude brighter than storage ring based synchrotron light sources for wide ranging applications. While FELs have already opened up experimental access to new regimes of x-ray science, their performance can be further enhanced with laser-based electron beam manipulation techniques. Here, guided by FEL scaling laws, we review some FEL basics and describe several of the more recent techniques designed to improve and expand FEL capabilities. Readers interested in more detailed discussions of FEL physics are directed to the references \cite{LaserHandbook, PhysRevSTAB.10.034801, SaldinBook, Schmuser, Brau} and the citations therein.

%
\subsection{Free-electron lasers}\label{sec:FELs}
%

FELs belong to a class of devices in which the electrons interact with the radiation field in vacuum. Unlike conventional lasers where photon energies are restricted to fixed transitions between atomic energy levels,  FELs use unbound electrons and have no such limitation on their output wavelength. This enables the FEL unique frequency tunability in that the wavelength scales with the electron beam energy as,
	\be
	\lambda_r
	=
	\frac{\lambda_u}{2\gamma^2}
	\left(1+\frac{K^2}{2}\right)
	.
	\ee
This also allows FELs to produce high power radiation without suffering from thermal lensing, birefringence or heat dissipation issues. Furthermore, generated in vacuum, the radiation is essentially diffraction limited, with a low angular divergence and narrow bandwidth.

High-gain FELs are characterized by a fundamental scaling parameter known as the Pierce parameter, $\rho$, ~\cite{Bonifacio} which relates the longitudinal plasma wavenumber $k_p=\sqrt{4\pi e^2n_0/\gamma^3mc^2}$ (where $n_0$ is the beam volume density) to the undulator wavenumber $k_u=2\pi/\lambda_u$,
	\be
	\rho=\left(\frac{K^2{\cal J}^2k_p^2}{32k_u^2}\right)^{1/3}.
	\ee
The longitudinal plasma oscillation period is typically much longer than the undulator period, so $\rho$ is typically on the order of $\sim 10^{-3}-10^{-4}$ for modern high-energy FELs. The parameter $\rho$ sets the characteristic exponential gain length of the radiation for a cold beam,
	\be
	L_g=\frac{\lambda_u}{4\pi\sqrt{3}\rho},
	\ee
as well as the output saturation power $P_\mathrm{sat}\simeq\rho {\cal E}I_e/e = \rho P_b$ ($P_b$ is electron beam power), saturation bandwidth $\Delta\omega/\omega_r\sim\rho$, and saturation length, $\sim 20 L_g \sim\lambda_u/\rho$ \cite{PhysRevLett.57.1871}. (See Fig.~\ref{FELgaincurve}.) Amplification of the radiation field is accompanied by the growth of the electron beam energy spread, so saturation occurs when the induced relative energy spread covers the FEL bandwidth, $\sigma_\eta\sim\rho$. This sets an upper limit not only on the intrinsic slice energy spread required to achieve high-gain, but also on the energy spread induced by laser seeding schemes in order that the coherent field can be amplified by the FEL process. Likewise, the geometric electron beam transverse emittance is limited to be $\epsilon_x<\lambda_r\beta/4\pi L_g$ in order to maintain high-gain conditions, where $\beta$ is the beta-function characterizing the strength of beam focusing in the undulators. The parameter $\rho$ also sets the radiation spot size of the gain-guided fundamental gaussian mode, $w_0\simeq2\sigma_x\eta_d^{1/4}$, in weakly diffracting systems where $\eta_d=L_g/2k\sigma_x^2\ll1$.

	\begin{figure}[h]
	\includegraphics[width = 0.48\textwidth]{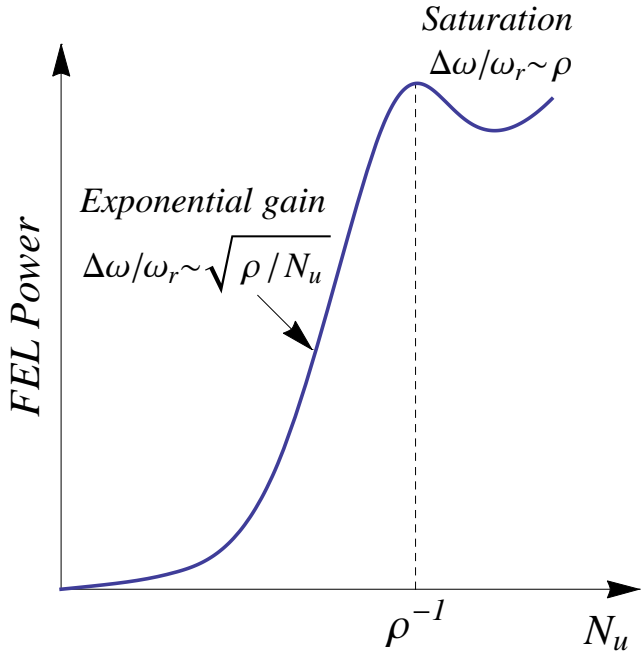}
	\caption{Characteristic power gain curve for a SASE FEL (adapted from \cite{PhysRevLett.57.1871}).}
	\label{FELgaincurve}
	\end{figure}

The regime when an FEL starts from random shot noise in the beam is called a self-amplified spontaneous emission (SASE).  In a simple 1-D model, the initial shot noise bunching factor is given by Eq.~\eqref{eq:coh-rad-6} and can be written as $b_\mathrm{sn} \sim \sqrt{1/N_c}$ where $N_c=I\lambda_r /ec\rho$ is the number of electrons in the length $\lambda_r/\rho$ of the beam~\cite{1367-2630-12-3-035010}). The shot noise has a broad bandwidth corresponding to an extremely short correlation length. In the FEL undulator (and in a laser modulator), resonant energy exchange with the electromagnetic field is maintained as electrons slip backward a distance $\lambda_r$ relative to the field each undulator period $\lambda_u$. After propagating together over a gain length, the electromagnetic wave slips through the electron beam one cooperation length $l_c=L_g\lambda_r/\lambda_u=\lambda_r/4\pi\sqrt{3}\rho$ \cite{PhysRevLett.73.70}. The temporal coherence length $l_\mathrm{coh}$ in a SASE FEL, established after the beam propagates a distance $L$ in the undulator, extends over a number of cooperation lengths and can be estimated as $l_\mathrm{coh} \sim (L/L_g)^{1/2}l_c$~\cite{krinsky02}. Accordingly, the number of temporal spikes in the SASE emission, given by $\sigma_s/l_\mathrm{coh}$, can be large if the electron beam length $\sigma_s$ is much longer than the cooperation length. The position and intensity of each spike depend strongly on initial shot noise bunching and therefore have large fluctuations from shot to shot \cite{PhysRevLett.73.70, Saldinstatistical}. In this case the SASE spectrum is also noisy, with a bandwidth much larger than the Fourier transform limit defined by the electron bunch length.

%
\subsection{Challenges in free-electron lasers}
%

Until recently, most of the modern high-gain FELs in the short wavelength (VUV to hard x-ray regime) have operated in SASE mode~\cite{FLASH, LCLS_2010, SACLA}, in which radiation from the electron beam shot noise is exponentially amplified to the multi-gigawatt level. While characterized by excellent transverse coherence, SASE FELs have limited temporal coherence and exhibit large statistical fluctuations in the output power spectrum. Thus, one of the most pressing needs is to make x-ray FELs fully coherent. Several such schemes are discussed in Section~\ref{SecIV:coherent}.

A second challenge is control over the x-ray pulse duration. For most of the FELs in operation the electron bunch length is longer than the cooperation length, and the pulse duration of the x-rays is typically on the order of tens of femtoseconds as limited by the electron bunch length. But in order to probe processes evolving on the time scale of electron motion in atoms, or to examine subtle details about how chemical bonds between atoms in molecules form or break, even shorter x-ray pulses down to attoseconds ($10^{-18}$ s) are required. Very recently, a novel scheme to push the x-ray pulse duration down to the zeptosecond ($10^{-21}$ s) regime has even been proposed~\cite{zs}. We discuss various approaches designed to address these challenges in Section~\ref{Section:IV.D}.

A third challenge is to generate ultra-high power x-ray pulses, primarily for studies of matter in extreme conditions and for advanced imaging of biological samples using the novel concept of ``diffraction before destruction''~\cite{DBD}. Because simplified FEL configurations typically saturate when beam energy spread becomes comparable to $\rho$ and beam energy goes out of resonance with the amplified light, the FEL power can be enhanced both by increasing the beam peak current (and hence $P_b$ and $\rho$), and by tapering the undulator tuning to continue extracting power from the electron beam. In recently proposed laser assisted schemes, for instance, such enhancements can be achieved by creating a train of periodic current-enhanced electron mircobunches and then applying a temporal shift between the FEL radiation and the micro-bunch train. This way, coherent x-rays up to 5 TW of peak power may be produced~\cite{5TW}. At such short wavelengths, the radiation can be focused to a tiny spot approaching the diffraction limit, thus producing high power densities useful for studies of matter in extreme conditions.

In addition to making brighter, shorter, and fully coherent x-ray pulses, there is also growing interest in adapting the experimental techniques developed for conventional lasers to the FEL by tailoring the properties of the radiation. For instance, mode-locked x-ray FELs (discussed in Section~\ref{Section:IV.E}) are envisioned to generate single radiation pulses that span a wide frequency range and consist of a series of equally spaced sharp spectral lines that may enable single-shot resonant inelastic x-ray scattering spectroscopy. Several schemes have recently demonstrated sequential x-ray pulses with independently tunable wavelengths~\cite{FERMI2color, Lutman}. These and other techniques (discussed in Section~\ref{Section:IV.F}) may also serve as new tools for x-ray spectroscopy, such as time-resolved three-wave mixing, stimulated Raman scattering, and so on.

%
\subsection{Generation of fully coherent x-ray pulses}\label{SecIV:coherent}
%

Several schemes geared towards improving the temporal coherence of SASE FELs have been proposed, and among those recently demonstrated, several are coming into common use for research.  One way to generate a fully coherent x-ray pulse in high-gain single-pass SASE FELs is to make the FEL cooperation length comparable to the electron bunch length. For SASE FELs driven by low charge beams on the order of 1 pC, the bunch after compression can be made as short as $\sim1~\mu $m to produce a single longitudinal radiation mode \cite{Rosenzweig200839, Reiche200845, LCLSlowcharge}.
In more general cases where the charge is on the order of 100 pC and the bunch length is significantly longer, the slippage length can be boosted either with chicanes between undulator sections, or by sequentially detuned undulator sections \cite{PhysRevLett.100.203901, HBSASE, iSASEFEL12}. The latter scheme has recently been demonstrated at the LCLS at SLAC~\cite{iSASE}. The former configuration may also enable x-ray FELs to reach shorter wavelengths via `harmonic lasing' as suggested in Ref~\cite{PhysRevLett.96.084801}. There, periodic temporal shifts of the electron beam along the FEL undulator from small chicanes (typically called phase shifters) are used to suppress lasing at the fundamental frequency while allowing harmonics to grow to saturation. Yet another proposed method is designed to increase the slippage with a short section of undulators that are tuned to a sub-harmonic of the fundamental FEL radiation wavelength~\cite{HLSchneidmiller, pSASE}. In this case, the FEL slippage length is increased by the sub-harmonic number to allow initially separated regions of radiation to overlap, leading to a significant improvement in FEL temporal coherence. In ideal conditions, the slippage length can be made comparable to the bunch length to produce fully coherent x-ray pulses. While some such schemes have shown promising results, in realistic conditions, however, the beam energy spread puts limits on the largest temporal shift that can be applied, which makes it difficult to achieve fully coherent x-ray pulses by such means \cite{HBSASE}. 

Fully coherent x-rays may also be produced in a low-gain multi-pass FEL oscillator configuration (see, e.g.,\cite{Gandhi, XrayOCL}) with the cavity resonator is composed of either several high reflective multilayer mirrors in the case of the soft x-rays, or of diamond crystals in the case of the hard x-rays.  The use of the regenerative amplifiers has also been proposed (see, e.g., \cite{XrayRGA}.

Alternatively, coherent FEL pulses can be produced if the FEL starts from a fully coherent seed that has sufficient power to dominate over the electron beam shot noise, thereby overcoming the SASE startup process. In the self-seeding scheme \cite{Feldhaus1997341, doi:10.1080/09500340.2011.586473}, such a seed is obtained from the monochromatized output of an upstream high-gain FEL stage that operates in SASE mode. The monochromatic radiation is then driven to saturation in the second stage FEL that operates as an amplifier. In principle, the self-seeding technique works at all wavelengths accessible to SASE FELs, and has recently been demonstrated at hard x-ray wavelengths \cite{2012NaPho...6..693A}. At the time of this writing, efforts are underway to minimize fluctuations in the final output and to implement this scheme at soft x-rays.

Seeding FELs with external lasers allows the generation of fully coherent radiation that is also well timed with respect to the seed laser for pump-probe experiments. One way to directly seed an FEL in the UV portion of the spectrum (400 nm to 20 nm) is to use a high harmonic generation (HHG) source produced from a high power laser injected into a noble gas (see, e.g.,~\cite{HHGreview}). In this scheme, the FEL operates as a high-gain amplifier of the low-power HHG source, amplifying it by several orders of magnitude. Direct FEL seeding at 160 nm \cite{HHG160nm}, 61 nm \cite{Togashi:11} and 38 nm \cite{HHG382013} from HHG sources has been demonstrated in recent years. Limited by the $\sim10^{-6}$ conversion efficiency, however, present HHG sources typically have a cutoff wavelength around $\sim$20 nm due to their relatively low pulse energy. While there has been progress from the laser community in extending HHG wavelengths to a few nm, FEL seeding with an HHG source at x-ray wavelengths has yet to be achieved because the HHG power needs to be at least one order of magnitude larger than the electron beam shot noise power in order to preserve the temporal coherence. This constraint is exacerbated by the fact that the shot noise power grows as the FEL wavelength is decreased. An interesting and developing technique that holds promise to lower requirements on the seed laser power is referred to as shot noise suppression. In such schemes, the intrinsic shot noise in the electron beam is reduced by tuning the effective longitudinal particle dynamics in the relativistic beam prior to injection into the FEL undulator~\cite{gover09d,ratner11hs}. Recently the noise suppression was experimentally demonstrated at the optical wavelengths~\cite{ratner_stupakov_12,gover_nature}, with some studies showing its prospects and challenges at shorter wavelengths~\cite{COOL13,FEL11_kim_lindberg}.

To circumvent the need for a high power electromagnetic seed at short wavelengths, many frequency up-conversion techniques that rely on using lasers to manipulate the beam's longitudinal phase space (e.g. HGHG, EEHG and their variants) have been proposed to seed x-ray FELs. These methods are designed to provide bunching at the FEL wavelength so that the coherent emission from the electron beam seeds the FEL amplification process. The theoretical aspects of these techniques are presented in the preceding Section~\ref{sec:3-1}. In these schemes, coherent bunching significantly above the shot noise level leads to emission of coherent radiation in the beginning of the FEL, which is then amplified to saturation. The first experiment to demonstrate the HGHG technique was performed at BNL \cite{HGHGScience}. There, a CO$_2$ laser with a wavelength of 10.6 $\mu$m was used to modulate the electron beam, and the induced 2nd harmonic bunching at 5.3 $\mu$m produced radiation with excellent temporal coherence that was amplified to saturation. Later, the 3rd and 4th harmonics of a Ti:Sapphire laser (800 nm) were amplified by FELs in 2003 \cite{HGHG3rd} and 2006 \cite{HGHG4th}. Recently, HGHG with a seed at 266 nm from an HHG source has also been demonstrated \cite{SPARCHHG}. So far, the highest harmonic achieved with the HGHG technique in a single stage is the 13th harmonic at 20 nm using a 1.2 GeV beam at FERMI FEL in 2011 \cite{FERMI}, which was limited by the requirement that the induced energy spread be less than $\rho$ to achieve high-gain. These experiments confirmed the advantages of seeding FELs with external lasers; namely, a stable central wavelength and excellent temporal coherence of the output radiation. For instance, as shown in Fig.~\ref{FERMIseeding}, the normalized central wavelength stability of 8th harmonic radiation seeded by a 260 nm UV laser is about $7\times10^{-5}$, a significant improvement over SASE FEL results obtained in the same wavelength range. Higher harmonics can be obtained by cascading multiple HGHG stages. Very recently, 4.3 nm radiation (60th harmonic of a 260 nm seed laser) has been achieved with a two-stage HGHG configuration at FERMI \cite{FERMI2stage}.
    \begin{figure}[h]
    \includegraphics[width = 0.48\textwidth]{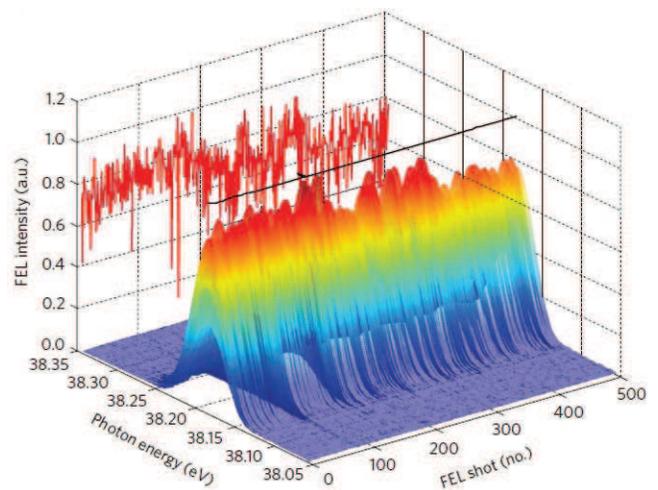}
    \caption{Measured FEL intensity and spectrum for 500 consecutive shots at 32.5 nm seeded with a 260 nm laser. From \cite{FERMI}.
    \label{FERMIseeding}}
    \end{figure}

Another harmonic bunching technique, EEHG (see Section \ref{sec:3-1-3}), holding the promise of generating fully coherent radiation in the soft x-ray wavelength in a single stage, is still in early experimental development. The first proof-of-principle experimental demonstration of the EEHG technique was achieved in 2010 at SLAC's NLCTA \cite{Echo3}. In this experiment, the 3rd and 4th harmonics of the second laser were generated. The experiment confirmed the basic physics behind the EEHG technique and indicated that the highly nonlinear phase space correlations were preserved and controlled with current technologies. Later, the 7th harmonic was also successfully produced by the same group, using induced energy modulations approximately twice that of the beam energy spread \cite{Echo7}. This experiment provided the first evidence that high harmonic radiation with harmonic number much higher than the ratio of energy modulation to energy spread could indeed be produced through EEHG. In 2011, a group at the Shanghai Institute of Applied Physics (SINAP) also observed the 3rd harmonic from EEHG, which was further amplified to saturation \cite{ZhaoZ12:first}. Currently, experimental efforts are on-going at several laboratories to extend EEHG to much higher harmonics and much shorter wavelengths. Very recently, coherent radiation at 160 nm (15th harmonic of a 2400 nm seed laser) has been produced with EEHG at SLAC~\cite{echo14}.

It is important to emphasize here that harmonic multiplication in the seeding increases the electron shot noise in the beam~\cite{saldin02sy,FEL10stupakov_4}. In addition, it also amplifies the phase errors in the seeding laser beam. As follows from analysis of~\cite{ratner_et_al_12} seeding with nearly transform-limited pulses in the soft x-ray regime will require development of new methods for phase measurement and control of short wavelength lasers or HHG sources.

%
\subsection{Generation of attosecond x-ray pulses}\label{Section:IV.D}
%

There are many ways to manipulate the electron beam properties to control the x-ray pulse length in FELs. As mentioned, a conceptually simple method is to lower the beam charge to reduce the electron bunch length, and thus radiation pulse length, by two orders of magnitude~\cite{XJWang2003, Rosenzweig200839, Reiche200845, LCLSlowcharge}. Alternately, one may use a ``slotted foil" in the center of a dispersive chicane to selectively spoil the transverse emittance in specific portions of the electron beam such that an ultrashort x-ray pulse is produced only from the short slice of unspoiled beam that passes cleanly through the slot~\cite{PE2004}. In a different technique, a few-cycle laser pulse can be used to manipulate the beam phase space to select only a short slice for lasing. This technique has another advantage in that the x-ray pulse is naturally synchronized with the laser, and therefore is well-suited for pump-probe experiments. The authors in~\cite{Martin} compare the pros and cons of these various proposals for producing attosecond x-rays.

One possible implementation of laser based approaches for the generation of attosecond x-ray pulses is shown in Fig.~\ref{Fig.IV.D.1} (see also,~\cite{Zhol4, Zhol5}).
   \begin{figure}[!htb]
    \bc
    \includegraphics[draft=false, width=0.5\textwidth]{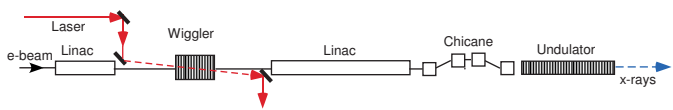}
    \caption{Schematic of a current enhanced SASE x-ray FEL~\cite{Zhol4}.
    \label{Fig.IV.D.1}}
    \ec
    \end{figure}
Here the electron bunch exits the upstream linac and enters a wiggler magnet. At the same time, a short laser pulse enters the wiggler and co-propagates with the electrons. The laser pulse overlaps only a short section of the bunch which we call the working section (WS). Electrons in the WS interact with the laser field and emerge from the wiggler with an energy modulation. The laser pulse energy is chosen such that the amplitude of the energy modulation exceeds the uncorrelated energy spread of the electrons by a factor of 5 to 10. Next, the electron beam enters a second linac and is accelerated to the final energy. Modulating lower energy electron beams is preferred because the required undulator period and $K$ value for resonance at optical to infrared wavelengths quickly grow with the beam energy. This second acceleration does not affect the energy modulation introduced in the wiggler and does not produce noticeable relative longitudinal motion of electrons because of the ultra-relativistic electron energies. Following acceleration, the electron beam passes through a dispersive magnetic chicane that produces bunching in the WS at the laser wavelength and thus a periodic enhancement of the electron peak current. Finally, the electron beam passes through a long undulator where radiation emitted by these electrons inside the WS is amplified with a shorter gain length because of the localized current enhancement, and therefore produces the dominant x-ray radiation that consists of short pulses equally separated at the laser wavelength. The x-ray radiation produced by electrons outside of the WS has significantly less intensity because of the longer FEL gain length due to their significantly lower peak current. Thus, there is precise synchronization between the output x-ray pulse and the laser pulse since only electrons from the WS produce intense x-rays. Moreover, by changing the duration of the laser pulse and adjusting the number of active wiggler periods, one can regulate the length of the WS and therefore the duration of the x-ray output.

Besides generating powerful x-rays in the FEL, electrons from the WS can also produce strong coherent synchrotron radiation at the modulating laser frequency (which is automatically temporally synchronized with the x-ray pulse) in a one-period wiggler that can be placed at the end of the FEL. This signal can be cross-correlated with the laser pulse to provide an accurate measure of the timing jitter between the laser pump pulse and the x-ray probe pulse. Note also that either edge radiation from a bending magnet or transition radiation from a thin metallic foil can be used if a wiggler is not available.

The duration of the seed laser pulse directly affects the duration of the x-ray output. Thus, short x-ray pulses can be obtained using seed laser pulses with only a few optical cycles, as briefly discussed in Section~\ref{Section:I.A}. Such a laser pulse, which interacts with the electrons in a wiggler magnet consisting of just one or two periods, produces an energy modulation with a temporal profile that closely resembles the waveform of the laser electric field. For example, manipulating the carrier-envelope phase (CEP), one can obtain a cosine-like waveform of the energy modulation when the peak of the energy modulation is at the maximum of the envelope, or a sine-like waveform when a zero crossing of the energy modulation is at the maximum of the envelope (see Fig.~\ref{Fig.IV.D.2}).
    \begin{figure}[!htb]
    \bc
    \includegraphics[draft=false, width=0.4\textwidth]{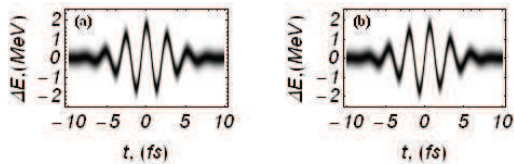}
    \caption{The phase space of the beam showing energy modulation of electrons produced in the interaction with a few-cycle, 800-nm-wavelength laser pulse with CEP stabilization interacting with the electron bunch in the wiggler magnet with two periods. (a) A cosine-like form, and (b) a sine-like form.
    \label{Fig.IV.D.2}}
    \ec
    \end{figure}
When the cosine-like form is used, the electron peak current obtained after the bunching at the laser wavelength has only three spikes, with the central spike producing the dominant x-ray radiation in the FEL with sub-femtosecond duration, as shown in Fig.~\ref{Fig.IV.D.3} (see, also \cite{Zhol4}). The dominance of the central peak over the side peaks can be further improved using two modulating lasers with slightly different frequencies \cite{attosecond1, Ding2}.
\begin{figure}[!htb]
    \bc
    \includegraphics[draft=false, width=0.4\textwidth]{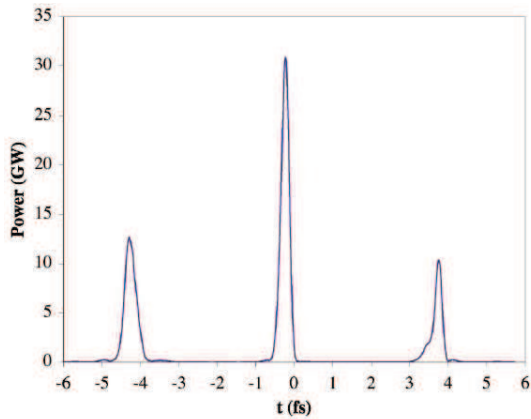}
    \caption{An example of the x-ray power profile produced in the FEL when using a few-cycle modulating laser. Only a small fragment of the entire x-ray pulse cut at $\pm$ 6 fs is shown. The typical FWHM pulse duration of the central spike is 250 attoseconds.
    \label{Fig.IV.D.3}}
    \ec
    \end{figure}
However,  the contrast of the central x-ray spike over the background radiation also depends on the total x-ray energy extracted from all of the unmodulated electrons, which increases with the electron bunch length and becomes comparable to the total x-ray energy in the central spike when the electron bunch length approaches 100 fs~\cite{attosecond1}.

Another technique, which combines undulator tapering with an electron beam chirp for enhancing the FEL output, can be used for obtaining sub-femtosecond x-ray pulses \cite{Sald3}. Recently demonstrated at the SPARC facility at optical wavelengths~\cite{GiannessiChirpedSeeded}, this scheme exploits the correlation between the electron energy and longitudinal position in the center of the sine-like energy modulation profile, as shown in Fig.~\ref{Fig.IV.D.2}b. Under normal conditions the energy chirp $d\gamma/ds$ in the electron bunch causes FEL gain degradation, but this reduction can be compensated and even reversed by means of undulator tapering. Tapering introduces a variation along the undulator of the parameter $K$, which changes the resonant beam energy $\gamma$ for the fixed FEL wavelength $\lambda _r$. Recall that the resonant FEL energy is given by
$\gamma ^2=(\lambda _u/2\lambda _r) \left(1+K^2/2\right)$. In the modulated beam, the energy $\gamma$ and longitudinal position $s$ are correlated. Therefore, in the region with just the right sign and slope of the chirp, electrons stay resonant with the changing value of $K$ as the emitted light slips forward, and the FEL interaction is sustained. For large $d\gamma/ds$ this requirement can be formulated with an approximate condition:
$(d\gamma /ds) (\bar{\beta }_{z}-1)\approx (d\gamma /dK) dK/dz$, where $\bar{\beta }_{z}$ is the scaled electron longitudinal velocity averaged over the undulator period.  Equivalently, one obtains \cite{Sald3, Fawley2008}:
	\begin{align} \label{eq:IV-D.1}
	\frac{d\ln K}{dz}
	=
	\frac{\lambda_r}{\lambda_u}
	\frac{1+K^2/2}{K^2/2}
	\frac{d\ln \gamma }{ds}
	.
	\end{align}
With the above-defined undulator taper, only a short slice of the electron bunch around the zero-crossing of the energy modulation in Fig.~\ref{Fig.IV.D.2}b will produce a powerful FEL pulse. The gain in the rest of the electron bunch (including both the unmodulated portions and the modulated regions with the wrong chirp) will be strongly reduced or even suppressed, as these electrons fall out of resonance. When optimized, as shown in Fig.~\ref{Fig.IV.D.4}, the calculated x-ray radiation has only one spike about 200 attoseconds (FWHM) in duration. It is worth mentioning that this scheme does not require a chicane to convert the energy modulation into a density modulation, therefore it also avoids the potentially strong longitudinal space charge forces that can exist in the ESASE scheme~\cite{SCGeloni}.
 \begin{figure}[!htb]
    \bc
    \includegraphics[draft=false, width=0.4\textwidth]{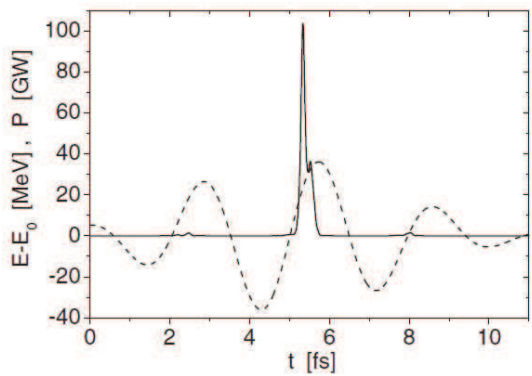}
    \caption{Energy modulation of the electron beam at the exit of the modulator undulator (dashed line), and a profile of the radiation pulse at the exit of the FEL (solid line) as shown in Ref.~\cite{Sald3}.
    \label{Fig.IV.D.4}}
    \ec
    \end{figure}

In contrast to modulating the beam energy, the proposal in~\cite{zholents2008} actually uses the modulation of the electron transverse momentum (angles) by employing a few-cycle CEP stabilized TEM$_{10}$ mode laser pulse (see Section~\ref{Section:I.A}). When interacting with the electrons in a one period undulator, this laser pulse deflects electrons (see Eq.~\eqref{eq:II-C.17}) located only in the ultra-short WS and directs them through the FEL on axis, while by design, all other electrons propagate with a large orbit distortion. As a result, the dominant FEL radiation comes only from the deflected on-axis electrons, while most of the other electrons produce only spontaneous emission. This technique can produce x-ray pulses approximately 100 attoseconds (FWHM) in length, with an excellent contrast over the background.

It should be pointed out that the preceding laser-based attosecond x-ray pulse generation schemes are all designed for SASE FELs. Therefore a long undulator is typically required to allow significant amplification. Because of the lengthened slippage associated with the long undulator, however, this limits the shortest pulse duration to the cooperation length, i.e., $\sim$100 as for hard x-rays with $\AA$ wavelengths, or $\sim$1 fs for soft x-rays with nm wavelengths. Alternatively, an intense attosecond x-ray pulse can be produced from a pre-bunched beam in a short undulator which may allow the x-ray pulse length to be pushed below the cooperation length. In~\cite{Zhol6} a short slice of a pre-bunched beam is proposed to produce a soft x-ray pulse down to 80 attoseconds (FWHM). There, a few-cycle laser pulse is used to create a WS only a fraction of the laser wavelength in duration that facilitates microbunching at x-rays. Then, electrons in the WS produce coherent undulator radiation that is frequency shifted and dominates over the spontaneous emission from the rest of the bunch. Another proposal~\cite{EEHG-as} aims to shorten a coherent undulator radiation pulse at 1 nm wavelength to 20 attoseconds (FWHM) by employing a combination of EEHG and a few-cycle laser pulse to generate microbunching at x-ray wavelengths only in an extremely short WS. In~\cite{CHGQiang1}, the authors studied a scheme that combined HGHG from a few-cycle laser pulse with bunch compression to produce attosecond x-ray pulses through modulation compression~\cite{MC1,MC2,MC3,MC4}. This scheme is also capable of generating ultrashort x-ray pulses with duration well below 100 as.

All of these techniques are designed to produce a single ultrashort x-ray pulse. The generation of a comb-like sequence of attosecond x-ray pulses with good temporal coherence is also possible, and is discussed in the next section.

%
\subsection{Generation of mode-locked x-ray pulse trains}\label{Section:IV.E}
%

Here, we discuss another important idea in adapting modern optical techniques to x-rays; namely, the generation of mode-locked x-ray pulses that may enable a new degree of control for x-ray laser applications.

The concept of mode locking in laser oscillators, where a fixed-phase relationship is established between all of the lasing longitudinal modes to generate short optical pulses, has been known for a long time (see, e.g.,~\cite{siegman,diels}). More recently, it was found to be particularly important in helping to establish a fixed phase relationship across a broad spectrum of frequencies for optical frequency metrology using frequency combs (see, e.g.,~\cite{Stenger_et_al:2002}) and for the generation of ultra-short EUV pulses (see, e.g.,~\cite{KraussIvanov:2009}). The mode locking concept is illustrated in Fig.~\ref{Fig.IV.E.1}. In mode-locked laser oscillators, the envelop bandwidth is determined by the gain medium, while the mode spacing is fixed by the cavity round trip time, which sets the temporal periodicity in the pulse train.
    \begin{figure}[!htb]
    \bc
    \includegraphics[draft=false, width=0.4\textwidth]{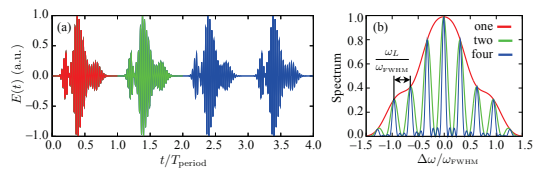}
    \caption{An example of mode-locking, where the time domain is shown in panel (a) and the frequency domain in panel (b).  The single pulse and its spectrum are shown in red.  Adding another pulse shown in green in panel (a) leads to the spectrum in panel (b) with additional substructure. The spectrum for the four pulses is shown in blue. (Figure courtesy of R. Lindberg).
    \label{Fig.IV.E.1}}
    \ec
    \end{figure}

Mode locking of x-ray pulses in FELs was first introduced in Ref.~\cite{PhysRevLett.100.203901}, wherein the basic physics of mode locking was shown to be achievable through the generation and coupling of a pulse train of FEL radiation.  X-ray pulse trains can be produced by modulating the FEL gain. For example, one can impose either a periodic variation in the electron beam energy as described in~\cite{PhysRevLett.100.203901}, or a periodic variation in the electron beam density (peak current) as in~\cite{Kur}. Both results can be obtained by modulating the energy of the electron beam in the undulator using a high-power optical laser. Establishing the necessary fixed phase relationship between radiation modes is accomplished by a series of small magnetic chicanes uniformly spaced along the FEL undulator (see Fig.~\ref{Fig.IV.E.2}). They are used to add precise delays between the electron beam and the radiation field, thereby extending the cooperation length of the FEL radiation. The mode spacing can be varied by changing the modulating laser wavelength, which requires that the strength of the chicane be changed accordingly in order that the total slippage length in each undulator-chicane module is equal to the laser wavelength. In the case described in ~\cite{Kur}, a magnetic chicane is added between the modulator and the first FEL undulator module, and is used to convert the periodic variation of the electron beam energy to a periodic variation of the electron beam current.
    \begin{figure}[!htb]
    \bc
    \includegraphics[draft=false, width=0.4\textwidth]{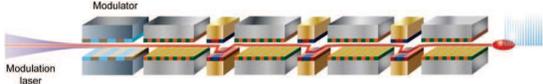}
    \caption{A schematic of an FEL design with mode-locking capabilities as shown in
    Ref.~\cite{PhysRevLett.100.203901}. Chicane magnets are placed between each undulator module to generate the desired slippage between the radiation and the electron beam and to create the radiation modes. The schematic includes a short beam-energy modulation undulator at the beginning of the FEL.
    \label{Fig.IV.E.2}}
    \ec
    \end{figure}

An simulated x-ray signal from a mode-locked SASE FEL is shown in Fig.~\ref{Fig.IV.E.3}, and the output spectrum is shown in the inset. One can see that the output signal of a mode-locked FEL in the time domain consists of many ultrashort pulses equally separated by the laser wavelength. In the frequency domain, the spectrum has many sharp lines over a wide bandwidth, a feature that is not easily attainable with other FELs and could be beneficial for examining the structure and dynamics of a large number of atomic states simultaneously. In an FEL seeded by an HHG source that consists of attosecond pulse trains, this feature may also allow one to maintain the temporal structure of the seed \cite{HHGML}.
    \begin{figure}[!htb]
    \bc
    \includegraphics[draft=false, width=0.4\textwidth]{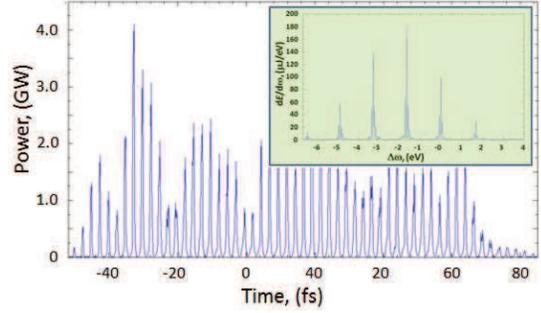}
    \caption{The output power of the mode-locked SASE FEL averaged over five simulation results to eliminate a relatively large shot-to-shot fluctuations. Each individual pulse in the pulse train has a sub-femtosecond duration.
    The insert shows the output spectrum averaged over five simulation results. The central photon energy is 200 eV and mode separation is 1.66 eV defined by the modulating laser \cite{Kur}.
     \label{Fig.IV.E.3}}
    \ec
    \end{figure}

In this example, the x-ray signal grows from shot noise, as in the SASE FEL. This results in power fluctuations across the output signal, as well as irregular variations within each spectral spike. Introducing a temporally coherent seed to dominate over the shot noise can lead to better stability and a cleaner spectrum. This has been studied in \cite{EEHGML, ML4} where the EEHG technique is used for generating microbunching in the electron beam, and a radiator in mode-locking configuration is used to produce a train of short pulses with excellent temporal coherence.

A different approach for creating a mode-locked x-ray pulse train is described in  ~\cite{ML3}. This proposal makes full use of the self-seeding technique~\cite{Feldhaus1997341} to impose temporal coherence over the entire electron bunch, thus eliminating the need for multiple chicanes. A possible implementation is shown in Fig.~\ref{Fig.IV.E.5}.
    \begin{figure}[!htb]
    \bc
    \includegraphics[draft=false, width=0.4\textwidth]{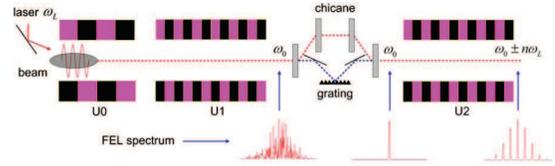}
    \caption{Schematic for generation of a mode-locked soft x ray pulse train in a self-seeded FEL \cite{ML3}.
    \label{Fig.IV.E.5}}
    \ec
    \end{figure}
Here, a short undulator (U0) is used to create a sinusoidal modulation of the electron beam energy and the chicane between undulators U1 and U2 converts it to a comb of electron microbunches. A grating-based monochromator is used at the exit of U1 to select narrowband radiation (with central angular frequency at $\omega_0$) from the broadband SASE radiation after U1. The chicane introduces additional path length for the electron beam to compensate for the delay of the x-rays in the monochromator (on the order of one picosecond). As a result, the monochromatic signal overlaps with the comb of electron microbunches and seeds FEL amplification in U2. Subsequent amplification in U2 produces a comb of mode-locked x-ray pulses. 

The physics of this scheme is analogous to amplitude modulation in telecommunications. Here, by modulating the lasing medium (the relativistic electron beam), the amplitude of the coherent seed with carrier frequency at $\omega_0$ is also modulated at the laser frequency $\omega_L$. This leads to the development of sidebands at the frequencies $\omega_0\pm\omega_L$. As the amplification continues, the amplitudes of the modes at these sideband frequencies will also be modulated by the electron beam, which further leads to development of new modes at $\omega_0\pm 2\omega_L$. This process repeats, with the single-frequency coherent seed being finally translated to a mode-locked output at the exit of U2 with sharp lines at the frequencies $\omega_0 \pm n \omega_L$.

Another modification of the mode-locked technique from Ref.~\cite{PhysRevLett.100.203901} was proposed in~\cite{ThompsonMcNeilModeZepta:2013} and aims at the production of mode-locked trains of few-cycle x-ray pulses. The technique involves preparing an electron beam with a large energy modulation in U0, which creates lower energy regions of the modulated beam that dominate the FEL interaction in the amplifier. This leads to periodic regions with microbunching at x-ray wavelengths at the end of U1. Once the comb structure of the microbunched x-ray regions is sufficiently well developed, and before saturation of the FEL process in U1, the electron beam proceeds into U2 without additional delay. (Also, no monochromator is used). In the case of \cite{ThompsonMcNeilModeZepta:2013}, U2 is comprised of a series of few-period undulators separated by chicanes that delay the electron bunch, similar to that used in~\cite{PhysRevLett.100.203901}. These undulator-chicane modules assist in maintaining overlap between regions where microbunching has developed and the train of few-cycle x-ray pulses produced as a result of the FEL amplification process. Simulations show that the train of hard x-ray pulses emitted from U2 reaches peak power levels approaching gigawatts, and that the rms pulse duration of each individual pulse is reduced to approximately 700 zeptoseconds ($10^{-21}$s). This is more than two orders of magnitude shorter than what is possible in a hard x-ray SASE FEL. Further, the envelope bandwidth of this mode-locked train of x-ray pulses is approximately two orders of magnitude wider than the one obtained from the standard SASE FEL.

%
\subsection{Generation of multicolor x-ray pulses}\label{Section:IV.F}
%

Similar to the spectroscopic techniques based on sample illumination from sequences of tailored optical laser pulses that provide multidimensional views of molecular and electronic processes (see, for example \cite{Mukamel}), sequential (or simultaneous) x-ray pulses with varying (or the same) wavelengths are foreseen to provide new tools for performing time-resolved x-ray spectroscopy (see, for example \cite{PattersonLCLS}) such as time-resolved x-ray four-wave mixing \cite{Bencivenga} and stimulated Raman scattering \cite{Schweigert, ZholTC}.

Two of the basic scenarios relevant to two-color x-ray applications are: The generation of a pair of sequential x-ray pulses with independent control of timing and spectrum, and the generation of one x-ray pulse with two discrete wavelengths. For instance, at SLAC's LCLS, two sequential x-ray pulses with slightly different wavelengths are produced by setting the first 16 undulators and the last 16 undulators to different strengths \cite{Lutman}. In this scheme, while the relative timing of the x-ray pulses can be varied with a chicane between the two undulator sections, the two pulses have comparable pulse lengths because they are generated by the same particles. In another method, the undulator strengths can also be set in an alternating fashion (e.g., with all the odd-numbered undulators set at $K_1$ and all the even-numbered undulators at $K_2$) to produce an x-ray pulse with two discrete spectral lines \cite{PhysRevLett.111.134801}. On the other hand, in a seeded or self-seeded arrangement, one can also introduce two seed signals at closely spaced wavelengths \cite{FreundTC, GeloniTC} and amplify them simultaneously to generate two colors. However, in this scheme the tunability of the photon energy difference is limited to the FEL gain bandwidth. Other schemes that exploit multipeaked electron beam energy spectra (e.g., two simultaneous beams with different energy) can also produce multifrequency and multi-pulse FEL output. Such a technique has been demonstrated in the IR~\cite{PetrilloTwoBeam}, and is currently under investigation in the x-ray regime at SLAC.

Yet there are many applications that require even more complicated arrangements of multiple x-ray pulses with carefully shaped time, frequency and polarization characteristics that may greatly benefit from laser-based beam manipulation techniques. For instance, in a proposed high-repetition rate seeded FEL, the baseline design of the two-color x-ray beam line~\cite{NGLSTN} uses two synchronized few-cycle mid-IR lasers to generate a strong energy chirp in a localized region of the beam. A tapered undulator is then used to compensate for the energy chirp to generate two ultrashort x-ray pulses with variable delay in a wide range from zero to $\sim$100 fs, variable angle between x-ray wave vectors, and independent photon energy tunability from 200 eV to 1 keV suitable for stimulated Raman scattering and three wave mixing applications.

Another example is the FERMI FEL where a chirped, high intensity seed laser is used to generate two narrow-band pulses separated both in time and frequency \cite{FERMI2color}. This idea is based on the fact that the bunching factor in FELs seeded by external lasers depends on the energy modulation amplitude. For a laser pulse with a finite length, this amplitude depends on time, and one can therefore use dispersion to manipulate the distribution of the time-dependent bunching (see, e.g. the measurement in~\cite{EMmeasurement}) to generate radiation pulses with the desired time and frequency shapes. To see how, we refer to~Eq. \eqref{eq:II-A.4} and consider bunching at the 8th harmonic of the modulating laser as an example. We rewrite this equation to indicate the time dependence of the energy modulation in the bunch due to the time-varying laser pulse amplitude $A(t)$,
    \begin{align}\label{eq:II-A.4-1}
    b_n(t)
    =
    e^{-\frac{1}{2} B^2 n^2}
   J_n(-A(t) B n)
   .
   \end{align}
Given a fixed normalized momentum compaction (here $B=0.2$), the time-dependent bunching factor $b_8(t)$ for various peak modulation amplitudes $A_\max = \max [A(t)]$ are shown in Fig.~\ref{fig2color}. With the peak energy modulation matching the optimal value ($A_\max=6$, red line in Fig.~\ref{fig2color})), the bunching distribution has a similar shape as the laser profile and one generates a single pulse with a single frequency. As the peak energy modulation amplitude is increased to beyond the optimal value, the bunching factor starts to oscillate, as dictated by the Bessel function dependence in Eq. \eqref{eq:II-A.4}. With $A_\max=7.6$ (blue line in Fig.~\ref{fig2color})) the bunching has a dip in the center of the laser where the beam is overbunched by the large energy modulation, while the energy modulation at the head and tail of the laser are optimal (with the local modulation amplitude equal to 6) to generate two peaks in the bunching distribution. Sending this beam through an undulator would yield two sequential pulses with a time separation equal to that of the two peaks (assuming the slippage is small compared to the separation). If the seed laser has a negligible chirp, the spectrum of the radiation shows the interference pattern of the two pulses \cite{HGHGSINAP}. On the contrary, if the laser has a strong frequency chirp, the output is two radiation pulses with different frequencies. Further increasing the energy modulation to $A_\max=9$ (magenta line in Fig.~\ref{fig2color})) increases the separation between the two peaks, and more bunching peaks appear in the central part of the beam. However, since the bunching in the central part is smaller and the beam energy spread is larger, simulations indicate that the reduced FEL growth rate in this region can still result in a simple two pulse structure, assuming a constant beam current.
    \begin{figure}[h]
    \includegraphics[width = 0.48\textwidth]{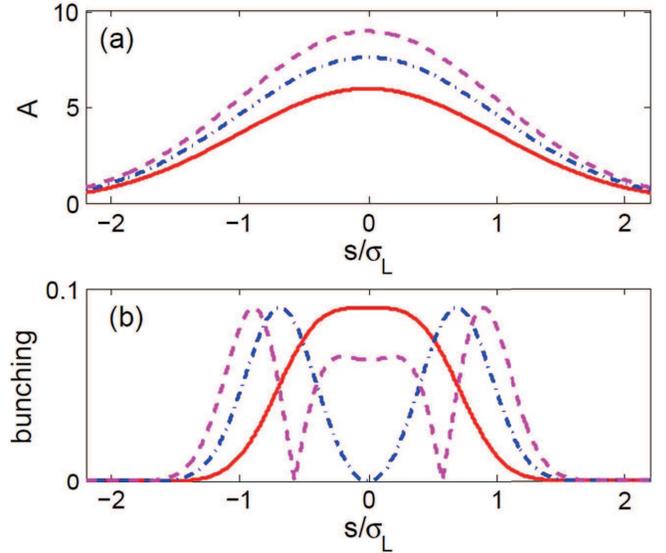}
        \caption{Time-dependent energy modulation $A(t)$ (a) and bunching distributions $b_8(t)$ (b). The horizontal axis is the longitudinal position normalized to the rms duration of the laser pulse $\sigma_L$. Red (solid), blue (dashed-dot), and magenta dashed) are for $A_\max=6$, $A_\max=7.6$, and $A_\max=9$, respectively.
    \label{fig2color}}
    \end{figure}

Experimental results of two-color EUV pulses produced by this method at the FERMI FEL are shown in Fig.~\ref{fig2colorfermi}. By controlling the laser energy modulation amplitude, chicane momentum compaction, and the laser chirp parameter, one can switch from one-pulse to two-pulse operation mode and also adjust the temporal separation and frequency difference. However, it should be pointed out that the tuning range of the frequency difference is still limited to the FEL gain bandwidth $\simeq\rho$. For a large frequency chirp, this can be overcome by imposing an associated energy chirp in the beam longitudinal phase space to match the resonant condition of the FEL wavelength \cite{CPACHG2}.
    \begin{figure}[h]
    \includegraphics[width = 0.48\textwidth]{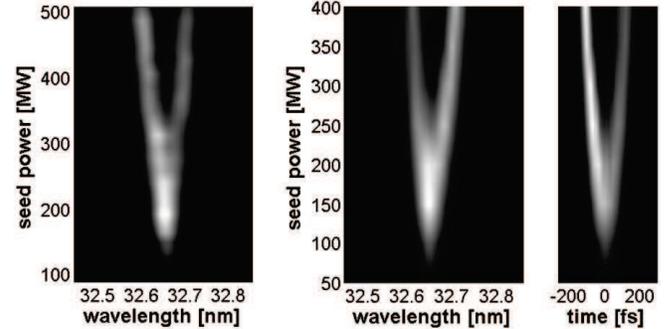}
    \caption{Spectral and temporal FEL intensities for different seed laser powers. Left: experimental spectral splitting; right: simulated spectral and temporal splitting. From \cite{FERMI2color}.
    \label{fig2colorfermi}}
    \end{figure}
    
%
\subsection{Laser beam as an undulator}\label{Section:IV.G}
%
    
In the push for compact x-ray sources, an attractive alternative to permanent magnet undulators (which have periods on the order of centimeters) are electromagnetic undulators, which use much shorter periods and thereby allow the electron beam energy to be significantly reduced in order to radiate at short wavelengths. Powerful lasers have been proposed for this purpose, and simple analysis shows that in the head-on collision of a laser pulse and a relativistic electron beam, the electrons radiate electromagnetic fields on-axis at a wavelength $\lambda_r$ (cf. with~\eqref{und_wavelength}),
    \begin{align}\label{und_wavelength1}
    \lambda_{r} 
    = 
    \lambda_{L}\frac {1+K^2/2} {4\gamma^{2}}
    ,
    \end{align}
where the undulator parameter $K$ is now computed using~\eqref{und_parameter} with $B_0$ the amplitude of the laser magnetic field. By comparison with Eq. \eqref{und_wavelength}, the laser acts as an undulator with period $\lambda_u=\lambda_L/2$. Of course, this ``undulator radiation'' is described equivalently by the inverse Compton scattering process, which is widely used as a source of incoherent x-rays and gamma radiation for various applications (see {\it e.g.}~\cite{compton_scattering} and references therein).  

Laser undulators for FELs have been analyzed in several papers~\cite{Gallardo,Gordon2001190,Bonifacio2007745,Sprangle}. In certain cases, a quantum description of the FEL is required (QFEL) because the photon momenta $\hbar k$ of the very short wavelengths can be comparable to the classical spread in the electron momentum $\gamma m c \rho$ (see, e.g., Ref. \cite{Bonifacio20111004} and references therein). For a soft x-ray of $\lambda_r\sim 1$ nm wavelength and a laser undulator from  a CO$_2$ laser with the wavelength of $\lambda_L = 10.6\,\,\mu$m, the required beam energy which follows from~\eqref{und_wavelength1} is only 26 MeV. While by itself such a small energy is advantageous, in practice it is difficult to generate a low-energy electron beam with the small geometric emittance, small energy spread, and high current needed for FEL gain at short wavelengths (a more detailed analysis of scalings involved with such FELs can be found in Ref.~\cite{Zolotorev2002445}).  This is the main reason why high gain FELs with laser undulators have not yet been demonstrated. 
 
Recently, in a renewed effort to overcome the challenges associated with laser undulators, the concept described in ~\cite{PhysRevLett.110.064802} proposed to use two ``tilted optical wave'' pulses from a CO$_2$ laser propagating at right angles to a relativistic electron bunch. Such an arrangement allows for an extended interaction length between the electron beam and the laser field, and solves some of the issues of the ``head-on'' laser-beam scheme. A different, although similar, approach was proposed in~\cite{Bisognano}.
    \begin{figure}[h]
    \includegraphics[width = 0.48\textwidth]{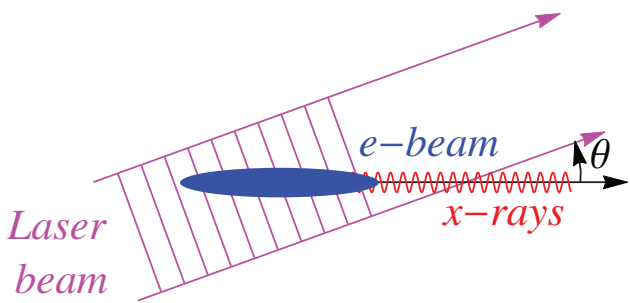}
    \caption{Geometry of the laser interaction with the relativistic beam.
    \label{laser_und}}
    \end{figure}
There, the authors suggested that the laser be sent at a small angle $\theta$ to the direction of the beam propagation, such that the on-axis radiation wavelength becomes~$\lambda_{r}   \approx  \lambda_{L}(1+\theta^2\gamma^{2})^{-1}$ (assuming $\theta \ll 1$ and $K\ll 1$). For a given $\lambda_r$ and $\lambda_L$ this allows the beam energy to be increased such that the beam quality is sufficient to drive FEL gain. To extend the length of the interaction between the laser and the electron beam the authors proposed to use a sheared laser pulse in which the wavefronts are tilted relative to the direction of propagation. As a representative set of parameters, they consider an electron beam of modest energy 170 MeV, normalized emittance of $6 \times 10^{-8}$ m at 20 pC charge, to generate $2.5$ nm radiation in a laser based undulator FEL with $\lambda_L = 750$ nm. While these are still challenging parameters, they may not be out of reach for the next generation of electron sources and laser systems.


%
\section{Beam conditioning and diagnostics with lasers}\label{sec:6}
%
In addition to tailoring the properties of radiation, lasers can be used to prepare optimal beams for driving accelerator-based light sources and for retrieving information on the electron beams themselves.
%
\subsection{Laser heater}
%

In modern FELs the electron beam generated in the electron gun has a small energy spread, typically 1$\sim$2 kiloelectron-volts. It was shown by several groups~\cite{borland02etal,saldin02-1sy,heifets02sk,huang02k} through theoretical analysis and computer simulations that transport of such a beam through a long linac (several hundred of meters in length) equipped with bunch compressors can lead to a so-called microbunching instability. Density variations in the beam drive energy modulations that both increase the energy spread and are converted into larger density modulations after dispersion. This process deteriorates the beam quality and reduces its efficacy as a lasing medium.

An effective method to suppress the instability was proposed in Ref.~\cite{Saldin-laser-heater} and was later given the name of a ``laser heater''. The idea is to ``heat-up'' the beam by increasing its uncorrelated energy spread using a laser-beam interaction in an undulator. The required slice energy spread of the beam is set to the level which, on the one hand, suppresses development of the microbunching instability through the mechanism of Landau damping, but, on the other hand, is small enough to not impede lasing in the FEL.  The laser heater works by introducing a correlated microstructure in the phase space of the beam on the scale of the laser wavelength that is effectively washed out through transport, resulting an increase in the uncorrelated energy spread. The laser heated beam is now considered as a necessary element in practically all designs of modern x-ray FELs~\cite{laser_heater_PAL-FEL,laser_heater_FERMI,laser_heater_XFEL}.

The laser-beam interaction was considered in Section~\ref{Section:II.C}. The maximum energy change $\Delta \gamma(r)$ of a particle in the beam as a function of radial position $r$ is given by
    \begin{align}\label{eq:laser_heater}
    \Delta \gamma(r)
    =
    \sqrt{
    \frac{P_{L}}{P_{0}}
    }
    \frac{2KL_{u}}{\gamma_{0}w_0}
    {\cal J}
    e^{-r^{2}/w_0^{2}}
    ,
    \end{align}
where $P_{L}$ is the laser power, $P_{0} = I_{A}mc^{2}/e \approx 8.7$ GW (cf. Eq.~\eqref{eq:II-C.13}; the factor $e^{-r^{2}/w_0^{2}}$ describes the radial profile of the laser beam, which now is on the order of the electron beam size). To find the energy distribution of the electron beam this formula should be combined with the beam distribution (see details in~\cite{huang_laser_heater}). The resulting energy distribution function depends on the input laser energy and the ratio of the rms transverse size of the beam $\sigma_{x}$ ($\sigma_{x} \approx \sigma_{y}$) and laser $\sigma_{r}=w_0/2$. In the limit $\sigma_{r} \gg \sigma_{x}$ one finds a double-horn distribution in energy (much like the dashed blue line projection in Fig. \ref{HGHGS}). In the case when the laser pulse is matched to the beam size, $\sigma_{x} \approx \sigma_{r}$, the distribution function becomes Gaussian-like.

The schematic of the laser heater for LCLS is shown in Fig.~\ref{fig:laser_heater}.
	\begin{figure}[htb]
	\centering
	\includegraphics[width=0.45\textwidth, trim=0mm 0mm 0mm 0mm, clip]{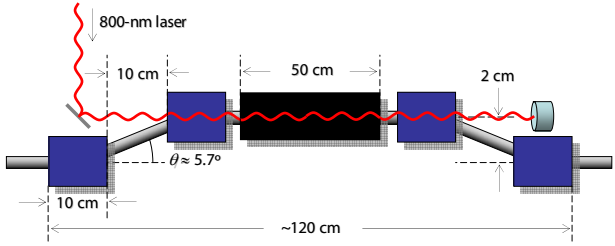}
	\caption{Layout of the LCLS laser heater inside a magnetic chicane (from~\cite{huang_laser_heater}).}
	\label{fig:laser_heater}
	\end{figure}
The laser heater consists of a 50 cm long, 5 cm period undulator located at the center of a small horizontal magnetic chicane in order to allow a convenient laser-electron interaction with no crossing angle. In addition to easy optical access, the chicane produces smearing of the laser-induced energy modulation, resulting in a random, uncorrelated energy spread with no temporal structure. This smearing occurs because the path length from chicane center (where the energy modulation is induced) to the end of the chicane depends on the electron�s horizontal angle, $x'$. The induced energy spread at the center of the chicane also causes some horizontal emittance growth which is typically negligible.

LCLS heater results are documented in Ref.~\cite{lcls_heater}. Figure~\ref{fig:laser_heater_exp} shows the longitudinal phase space of the beam measured with the help of an rf deflecting cavity.
	\begin{figure}[htb]
	\centering
	\includegraphics[width=0.3\textwidth, trim=0mm 0mm 0mm 0mm, clip]{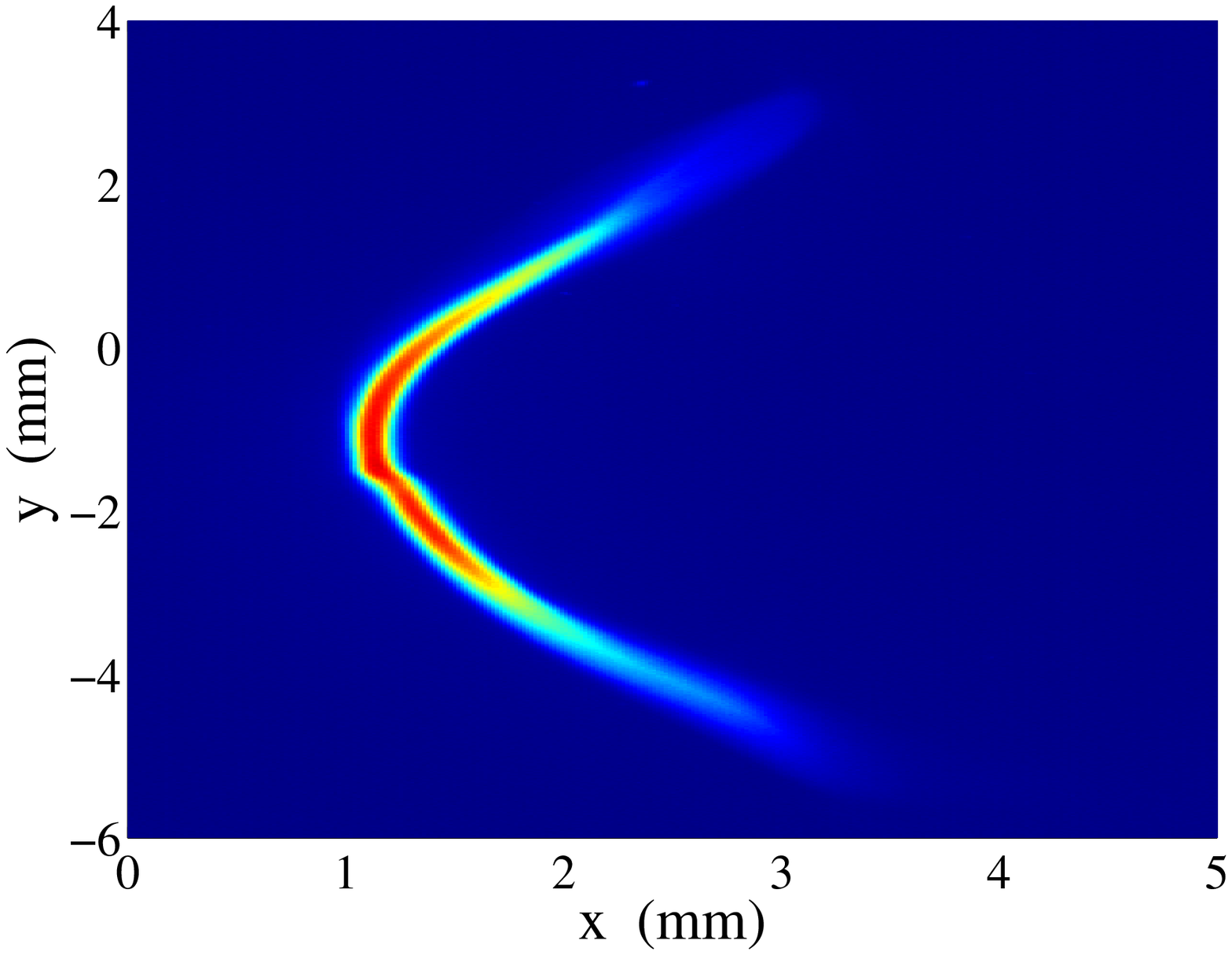}
	\includegraphics[width=0.3\textwidth, trim=0mm 0mm 0mm 0mm, clip]{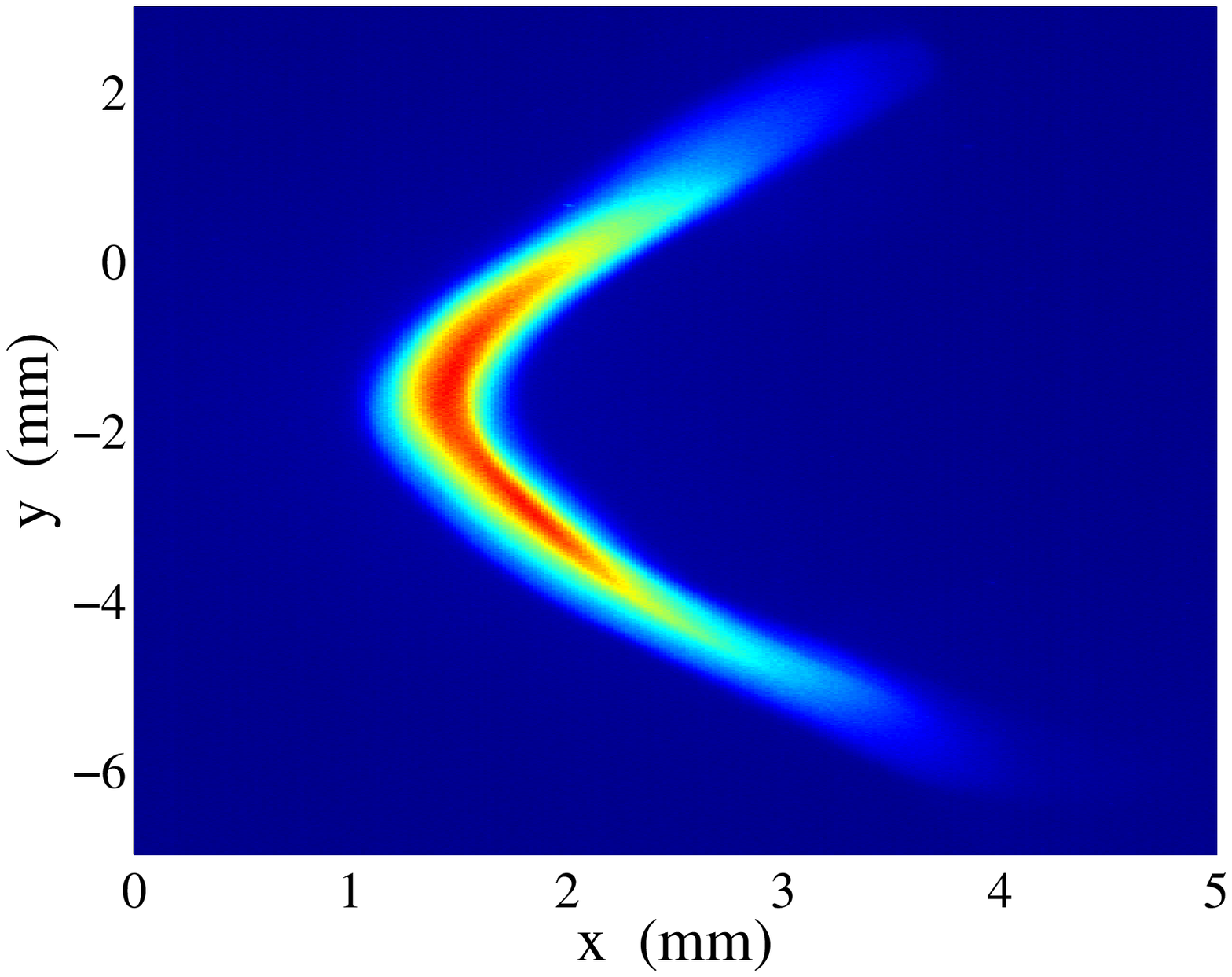}
	\includegraphics[width=0.3\textwidth, trim=0mm 0mm 0mm 0mm, clip]{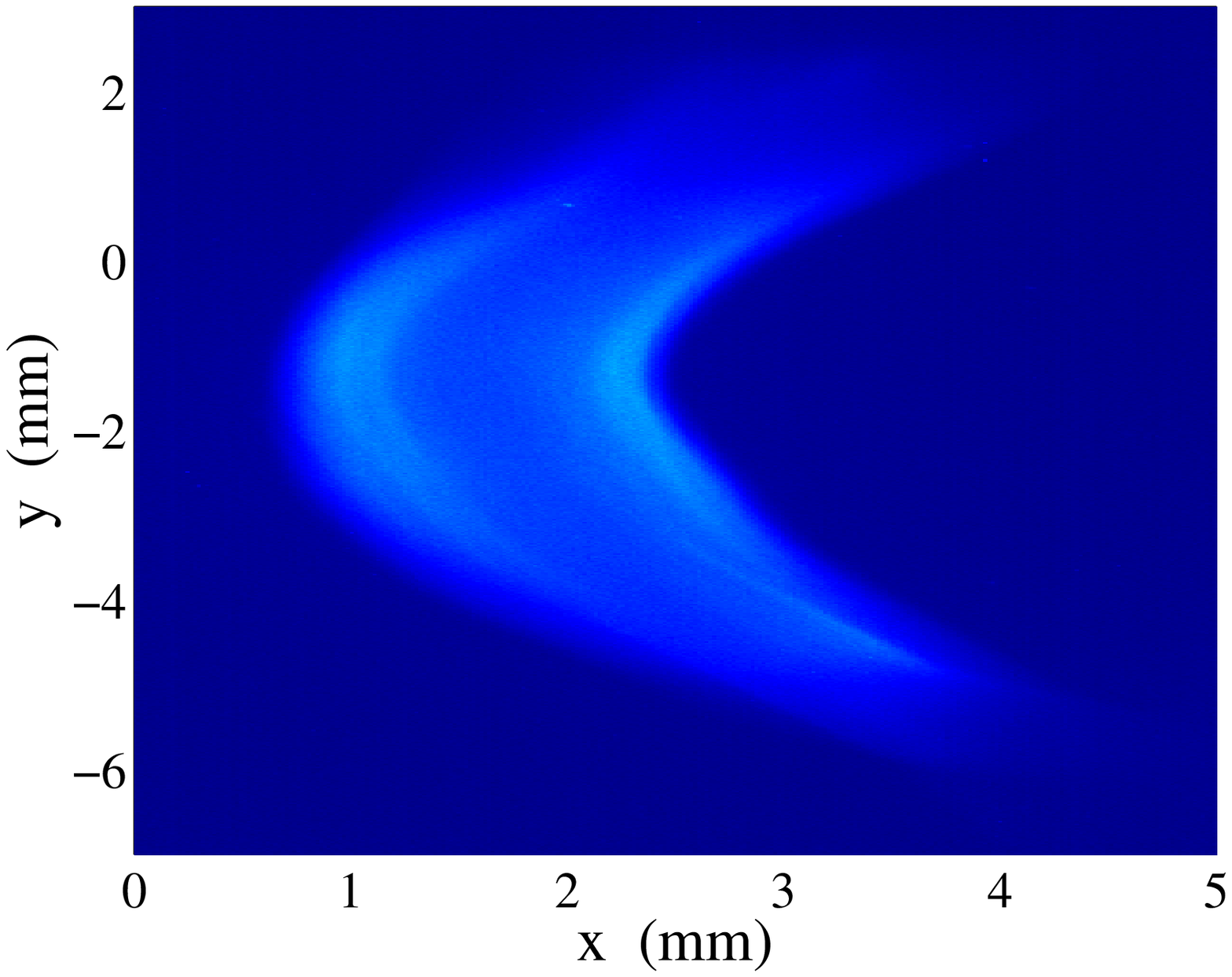}
	\caption{Measured longitudinal phase space at 135 MeV with (a) laser heater off, (b) IR-laser energy at 10 $\mu$J, and (c) at 220 $\mu$J. The vertical axis shows beam longitudinal position and the horizonal axis shows the beam energy. [from Ref.~\cite{lcls_heater}].}
	\label{fig:laser_heater_exp}
	\end{figure}
The figure clearly shows the increase of the beam energy spread with the energy of the laser pulse.

An interesting and surprising effect which the authors called ``trickle heating'' was discovered in~\cite{lcls_heater}. For small laser pulse energy $\sim 1\,\,\mu$J, the authors observed an increased heating effect compared with the theoretical value based on Eq.~\eqref{eq:laser_heater}. It turns out that, for this laser energy, the energy modulation periodic structures are not completely removed by the smearing effect in the chicane magnets, but are transformed into $x-z$ correlations in the phase space of the beam. These correlations excite space charge wakefields that apparently lead to additional heating.

%
       \subsection{Emittance exchange}\label{V.H}
%

The performance of an FEL depends critically on the Pierce parameter $\rho$, which is determined by the transverse electron beam size and the peak current. While the 6-D phase space volume of a beam is conserved along phase space trajectories, the phase space area of individual planes may be rearranged in different ways to maximize $\rho$ without violating Liouville's theorem (see, e.g.~\cite{EP}). One of the effective ways to rearrange beam's distribution in 6-D phase space is emittance exchange (EEX) which exchanges the projected emittance between different planes. Of particular interest is the transverse-to-longitudinal EEX that exchanges the emittance between one of the transverse planes ($x$ or $y$) with the assumed  much smaller emittance of the longitudinal plane ($s$).

Hereafter we will only consider beam dynamics in $x$ and $s$ planes and ignore the uncoupled motions in $y$ plane. An electron's initial state in phase space is denoted as $\vec{X_0}=(x_0,x_0',s_0,\eta_0)^T$, where $x_0$ is the horizontal position, $x_0'$ is the horizontal angle, $s_0$ and $\eta_0$ are the longitudinal position and the relative energy deviation with respect to the reference particle, respectively\footnote{We remind the reader that we use variable $s$ as a coordinate directed along $z$ and associated with particle's position inside the beam.}. After passing through a linear Hamiltonian system, the electron's state $\vec{X_1}$ is related to its initial state $\vec{X_0}$ by $\vec{X_1}=R\vec{X_0}$, where $R$ is a symplectic 4$\times$4 transport matrix that describes the beam dynamics associated with the system. The 4$\times$4 transfer matrix may be written as four 2$\times$2 blocks,
\begin{equation} \label{eexs}
R=
\left[
\begin{array}{cc}
 A & B \\
C & D\\
\end{array}
\right].
\end{equation}
If all elements in $A$ and $D$ are zero, then an electron's final transverse coordinates will only depend on its initial longitudinal coordinates, and vice versa. As a result, beam's transverse and longitudinal degrees of freedom will be exchanged after the beam passes through the beam line.

Construction of a beam line with such an exotic transfer matrix certainly requires elements that couple the beam dynamics in $x$ and $s$ planes. Two representative elements are the dogleg magnet arrangement and the rf deflecting cavity. A dogleg magnet is just half of a chicane that correlates a particle's final transverse position with its initial energy, and final longitudinal position with its initial transverse angle. A deflecting cavity is an rf structure operating in the transverse mode (e.g. TEM$_{10}$) which couples a particle's final transverse deflection with its initial longitudinal position, and final energy with its initial transverse position. Combined, these elements provide the desired coordinate mapping. Currently most of the EEX schemes rely on proper arrangement of dogleg magnets and deflecting cavities, with the simplest scheme being two identical dogleg magnets with a deflecting cavity in between \cite{EEX1, EEX2, EEX3}.

Transverse-to-longitudinal EEX was originally proposed to reduce beam transverse emittance for enhancing FEL gain \cite{EEX1}. As shown in \cite{EEX2} one may use an ultra-short laser pulse in a photocathode gun to generate a beam with small longitudinal emittance, and then exchange the transverse and longitudinal emittance to obtain a beam with small transverse emittance. The increased longitudinal emittance does not affect FEL gain as long as it is kept below the threshold set by $\rho$. Furthermore, the increased longitudinal emittance after EEX is helpful for suppressing the microbunching instability, which can also degrade the FEL performance. So there are combined advantages in swapping beam phase space areas between different planes while keeping the 6-D phase space volume unchanged.

In addition to reducing beam transverse emittance, the fact that beam's transverse and longitudinal degrees of freedom are exchanged in EEX implies that one can tailor beam's longitudinal distribution by shaping beam's initial transverse phase space. For example, a simple mask has been used to generate transversely modulated beamlets that are then converted into longitudinal sub-ps bunch trains through EEX \cite{EEX4} useful for generating narrow-band THz radiation. With a special mask, one may also generate a beam with linearly ramped current profile \cite{EEX5} that is crucial for obtaining a high transformer ratio in advanced beam driven accelerators. It is also possible to convert these transverse beamlets into separated energy bands which may be used for generation of high harmonics in FELs \cite{EEX6}, similar to the EEHG technique.

Another advanced application enabled by transverse-to-longitudinal EEX is chirp-free bunch compression \cite{EEX7}. Typically an energy chirp is required for bunch compression, which puts certain constraints on the linearity of the beam longitudinal phase space. In the chirp-free bunch compression scheme, an EEX beam line is first used to convert $s$ to $x$ (or $x'$). Then a telescope beam line is used to demagnify the beam in $x$ (or $x'$), after which a second EEX beam line converts $x$ (or $x'$) back to $s$ and the beam is compressed. This application may be particularly useful for compact x-ray FELs based on x-band linac structures where the power source for a third harmonic cavity required to linearize the beam longitudinal phase space is not available. This cascaded EEX scheme may also be used to convert a laser modulation to a higher frequency for FEL seeding, or to convert a laser modulation to a lower frequency for the generation of THz radiation.

Another interesting scenario in transverse-to-longitudinal EEX is to replace the rf deflecting cavity with a TEM$_{10}$ mode laser; the so-called laser assisted EEX \cite{LAEE}. The advantage of using a laser to perform EEX is that instead of exchanging the emittance for the whole bunch, one can exchange the emittance for only part of the electrons. This is because the TEM$_{10}$ laser gives beam a sinusoidal kick (Fig.~\ref{figeex}b) while the EEX requires a linear kick. So only the electrons around the laser field zero-crossing will experience EEX (red particles in Fig.~\ref{figeex}). For these particles, the rms length is only a fraction of the laser wavelength so they are characterized by an extremely small longitudinal emittance. After EEX, these portions of the beam will have an extremely small transverse emittance (Fig.~\ref{figeex}c). Such small transverse emittances enable operation of an x-ray FEL at lower energy, which can greatly reduce the size and cost of the facility.
    \begin{figure}[h]
    \includegraphics[width = 0.49\textwidth]{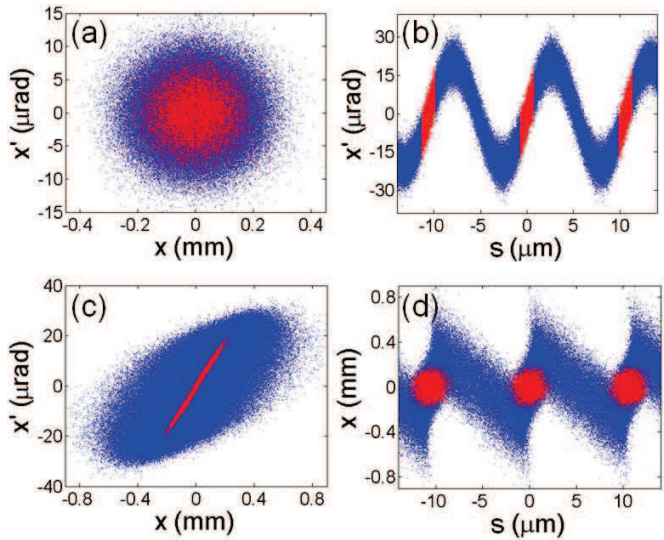}
    \caption{Representative beam phase space evolution in EEX with a laser. (a) Initial transverse phase space; (b) $x'-s$ distribution after interacting with the laser with wavelength at 10.6 $\mu$m; (c) Final transverse phase space; (d) $x-s$ distribution after EEX. The particles that have effective EEX are shown with red dots. From \cite{LAEE}.
    \label{figeex}}
    \end{figure}

It should be pointed out that EEX between $x$ and $y$ planes is also possible, and is conceptually simpler than the transverse-to-longitudinal EEX. For instance, a solenoid with suitable strength that rotates the beam by $n\pi/2$ ($n$ is an odd number) naturally exchanges the beam emittance and coordinates in the $x$ and $y$ planes. A series of skew quadrupoles may be used to exchange the beam emittance in $x$ and $y$ planes as well \cite{beamrotator, Mobius}.

Note, in discussions above, we have assumed that the beam is uncoupled before and after its passage through some elements. As discussed in Sec.II.A, in this case the emittances of the subspaces can only be completely exchanged. By contrast, if the beam is allowed to have coupling in different planes, the emittance of one plane can actually be made smaller at the cost of significantly increasing the emittance in another plane (the product of the final emittances is larger than the initial value). Recently, a scheme that uses a dogleg to reduce beam slice energy spread at the cost of an increase in beam transverse emittance has been proposed to enhance the frequency multiplication efficiency of HGHG technique~\cite{CooledHGHG}. With the beam coupled in transverse and longitudinal planes, however, this scheme has demanding requirements on the transverse emittance.

It is also worth mentioning that EEX can only exchange the emittances among the different planes, but it does not change them. With the electron beam born with coupling (e.g. from a tilted laser that introduces $x-s$ coupling, a magnetized cathode that introduces $x-y$ coupling, etc.), or with the use of non-symplectic elements in the beam line, emittance may be partitioned in different planes (see. e.g.~\cite{FB, EP} and references therein).

%
\subsection{Beam conditioning}\label{sec:IV-H}
%

As was mentioned in Section~\ref{sec:FELs}, modern x-ray FELs require relativistic beams with small emittance and energy spread. The physical mechanism behind such a requirement is the need to maintain resonance between the wiggling motion of electrons in the undulator and the FEL radiation propagating with the speed of light. Electrons with different amplitudes of betatron oscillations and different energies propagate in the undulator with unequal longitudinal velocities, slipping relative to the radiation phase and leading to a smearing of the resonance. If the smearing is strong enough it suppresses the lasing of the beam.

In 1992, Sessler, Whittum and Yu~\cite{sessler92wy} noticed that the velocity spread due to the beam transverse emittance can be compensated by correlating the particle energy with its amplitude of the betatron oscillations. Specifically, for equal focussing in both transverse planes, the requirement for such a compensation reads
    \begin{align}\label{eq:IV-H.1}
    \frac{\Delta \gamma}{\gamma}
    =
    \kappa
    (J_{x}+J_{y})
    ,
    \end{align}
with
    \begin{align}\label{eq:IV-H.2}
    \kappa
    =
    \frac{1}{2\bar\beta}
    \frac{\lambda_{u}}{\lambda_{r}}
    ,
    \end{align}
where $\Delta \gamma$ is the energy deviation of the particle (in units of $mc^{2}$),  $\bar\beta$ is the averaged beta function in the undulator, $\lambda_{u}$ is the undulator period, $\lambda_{r}$ is the wavelength of the undulator emission, and $J_{x}$ and $J_{y}$ are the actions for betatron oscillations in $x$ and $y$ directions, respectively, normalized so that $\langle J_{x}\rangle = \langle J_{y}\rangle= \epsilon$, where $ \epsilon$ is the geometric transverse emittance of the beam. Preparing a beam which satisfies the condition~\eqref{eq:IV-H.1} is called ``beam conditioning''. The physical picture of beam conditioning is that particles with larger betatron amplitudes are given extra energy (and therefore larger velocity) to compensate for the longer path length. Values of $\kappa$ required for proper conditioning are typically on the order of 10--100 $\mu$m$^{-1}$.

Several methods were proposed in the literature to condition a beam~\cite{sessler92wy,Sprangle19936,beam_cond_neil,Papadichev1995ABS79,vinokurov96,emma03s,zholents_conditioner,schroeder_conditioner,wolsky_conditioner}. Of special interest for this article is a laser based approach developed in Ref.~\cite{zholents_conditioner} which we review in this section. The idea of the method is based on the scheme proposed by Vinokurov~\cite{vinokurov96} who suggested a conditioner setup that first uses an rf cavity to ``chirp'' the beam (i.e., introduce a correlation between longitudinal position and energy), then passes the beam through a focusing channel with some chromaticity, and finally uses a second rf cavity to remove the chirp.

If the frequency of the first rf accelerating section is $\omega_\mathrm{rf}$, electrons near the zero phase crossing change the energy by
    \begin{align}\label{eq:IV-H.3}
    \Delta \gamma_{1}
    =
    \frac{eU}{mc^{2}}
    \sin
    \left(
    \frac{\omega_\mathrm{rf}s}{c}
    \right)
    ,
    \end{align}
where $s$ is the longitudinal coordinate of particles in the bunch relative to its center. The focusing beam line then delays the particles with large betatron oscillation amplitudes and longer orbits by
    \begin{align}\label{eq:IV-H.4}
    \Delta s
    =
    -
    \frac{1}{2}
    \int
    \left[
    \left(
    \frac{dx}{dz}
    \right)^{2}
    +
    \left(
    \frac{dy}{dz}
    \right)^{2}
    \right]
    dz
    .
    \end{align}
Assuming that the two transverse directions in the focusing channel are identical, $\Delta s$ can be expressed through the action variables $J_{x}$ and $J_{y}$
    \begin{align}\label{eq:IV-H.5}
    \Delta s
    =
    2\pi\xi
    (J_{x}+J_{y})
    ,
    \end{align}
where $\xi$ is the chromaticity of the channel.

The second rf accelerating section is phased so as to remove the energy modulation introduced by the first one. However, the energy correction does not exactly cancel the energy deviation introduced by the first section because of the delay $\Delta s$, and there will be a residual energy deviation depending on the particle's betatron amplitude and the chromaticity of the beam line. The residual energy is
    \begin{align}\label{eq:IV-H.6}
    \Delta \gamma
    &=
    \frac{eU}{mc^{2}}
    \left[
    \sin
    \left(
    \frac{\omega_\mathrm{rf}s}{c}
    \right)
    -
    \sin
    \left(
    \frac{\omega_\mathrm{rf}}{c}
    (s-\Delta s)
    \right)
    \right]
    \nonumber\\
    &=
    2\pi\xi
    (J_{x}+J_{y})
    \frac{eU}{mc^{2}}
    \frac{\omega_\mathrm{rf}}{c}
    \cos
    \left(
    \frac{\omega_\mathrm{rf}s}{c}
    \right)
    ,
    \end{align}
where $\omega_\mathrm{rf}|\Delta s|/{c} \ll 1$ was assumed. Typically the rf wavelength is much larger than the bunch length, $\omega_\mathrm{rf}|s|/{c} \ll 1$, so ~\eqref{eq:IV-H.6} reduces to~\eqref{eq:IV-H.1} with
    \begin{align}\label{eq:IV-H.7}
    \kappa
    =
    \kappa_{0}
    \equiv
    2\pi\xi
    \frac{eU}{\gamma mc^{2}}
    \frac{\omega_\mathrm{rf}}{c}
    .
    \end{align}
Unfortunately, as detailed analysis shows~\cite{wolsky_conditioner}, realistic values for the rf voltage $U$ and the length of the focusing channel give values of $\kappa$ that are several orders of magnitude smaller than what is required to condition a beam in a modern x-ray FEL.

As pointed out in~\cite{zholents_conditioner}, however, with $\kappa$ being proportional to $\omega_{\mathrm{rf}}$, the conditioning can be made much stronger if a laser beam is used to modulate the particles energy, instead of the accelerating structures, thereby replacing $\omega_{\mathrm{rf}}$ with a much larger laser frequency $\omega_{L}$. The setup of the laser assisted conditioner is shown in Fig.~\ref{laser_conditioner}.
    \begin{figure}[h]
    \includegraphics[width = 0.48\textwidth]{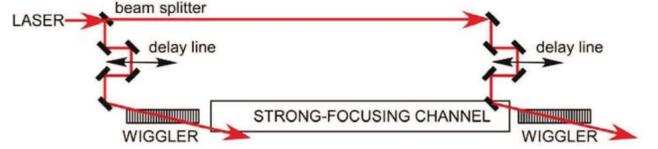}
    \caption{A schematic of the laser assisted conditioner. From \cite{zholents_conditioner}.
    \label{laser_conditioner}}
    \end{figure}
The laser pulse co-propagates in the wiggler magnet at a small angle with the electron beam and produces an energy modulation of the electrons at the laser wavelength $\lambda_{L}$. The electrons then pass through a focusing channel. Finally, the reverse modulation is applied to the beam with a second laser pulse in the second wiggler magnet. The arrival times of the electron beam and laser pulses in the wigglers are controlled by providing tight synchronization between the laser pulses and interferometric control of all path lengths with active feedback.

With the laser wavelength $\lambda_{L}$ typically much shorter than the bunch length the approximation~\eqref{eq:IV-H.7} is invalid and should be replaced by the conditioning factor $\kappa(s)$ which oscillates over the bunch length
    \begin{align}\label{eq:IV-H.8}
    \kappa(s)
    =
    \kappa_{0}
    \cos
    \left(
    \frac{2\pi s}{\lambda_{L}}
    \right)
    ,
    \end{align}
where $\kappa_{0}$ is given by~\eqref{eq:IV-H.7} with $\omega_{\mathrm{rf}}$ replaced by $\omega_{L}$. Assuming that $\kappa_{0}$ is equal to the required value in~\eqref{eq:IV-H.2}, we see that the conditioning is achieved near discrete points in the beam $s=n\lambda_{L}$ where $n$ is an integer number with $\kappa$ taking all possible values between $-\kappa_{0}$ and $\kappa_{0}$. The entire electron beam can now be viewed as a sequence of alternating slices of electrons with a variable degree (and the sign) of conditioning. As long as the length of the slice (which is approximately equal to $\lambda_{L}/2$) is longer than the slippage length, the correctly conditioned slices will generate the FEL radiation, while the incorrectly conditioned regions will not radiate. Simulations in \cite{zholents_conditioner} showed that the overall performance of the laser conditioned beam could be considerably improved in the regime when the transverse emittance is large.

In another approach~\cite{schroeder_conditioner}, FEL conditioning via Thomson backscattering of an intense laser pulse was proposed. The number of scattered photons, and hence the electron energy loss, is proportional to the laser intensity which decreases off-axis for a focused beam. Therefore, Thomson backscattering produces a correlation between an electron's energy loss and its transverse location in the laser field, thus allowing beam conditioning. However, according to~\cite{schroeder_conditioner}, a rather large laser energy on the order of $\sim 10^2$ Joules in $\sim 10$ ps is required for conditioning a beam with realistic parameters.

%
\subsection{Measurement of ultra-short beams}\label{Sec:MeasuringUltraShortBeams}
%
Using lasers to measure electron beam parameters dates all the way back to the early days of lasers. In the 1960's Thomson backscattering~\cite{1963} was suggested as a method to measure plasma parameters. In the 1990's Thomson backscattering was used to measure both transverse beam sizes in the nanometer range \cite{Kezerashvili, Shintake} and the length of a relativistic electron beam with sub-picosecond time resolution~\cite{Wim1996}. Here we discuss several recent techniques that use lasers to manipulate the electron beam phase space for the determination of the temporal structure with femtosecond time resolution.

%
     \subsubsection{Optical replica synthesizer}
%

In the so-called optical replica synthesizer (ORS) technique \cite{ORS}, an optical laser with a pulse duration much longer than the electron beam is first used to generate a density modulation in the electron beam with a modulator-chicane system. The density-modulated beam is then sent through an undulator tuned to the laser frequency to generate intense coherent radiation. Analysis shows that under proper conditions (i.e. the modulation amplitude and beam size do not vary in time, and the slippage is sufficiently small) the  radiated fields are a replica of the electron beam temporal profile. The electron beam information is thus encoded into the optical light, whereby well-established techniques borrowed from the ultrafast laser community such as frequency-resolved optical gating (FROG \cite{FROG}) can be used to obtain the electron beam temporal information, including both the temporal shape and energy chirp.

While the FROG technique has been successfully used to measure a 4.5 fs laser pulse \cite{4fslaser}, the slippage length in the radiator typically limits the resolution of ORS to $>10~$fs. This is because implementation of the FROG technique in single-shot mode typically requires $>1~\mu$J of radiation energy, which in turn requires an undulator with many periods to generate enough signal. Increasing the number of periods also increases the total slippage, however, which can quickly exceed $10~$fs at optical wavelengths. A proof-of-principle experiment at FLASH provided encouraging results \cite{ORS1}, but more work is needed to demonstrate the potential of this technique in realistic conditions where the coherent radiation is also contaminated by the contribution from microbunching instabilities.

A simplified version of ORS in which the radiator undulator is replaced with an optical transition radiation (OTR) screen and only the radiation spectrum needs to be measured has also been proposed \cite{ICFA-Xiang}. This scheme is easy to implement and does not require generation of high-power radiation for a FROG measurement, yet still allows the rms length of an ultrashort bunch to be obtained. The idea to extract bunch length information by measuring the coherent OTR (COTR) spectrum is rather simple, since, for an idealized Gaussian beam with rms length $\Delta t$ the rms frequency spread is $\Delta f=1/(2\Delta t)$.

Consider a Gaussian electron beam with rms length of 5 fs. The longitudinal phase space is shown in Fig.~\ref{ORS}a. After interacting with an 800 nm laser and passing through a chicane, the beam longitudinal phase space evolves to that in Fig.~\ref{ORS}b. The corresponding beam current is shown with dashed red line in Fig.~\ref{ORS}c (the initial beam current is shown with blue-solid line). It is justified to assume a sufficiently prompt response for generation of OTR such that the pulse shape of the COTR generated when beam strikes an OTR screen will be a replica of that of the beam current distribution. Accordingly, the COTR spectrum generated by this modulated beam will carry information about the beam current distribution. The COTR spectrum for the beam in Fig.~\ref{ORS}c is shown in Fig.~\ref{ORS}d where one can see that the quasi-monochromatic COTR has a central frequency peak at $f=375$ THz (corresponding to a central wavelength of 800 nm) with an rms frequency spread of $\Delta f=33$~THz, which corresponds to a transform limited rms pulse width of 5 fs, which is the same as that of the electron bunch.
    \begin{figure}[h]
    \includegraphics[width = 0.23\textwidth]{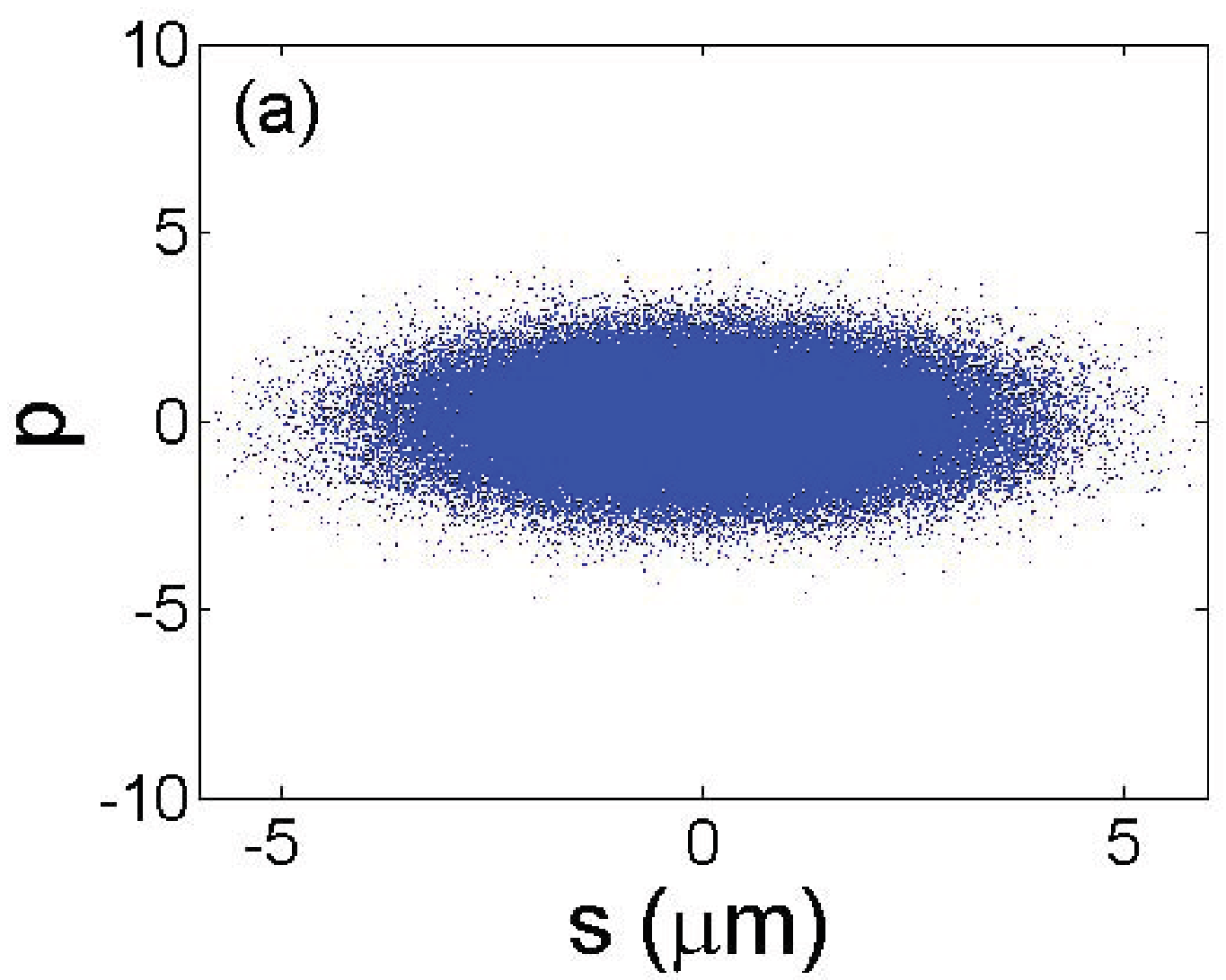}
    \includegraphics[width = 0.23\textwidth]{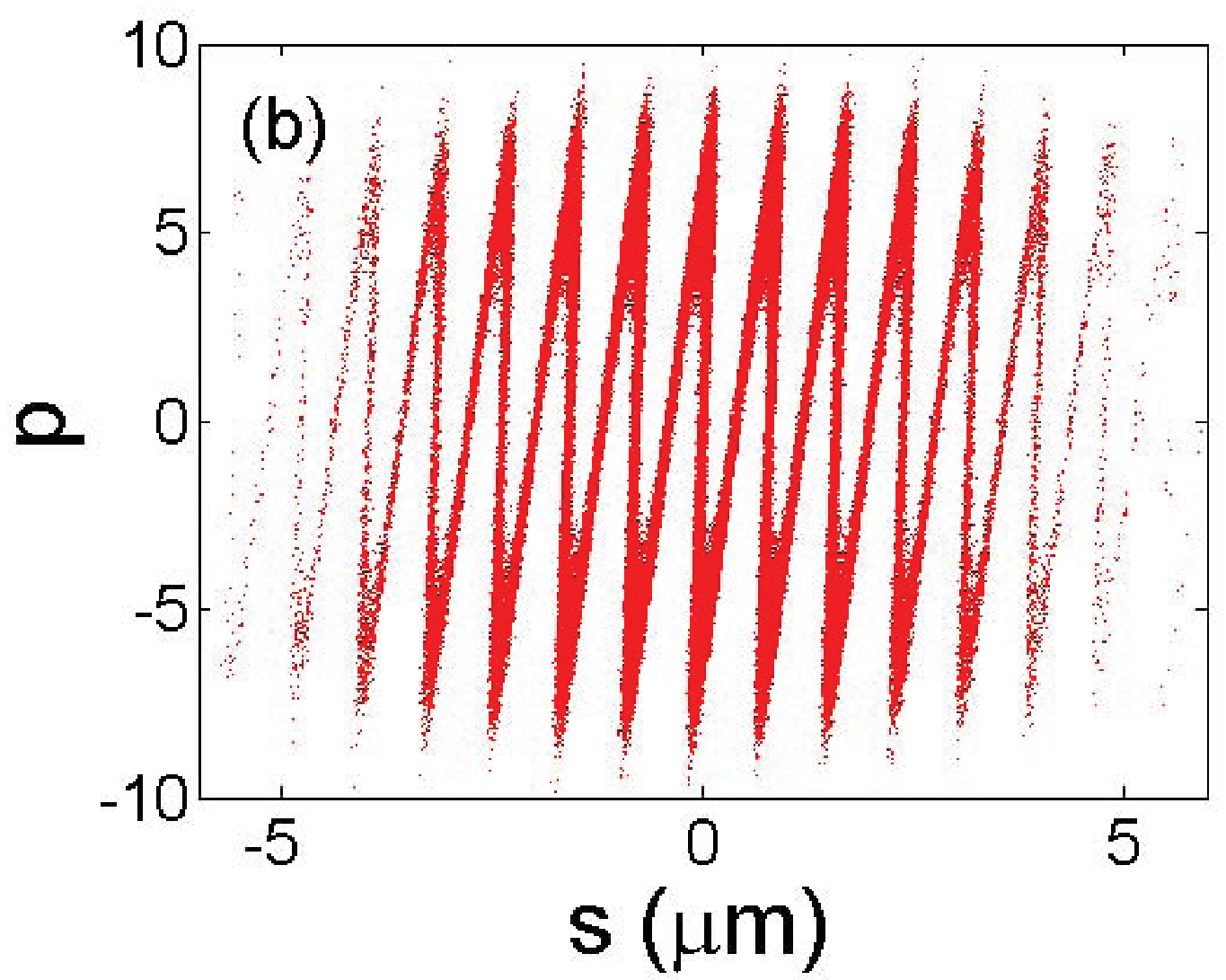}
    \includegraphics[width = 0.23\textwidth]{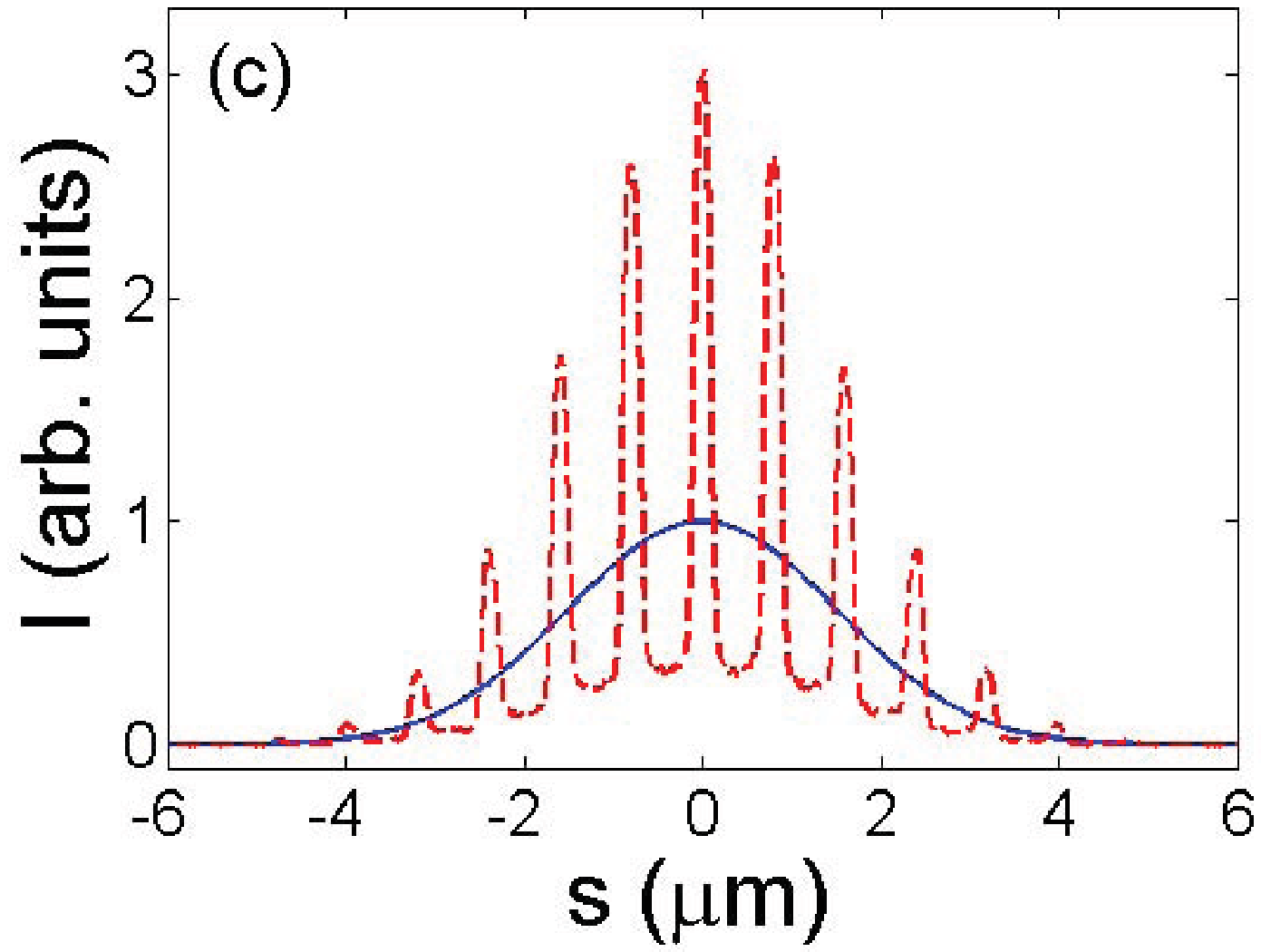}
    \includegraphics[width = 0.23\textwidth]{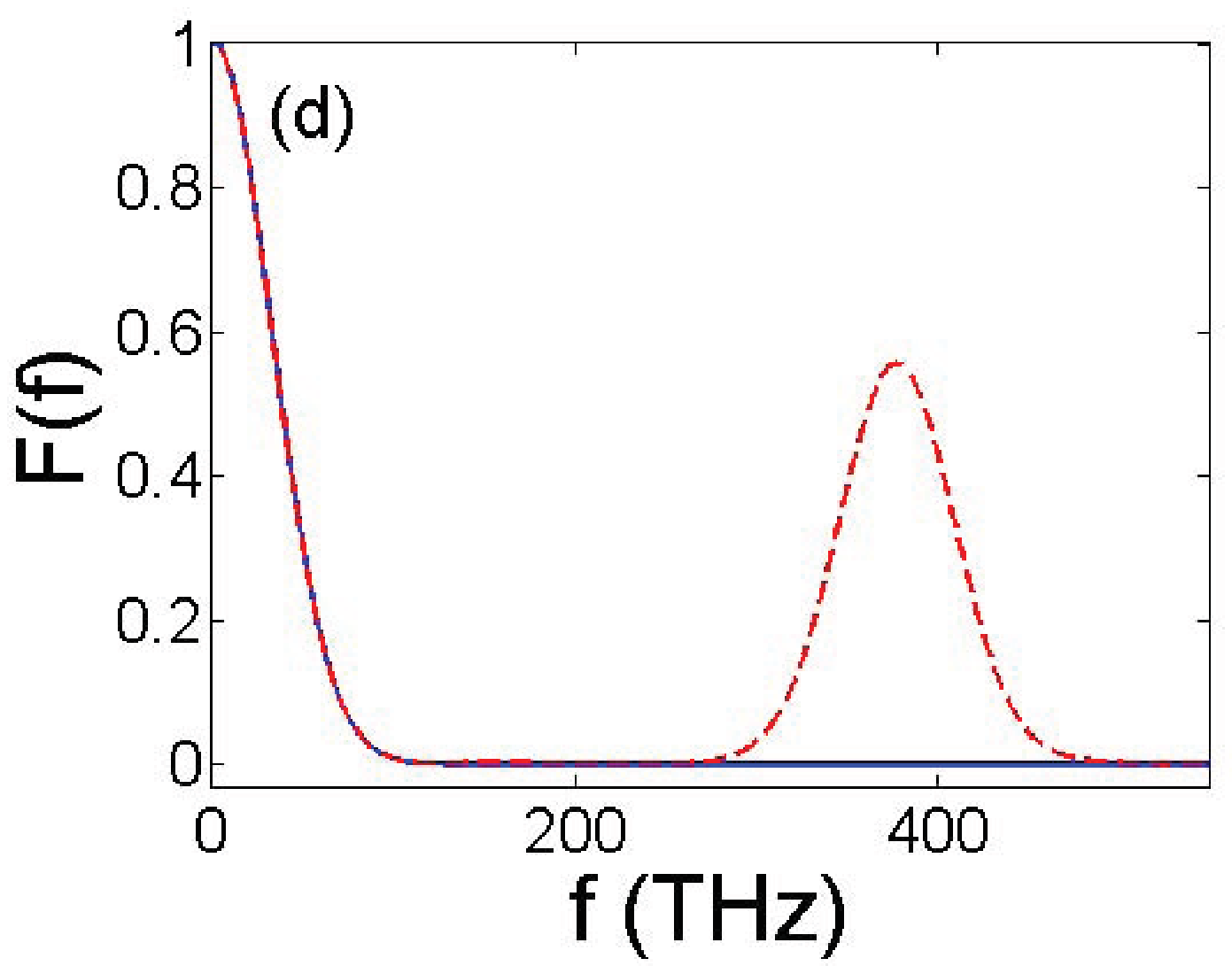}
    \caption{Evolution of the longitudinal phase space in a simplified ORS scheme: (a) initial beam phase space; (b) beam phase space after interaction with an optical laser; (c) beam current with (dashed red) and without (solid blue) the laser modulation; (d) corresponding radiation spectra.
    \label{ORS}}
    \end{figure}

This method can be applied to extract rms bunch length in more general cases, such as beams with asymmetric Gaussian and flat-top distributions. However, it should be noted that it may yield inaccurate results if the beam temporal profile has a complicated shape.

%
\subsubsection{Optical streaking}
%

In the ORS method, the laser pulse duration needs to be much longer than electron bunch length in order to provide a constant modulation across the bunch. On the contrary, if a relatively short laser pulse (with duration a few times longer than electron bunch length) is used to modulate the beam, the energy modulation amplitude varies in time. This leads to an effect that can be exploited to measure the electron bunch length. The physics behind this so-called optical streaking method~\cite{OS} is illustrated in Fig.~\ref{OS}.  If the electron beam is synchronized with the laser in such a way that the beam interacts with the sloped region of the laser pulse as shown in Fig.~\ref{OS}, the beam energy distribution after the laser modulation can be used to reconstruct the beam temporal profile. Consider as an example a Gaussian electron beam with an rms duration of 4 fs. The initial longitudinal phase space is shown with red dots and the projected beam temporal profile with the solid red line. After interacting with a high power 800 nm laser, the beam phase space is shown with green dots and its energy distribution with the blue line. Here, the laser FWHM duration is assumed to be 40 fs and the temporal offset between the electron beam and laser intensity peak is 25 fs (part of the laser field profile is shown with dashed magenta line). This generates an energy modulation in the beam that grows in amplitude.
    \begin{figure}[h]
    \includegraphics[width = 0.48\textwidth]{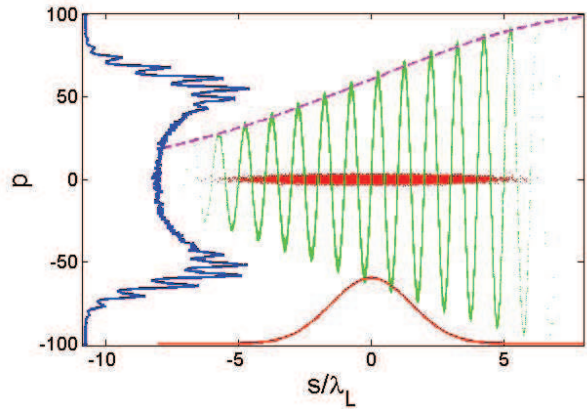}
    \caption{Illustration of the optical streaking method to measure electron bunch temporal profile. The horizontal axis is the beam longitudinal position normalized to the laser wavelength and the vertical axis is particle's energy deviation normalized to the rms slice energy spread of the beam.
    \label{OS}}
    \end{figure}
When the laser power is sufficiently high, it is possible to make the energy difference between two adjacent cycles larger than beam's intrinsic energy spread (see blue line in Fig.~\ref{OS}) so that a clear modulation in the projected energy distribution can be observed. Because the oscillation period of beam energy is correlated with the laser wavelength, the projected energy modulation can be measured with a high resolution magnetic energy spectrometer and used to determine the electron beam's temporal shape using the laser wavelength as the ruler~\cite{OS}. This method is particularly suited for electron beams with small intrinsic energy spreads such that modest laser powers can be used.

%
\subsubsection{Optical oscilloscope}
%

The rf deflecting cavity is typically used for time resolved diagnosis of the electron bunches in linear accelerators~\cite{XJWang1999, Akre}. For a beam with given emittance and intrinsic divergence at the cavity, the temporal resolution provided by this cavity is limited by the available kick strength. A stronger kick can be achieved by increasing the voltage and/or the frequency of the cavity. For instance, given the same voltage, an X-band deflecting cavity provides 4 times higher resolution than an S-band cavity. Thus, lasers operating in transverse mode have potential to dramatically improve temporal resolution by virtue of their much higher frequencies (see Sec.~\ref{Section:II.C.AM} ).

A drawback of using lasers to generate the required angular kick is that the electron beam is typically much longer than the wavelength. Because the desired linear kick occurs only for electrons around the zero field crossing, this  limits application of such ``optical deflectors" to cases where the bunch is much shorter than the laser wavelength. However, to avoid such limitations and to improve the dynamic range of the measurement, a so-called ``optical oscilloscope'' has been recently proposed~\cite{OO}. Here, in addition to an optical laser operating in transverse mode, an rf deflecting cavity is used to streak the beam in an orthogonal direction so that the beam distribution at different laser cycles can be separated as well. The outcome of such a configuration is a 2-D map which shows the overall bunch shape (in the kicking direction of rf cavity) as well as the fine structures (in the kicking direction of the laser). Note, in this scheme the resolution of the rf deflecting cavity needs to be smaller than the wavelength of the laser, therefore long wavelength lasers on the order of a few microns may be favored for this application. In principle, with a high power laser the resolution of this method can be pushed to the sub-femtosecond regime.


%
\section{Summary}\label{sec:7}
%

Over the last several decades, relativistic electron beams have proven to be effective and enormously versatile sources of intense radiation used to probe the structure and dynamical behavior of matter. Continued improvements in electron beam quality and brightness have led to corresponding enhancements in the radiation they produce, with higher intensities, shorter wavelengths, and shorter pulse durations emerging with each successive generation in source technology. Over the last decade in particular, significant progress has been made with the birth of 
free-electron lasers that are capable of producing hard x-ray beams with unprecedented brightness. These devices enable the investigation of processes on the femtosecond time scale with x-rays at 10 keV energies in a single shot, opening new opportunities to capture the ultrafast and probe the ultrasmall.

Parallel advances in laser technology have led to new opportunities to further improve the performance of accelerator based light source facilities, from the development of now ubiquitous photocathode injectors that deliver high brightness beams with low energy spread and emittance, to more recent seeding techniques that rely on precision laser manipulation of the high energy electron beams for tailored radiation production. Together, these innovations allow the fully coherent field distribution from a laser pulse to be imprinted onto the electron beam, and then transferred to radiation at much shorter wavelengths. Through the variety of techniques described in this review, this provides additional flexibility in controlling the pulse duration, spectral bandwidth, and even transverse distribution of the radiation. Associated improvements in timing synchronization and stabilization between the laser and electron beams at the tens of femtoseconds level have also 
enabled the merging of these technologies to access new regimes in, for example, x-ray pump-probe experiments.

Presented in this article is a broad mixture of both new ideas and recently demonstrated methods that aims to highlight current progress and to stimulate future research. We describe the simple fundamental principles on which many of these methods are based. In most cases they involve an interaction of the laser beam with a relativistic electron bunch in a resonantly tuned undulator, and a subsequent evolution of the beam distribution function when it passes through specially arranged magnets or other components of the beam line. While such manipulations create modulations on the scale of the laser wavelength, in more sophisticated arrangements, one can generate the desired structures in the beam on much shorter scales, leading to an effective laser frequency upshifting and access to nanometer scales. Various other emergent research opportunities are also discussed in this article, including the generation of ultrashort x-ray pulses both in light source storage rings and FELs, the production of beams with orbital angular momentum, the production of mode-locked x-ray pulse trains, and numerous sophisticated diagnostic techniques.

As is typical with emerging technologies, the number of newly proposed techniques has outpaced the number that have been experimentally demonstrated and adapted in practice.  Accordingly, several of the more recent concepts are in the research stage of development, but are included here for their value in leading to novel insights and stimulating new ideas.

%
\section{Acknowledgements}\label{sec:8}
%

This work was supported by the U.S. Department of Energy contracts DE-AC02-76SF00515 and DE-AC02-06CH11357.

\bibliography{beam_by_design_rev6}

\begin{thebibliography}{278}%
\makeatletter
\providecommand \@ifxundefined [1]{%
 \@ifx{#1\undefined}
}%
\providecommand \@ifnum [1]{%
 \ifnum #1\expandafter \@firstoftwo
 \else \expandafter \@secondoftwo
 \fi
}%
\providecommand \@ifx [1]{%
 \ifx #1\expandafter \@firstoftwo
 \else \expandafter \@secondoftwo
 \fi
}%
\providecommand \natexlab [1]{#1}%
\providecommand \enquote  [1]{``#1''}%
\providecommand \bibnamefont  [1]{#1}%
\providecommand \bibfnamefont [1]{#1}%
\providecommand \citenamefont [1]{#1}%
\providecommand \href@noop [0]{\@secondoftwo}%
\providecommand \href [0]{\begingroup \@sanitize@url \@href}%
\providecommand \@href[1]{\@@startlink{#1}\@@href}%
\providecommand \@@href[1]{\endgroup#1\@@endlink}%
\providecommand \@sanitize@url [0]{\catcode `\\12\catcode `\$12\catcode
  `\&12\catcode `\#12\catcode `\^12\catcode `\_12\catcode `\%12\relax}%
\providecommand \@@startlink[1]{}%
\providecommand \@@endlink[0]{}%
\providecommand \url  [0]{\begingroup\@sanitize@url \@url }%
\providecommand \@url [1]{\endgroup\@href {#1}{\urlprefix }}%
\providecommand \urlprefix  [0]{URL }%
\providecommand \Eprint [0]{\href }%
\providecommand \doibase [0]{http://dx.doi.org/}%
\providecommand \selectlanguage [0]{\@gobble}%
\providecommand \bibinfo  [0]{\@secondoftwo}%
\providecommand \bibfield  [0]{\@secondoftwo}%
\providecommand \translation [1]{[#1]}%
\providecommand \BibitemOpen [0]{}%
\providecommand \bibitemStop [0]{}%
\providecommand \bibitemNoStop [0]{.\EOS\space}%
\providecommand \EOS [0]{\spacefactor3000\relax}%
\providecommand \BibitemShut  [1]{\csname bibitem#1\endcsname}%
\let\auto@bib@innerbib\@empty
\bibitem [{\citenamefont {Ackermann}\ \emph {et~al.}(2013)\citenamefont
  {Ackermann}, \citenamefont {Azima}, \citenamefont {Bajt}, \citenamefont
  {B\"odewadt}, \citenamefont {Curbis}, \citenamefont {Dachraoui},
  \citenamefont {Delsim-Hashemi}, \citenamefont {Drescher}, \citenamefont
  {D\"usterer}, \citenamefont {Faatz}, \citenamefont {Felber}, \citenamefont
  {Feldhaus}, \citenamefont {Hass}, \citenamefont {Hipp}, \citenamefont
  {Honkavaara}, \citenamefont {Ischebeck}, \citenamefont {Khan}, \citenamefont
  {Laarmann}, \citenamefont {Lechner}, \citenamefont {Maltezopoulos},
  \citenamefont {Miltchev}, \citenamefont {Mittenzwey}, \citenamefont
  {Rehders}, \citenamefont {R\"onsch-Schulenburg}, \citenamefont {Rossbach},
  \citenamefont {Schlarb}, \citenamefont {Schreiber}, \citenamefont
  {Schroedter}, \citenamefont {Schulz}, \citenamefont {Schulz}, \citenamefont
  {Tarkeshian}, \citenamefont {Tischer}, \citenamefont {Wacker},\ and\
  \citenamefont {Wieland}}]{HHG382013}%
  \BibitemOpen
  \bibfield  {author} {\bibinfo {author} {\bibnamefont {Ackermann},
  \bibfnamefont {S.}}, \bibinfo {author} {\bibfnamefont {A.}~\bibnamefont
  {Azima}}, \bibinfo {author} {\bibfnamefont {S.}~\bibnamefont {Bajt}},
  \bibinfo {author} {\bibfnamefont {J.}~\bibnamefont {B\"odewadt}}, \bibinfo
  {author} {\bibfnamefont {F.}~\bibnamefont {Curbis}}, \bibinfo {author}
  {\bibfnamefont {H.}~\bibnamefont {Dachraoui}}, \bibinfo {author}
  {\bibfnamefont {H.}~\bibnamefont {Delsim-Hashemi}}, \bibinfo {author}
  {\bibfnamefont {M.}~\bibnamefont {Drescher}}, \bibinfo {author}
  {\bibfnamefont {S.}~\bibnamefont {D\"usterer}}, \bibinfo {author}
  {\bibfnamefont {B.}~\bibnamefont {Faatz}}, \bibinfo {author} {\bibfnamefont
  {M.}~\bibnamefont {Felber}}, \bibinfo {author} {\bibfnamefont
  {J.}~\bibnamefont {Feldhaus}}, \bibinfo {author} {\bibfnamefont
  {E.}~\bibnamefont {Hass}}, \bibinfo {author} {\bibfnamefont {U.}~\bibnamefont
  {Hipp}}, \bibinfo {author} {\bibfnamefont {K.}~\bibnamefont {Honkavaara}},
  \bibinfo {author} {\bibfnamefont {R.}~\bibnamefont {Ischebeck}}, \bibinfo
  {author} {\bibfnamefont {S.}~\bibnamefont {Khan}}, \bibinfo {author}
  {\bibfnamefont {T.}~\bibnamefont {Laarmann}}, \bibinfo {author}
  {\bibfnamefont {C.}~\bibnamefont {Lechner}}, \bibinfo {author} {\bibfnamefont
  {T.}~\bibnamefont {Maltezopoulos}}, \bibinfo {author} {\bibfnamefont
  {V.}~\bibnamefont {Miltchev}}, \bibinfo {author} {\bibfnamefont
  {M.}~\bibnamefont {Mittenzwey}}, \bibinfo {author} {\bibfnamefont
  {M.}~\bibnamefont {Rehders}}, \bibinfo {author} {\bibfnamefont
  {J.}~\bibnamefont {R\"onsch-Schulenburg}}, \bibinfo {author} {\bibfnamefont
  {J.}~\bibnamefont {Rossbach}}, \bibinfo {author} {\bibfnamefont
  {H.}~\bibnamefont {Schlarb}}, \bibinfo {author} {\bibfnamefont
  {S.}~\bibnamefont {Schreiber}}, \bibinfo {author} {\bibfnamefont
  {L.}~\bibnamefont {Schroedter}}, \bibinfo {author} {\bibfnamefont
  {M.}~\bibnamefont {Schulz}}, \bibinfo {author} {\bibfnamefont
  {S.}~\bibnamefont {Schulz}}, \bibinfo {author} {\bibfnamefont
  {R.}~\bibnamefont {Tarkeshian}}, \bibinfo {author} {\bibfnamefont
  {M.}~\bibnamefont {Tischer}}, \bibinfo {author} {\bibfnamefont
  {V.}~\bibnamefont {Wacker}}, \ and\ \bibinfo {author} {\bibfnamefont
  {M.}~\bibnamefont {Wieland}}} (\bibinfo {year} {2013}),\ \href
  {http://link.aps.org/doi/10.1103/PhysRevLett.111.114801} {\bibfield
  {journal} {\bibinfo  {journal} {Phys. Rev. Lett.}\ }\textbf {\bibinfo
  {volume} {111}},\ \bibinfo {pages} {114801}}\BibitemShut {NoStop}%
\bibitem [{\citenamefont {Ackermann}\ \emph {et~al.}(2007)\citenamefont
  {Ackermann}, \citenamefont {Asova}, \citenamefont {Ayvazyan}, \citenamefont
  {Azima}, \citenamefont {Baboi}, \citenamefont {B\"ahr}, \citenamefont
  {Balandin}, \citenamefont {Beutner}, \citenamefont {Brandt}, \citenamefont
  {Bolzmann}, \citenamefont {Brinkmann}, \citenamefont {Brovko}, \citenamefont
  {Castellano}, \citenamefont {Castro}, \citenamefont {Catani}, \citenamefont
  {Chiadroni}, \citenamefont {Choroba}, \citenamefont {Cianchi}, \citenamefont
  {Costello}, \citenamefont {Cubaynes}, \citenamefont {Dardis}, \citenamefont
  {Decking}, \citenamefont {Delsim-Hashemi}, \citenamefont {Delserieys},
  \citenamefont {Pirro}, \citenamefont {Dohlus}, \citenamefont {D\"usterer},
  \citenamefont {Eckhardt}, \citenamefont {Edwards}, \citenamefont {Faatz},
  \citenamefont {Feldhaus}, \citenamefont {Fl\"ottmann}, \citenamefont
  {Frisch}, \citenamefont {Fr\"ohlich}, \citenamefont {Garvey}, \citenamefont
  {Gensch}, \citenamefont {Gerth}, \citenamefont {G\"oler}, \citenamefont
  {Golubeva}, \citenamefont {Grabosch}, \citenamefont {Grecki}, \citenamefont
  {Grimm}, \citenamefont {Hacker}, \citenamefont {Hahn}, \citenamefont {Han},
  \citenamefont {Honkavaara}, \citenamefont {Hott}, \citenamefont {H\"uning},
  \citenamefont {Ivanisenko}, \citenamefont {Jaeschke}, \citenamefont
  {Jalmuzna}, \citenamefont {Jezynski}, \citenamefont {Kammering},
  \citenamefont {Katalev}, \citenamefont {Kavanagh}, \citenamefont {Kennedy},
  \citenamefont {Khodyachykh}, \citenamefont {Klose}, \citenamefont
  {Kocharyan}, \citenamefont {K\"orfer}, \citenamefont {Kollewe}, \citenamefont
  {Koprek}, \citenamefont {Korepanov}, \citenamefont {Kostin}, \citenamefont
  {Krassilnikov}, \citenamefont {Kube}, \citenamefont {Kuhlmann}, \citenamefont
  {Lewis}, \citenamefont {Lilje}, \citenamefont {Limberg}, \citenamefont
  {Lipka}, \citenamefont {L\"ohl}, \citenamefont {Luna}, \citenamefont {Luong},
  \citenamefont {Martins}, \citenamefont {Meyer}, \citenamefont {Michelato},
  \citenamefont {Miltchev}, \citenamefont {M\"oller}, \citenamefont {Monaco},
  \citenamefont {M\"uller}, \citenamefont {Napieralski}, \citenamefont
  {Napoly}, \citenamefont {Nicolosi}, \citenamefont {N\"olle}, \citenamefont
  {Nunez}, \citenamefont {Oppelt}, \citenamefont {Pagani}, \citenamefont
  {Paparella}, \citenamefont {Pchalek}, \citenamefont {Pedregosa-Gutierrez},
  \citenamefont {Petersen}, \citenamefont {Petrosyan}, \citenamefont
  {Petrosyan}, \citenamefont {Petrosyan}, \citenamefont {Pfl\"uger},
  \citenamefont {Pl\"onjes}, \citenamefont {Poletto}, \citenamefont {Pozniak},
  \citenamefont {Prat}, \citenamefont {Proch}, \citenamefont {Pucyk},
  \citenamefont {Radcliffe}, \citenamefont {Redlin}, \citenamefont {Rehlich},
  \citenamefont {Richter}, \citenamefont {Roehrs}, \citenamefont {Roensch},
  \citenamefont {Romaniuk}, \citenamefont {Ross}, \citenamefont {Rossbach},
  \citenamefont {Rybnikov}, \citenamefont {Sachwitz}, \citenamefont {Saldin},
  \citenamefont {Sandner}, \citenamefont {Schlarb}, \citenamefont {Schmidt},
  \citenamefont {Schmitz}, \citenamefont {Schm\"user}, \citenamefont
  {Schneider}, \citenamefont {Schneidmiller}, \citenamefont {Schnepp},
  \citenamefont {Schreiber}, \citenamefont {Seidel}, \citenamefont {Sertore},
  \citenamefont {Shabunov}, \citenamefont {Simon}, \citenamefont {Simrock},
  \citenamefont {Sombrowski}, \citenamefont {Sorokin}, \citenamefont
  {Spanknebel}, \citenamefont {Spesyvtsev}, \citenamefont {Staykov},
  \citenamefont {Steffen}, \citenamefont {Stephan}, \citenamefont {Stulle},
  \citenamefont {Thom}, \citenamefont {Tiedtke}, \citenamefont {Tischer},
  \citenamefont {Toleikis}, \citenamefont {Treusch}, \citenamefont {Trines},
  \citenamefont {Tsakov}, \citenamefont {Vogel}, \citenamefont {Weiland},
  \citenamefont {Weise}, \citenamefont {Wellh\"ofer}, \citenamefont {Wendt},
  \citenamefont {Will}, \citenamefont {Winter}, \citenamefont {Wittenburg},
  \citenamefont {Wurth}, \citenamefont {Yeates}, \citenamefont {Yurkov},
  \citenamefont {Zagorodnov},\ and\ \citenamefont {Zapfe}}]{FLASH}%
  \BibitemOpen
  \bibfield  {author} {\bibinfo {author} {\bibnamefont {Ackermann},
  \bibfnamefont {W.}}, \bibinfo {author} {\bibfnamefont {G.}~\bibnamefont
  {Asova}}, \bibinfo {author} {\bibfnamefont {V.}~\bibnamefont {Ayvazyan}},
  \bibinfo {author} {\bibfnamefont {A.}~\bibnamefont {Azima}}, \bibinfo
  {author} {\bibfnamefont {N.}~\bibnamefont {Baboi}}, \bibinfo {author}
  {\bibfnamefont {J.}~\bibnamefont {B\"ahr}}, \bibinfo {author} {\bibfnamefont
  {V.}~\bibnamefont {Balandin}}, \bibinfo {author} {\bibfnamefont
  {B.}~\bibnamefont {Beutner}}, \bibinfo {author} {\bibfnamefont
  {A.}~\bibnamefont {Brandt}}, \bibinfo {author} {\bibfnamefont
  {A.}~\bibnamefont {Bolzmann}}, \bibinfo {author} {\bibfnamefont
  {R.}~\bibnamefont {Brinkmann}}, \bibinfo {author} {\bibfnamefont {O.~I.}\
  \bibnamefont {Brovko}}, \bibinfo {author} {\bibfnamefont {M.}~\bibnamefont
  {Castellano}}, \bibinfo {author} {\bibfnamefont {P.}~\bibnamefont {Castro}},
  \bibinfo {author} {\bibfnamefont {L.}~\bibnamefont {Catani}}, \bibinfo
  {author} {\bibfnamefont {E.}~\bibnamefont {Chiadroni}}, \bibinfo {author}
  {\bibfnamefont {S.}~\bibnamefont {Choroba}}, \bibinfo {author} {\bibfnamefont
  {A.}~\bibnamefont {Cianchi}}, \bibinfo {author} {\bibfnamefont {J.~T.}\
  \bibnamefont {Costello}}, \bibinfo {author} {\bibfnamefont {D.}~\bibnamefont
  {Cubaynes}}, \bibinfo {author} {\bibfnamefont {J.}~\bibnamefont {Dardis}},
  \bibinfo {author} {\bibfnamefont {W.}~\bibnamefont {Decking}}, \bibinfo
  {author} {\bibfnamefont {H.}~\bibnamefont {Delsim-Hashemi}}, \bibinfo
  {author} {\bibfnamefont {A.}~\bibnamefont {Delserieys}}, \bibinfo {author}
  {\bibfnamefont {G.~D.}\ \bibnamefont {Pirro}}, \bibinfo {author}
  {\bibfnamefont {M.}~\bibnamefont {Dohlus}}, \bibinfo {author} {\bibfnamefont
  {S.}~\bibnamefont {D\"usterer}}, \bibinfo {author} {\bibfnamefont
  {A.}~\bibnamefont {Eckhardt}}, \bibinfo {author} {\bibfnamefont {H.~T.}\
  \bibnamefont {Edwards}}, \bibinfo {author} {\bibfnamefont {B.}~\bibnamefont
  {Faatz}}, \bibinfo {author} {\bibfnamefont {J.}~\bibnamefont {Feldhaus}},
  \bibinfo {author} {\bibfnamefont {K.}~\bibnamefont {Fl\"ottmann}}, \bibinfo
  {author} {\bibfnamefont {J.}~\bibnamefont {Frisch}}, \bibinfo {author}
  {\bibfnamefont {L.}~\bibnamefont {Fr\"ohlich}}, \bibinfo {author}
  {\bibfnamefont {T.}~\bibnamefont {Garvey}}, \bibinfo {author} {\bibfnamefont
  {U.}~\bibnamefont {Gensch}}, \bibinfo {author} {\bibfnamefont
  {C.}~\bibnamefont {Gerth}}, \bibinfo {author} {\bibfnamefont
  {M.}~\bibnamefont {G\"oler}}, \bibinfo {author} {\bibfnamefont
  {N.}~\bibnamefont {Golubeva}}, \bibinfo {author} {\bibfnamefont {H.-J.}\
  \bibnamefont {Grabosch}}, \bibinfo {author} {\bibfnamefont {M.}~\bibnamefont
  {Grecki}}, \bibinfo {author} {\bibfnamefont {O.}~\bibnamefont {Grimm}},
  \bibinfo {author} {\bibfnamefont {K.}~\bibnamefont {Hacker}}, \bibinfo
  {author} {\bibfnamefont {U.}~\bibnamefont {Hahn}}, \bibinfo {author}
  {\bibfnamefont {J.~H.}\ \bibnamefont {Han}}, \bibinfo {author} {\bibfnamefont
  {K.}~\bibnamefont {Honkavaara}}, \bibinfo {author} {\bibfnamefont
  {T.}~\bibnamefont {Hott}}, \bibinfo {author} {\bibfnamefont {M.}~\bibnamefont
  {H\"uning}}, \bibinfo {author} {\bibfnamefont {Y.}~\bibnamefont
  {Ivanisenko}}, \bibinfo {author} {\bibfnamefont {E.}~\bibnamefont
  {Jaeschke}}, \bibinfo {author} {\bibfnamefont {W.}~\bibnamefont {Jalmuzna}},
  \bibinfo {author} {\bibfnamefont {T.}~\bibnamefont {Jezynski}}, \bibinfo
  {author} {\bibfnamefont {R.}~\bibnamefont {Kammering}}, \bibinfo {author}
  {\bibfnamefont {V.}~\bibnamefont {Katalev}}, \bibinfo {author} {\bibfnamefont
  {K.}~\bibnamefont {Kavanagh}}, \bibinfo {author} {\bibfnamefont {E.~T.}\
  \bibnamefont {Kennedy}}, \bibinfo {author} {\bibfnamefont {S.}~\bibnamefont
  {Khodyachykh}}, \bibinfo {author} {\bibfnamefont {K.}~\bibnamefont {Klose}},
  \bibinfo {author} {\bibfnamefont {V.}~\bibnamefont {Kocharyan}}, \bibinfo
  {author} {\bibfnamefont {M.}~\bibnamefont {K\"orfer}}, \bibinfo {author}
  {\bibfnamefont {M.}~\bibnamefont {Kollewe}}, \bibinfo {author} {\bibfnamefont
  {W.}~\bibnamefont {Koprek}}, \bibinfo {author} {\bibfnamefont
  {S.}~\bibnamefont {Korepanov}}, \bibinfo {author} {\bibfnamefont
  {D.}~\bibnamefont {Kostin}}, \bibinfo {author} {\bibfnamefont
  {M.}~\bibnamefont {Krassilnikov}}, \bibinfo {author} {\bibfnamefont
  {G.}~\bibnamefont {Kube}}, \bibinfo {author} {\bibfnamefont {M.}~\bibnamefont
  {Kuhlmann}}, \bibinfo {author} {\bibfnamefont {C.~L.~S.}\ \bibnamefont
  {Lewis}}, \bibinfo {author} {\bibfnamefont {L.}~\bibnamefont {Lilje}},
  \bibinfo {author} {\bibfnamefont {T.}~\bibnamefont {Limberg}}, \bibinfo
  {author} {\bibfnamefont {D.}~\bibnamefont {Lipka}}, \bibinfo {author}
  {\bibfnamefont {F.}~\bibnamefont {L\"ohl}}, \bibinfo {author} {\bibfnamefont
  {H.}~\bibnamefont {Luna}}, \bibinfo {author} {\bibfnamefont {M.}~\bibnamefont
  {Luong}}, \bibinfo {author} {\bibfnamefont {M.}~\bibnamefont {Martins}},
  \bibinfo {author} {\bibfnamefont {M.}~\bibnamefont {Meyer}}, \bibinfo
  {author} {\bibfnamefont {P.}~\bibnamefont {Michelato}}, \bibinfo {author}
  {\bibfnamefont {V.}~\bibnamefont {Miltchev}}, \bibinfo {author}
  {\bibfnamefont {W.~D.}\ \bibnamefont {M\"oller}}, \bibinfo {author}
  {\bibfnamefont {L.}~\bibnamefont {Monaco}}, \bibinfo {author} {\bibfnamefont
  {W.~F.~O.}\ \bibnamefont {M\"uller}}, \bibinfo {author} {\bibfnamefont
  {O.}~\bibnamefont {Napieralski}}, \bibinfo {author} {\bibfnamefont
  {O.}~\bibnamefont {Napoly}}, \bibinfo {author} {\bibfnamefont
  {P.}~\bibnamefont {Nicolosi}}, \bibinfo {author} {\bibfnamefont
  {D.}~\bibnamefont {N\"olle}}, \bibinfo {author} {\bibfnamefont
  {T.}~\bibnamefont {Nunez}}, \bibinfo {author} {\bibfnamefont
  {A.}~\bibnamefont {Oppelt}}, \bibinfo {author} {\bibfnamefont
  {C.}~\bibnamefont {Pagani}}, \bibinfo {author} {\bibfnamefont
  {R.}~\bibnamefont {Paparella}}, \bibinfo {author} {\bibfnamefont
  {N.}~\bibnamefont {Pchalek}}, \bibinfo {author} {\bibfnamefont
  {J.}~\bibnamefont {Pedregosa-Gutierrez}}, \bibinfo {author} {\bibfnamefont
  {B.}~\bibnamefont {Petersen}}, \bibinfo {author} {\bibfnamefont
  {B.}~\bibnamefont {Petrosyan}}, \bibinfo {author} {\bibfnamefont
  {G.}~\bibnamefont {Petrosyan}}, \bibinfo {author} {\bibfnamefont
  {L.}~\bibnamefont {Petrosyan}}, \bibinfo {author} {\bibfnamefont
  {J.}~\bibnamefont {Pfl\"uger}}, \bibinfo {author} {\bibfnamefont
  {E.}~\bibnamefont {Pl\"onjes}}, \bibinfo {author} {\bibfnamefont
  {L.}~\bibnamefont {Poletto}}, \bibinfo {author} {\bibfnamefont
  {K.}~\bibnamefont {Pozniak}}, \bibinfo {author} {\bibfnamefont
  {E.}~\bibnamefont {Prat}}, \bibinfo {author} {\bibfnamefont {D.}~\bibnamefont
  {Proch}}, \bibinfo {author} {\bibfnamefont {P.}~\bibnamefont {Pucyk}},
  \bibinfo {author} {\bibfnamefont {P.}~\bibnamefont {Radcliffe}}, \bibinfo
  {author} {\bibfnamefont {H.}~\bibnamefont {Redlin}}, \bibinfo {author}
  {\bibfnamefont {K.}~\bibnamefont {Rehlich}}, \bibinfo {author} {\bibfnamefont
  {M.}~\bibnamefont {Richter}}, \bibinfo {author} {\bibfnamefont
  {M.}~\bibnamefont {Roehrs}}, \bibinfo {author} {\bibfnamefont
  {J.}~\bibnamefont {Roensch}}, \bibinfo {author} {\bibfnamefont
  {R.}~\bibnamefont {Romaniuk}}, \bibinfo {author} {\bibfnamefont
  {M.}~\bibnamefont {Ross}}, \bibinfo {author} {\bibfnamefont {J.}~\bibnamefont
  {Rossbach}}, \bibinfo {author} {\bibfnamefont {V.}~\bibnamefont {Rybnikov}},
  \bibinfo {author} {\bibfnamefont {M.}~\bibnamefont {Sachwitz}}, \bibinfo
  {author} {\bibfnamefont {E.~L.}\ \bibnamefont {Saldin}}, \bibinfo {author}
  {\bibfnamefont {W.}~\bibnamefont {Sandner}}, \bibinfo {author} {\bibfnamefont
  {H.}~\bibnamefont {Schlarb}}, \bibinfo {author} {\bibfnamefont
  {B.}~\bibnamefont {Schmidt}}, \bibinfo {author} {\bibfnamefont
  {M.}~\bibnamefont {Schmitz}}, \bibinfo {author} {\bibfnamefont
  {P.}~\bibnamefont {Schm\"user}}, \bibinfo {author} {\bibfnamefont {J.~R.}\
  \bibnamefont {Schneider}}, \bibinfo {author} {\bibfnamefont {E.~A.}\
  \bibnamefont {Schneidmiller}}, \bibinfo {author} {\bibfnamefont
  {S.}~\bibnamefont {Schnepp}}, \bibinfo {author} {\bibfnamefont
  {S.}~\bibnamefont {Schreiber}}, \bibinfo {author} {\bibfnamefont
  {M.}~\bibnamefont {Seidel}}, \bibinfo {author} {\bibfnamefont
  {D.}~\bibnamefont {Sertore}}, \bibinfo {author} {\bibfnamefont {A.~V.}\
  \bibnamefont {Shabunov}}, \bibinfo {author} {\bibfnamefont {C.}~\bibnamefont
  {Simon}}, \bibinfo {author} {\bibfnamefont {S.}~\bibnamefont {Simrock}},
  \bibinfo {author} {\bibfnamefont {E.}~\bibnamefont {Sombrowski}}, \bibinfo
  {author} {\bibfnamefont {A.~A.}\ \bibnamefont {Sorokin}}, \bibinfo {author}
  {\bibfnamefont {P.}~\bibnamefont {Spanknebel}}, \bibinfo {author}
  {\bibfnamefont {R.}~\bibnamefont {Spesyvtsev}}, \bibinfo {author}
  {\bibfnamefont {L.}~\bibnamefont {Staykov}}, \bibinfo {author} {\bibfnamefont
  {B.}~\bibnamefont {Steffen}}, \bibinfo {author} {\bibfnamefont
  {F.}~\bibnamefont {Stephan}}, \bibinfo {author} {\bibfnamefont
  {F.}~\bibnamefont {Stulle}}, \bibinfo {author} {\bibfnamefont
  {H.}~\bibnamefont {Thom}}, \bibinfo {author} {\bibfnamefont {K.}~\bibnamefont
  {Tiedtke}}, \bibinfo {author} {\bibfnamefont {M.}~\bibnamefont {Tischer}},
  \bibinfo {author} {\bibfnamefont {S.}~\bibnamefont {Toleikis}}, \bibinfo
  {author} {\bibfnamefont {R.}~\bibnamefont {Treusch}}, \bibinfo {author}
  {\bibfnamefont {D.}~\bibnamefont {Trines}}, \bibinfo {author} {\bibfnamefont
  {I.}~\bibnamefont {Tsakov}}, \bibinfo {author} {\bibfnamefont
  {E.}~\bibnamefont {Vogel}}, \bibinfo {author} {\bibfnamefont
  {T.}~\bibnamefont {Weiland}}, \bibinfo {author} {\bibfnamefont
  {H.}~\bibnamefont {Weise}}, \bibinfo {author} {\bibfnamefont
  {M.}~\bibnamefont {Wellh\"ofer}}, \bibinfo {author} {\bibfnamefont
  {M.}~\bibnamefont {Wendt}}, \bibinfo {author} {\bibfnamefont
  {I.}~\bibnamefont {Will}}, \bibinfo {author} {\bibfnamefont {A.}~\bibnamefont
  {Winter}}, \bibinfo {author} {\bibfnamefont {K.}~\bibnamefont {Wittenburg}},
  \bibinfo {author} {\bibfnamefont {W.}~\bibnamefont {Wurth}}, \bibinfo
  {author} {\bibfnamefont {P.}~\bibnamefont {Yeates}}, \bibinfo {author}
  {\bibfnamefont {M.~V.}\ \bibnamefont {Yurkov}}, \bibinfo {author}
  {\bibfnamefont {I.}~\bibnamefont {Zagorodnov}}, \ and\ \bibinfo {author}
  {\bibfnamefont {K.}~\bibnamefont {Zapfe}}} (\bibinfo {year} {2007}),\
  \href@noop {} {\bibfield  {journal} {\bibinfo  {journal} {Nature Photon.}\
  }\textbf {\bibinfo {volume} {1}},\ \bibinfo {pages} {336}}\BibitemShut
  {NoStop}%
\bibitem [{\citenamefont {Akre}\ \emph {et~al.}(2002)\citenamefont {Akre},
  \citenamefont {Bentson}, \citenamefont {Emma},\ and\ \citenamefont
  {Krejcik}}]{Akre}%
  \BibitemOpen
  \bibfield  {author} {\bibinfo {author} {\bibnamefont {Akre}, \bibfnamefont
  {R.}}, \bibinfo {author} {\bibfnamefont {L.}~\bibnamefont {Bentson}},
  \bibinfo {author} {\bibfnamefont {P.}~\bibnamefont {Emma}}, \ and\ \bibinfo
  {author} {\bibfnamefont {P.}~\bibnamefont {Krejcik}}} (\bibinfo {year}
  {2002}),\ in\ \href@noop {} {\emph {\bibinfo {booktitle} {Proceedings of EPAC
  2002}}},\ p.\ \bibinfo {pages} {1882}\BibitemShut {NoStop}%
\bibitem [{\citenamefont {Allaria}\ \emph {et~al.}(2012)\citenamefont
  {Allaria}, \citenamefont {Appio}, \citenamefont {Badano}, \citenamefont
  {Barletta}, \citenamefont {Bassanese}, \citenamefont {Biedron}, \citenamefont
  {Borga}, \citenamefont {Busetto}, \citenamefont {Castronovo}, \citenamefont
  {Cinquegrana}, \citenamefont {Cleva}, \citenamefont {Cocco}, \citenamefont
  {Cornacchia}, \citenamefont {Craievich}, \citenamefont {Cudin}, \citenamefont
  {D'Auria}, \citenamefont {Forno}, \citenamefont {Danailov}, \citenamefont
  {Monte}, \citenamefont {Ninno}, \citenamefont {Delgiusto}, \citenamefont
  {Demidovich}, \citenamefont {Mitri}, \citenamefont {Diviacco}, \citenamefont
  {Fabris}, \citenamefont {Fabris}, \citenamefont {Fawley}, \citenamefont
  {Ferianis}, \citenamefont {Ferrari}, \citenamefont {Ferry}, \citenamefont
  {Froehlich}, \citenamefont {Furlan}, \citenamefont {Gaio}, \citenamefont
  {Gelmetti}, \citenamefont {Giannessi}, \citenamefont {Giannini},
  \citenamefont {Gobessi}, \citenamefont {Ivanov}, \citenamefont
  {Karantzoulis}, \citenamefont {Lonza}, \citenamefont {Lutman}, \citenamefont
  {Mahieu}, \citenamefont {Milloch}, \citenamefont {Milton}, \citenamefont
  {Musardo}, \citenamefont {Nikolov}, \citenamefont {Noe}, \citenamefont
  {Parmigiani}, \citenamefont {Penco}, \citenamefont {Petronio}, \citenamefont
  {Pivetta}, \citenamefont {Predonzani}, \citenamefont {Rossi}, \citenamefont
  {Rumiz}, \citenamefont {Salom}, \citenamefont {Scafuri}, \citenamefont
  {Serpico}, \citenamefont {Sigalotti}, \citenamefont {Spampinati},
  \citenamefont {Spezzani}, \citenamefont {Svandrlik}, \citenamefont {Svetina},
  \citenamefont {Tazzari}, \citenamefont {Trovo}, \citenamefont {Umer},
  \citenamefont {Vascotto}, \citenamefont {Veronese}, \citenamefont
  {Visintini}, \citenamefont {Zaccaria}, \citenamefont {Zangrando},\ and\
  \citenamefont {Zangrando}}]{FERMI}%
  \BibitemOpen
  \bibfield  {author} {\bibinfo {author} {\bibnamefont {Allaria}, \bibfnamefont
  {E.}}, \bibinfo {author} {\bibfnamefont {R.}~\bibnamefont {Appio}}, \bibinfo
  {author} {\bibfnamefont {L.}~\bibnamefont {Badano}}, \bibinfo {author}
  {\bibfnamefont {W.}~\bibnamefont {Barletta}}, \bibinfo {author}
  {\bibfnamefont {S.}~\bibnamefont {Bassanese}}, \bibinfo {author}
  {\bibfnamefont {S.}~\bibnamefont {Biedron}}, \bibinfo {author} {\bibfnamefont
  {A.}~\bibnamefont {Borga}}, \bibinfo {author} {\bibfnamefont
  {E.}~\bibnamefont {Busetto}}, \bibinfo {author} {\bibfnamefont
  {D.}~\bibnamefont {Castronovo}}, \bibinfo {author} {\bibfnamefont
  {P.}~\bibnamefont {Cinquegrana}}, \bibinfo {author} {\bibfnamefont
  {S.}~\bibnamefont {Cleva}}, \bibinfo {author} {\bibfnamefont
  {D.}~\bibnamefont {Cocco}}, \bibinfo {author} {\bibfnamefont
  {M.}~\bibnamefont {Cornacchia}}, \bibinfo {author} {\bibfnamefont
  {P.}~\bibnamefont {Craievich}}, \bibinfo {author} {\bibfnamefont
  {I.}~\bibnamefont {Cudin}}, \bibinfo {author} {\bibnamefont {D'Auria}},
  \bibinfo {author} {\bibfnamefont {M.~D.}\ \bibnamefont {Forno}}, \bibinfo
  {author} {\bibfnamefont {M.}~\bibnamefont {Danailov}}, \bibinfo {author}
  {\bibfnamefont {R.~D.}\ \bibnamefont {Monte}}, \bibinfo {author}
  {\bibfnamefont {G.~D.}\ \bibnamefont {Ninno}}, \bibinfo {author}
  {\bibfnamefont {P.}~\bibnamefont {Delgiusto}}, \bibinfo {author}
  {\bibfnamefont {A.}~\bibnamefont {Demidovich}}, \bibinfo {author}
  {\bibfnamefont {S.~D.}\ \bibnamefont {Mitri}}, \bibinfo {author}
  {\bibfnamefont {B.}~\bibnamefont {Diviacco}}, \bibinfo {author}
  {\bibfnamefont {A.}~\bibnamefont {Fabris}}, \bibinfo {author} {\bibfnamefont
  {R.}~\bibnamefont {Fabris}}, \bibinfo {author} {\bibfnamefont
  {W.}~\bibnamefont {Fawley}}, \bibinfo {author} {\bibfnamefont
  {M.}~\bibnamefont {Ferianis}}, \bibinfo {author} {\bibfnamefont
  {E.}~\bibnamefont {Ferrari}}, \bibinfo {author} {\bibfnamefont
  {S.}~\bibnamefont {Ferry}}, \bibinfo {author} {\bibfnamefont
  {L.}~\bibnamefont {Froehlich}}, \bibinfo {author} {\bibfnamefont
  {P.}~\bibnamefont {Furlan}}, \bibinfo {author} {\bibfnamefont
  {G.}~\bibnamefont {Gaio}}, \bibinfo {author} {\bibfnamefont {F.}~\bibnamefont
  {Gelmetti}}, \bibinfo {author} {\bibfnamefont {L.}~\bibnamefont {Giannessi}},
  \bibinfo {author} {\bibfnamefont {M.}~\bibnamefont {Giannini}}, \bibinfo
  {author} {\bibfnamefont {R.}~\bibnamefont {Gobessi}}, \bibinfo {author}
  {\bibfnamefont {R.}~\bibnamefont {Ivanov}}, \bibinfo {author} {\bibfnamefont
  {E.}~\bibnamefont {Karantzoulis}}, \bibinfo {author} {\bibfnamefont
  {M.}~\bibnamefont {Lonza}}, \bibinfo {author} {\bibfnamefont
  {A.}~\bibnamefont {Lutman}}, \bibinfo {author} {\bibfnamefont
  {B.}~\bibnamefont {Mahieu}}, \bibinfo {author} {\bibfnamefont
  {M.}~\bibnamefont {Milloch}}, \bibinfo {author} {\bibfnamefont
  {S.}~\bibnamefont {Milton}}, \bibinfo {author} {\bibfnamefont
  {M.}~\bibnamefont {Musardo}}, \bibinfo {author} {\bibfnamefont
  {I.}~\bibnamefont {Nikolov}}, \bibinfo {author} {\bibfnamefont
  {S.}~\bibnamefont {Noe}}, \bibinfo {author} {\bibfnamefont {F.}~\bibnamefont
  {Parmigiani}}, \bibinfo {author} {\bibfnamefont {G.}~\bibnamefont {Penco}},
  \bibinfo {author} {\bibfnamefont {M.}~\bibnamefont {Petronio}}, \bibinfo
  {author} {\bibfnamefont {L.}~\bibnamefont {Pivetta}}, \bibinfo {author}
  {\bibfnamefont {M.}~\bibnamefont {Predonzani}}, \bibinfo {author}
  {\bibfnamefont {F.}~\bibnamefont {Rossi}}, \bibinfo {author} {\bibfnamefont
  {L.}~\bibnamefont {Rumiz}}, \bibinfo {author} {\bibfnamefont
  {A.}~\bibnamefont {Salom}}, \bibinfo {author} {\bibfnamefont
  {C.}~\bibnamefont {Scafuri}}, \bibinfo {author} {\bibfnamefont
  {C.}~\bibnamefont {Serpico}}, \bibinfo {author} {\bibfnamefont
  {P.}~\bibnamefont {Sigalotti}}, \bibinfo {author} {\bibfnamefont
  {S.}~\bibnamefont {Spampinati}}, \bibinfo {author} {\bibfnamefont
  {C.}~\bibnamefont {Spezzani}}, \bibinfo {author} {\bibfnamefont
  {M.}~\bibnamefont {Svandrlik}}, \bibinfo {author} {\bibfnamefont
  {C.}~\bibnamefont {Svetina}}, \bibinfo {author} {\bibfnamefont
  {S.}~\bibnamefont {Tazzari}}, \bibinfo {author} {\bibfnamefont
  {M.}~\bibnamefont {Trovo}}, \bibinfo {author} {\bibfnamefont
  {R.}~\bibnamefont {Umer}}, \bibinfo {author} {\bibfnamefont {A.}~\bibnamefont
  {Vascotto}}, \bibinfo {author} {\bibfnamefont {M.}~\bibnamefont {Veronese}},
  \bibinfo {author} {\bibfnamefont {R.}~\bibnamefont {Visintini}}, \bibinfo
  {author} {\bibfnamefont {M.}~\bibnamefont {Zaccaria}}, \bibinfo {author}
  {\bibfnamefont {D.}~\bibnamefont {Zangrando}}, \ and\ \bibinfo {author}
  {\bibfnamefont {M.}~\bibnamefont {Zangrando}}} (\bibinfo {year} {2012}),\
  \href@noop {} {\bibfield  {journal} {\bibinfo  {journal} {Nature Photon.}\
  }\textbf {\bibinfo {volume} {6}},\ \bibinfo {pages} {699}}\BibitemShut
  {NoStop}%
\bibitem [{\citenamefont {Allaria}\ \emph {et~al.}(2013)\citenamefont
  {Allaria}, \citenamefont {Castronovo}, \citenamefont {Cinquegrana},
  \citenamefont {Craievich}, \citenamefont {Dal~Forno}, \citenamefont
  {Danailov}, \citenamefont {D'Auria}, \citenamefont {Demidovich},
  \citenamefont {De~Ninno}, \citenamefont {Di~Mitri}, \citenamefont {Diviacco},
  \citenamefont {Fawley}, \citenamefont {Ferianis}, \citenamefont {Ferrari},
  \citenamefont {Froehlich}, \citenamefont {Gaio}, \citenamefont {Gauthier},
  \citenamefont {Giannessi}, \citenamefont {Ivanov}, \citenamefont {Mahieu},
  \citenamefont {Mahne}, \citenamefont {Nikolov}, \citenamefont {Parmigiani},
  \citenamefont {Penco}, \citenamefont {Raimondi}, \citenamefont {Scafuri},
  \citenamefont {Serpico}, \citenamefont {Sigalotti}, \citenamefont
  {Spampinati}, \citenamefont {Spezzani}, \citenamefont {Svandrlik},
  \citenamefont {Svetina}, \citenamefont {Trovo}, \citenamefont {Veronese},
  \citenamefont {Zangrando},\ and\ \citenamefont {Zangrando}}]{FERMI2stage}%
  \BibitemOpen
  \bibfield  {author} {\bibinfo {author} {\bibnamefont {Allaria}, \bibfnamefont
  {E.}}, \bibinfo {author} {\bibfnamefont {D.}~\bibnamefont {Castronovo}},
  \bibinfo {author} {\bibfnamefont {P.}~\bibnamefont {Cinquegrana}}, \bibinfo
  {author} {\bibfnamefont {P.}~\bibnamefont {Craievich}}, \bibinfo {author}
  {\bibfnamefont {M.}~\bibnamefont {Dal~Forno}}, \bibinfo {author}
  {\bibfnamefont {M.}~\bibnamefont {Danailov}}, \bibinfo {author}
  {\bibfnamefont {G.}~\bibnamefont {D'Auria}}, \bibinfo {author} {\bibfnamefont
  {A.}~\bibnamefont {Demidovich}}, \bibinfo {author} {\bibfnamefont
  {G.}~\bibnamefont {De~Ninno}}, \bibinfo {author} {\bibfnamefont
  {S.}~\bibnamefont {Di~Mitri}}, \bibinfo {author} {\bibfnamefont
  {B.}~\bibnamefont {Diviacco}}, \bibinfo {author} {\bibfnamefont
  {W.}~\bibnamefont {Fawley}}, \bibinfo {author} {\bibfnamefont
  {M.}~\bibnamefont {Ferianis}}, \bibinfo {author} {\bibfnamefont
  {E.}~\bibnamefont {Ferrari}}, \bibinfo {author} {\bibfnamefont
  {L.}~\bibnamefont {Froehlich}}, \bibinfo {author} {\bibfnamefont
  {G.}~\bibnamefont {Gaio}}, \bibinfo {author} {\bibfnamefont {D.}~\bibnamefont
  {Gauthier}}, \bibinfo {author} {\bibfnamefont {L.}~\bibnamefont {Giannessi}},
  \bibinfo {author} {\bibfnamefont {R.}~\bibnamefont {Ivanov}}, \bibinfo
  {author} {\bibfnamefont {B.}~\bibnamefont {Mahieu}}, \bibinfo {author}
  {\bibfnamefont {N.}~\bibnamefont {Mahne}}, \bibinfo {author} {\bibfnamefont
  {I.}~\bibnamefont {Nikolov}}, \bibinfo {author} {\bibfnamefont
  {F.}~\bibnamefont {Parmigiani}}, \bibinfo {author} {\bibfnamefont
  {G.}~\bibnamefont {Penco}}, \bibinfo {author} {\bibfnamefont
  {L.}~\bibnamefont {Raimondi}}, \bibinfo {author} {\bibfnamefont
  {C.}~\bibnamefont {Scafuri}}, \bibinfo {author} {\bibfnamefont
  {C.}~\bibnamefont {Serpico}}, \bibinfo {author} {\bibfnamefont
  {P.}~\bibnamefont {Sigalotti}}, \bibinfo {author} {\bibfnamefont
  {S.}~\bibnamefont {Spampinati}}, \bibinfo {author} {\bibfnamefont
  {C.}~\bibnamefont {Spezzani}}, \bibinfo {author} {\bibfnamefont
  {M.}~\bibnamefont {Svandrlik}}, \bibinfo {author} {\bibfnamefont
  {C.}~\bibnamefont {Svetina}}, \bibinfo {author} {\bibfnamefont
  {M.}~\bibnamefont {Trovo}}, \bibinfo {author} {\bibfnamefont
  {M.}~\bibnamefont {Veronese}}, \bibinfo {author} {\bibfnamefont
  {D.}~\bibnamefont {Zangrando}}, \ and\ \bibinfo {author} {\bibfnamefont
  {M.}~\bibnamefont {Zangrando}}} (\bibinfo {year} {2013}),\ \href@noop {}
  {\bibfield  {journal} {\bibinfo  {journal} {Nature Photon.}\ }\textbf
  {\bibinfo {volume} {7}},\ \bibinfo {pages} {913}}\BibitemShut {NoStop}%
\bibitem [{\citenamefont {Allaria}\ and\ \citenamefont
  {De~Ninno}(2007)}]{HGHGS2}%
  \BibitemOpen
  \bibfield  {author} {\bibinfo {author} {\bibnamefont {Allaria}, \bibfnamefont
  {E.}}, \ and\ \bibinfo {author} {\bibfnamefont {G.}~\bibnamefont {De~Ninno}}}
  (\bibinfo {year} {2007}),\ \href@noop {} {\bibfield  {journal} {\bibinfo
  {journal} {Phys. Rev. Lett.}\ }\textbf {\bibinfo {volume} {99}},\ \bibinfo
  {pages} {014801}}\BibitemShut {NoStop}%
\bibitem [{\citenamefont {Allen}\ \emph {et~al.}(2003)\citenamefont {Allen},
  \citenamefont {Barnett},\ and\ \citenamefont {Padgett}}]{AllenBook}%
  \BibitemOpen
  \bibfield  {author} {\bibinfo {author} {\bibnamefont {Allen}, \bibfnamefont
  {L.}}, \bibinfo {author} {\bibfnamefont {S.~M.}\ \bibnamefont {Barnett}}, \
  and\ \bibinfo {author} {\bibfnamefont {M.~J.}\ \bibnamefont {Padgett}}}
  (\bibinfo {year} {2003}),\ \href@noop {} {\emph {\bibinfo {title} {Optical
  angular momentum}}}\ (\bibinfo  {publisher} {Institute of Physics
  Pub.})\BibitemShut {NoStop}%
\bibitem [{\citenamefont {Allen}\ \emph {et~al.}(1992)\citenamefont {Allen},
  \citenamefont {Beijersbergen}, \citenamefont {Spreeuw},\ and\ \citenamefont
  {Woerdman}}]{Allen}%
  \BibitemOpen
  \bibfield  {author} {\bibinfo {author} {\bibnamefont {Allen}, \bibfnamefont
  {L.}}, \bibinfo {author} {\bibfnamefont {M.~W.}\ \bibnamefont
  {Beijersbergen}}, \bibinfo {author} {\bibfnamefont {R.~J.~C.}\ \bibnamefont
  {Spreeuw}}, \ and\ \bibinfo {author} {\bibfnamefont {J.~P.}\ \bibnamefont
  {Woerdman}}} (\bibinfo {year} {1992}),\ \href
  {http://dx.doi.org/10.1103/PhysRevA.45.8185} {\bibfield  {journal} {\bibinfo
  {journal} {Phys. Rev. A}\ }\textbf {\bibinfo {volume} {45}}~(\bibinfo
  {number} {11}),\ \bibinfo {pages} {8185}}\BibitemShut {NoStop}%
\bibitem [{\citenamefont {Amann}\ \emph {et~al.}(2012)\citenamefont {Amann},
  \citenamefont {Berg}, \citenamefont {Blank}, \citenamefont {Decker},
  \citenamefont {Ding}, \citenamefont {Emma}, \citenamefont {Feng},
  \citenamefont {Frisch}, \citenamefont {Fritz}, \citenamefont {Hastings},
  \citenamefont {Huang}, \citenamefont {Krzywinski}, \citenamefont {Lindberg},
  \citenamefont {Loos}, \citenamefont {Lutman}, \citenamefont {Nuhn},
  \citenamefont {Ratner}, \citenamefont {Rzepiela}, \citenamefont {Shu},
  \citenamefont {Shvyd'ko}, \citenamefont {Spampinati}, \citenamefont
  {Stoupin}, \citenamefont {Terentyev}, \citenamefont {Trakhtenberg},
  \citenamefont {Walz}, \citenamefont {Welch}, \citenamefont {Wu},
  \citenamefont {Zholents},\ and\ \citenamefont {Zhu}}]{2012NaPho...6..693A}%
  \BibitemOpen
  \bibfield  {author} {\bibinfo {author} {\bibnamefont {Amann}, \bibfnamefont
  {J.}}, \bibinfo {author} {\bibfnamefont {W.}~\bibnamefont {Berg}}, \bibinfo
  {author} {\bibfnamefont {V.}~\bibnamefont {Blank}}, \bibinfo {author}
  {\bibfnamefont {F.-J.}\ \bibnamefont {Decker}}, \bibinfo {author}
  {\bibfnamefont {Y.}~\bibnamefont {Ding}}, \bibinfo {author} {\bibfnamefont
  {P.}~\bibnamefont {Emma}}, \bibinfo {author} {\bibfnamefont {Y.}~\bibnamefont
  {Feng}}, \bibinfo {author} {\bibfnamefont {J.}~\bibnamefont {Frisch}},
  \bibinfo {author} {\bibfnamefont {D.}~\bibnamefont {Fritz}}, \bibinfo
  {author} {\bibfnamefont {J.}~\bibnamefont {Hastings}}, \bibinfo {author}
  {\bibfnamefont {Z.}~\bibnamefont {Huang}}, \bibinfo {author} {\bibfnamefont
  {J.}~\bibnamefont {Krzywinski}}, \bibinfo {author} {\bibfnamefont
  {R.}~\bibnamefont {Lindberg}}, \bibinfo {author} {\bibfnamefont
  {H.}~\bibnamefont {Loos}}, \bibinfo {author} {\bibfnamefont {A.}~\bibnamefont
  {Lutman}}, \bibinfo {author} {\bibfnamefont {H.-D.}\ \bibnamefont {Nuhn}},
  \bibinfo {author} {\bibfnamefont {D.}~\bibnamefont {Ratner}}, \bibinfo
  {author} {\bibfnamefont {J.}~\bibnamefont {Rzepiela}}, \bibinfo {author}
  {\bibfnamefont {D.}~\bibnamefont {Shu}}, \bibinfo {author} {\bibfnamefont
  {Y.}~\bibnamefont {Shvyd'ko}}, \bibinfo {author} {\bibfnamefont
  {S.}~\bibnamefont {Spampinati}}, \bibinfo {author} {\bibfnamefont
  {S.}~\bibnamefont {Stoupin}}, \bibinfo {author} {\bibfnamefont
  {S.}~\bibnamefont {Terentyev}}, \bibinfo {author} {\bibfnamefont
  {E.}~\bibnamefont {Trakhtenberg}}, \bibinfo {author} {\bibfnamefont
  {D.}~\bibnamefont {Walz}}, \bibinfo {author} {\bibfnamefont {J.}~\bibnamefont
  {Welch}}, \bibinfo {author} {\bibfnamefont {J.}~\bibnamefont {Wu}}, \bibinfo
  {author} {\bibfnamefont {A.}~\bibnamefont {Zholents}}, \ and\ \bibinfo
  {author} {\bibfnamefont {D.}~\bibnamefont {Zhu}}} (\bibinfo {year} {2012}),\
  \href@noop {} {\bibfield  {journal} {\bibinfo  {journal} {Nature Photon.}\
  }\textbf {\bibinfo {volume} {6}},\ \bibinfo {pages} {693}}\BibitemShut
  {NoStop}%
\bibitem [{\citenamefont {Andonian}\ \emph {et~al.}(2011)\citenamefont
  {Andonian}, \citenamefont {Hemsing}, \citenamefont {Xiang}, \citenamefont
  {Musumeci}, \citenamefont {Murokh}, \citenamefont {Tochitsky},\ and\
  \citenamefont {Rosenzweig}}]{OO}%
  \BibitemOpen
  \bibfield  {author} {\bibinfo {author} {\bibnamefont {Andonian},
  \bibfnamefont {G.}}, \bibinfo {author} {\bibfnamefont {E.}~\bibnamefont
  {Hemsing}}, \bibinfo {author} {\bibfnamefont {D.}~\bibnamefont {Xiang}},
  \bibinfo {author} {\bibfnamefont {P.}~\bibnamefont {Musumeci}}, \bibinfo
  {author} {\bibfnamefont {A.}~\bibnamefont {Murokh}}, \bibinfo {author}
  {\bibfnamefont {S.}~\bibnamefont {Tochitsky}}, \ and\ \bibinfo {author}
  {\bibfnamefont {J.~B.}\ \bibnamefont {Rosenzweig}}} (\bibinfo {year}
  {2011}),\ \href {http://link.aps.org/doi/10.1103/PhysRevSTAB.14.072802}
  {\bibfield  {journal} {\bibinfo  {journal} {Phys. Rev. ST Accel. Beams}\
  }\textbf {\bibinfo {volume} {14}},\ \bibinfo {pages} {072802}}\BibitemShut
  {NoStop}%
\bibitem [{\citenamefont {Angelova}\ \emph {et~al.}(2008)\citenamefont
  {Angelova}, \citenamefont {Ziemann}, \citenamefont {Dohlus},\ and\
  \citenamefont {Kot}}]{laser_heater_XFEL}%
  \BibitemOpen
  \bibfield  {author} {\bibinfo {author} {\bibnamefont {Angelova},
  \bibfnamefont {G.}}, \bibinfo {author} {\bibfnamefont {V.}~\bibnamefont
  {Ziemann}}, \bibinfo {author} {\bibfnamefont {M.}~\bibnamefont {Dohlus}}, \
  and\ \bibinfo {author} {\bibfnamefont {Y.}~\bibnamefont {Kot}}} (\bibinfo
  {year} {2008}),\ in\ \href@noop {} {\emph {\bibinfo {booktitle} {2008
  European Particle Accelerator Conference}}},\ p.\ \bibinfo {pages}
  {2695}\BibitemShut {NoStop}%
\bibitem [{\citenamefont {Attwood}(2000)}]{attwood}%
  \BibitemOpen
  \bibfield  {author} {\bibinfo {author} {\bibnamefont {Attwood}, \bibfnamefont
  {D.}}} (\bibinfo {year} {2000}),\ \href@noop {} {\emph {\bibinfo {title}
  {Soft X-rays and Extreme Ultraviolet Radiation: Principles and
  Applications}}}\ (\bibinfo  {publisher} {Cambridge University
  Press})\BibitemShut {NoStop}%
\bibitem [{\citenamefont {Bahrdt}\ \emph {et~al.}(2013)\citenamefont {Bahrdt},
  \citenamefont {Holldack}, \citenamefont {Kuske}, \citenamefont {M\"uller},
  \citenamefont {Scheer},\ and\ \citenamefont
  {Schmid}}]{PhysRevLett.111.034801}%
  \BibitemOpen
  \bibfield  {author} {\bibinfo {author} {\bibnamefont {Bahrdt}, \bibfnamefont
  {J.}}, \bibinfo {author} {\bibfnamefont {K.}~\bibnamefont {Holldack}},
  \bibinfo {author} {\bibfnamefont {P.}~\bibnamefont {Kuske}}, \bibinfo
  {author} {\bibfnamefont {R.}~\bibnamefont {M\"uller}}, \bibinfo {author}
  {\bibfnamefont {M.}~\bibnamefont {Scheer}}, \ and\ \bibinfo {author}
  {\bibfnamefont {P.}~\bibnamefont {Schmid}}} (\bibinfo {year} {2013}),\ \href
  {http://link.aps.org/doi/10.1103/PhysRevLett.111.034801} {\bibfield
  {journal} {\bibinfo  {journal} {Phys. Rev. Lett.}\ }\textbf {\bibinfo
  {volume} {111}},\ \bibinfo {pages} {034801}}\BibitemShut {NoStop}%
\bibitem [{\citenamefont {Baltu\v{s}ka}\ \emph {et~al.}(1998)\citenamefont
  {Baltu\v{s}ka}, \citenamefont {Pshenichnikov},\ and\ \citenamefont
  {Wiersma}}]{4fslaser}%
  \BibitemOpen
  \bibfield  {author} {\bibinfo {author} {\bibnamefont {Baltu\v{s}ka},
  \bibfnamefont {A.}}, \bibinfo {author} {\bibfnamefont {M.~S.}\ \bibnamefont
  {Pshenichnikov}}, \ and\ \bibinfo {author} {\bibfnamefont {D.~A.}\
  \bibnamefont {Wiersma}}} (\bibinfo {year} {1998}),\ \href
  {http://ol.osa.org/abstract.cfm?URI=ol-23-18-1474} {\bibfield  {journal}
  {\bibinfo  {journal} {Opt. Lett.}\ }\textbf {\bibinfo {volume}
  {23}}~(\bibinfo {number} {18}),\ \bibinfo {pages} {1474}}\BibitemShut
  {NoStop}%
\bibitem [{\citenamefont {Bane}\ and\ \citenamefont {Wilson}(1977)}]{Bane}%
  \BibitemOpen
  \bibfield  {author} {\bibinfo {author} {\bibnamefont {Bane}, \bibfnamefont
  {K.}}, \ and\ \bibinfo {author} {\bibfnamefont {P.~B.}\ \bibnamefont
  {Wilson}}} (\bibinfo {year} {1977}),\ \href@noop {} {\bibfield  {journal}
  {\bibinfo  {journal} {IEEE Trans. Nucl. Sci}\ }\textbf {\bibinfo {volume}
  {24}},\ \bibinfo {pages} {1977}}\BibitemShut {NoStop}%
\bibitem [{\citenamefont {Beaud}\ \emph {et~al.}(2007)\citenamefont {Beaud},
  \citenamefont {Johnson}, \citenamefont {Streun}, \citenamefont {Abela},
  \citenamefont {Abramsohn}, \citenamefont {Grolimund}, \citenamefont
  {Krasniqi}, \citenamefont {Schmidt}, \citenamefont {Schlott},\ and\
  \citenamefont {Ingold}}]{laser-slicingSLS}%
  \BibitemOpen
  \bibfield  {author} {\bibinfo {author} {\bibnamefont {Beaud}, \bibfnamefont
  {P.}}, \bibinfo {author} {\bibfnamefont {S.~L.}\ \bibnamefont {Johnson}},
  \bibinfo {author} {\bibfnamefont {A.}~\bibnamefont {Streun}}, \bibinfo
  {author} {\bibfnamefont {R.}~\bibnamefont {Abela}}, \bibinfo {author}
  {\bibfnamefont {D.}~\bibnamefont {Abramsohn}}, \bibinfo {author}
  {\bibfnamefont {D.}~\bibnamefont {Grolimund}}, \bibinfo {author}
  {\bibfnamefont {F.}~\bibnamefont {Krasniqi}}, \bibinfo {author}
  {\bibfnamefont {T.}~\bibnamefont {Schmidt}}, \bibinfo {author} {\bibfnamefont
  {V.}~\bibnamefont {Schlott}}, \ and\ \bibinfo {author} {\bibfnamefont
  {G.}~\bibnamefont {Ingold}}} (\bibinfo {year} {2007}),\ \href@noop {}
  {\bibfield  {journal} {\bibinfo  {journal} {Phys. Rev. Lett.}\ }\textbf
  {\bibinfo {volume} {99}},\ \bibinfo {pages} {174801}}\BibitemShut {NoStop}%
\bibitem [{\citenamefont {Ben-Zvi}\ \emph {et~al.}(1991)\citenamefont
  {Ben-Zvi}, \citenamefont {Mauro}, \citenamefont {Krinsky}, \citenamefont
  {White},\ and\ \citenamefont {Yu}}]{BenZvi1991181}%
  \BibitemOpen
  \bibfield  {author} {\bibinfo {author} {\bibnamefont {Ben-Zvi}, \bibfnamefont
  {I.}}, \bibinfo {author} {\bibfnamefont {L.~D.}\ \bibnamefont {Mauro}},
  \bibinfo {author} {\bibfnamefont {S.}~\bibnamefont {Krinsky}}, \bibinfo
  {author} {\bibfnamefont {M.}~\bibnamefont {White}}, \ and\ \bibinfo {author}
  {\bibfnamefont {L.}~\bibnamefont {Yu}}} (\bibinfo {year} {1991}),\ \href
  {http://dx.doi.org/10.1016/0168-9002(91)90845-H} {\bibfield  {journal}
  {\bibinfo  {journal} {Nucl. Instrum. Methods Phys. Res. A}\ }\textbf
  {\bibinfo {volume} {304}}~(\bibinfo {number} {1-€"3}),\ \bibinfo {pages}
  {181}}\BibitemShut {NoStop}%
\bibitem [{\citenamefont {Bencivenga}\ \emph {et~al.}(2013)\citenamefont
  {Bencivenga}, \citenamefont {Raimondi}, \citenamefont {Svetina},\ and\
  \citenamefont {Masciovecchio}}]{Bencivenga}%
  \BibitemOpen
  \bibfield  {author} {\bibinfo {author} {\bibnamefont {Bencivenga},
  \bibfnamefont {F.}}, \bibinfo {author} {\bibfnamefont {L.}~\bibnamefont
  {Raimondi}}, \bibinfo {author} {\bibfnamefont {C.}~\bibnamefont {Svetina}}, \
  and\ \bibinfo {author} {\bibfnamefont {C.}~\bibnamefont {Masciovecchio}}}
  (\bibinfo {year} {2013}),\ \href@noop {} {\bibfield  {journal} {\bibinfo
  {journal} {Proc. SPIE}\ }\textbf {\bibinfo {volume} {8778}},\ \bibinfo
  {pages} {877807}}\BibitemShut {NoStop}%
\bibitem [{\citenamefont {Biedron}\ \emph {et~al.}(2001)\citenamefont
  {Biedron}, \citenamefont {Milton},\ and\ \citenamefont {Freund}}]{MC1}%
  \BibitemOpen
  \bibfield  {author} {\bibinfo {author} {\bibnamefont {Biedron}, \bibfnamefont
  {S.}}, \bibinfo {author} {\bibfnamefont {S.}~\bibnamefont {Milton}}, \ and\
  \bibinfo {author} {\bibfnamefont {H.}~\bibnamefont {Freund}}} (\bibinfo
  {year} {2001}),\ \href@noop {} {\bibfield  {journal} {\bibinfo  {journal}
  {Nucl. Instrum. Methods Phys. Res. A}\ }\textbf {\bibinfo {volume} {475}},\
  \bibinfo {pages} {401}}\BibitemShut {NoStop}%
\bibitem [{\citenamefont {Bielawski}\ \emph {et~al.}(2008)\citenamefont
  {Bielawski}, \citenamefont {Evain}, \citenamefont {Hara}, \citenamefont
  {Hosaka}, \citenamefont {Katoh}, \citenamefont {Kimura}, \citenamefont
  {Mochihashi}, \citenamefont {Shimada}, \citenamefont {Szwaj}, \citenamefont
  {Takahashi},\ and\ \citenamefont {Takashima}}]{Bielawski}%
  \BibitemOpen
  \bibfield  {author} {\bibinfo {author} {\bibnamefont {Bielawski},
  \bibfnamefont {S.}}, \bibinfo {author} {\bibfnamefont {C.}~\bibnamefont
  {Evain}}, \bibinfo {author} {\bibfnamefont {T.}~\bibnamefont {Hara}},
  \bibinfo {author} {\bibfnamefont {M.}~\bibnamefont {Hosaka}}, \bibinfo
  {author} {\bibfnamefont {M.}~\bibnamefont {Katoh}}, \bibinfo {author}
  {\bibfnamefont {S.}~\bibnamefont {Kimura}}, \bibinfo {author} {\bibfnamefont
  {A.}~\bibnamefont {Mochihashi}}, \bibinfo {author} {\bibfnamefont
  {M.}~\bibnamefont {Shimada}}, \bibinfo {author} {\bibfnamefont
  {C.}~\bibnamefont {Szwaj}}, \bibinfo {author} {\bibfnamefont
  {T.}~\bibnamefont {Takahashi}}, \ and\ \bibinfo {author} {\bibfnamefont
  {Y.}~\bibnamefont {Takashima}}} (\bibinfo {year} {2008}),\ \href@noop {}
  {\bibfield  {journal} {\bibinfo  {journal} {Nature Phys.}\ }\textbf {\bibinfo
  {volume} {4}},\ \bibinfo {pages} {390}}\BibitemShut {NoStop}%
\bibitem [{\citenamefont {Bonifacio}\ \emph {et~al.}(1994)\citenamefont
  {Bonifacio}, \citenamefont {De~Salvo}, \citenamefont {Pierini}, \citenamefont
  {Piovella},\ and\ \citenamefont {Pellegrini}}]{PhysRevLett.73.70}%
  \BibitemOpen
  \bibfield  {author} {\bibinfo {author} {\bibnamefont {Bonifacio},
  \bibfnamefont {R.}}, \bibinfo {author} {\bibfnamefont {L.}~\bibnamefont
  {De~Salvo}}, \bibinfo {author} {\bibfnamefont {P.}~\bibnamefont {Pierini}},
  \bibinfo {author} {\bibfnamefont {N.}~\bibnamefont {Piovella}}, \ and\
  \bibinfo {author} {\bibfnamefont {C.}~\bibnamefont {Pellegrini}}} (\bibinfo
  {year} {1994}),\ \href {http://link.aps.org/doi/10.1103/PhysRevLett.73.70}
  {\bibfield  {journal} {\bibinfo  {journal} {Phys. Rev. Lett.}\ }\textbf
  {\bibinfo {volume} {73}},\ \bibinfo {pages} {70}}\BibitemShut {NoStop}%
\bibitem [{\citenamefont {{Bonifacio}}\ \emph {et~al.}(1984)\citenamefont
  {{Bonifacio}}, \citenamefont {{Pellegrini}},\ and\ \citenamefont
  {{Narducci}}}]{Bonifacio}%
  \BibitemOpen
  \bibfield  {author} {\bibinfo {author} {\bibnamefont {{Bonifacio}},
  \bibfnamefont {R.}}, \bibinfo {author} {\bibfnamefont {C.}~\bibnamefont
  {{Pellegrini}}}, \ and\ \bibinfo {author} {\bibfnamefont {L.~M.}\
  \bibnamefont {{Narducci}}}} (\bibinfo {year} {1984}),\ \href@noop {}
  {\bibfield  {journal} {\bibinfo  {journal} {Opt. Commun.}\ }\textbf {\bibinfo
  {volume} {50}},\ \bibinfo {pages} {373}}\BibitemShut {NoStop}%
\bibitem [{\citenamefont {Bonifacio}\ \emph {et~al.}(2007)\citenamefont
  {Bonifacio}, \citenamefont {Piovella}, \citenamefont {Cola},\ and\
  \citenamefont {Volpe}}]{Bonifacio2007745}%
  \BibitemOpen
  \bibfield  {author} {\bibinfo {author} {\bibnamefont {Bonifacio},
  \bibfnamefont {R.}}, \bibinfo {author} {\bibfnamefont {N.}~\bibnamefont
  {Piovella}}, \bibinfo {author} {\bibfnamefont {M.}~\bibnamefont {Cola}}, \
  and\ \bibinfo {author} {\bibfnamefont {L.}~\bibnamefont {Volpe}}} (\bibinfo
  {year} {2007}),\ \href
  {http://www.sciencedirect.com/science/article/pii/S0168900207005360}
  {\bibfield  {journal} {\bibinfo  {journal} {Nuclear Instruments and Methods
  in Physics Research Section A: Accelerators, Spectrometers, Detectors and
  Associated Equipment}\ }\textbf {\bibinfo {volume} {577}}~(\bibinfo {number}
  {3}),\ \bibinfo {pages} {745 }}\BibitemShut {NoStop}%
\bibitem [{\citenamefont {Bonifacio}\ \emph {et~al.}(2011)\citenamefont
  {Bonifacio}, \citenamefont {Robb},\ and\ \citenamefont
  {Piovella}}]{Bonifacio20111004}%
  \BibitemOpen
  \bibfield  {author} {\bibinfo {author} {\bibnamefont {Bonifacio},
  \bibfnamefont {R.}}, \bibinfo {author} {\bibfnamefont {G.}~\bibnamefont
  {Robb}}, \ and\ \bibinfo {author} {\bibfnamefont {N.}~\bibnamefont
  {Piovella}}} (\bibinfo {year} {2011}),\ \href {\doibase
  http://dx.doi.org/10.1016/j.optcom.2010.10.030} {\bibfield  {journal}
  {\bibinfo  {journal} {Optics Communications}\ }\textbf {\bibinfo {volume}
  {284}}~(\bibinfo {number} {4}),\ \bibinfo {pages} {1004 }}\BibitemShut
  {NoStop}%
\bibitem [{\citenamefont {Borchers}\ \emph {et~al.}(2011)\citenamefont
  {Borchers}, \citenamefont {Koke}, \citenamefont {Husakou}, \citenamefont
  {Herrmann},\ and\ \citenamefont {Steinmeyer}}]{Borchers}%
  \BibitemOpen
  \bibfield  {author} {\bibinfo {author} {\bibnamefont {Borchers},
  \bibfnamefont {B.}}, \bibinfo {author} {\bibfnamefont {S.}~\bibnamefont
  {Koke}}, \bibinfo {author} {\bibfnamefont {A.}~\bibnamefont {Husakou}},
  \bibinfo {author} {\bibfnamefont {J.}~\bibnamefont {Herrmann}}, \ and\
  \bibinfo {author} {\bibfnamefont {G.}~\bibnamefont {Steinmeyer}}} (\bibinfo
  {year} {2011}),\ \href {http://ol.osa.org/abstract.cfm?URI=ol-36-21-4146}
  {\bibfield  {journal} {\bibinfo  {journal} {Opt. Lett.}\ }\textbf {\bibinfo
  {volume} {36}}~(\bibinfo {number} {21}),\ \bibinfo {pages}
  {4146}}\BibitemShut {NoStop}%
\bibitem [{\citenamefont {Borland}(2005)}]{BorlandAPSU}%
  \BibitemOpen
  \bibfield  {author} {\bibinfo {author} {\bibnamefont {Borland}, \bibfnamefont
  {M.}}} (\bibinfo {year} {2005}),\ \href
  {http://link.aps.org/doi/10.1103/PhysRevSTAB.8.074001} {\bibfield  {journal}
  {\bibinfo  {journal} {Phys. Rev. ST Accel. Beams}\ }\textbf {\bibinfo
  {volume} {8}},\ \bibinfo {pages} {074001}}\BibitemShut {NoStop}%
\bibitem [{\citenamefont {Borland}(2012)}]{Borland2012}%
  \BibitemOpen
  \bibfield  {author} {\bibinfo {author} {\bibnamefont {Borland}, \bibfnamefont
  {M.}}} (\bibinfo {year} {2012}),\ in\ \href@noop {} {\emph {\bibinfo
  {booktitle} {Proceedings of Intern. Part. Acc. Conf., IPAC'2012, New
  Orlean}}},\ p.\ \bibinfo {pages} {1035}\BibitemShut {NoStop}%
\bibitem [{\citenamefont {Borland}\ \emph {et~al.}(2002)\citenamefont
  {Borland}, \citenamefont {Chae}, \citenamefont {Emma}, \citenamefont
  {Lewellen}, \citenamefont {Bharadwaj}, \citenamefont {Fawley}, \citenamefont
  {Krejcik}, \citenamefont {Limborg}, \citenamefont {Milton}, \citenamefont
  {Nuhn}, \citenamefont {Soliday},\ and\ \citenamefont
  {Woodley}}]{borland02etal}%
  \BibitemOpen
  \bibfield  {author} {\bibinfo {author} {\bibnamefont {Borland}, \bibfnamefont
  {M.}}, \bibinfo {author} {\bibfnamefont {Y.}~\bibnamefont {Chae}}, \bibinfo
  {author} {\bibfnamefont {P.}~\bibnamefont {Emma}}, \bibinfo {author}
  {\bibfnamefont {J.}~\bibnamefont {Lewellen}}, \bibinfo {author}
  {\bibfnamefont {V.}~\bibnamefont {Bharadwaj}}, \bibinfo {author}
  {\bibfnamefont {W.}~\bibnamefont {Fawley}}, \bibinfo {author} {\bibfnamefont
  {P.}~\bibnamefont {Krejcik}}, \bibinfo {author} {\bibfnamefont
  {C.}~\bibnamefont {Limborg}}, \bibinfo {author} {\bibfnamefont
  {S.}~\bibnamefont {Milton}}, \bibinfo {author} {\bibfnamefont {H.-D.}\
  \bibnamefont {Nuhn}}, \bibinfo {author} {\bibfnamefont {R.}~\bibnamefont
  {Soliday}}, \ and\ \bibinfo {author} {\bibfnamefont {M.}~\bibnamefont
  {Woodley}}} (\bibinfo {year} {2002}),\ \href@noop {} {\bibfield  {journal}
  {\bibinfo  {journal} {Nucl. Instrum. Methods Phys. Res. A}\ }\textbf
  {\bibinfo {volume} {483}},\ \bibinfo {pages} {268}}\BibitemShut {NoStop}%
\bibitem [{\citenamefont {Brau}(1990)}]{Brau}%
  \BibitemOpen
  \bibfield  {author} {\bibinfo {author} {\bibnamefont {Brau}, \bibfnamefont
  {C.}}} (\bibinfo {year} {1990}),\ \href@noop {} {\emph {\bibinfo {title}
  {Free-Electron Lasers}}}\ (\bibinfo  {publisher} {Academic Press,
  Oxford})\BibitemShut {NoStop}%
\bibitem [{\citenamefont {Bressler}\ \emph {et~al.}(2009)\citenamefont
  {Bressler}, \citenamefont {Milne}, \citenamefont {Pham}, \citenamefont
  {ElNahhas}, \citenamefont {van~der Veen}, \citenamefont {Gawelda},
  \citenamefont {Johnson}, \citenamefont {Beaud}, \citenamefont {Grolimund},
  \citenamefont {Kaiser}, \citenamefont {Borca}, \citenamefont {Ingold},
  \citenamefont {Abela},\ and\ \citenamefont {Chergui}}]{LSapplication6}%
  \BibitemOpen
  \bibfield  {author} {\bibinfo {author} {\bibnamefont {Bressler},
  \bibfnamefont {C.}}, \bibinfo {author} {\bibfnamefont {C.}~\bibnamefont
  {Milne}}, \bibinfo {author} {\bibfnamefont {V.-T.}\ \bibnamefont {Pham}},
  \bibinfo {author} {\bibfnamefont {A.}~\bibnamefont {ElNahhas}}, \bibinfo
  {author} {\bibfnamefont {R.~M.}\ \bibnamefont {van~der Veen}}, \bibinfo
  {author} {\bibfnamefont {W.}~\bibnamefont {Gawelda}}, \bibinfo {author}
  {\bibfnamefont {S.}~\bibnamefont {Johnson}}, \bibinfo {author} {\bibfnamefont
  {P.}~\bibnamefont {Beaud}}, \bibinfo {author} {\bibfnamefont
  {D.}~\bibnamefont {Grolimund}}, \bibinfo {author} {\bibfnamefont
  {M.}~\bibnamefont {Kaiser}}, \bibinfo {author} {\bibfnamefont {C.~N.}\
  \bibnamefont {Borca}}, \bibinfo {author} {\bibfnamefont {G.}~\bibnamefont
  {Ingold}}, \bibinfo {author} {\bibfnamefont {R.}~\bibnamefont {Abela}}, \
  and\ \bibinfo {author} {\bibfnamefont {M.}~\bibnamefont {Chergui}}} (\bibinfo
  {year} {2009}),\ \href@noop {} {\bibfield  {journal} {\bibinfo  {journal}
  {Science}\ }\textbf {\bibinfo {volume} {323}},\ \bibinfo {pages}
  {489}}\BibitemShut {NoStop}%
\bibitem [{\citenamefont {Breuer}\ and\ \citenamefont
  {Hommelhoff}(2013)}]{PhysRevLett.111.134803}%
  \BibitemOpen
  \bibfield  {author} {\bibinfo {author} {\bibnamefont {Breuer}, \bibfnamefont
  {J.}}, \ and\ \bibinfo {author} {\bibfnamefont {P.}~\bibnamefont
  {Hommelhoff}}} (\bibinfo {year} {2013}),\ \href@noop {} {\bibfield  {journal}
  {\bibinfo  {journal} {Phys. Rev. Lett.}\ }\textbf {\bibinfo {volume} {111}},\
  \bibinfo {pages} {134803}}\BibitemShut {NoStop}%
\bibitem [{\citenamefont {Brinkmann}\ \emph {et~al.}(2001)\citenamefont
  {Brinkmann}, \citenamefont {Derbenev},\ and\ \citenamefont
  {Fl\"ottmann}}]{FB}%
  \BibitemOpen
  \bibfield  {author} {\bibinfo {author} {\bibnamefont {Brinkmann},
  \bibfnamefont {R.}}, \bibinfo {author} {\bibfnamefont {Y.}~\bibnamefont
  {Derbenev}}, \ and\ \bibinfo {author} {\bibfnamefont {K.}~\bibnamefont
  {Fl\"ottmann}}} (\bibinfo {year} {2001}),\ \href
  {http://link.aps.org/doi/10.1103/PhysRevSTAB.4.053501} {\bibfield  {journal}
  {\bibinfo  {journal} {Phys. Rev. ST Accel. Beams}\ }\textbf {\bibinfo
  {volume} {4}},\ \bibinfo {pages} {053501}}\BibitemShut {NoStop}%
\bibitem [{\citenamefont {Byrd}\ \emph {et~al.}(2006)\citenamefont {Byrd},
  \citenamefont {Hao}, \citenamefont {Martin}, \citenamefont {Robin},
  \citenamefont {Sannibale}, \citenamefont {Schoenlein}, \citenamefont
  {Zholents},\ and\ \citenamefont {Zolotorev}}]{Byrd1}%
  \BibitemOpen
  \bibfield  {author} {\bibinfo {author} {\bibnamefont {Byrd}, \bibfnamefont
  {J.}}, \bibinfo {author} {\bibfnamefont {Z.}~\bibnamefont {Hao}}, \bibinfo
  {author} {\bibfnamefont {M.}~\bibnamefont {Martin}}, \bibinfo {author}
  {\bibfnamefont {D.}~\bibnamefont {Robin}}, \bibinfo {author} {\bibfnamefont
  {F.}~\bibnamefont {Sannibale}}, \bibinfo {author} {\bibfnamefont
  {R.}~\bibnamefont {Schoenlein}}, \bibinfo {author} {\bibfnamefont
  {A.}~\bibnamefont {Zholents}}, \ and\ \bibinfo {author} {\bibfnamefont
  {M.}~\bibnamefont {Zolotorev}}} (\bibinfo {year} {2006}),\ \href@noop {}
  {\bibfield  {journal} {\bibinfo  {journal} {Phys. Rev. Lett.}\ }\textbf
  {\bibinfo {volume} {96}},\ \bibinfo {pages} {164801}}\BibitemShut {NoStop}%
\bibitem [{\citenamefont {Cai}\ \emph {et~al.}(2012)\citenamefont {Cai},
  \citenamefont {Bane}, \citenamefont {Hettel}, \citenamefont {Nosochkov},
  \citenamefont {Wang},\ and\ \citenamefont {Borland}}]{Cai2012}%
  \BibitemOpen
  \bibfield  {author} {\bibinfo {author} {\bibnamefont {Cai}, \bibfnamefont
  {Y.}}, \bibinfo {author} {\bibfnamefont {K.}~\bibnamefont {Bane}}, \bibinfo
  {author} {\bibfnamefont {R.}~\bibnamefont {Hettel}}, \bibinfo {author}
  {\bibfnamefont {Y.}~\bibnamefont {Nosochkov}}, \bibinfo {author}
  {\bibfnamefont {M.-H.}\ \bibnamefont {Wang}}, \ and\ \bibinfo {author}
  {\bibfnamefont {M.}~\bibnamefont {Borland}}} (\bibinfo {year} {2012}),\ \href
  {http://link.aps.org/doi/10.1103/PhysRevSTAB.15.054002} {\bibfield  {journal}
  {\bibinfo  {journal} {Phys. Rev. ST Accel. Beams}\ }\textbf {\bibinfo
  {volume} {15}},\ \bibinfo {pages} {054002}}\BibitemShut {NoStop}%
\bibitem [{\citenamefont {Carlsten}\ \emph {et~al.}(2011)\citenamefont
  {Carlsten}, \citenamefont {Bishofberger}, \citenamefont {Duffy},
  \citenamefont {Russell}, \citenamefont {Ryne}, \citenamefont {Yampolsky},\
  and\ \citenamefont {Dragt}}]{EP}%
  \BibitemOpen
  \bibfield  {author} {\bibinfo {author} {\bibnamefont {Carlsten},
  \bibfnamefont {B.~E.}}, \bibinfo {author} {\bibfnamefont {K.~A.}\
  \bibnamefont {Bishofberger}}, \bibinfo {author} {\bibfnamefont {L.~D.}\
  \bibnamefont {Duffy}}, \bibinfo {author} {\bibfnamefont {S.~J.}\ \bibnamefont
  {Russell}}, \bibinfo {author} {\bibfnamefont {R.~D.}\ \bibnamefont {Ryne}},
  \bibinfo {author} {\bibfnamefont {N.~A.}\ \bibnamefont {Yampolsky}}, \ and\
  \bibinfo {author} {\bibfnamefont {A.~J.}\ \bibnamefont {Dragt}}} (\bibinfo
  {year} {2011}),\ \href
  {http://link.aps.org/doi/10.1103/PhysRevSTAB.14.050706} {\bibfield  {journal}
  {\bibinfo  {journal} {Phys. Rev. ST Accel. Beams}\ }\textbf {\bibinfo
  {volume} {14}},\ \bibinfo {pages} {050706}}\BibitemShut {NoStop}%
\bibitem [{\citenamefont {Cavalleri}\ \emph {et~al.}(2005)\citenamefont
  {Cavalleri}, \citenamefont {Rini}, \citenamefont {Chong}, \citenamefont
  {Fourmaux}, \citenamefont {Glover}, \citenamefont {Heimann}, \citenamefont
  {Kieffer},\ and\ \citenamefont {Schoenlein}}]{LSapplication4}%
  \BibitemOpen
  \bibfield  {author} {\bibinfo {author} {\bibnamefont {Cavalleri},
  \bibfnamefont {A.}}, \bibinfo {author} {\bibfnamefont {M.}~\bibnamefont
  {Rini}}, \bibinfo {author} {\bibfnamefont {H.~H.~W.}\ \bibnamefont {Chong}},
  \bibinfo {author} {\bibfnamefont {S.}~\bibnamefont {Fourmaux}}, \bibinfo
  {author} {\bibfnamefont {T.~E.}\ \bibnamefont {Glover}}, \bibinfo {author}
  {\bibfnamefont {P.~A.}\ \bibnamefont {Heimann}}, \bibinfo {author}
  {\bibfnamefont {J.~C.}\ \bibnamefont {Kieffer}}, \ and\ \bibinfo {author}
  {\bibfnamefont {R.~W.}\ \bibnamefont {Schoenlein}}} (\bibinfo {year}
  {2005}),\ \href@noop {} {\bibfield  {journal} {\bibinfo  {journal} {Phys.
  Rev. Lett.}\ }\textbf {\bibinfo {volume} {95}},\ \bibinfo {pages}
  {067405}}\BibitemShut {NoStop}%
\bibitem [{\citenamefont {Cavalleri}\ \emph {et~al.}(2006)\citenamefont
  {Cavalleri}, \citenamefont {Wall}, \citenamefont {Simpson}, \citenamefont
  {Statz}, \citenamefont {Ward}, \citenamefont {Nelson}, \citenamefont {Rini},\
  and\ \citenamefont {Schoenlein}}]{LSapplication2}%
  \BibitemOpen
  \bibfield  {author} {\bibinfo {author} {\bibnamefont {Cavalleri},
  \bibfnamefont {A.}}, \bibinfo {author} {\bibfnamefont {S.}~\bibnamefont
  {Wall}}, \bibinfo {author} {\bibfnamefont {C.}~\bibnamefont {Simpson}},
  \bibinfo {author} {\bibfnamefont {E.}~\bibnamefont {Statz}}, \bibinfo
  {author} {\bibfnamefont {D.~W.}\ \bibnamefont {Ward}}, \bibinfo {author}
  {\bibfnamefont {K.~A.}\ \bibnamefont {Nelson}}, \bibinfo {author}
  {\bibfnamefont {M.}~\bibnamefont {Rini}}, \ and\ \bibinfo {author}
  {\bibfnamefont {R.~W.}\ \bibnamefont {Schoenlein}}} (\bibinfo {year}
  {2006}),\ \href@noop {} {\bibfield  {journal} {\bibinfo  {journal} {Nature}\
  }\textbf {\bibinfo {volume} {442}},\ \bibinfo {pages} {664}}\BibitemShut
  {NoStop}%
\bibitem [{\citenamefont {Cerullo}\ and\ \citenamefont
  {De~Silverstri}(2003)}]{OPA}%
  \BibitemOpen
  \bibfield  {author} {\bibinfo {author} {\bibnamefont {Cerullo}, \bibfnamefont
  {G.}}, \ and\ \bibinfo {author} {\bibfnamefont {S.}~\bibnamefont
  {De~Silverstri}}} (\bibinfo {year} {2003}),\ \href@noop {} {\bibfield
  {journal} {\bibinfo  {journal} {Rev. Sci. Instrum.}\ }\textbf {\bibinfo
  {volume} {74}},\ \bibinfo {pages} {1}}\BibitemShut {NoStop}%
\bibitem [{\citenamefont {Chang}\ \emph {et~al.}(2013)\citenamefont {Chang},
  \citenamefont {Tang},\ and\ \citenamefont {Wu}}]{PhysRevLett.110.064802}%
  \BibitemOpen
  \bibfield  {author} {\bibinfo {author} {\bibnamefont {Chang}, \bibfnamefont
  {C.}}, \bibinfo {author} {\bibfnamefont {C.}~\bibnamefont {Tang}}, \ and\
  \bibinfo {author} {\bibfnamefont {J.}~\bibnamefont {Wu}}} (\bibinfo {year}
  {2013}),\ \href {http://link.aps.org/doi/10.1103/PhysRevLett.110.064802}
  {\bibfield  {journal} {\bibinfo  {journal} {Phys. Rev. Lett.}\ }\textbf
  {\bibinfo {volume} {110}},\ \bibinfo {pages} {064802}}\BibitemShut {NoStop}%
\bibitem [{\citenamefont {Chao}(1993)}]{Chao}%
  \BibitemOpen
  \bibfield  {author} {\bibinfo {author} {\bibnamefont {Chao}, \bibfnamefont
  {A.~W.}}} (\bibinfo {year} {1993}),\ \href@noop {} {\emph {\bibinfo {title}
  {Physics of Collective Beam Instabilities in High Energy Accelerators}}}\
  (\bibinfo  {publisher} {Wiley},\ \bibinfo {address} {New York})\BibitemShut
  {NoStop}%
\bibitem [{\citenamefont {Chen}\ \emph {et~al.}(1985)\citenamefont {Chen},
  \citenamefont {Dawson}, \citenamefont {Huff},\ and\ \citenamefont
  {Katsouleas}}]{PhysRevLett.54.693}%
  \BibitemOpen
  \bibfield  {author} {\bibinfo {author} {\bibnamefont {Chen}, \bibfnamefont
  {P.}}, \bibinfo {author} {\bibfnamefont {J.~M.}\ \bibnamefont {Dawson}},
  \bibinfo {author} {\bibfnamefont {R.~W.}\ \bibnamefont {Huff}}, \ and\
  \bibinfo {author} {\bibfnamefont {T.}~\bibnamefont {Katsouleas}}} (\bibinfo
  {year} {1985}),\ \href {http://link.aps.org/doi/10.1103/PhysRevLett.54.693}
  {\bibfield  {journal} {\bibinfo  {journal} {Phys. Rev. Lett.}\ }\textbf
  {\bibinfo {volume} {54}},\ \bibinfo {pages} {693}}\BibitemShut {NoStop}%
\bibitem [{\citenamefont {Colson}(1981)}]{Colson1}%
  \BibitemOpen
  \bibfield  {author} {\bibinfo {author} {\bibnamefont {Colson}, \bibfnamefont
  {W.}}} (\bibinfo {year} {1981}),\ \href@noop {} {\bibfield  {journal}
  {\bibinfo  {journal} {IEEE J. of Quantum Electronics}\ }\textbf {\bibinfo
  {volume} {QE-17}},\ \bibinfo {pages} {1417}}\BibitemShut {NoStop}%
\bibitem [{\citenamefont {Cornacchia}\ and\ \citenamefont {Emma}(2002)}]{EEX1}%
  \BibitemOpen
  \bibfield  {author} {\bibinfo {author} {\bibnamefont {Cornacchia},
  \bibfnamefont {M.}}, \ and\ \bibinfo {author} {\bibfnamefont
  {P.}~\bibnamefont {Emma}}} (\bibinfo {year} {2002}),\ \href
  {http://link.aps.org/doi/10.1103/PhysRevSTAB.5.084001} {\bibfield  {journal}
  {\bibinfo  {journal} {Phys. Rev. ST Accel. Beams}\ }\textbf {\bibinfo
  {volume} {5}},\ \bibinfo {pages} {084001}}\BibitemShut {NoStop}%
\bibitem [{\citenamefont {Courant}(1966)}]{Courant}%
  \BibitemOpen
  \bibfield  {author} {\bibinfo {author} {\bibnamefont {Courant}, \bibfnamefont
  {E.}}} (\bibinfo {year} {1966}),\ \enquote {\bibinfo {title} {Conservation of
  phase space in hamiltonian systems and particle beams},}\ in\ \href@noop {}
  {\emph {\bibinfo {booktitle} {Perspectives in Modern Physics, Essays in Honor
  of Hans A. Bethe}}},\ \bibinfo {editor} {edited by\ \bibinfo {editor}
  {\bibfnamefont {R.~E.}\ \bibnamefont {Marshak}}}\ (\bibinfo  {publisher}
  {Interscience Publishers, New York})\ p.\ \bibinfo {pages} {257}\BibitemShut
  {NoStop}%
\bibitem [{\citenamefont {Courant}\ \emph {et~al.}(1985)\citenamefont
  {Courant}, \citenamefont {Pellegrini},\ and\ \citenamefont
  {Zakowicz}}]{PhysRevA.32.2813}%
  \BibitemOpen
  \bibfield  {author} {\bibinfo {author} {\bibnamefont {Courant}, \bibfnamefont
  {E.~D.}}, \bibinfo {author} {\bibfnamefont {C.}~\bibnamefont {Pellegrini}}, \
  and\ \bibinfo {author} {\bibfnamefont {W.}~\bibnamefont {Zakowicz}}}
  (\bibinfo {year} {1985}),\ \href
  {http://link.aps.org/doi/10.1103/PhysRevA.32.2813} {\bibfield  {journal}
  {\bibinfo  {journal} {Phys. Rev. A}\ }\textbf {\bibinfo {volume} {32}},\
  \bibinfo {pages} {2813}}\BibitemShut {NoStop}%
\bibitem [{\citenamefont {De~Ninno}\ \emph {et~al.}(2008)\citenamefont
  {De~Ninno}, \citenamefont {Allaria}, \citenamefont {Coreno}, \citenamefont
  {Curbis}, \citenamefont {Danailov}, \citenamefont {Karantzoulis},
  \citenamefont {Locatelli}, \citenamefont {Mente\ifmmode~\mbox{\c{s}}\else
  \c{s}\fi{}}, \citenamefont {Nino}, \citenamefont {Spezzani},\ and\
  \citenamefont {Trov\`o}}]{Elettra}%
  \BibitemOpen
  \bibfield  {author} {\bibinfo {author} {\bibnamefont {De~Ninno},
  \bibfnamefont {G.}}, \bibinfo {author} {\bibfnamefont {E.}~\bibnamefont
  {Allaria}}, \bibinfo {author} {\bibfnamefont {M.}~\bibnamefont {Coreno}},
  \bibinfo {author} {\bibfnamefont {F.}~\bibnamefont {Curbis}}, \bibinfo
  {author} {\bibfnamefont {M.~B.}\ \bibnamefont {Danailov}}, \bibinfo {author}
  {\bibfnamefont {E.}~\bibnamefont {Karantzoulis}}, \bibinfo {author}
  {\bibfnamefont {A.}~\bibnamefont {Locatelli}}, \bibinfo {author}
  {\bibfnamefont {T.~O.}\ \bibnamefont {Mente\ifmmode~\mbox{\c{s}}\else
  \c{s}\fi{}}}, \bibinfo {author} {\bibfnamefont {M.~A.}\ \bibnamefont {Nino}},
  \bibinfo {author} {\bibfnamefont {C.}~\bibnamefont {Spezzani}}, \ and\
  \bibinfo {author} {\bibfnamefont {M.}~\bibnamefont {Trov\`o}}} (\bibinfo
  {year} {2008}),\ \href@noop {} {\bibfield  {journal} {\bibinfo  {journal}
  {Phys. Rev. Lett.}\ }\textbf {\bibinfo {volume} {101}},\ \bibinfo {pages}
  {053902}}\BibitemShut {NoStop}%
\bibitem [{\citenamefont {De~Ninno}\ \emph {et~al.}(2013)\citenamefont
  {De~Ninno}, \citenamefont {Mahieu}, \citenamefont {Allaria}, \citenamefont
  {Giannessi},\ and\ \citenamefont {Spampinati}}]{FERMI2color}%
  \BibitemOpen
  \bibfield  {author} {\bibinfo {author} {\bibnamefont {De~Ninno},
  \bibfnamefont {G.}}, \bibinfo {author} {\bibfnamefont {B.}~\bibnamefont
  {Mahieu}}, \bibinfo {author} {\bibfnamefont {E.}~\bibnamefont {Allaria}},
  \bibinfo {author} {\bibfnamefont {L.}~\bibnamefont {Giannessi}}, \ and\
  \bibinfo {author} {\bibfnamefont {S.}~\bibnamefont {Spampinati}}} (\bibinfo
  {year} {2013}),\ \href
  {http://link.aps.org/doi/10.1103/PhysRevLett.110.064801} {\bibfield
  {journal} {\bibinfo  {journal} {Phys. Rev. Lett.}\ }\textbf {\bibinfo
  {volume} {110}},\ \bibinfo {pages} {064801}}\BibitemShut {NoStop}%
\bibitem [{\citenamefont {Deacon}\ \emph {et~al.}(1977)\citenamefont {Deacon},
  \citenamefont {Elias}, \citenamefont {Madey}, \citenamefont {Ramian},
  \citenamefont {Schwettman},\ and\ \citenamefont {Smith}}]{Madey3}%
  \BibitemOpen
  \bibfield  {author} {\bibinfo {author} {\bibnamefont {Deacon}, \bibfnamefont
  {D.~A.~G.}}, \bibinfo {author} {\bibfnamefont {L.~R.}\ \bibnamefont {Elias}},
  \bibinfo {author} {\bibfnamefont {J.~M.~J.}\ \bibnamefont {Madey}}, \bibinfo
  {author} {\bibfnamefont {G.~J.}\ \bibnamefont {Ramian}}, \bibinfo {author}
  {\bibfnamefont {H.~A.}\ \bibnamefont {Schwettman}}, \ and\ \bibinfo {author}
  {\bibfnamefont {T.~I.}\ \bibnamefont {Smith}}} (\bibinfo {year} {1977}),\
  \href {http://link.aps.org/doi/10.1103/PhysRevLett.38.892} {\bibfield
  {journal} {\bibinfo  {journal} {Phys. Rev. Lett.}\ }\textbf {\bibinfo
  {volume} {38}},\ \bibinfo {pages} {892}}\BibitemShut {NoStop}%
\bibitem [{\citenamefont {Decker}(2005)}]{Decker}%
  \BibitemOpen
  \bibfield  {author} {\bibinfo {author} {\bibnamefont {Decker}, \bibfnamefont
  {G.}}} (\bibinfo {year} {2005}),\ in\ \href@noop {} {\emph {\bibinfo
  {booktitle} {Proceedings of DIPAC 2005, Lyon, France}}},\ p.\ \bibinfo
  {pages} {233}\BibitemShut {NoStop}%
\bibitem [{\citenamefont {Deng}\ and\ \citenamefont {Feng}(2013)}]{CooledHGHG}%
  \BibitemOpen
  \bibfield  {author} {\bibinfo {author} {\bibnamefont {Deng}, \bibfnamefont
  {H.}}, \ and\ \bibinfo {author} {\bibfnamefont {C.}~\bibnamefont {Feng}}}
  (\bibinfo {year} {2013}),\ \href
  {http://link.aps.org/doi/10.1103/PhysRevLett.111.084801} {\bibfield
  {journal} {\bibinfo  {journal} {Phys. Rev. Lett.}\ }\textbf {\bibinfo
  {volume} {111}},\ \bibinfo {pages} {084801}}\BibitemShut {NoStop}%
\bibitem [{\citenamefont {Diels}\ and\ \citenamefont {Rudolph}(1996)}]{diels}%
  \BibitemOpen
  \bibfield  {author} {\bibinfo {author} {\bibnamefont {Diels}, \bibfnamefont
  {J.-C.}}, \ and\ \bibinfo {author} {\bibfnamefont {W.}~\bibnamefont
  {Rudolph}}} (\bibinfo {year} {1996}),\ \href@noop {} {\emph {\bibinfo {title}
  {Ultrashort laser pulse phenomena}}}\ (\bibinfo  {publisher} {Academic
  Press})\BibitemShut {NoStop}%
\bibitem [{\citenamefont {Ding}\ \emph {et~al.}(2011)\citenamefont {Ding},
  \citenamefont {Bane},\ and\ \citenamefont {Huang}}]{OS}%
  \BibitemOpen
  \bibfield  {author} {\bibinfo {author} {\bibnamefont {Ding}, \bibfnamefont
  {Y.}}, \bibinfo {author} {\bibfnamefont {K.}~\bibnamefont {Bane}}, \ and\
  \bibinfo {author} {\bibfnamefont {Z.}~\bibnamefont {Huang}}} (\bibinfo {year}
  {2011}),\ in\ \href@noop {} {\emph {\bibinfo {booktitle} {2011 Free-electron
  laser Conference}}},\ p.\ \bibinfo {pages} {431}\BibitemShut {NoStop}%
\bibitem [{\citenamefont {Ding}\ \emph
  {et~al.}(2009{\natexlab{a}})\citenamefont {Ding}, \citenamefont {Brachmann},
  \citenamefont {Decker}, \citenamefont {Dowell}, \citenamefont {Emma},
  \citenamefont {Frisch}, \citenamefont {Gilevich}, \citenamefont {Hays},
  \citenamefont {Hering}, \citenamefont {Huang}, \citenamefont {Iverson},
  \citenamefont {Loos}, \citenamefont {Miahnahri}, \citenamefont {Nuhn},
  \citenamefont {Ratner}, \citenamefont {Turner}, \citenamefont {Welch},
  \citenamefont {White},\ and\ \citenamefont {Wu}}]{LCLSlowcharge}%
  \BibitemOpen
  \bibfield  {author} {\bibinfo {author} {\bibnamefont {Ding}, \bibfnamefont
  {Y.}}, \bibinfo {author} {\bibfnamefont {A.}~\bibnamefont {Brachmann}},
  \bibinfo {author} {\bibfnamefont {F.-J.}\ \bibnamefont {Decker}}, \bibinfo
  {author} {\bibfnamefont {D.}~\bibnamefont {Dowell}}, \bibinfo {author}
  {\bibfnamefont {P.}~\bibnamefont {Emma}}, \bibinfo {author} {\bibfnamefont
  {J.}~\bibnamefont {Frisch}}, \bibinfo {author} {\bibfnamefont
  {S.}~\bibnamefont {Gilevich}}, \bibinfo {author} {\bibfnamefont
  {G.}~\bibnamefont {Hays}}, \bibinfo {author} {\bibfnamefont {P.}~\bibnamefont
  {Hering}}, \bibinfo {author} {\bibfnamefont {Z.}~\bibnamefont {Huang}},
  \bibinfo {author} {\bibfnamefont {R.}~\bibnamefont {Iverson}}, \bibinfo
  {author} {\bibfnamefont {H.}~\bibnamefont {Loos}}, \bibinfo {author}
  {\bibfnamefont {A.}~\bibnamefont {Miahnahri}}, \bibinfo {author}
  {\bibfnamefont {H.-D.}\ \bibnamefont {Nuhn}}, \bibinfo {author}
  {\bibfnamefont {D.}~\bibnamefont {Ratner}}, \bibinfo {author} {\bibfnamefont
  {J.}~\bibnamefont {Turner}}, \bibinfo {author} {\bibfnamefont
  {J.}~\bibnamefont {Welch}}, \bibinfo {author} {\bibfnamefont
  {W.}~\bibnamefont {White}}, \ and\ \bibinfo {author} {\bibfnamefont
  {J.}~\bibnamefont {Wu}}} (\bibinfo {year} {2009}{\natexlab{a}}),\ \href
  {http://link.aps.org/doi/10.1103/PhysRevLett.102.254801} {\bibfield
  {journal} {\bibinfo  {journal} {Phys. Rev. Lett.}\ }\textbf {\bibinfo
  {volume} {102}},\ \bibinfo {pages} {254801}}\BibitemShut {NoStop}%
\bibitem [{\citenamefont {Ding}\ \emph
  {et~al.}(2009{\natexlab{b}})\citenamefont {Ding}, \citenamefont {Huang},
  \citenamefont {Ratner}, \citenamefont {Bucksbaum},\ and\ \citenamefont
  {Merdji}}]{Ding2}%
  \BibitemOpen
  \bibfield  {author} {\bibinfo {author} {\bibnamefont {Ding}, \bibfnamefont
  {Y.}}, \bibinfo {author} {\bibfnamefont {Z.}~\bibnamefont {Huang}}, \bibinfo
  {author} {\bibfnamefont {D.}~\bibnamefont {Ratner}}, \bibinfo {author}
  {\bibfnamefont {P.}~\bibnamefont {Bucksbaum}}, \ and\ \bibinfo {author}
  {\bibfnamefont {H.}~\bibnamefont {Merdji}}} (\bibinfo {year}
  {2009}{\natexlab{b}}),\ \href@noop {} {\bibfield  {journal} {\bibinfo
  {journal} {Phys. Rev. Spec. Topics Accel. and Beams}\ }\textbf {\bibinfo
  {volume} {12}},\ \bibinfo {pages} {060703}}\BibitemShut {NoStop}%
\bibitem [{\citenamefont {Dunning}\ \emph
  {et~al.}(2013{\natexlab{a}})\citenamefont {Dunning}, \citenamefont {McNeil},\
  and\ \citenamefont {Thompson}}]{zs}%
  \BibitemOpen
  \bibfield  {author} {\bibinfo {author} {\bibnamefont {Dunning}, \bibfnamefont
  {D.~J.}}, \bibinfo {author} {\bibfnamefont {B.~W.~J.}\ \bibnamefont
  {McNeil}}, \ and\ \bibinfo {author} {\bibfnamefont {N.~R.}\ \bibnamefont
  {Thompson}}} (\bibinfo {year} {2013}{\natexlab{a}}),\ \href@noop {}
  {\bibfield  {journal} {\bibinfo  {journal} {Phys. Rev. Lett.}\ }\textbf
  {\bibinfo {volume} {110}},\ \bibinfo {pages} {104801}}\BibitemShut {NoStop}%
\bibitem [{\citenamefont {Dunning}\ \emph {et~al.}(2012)\citenamefont
  {Dunning}, \citenamefont {Hast}, \citenamefont {Hemsing}, \citenamefont
  {Jobe}, \citenamefont {McCormick}, \citenamefont {Nelson}, \citenamefont
  {Raubenheimer}, \citenamefont {Soong}, \citenamefont {Szalata}, \citenamefont
  {Walz}, \citenamefont {Weathersby},\ and\ \citenamefont {Xiang}}]{THzNLCTA}%
  \BibitemOpen
  \bibfield  {author} {\bibinfo {author} {\bibnamefont {Dunning}, \bibfnamefont
  {M.}}, \bibinfo {author} {\bibfnamefont {C.}~\bibnamefont {Hast}}, \bibinfo
  {author} {\bibfnamefont {E.}~\bibnamefont {Hemsing}}, \bibinfo {author}
  {\bibfnamefont {K.}~\bibnamefont {Jobe}}, \bibinfo {author} {\bibfnamefont
  {D.}~\bibnamefont {McCormick}}, \bibinfo {author} {\bibfnamefont
  {J.}~\bibnamefont {Nelson}}, \bibinfo {author} {\bibfnamefont {T.~O.}\
  \bibnamefont {Raubenheimer}}, \bibinfo {author} {\bibfnamefont
  {K.}~\bibnamefont {Soong}}, \bibinfo {author} {\bibfnamefont
  {Z.}~\bibnamefont {Szalata}}, \bibinfo {author} {\bibfnamefont
  {D.}~\bibnamefont {Walz}}, \bibinfo {author} {\bibfnamefont {S.}~\bibnamefont
  {Weathersby}}, \ and\ \bibinfo {author} {\bibfnamefont {D.}~\bibnamefont
  {Xiang}}} (\bibinfo {year} {2012}),\ \href
  {http://link.aps.org/doi/10.1103/PhysRevLett.109.074801} {\bibfield
  {journal} {\bibinfo  {journal} {Phys. Rev. Lett.}\ }\textbf {\bibinfo
  {volume} {109}},\ \bibinfo {pages} {074801}}\BibitemShut {NoStop}%
\bibitem [{\citenamefont {Dunning}\ \emph
  {et~al.}(2013{\natexlab{b}})\citenamefont {Dunning}, \citenamefont {Hemsing},
  \citenamefont {Hast}, \citenamefont {Raubenheimer}, \citenamefont
  {Weathersby}, \citenamefont {Xiang},\ and\ \citenamefont
  {Fu}}]{PhysRevLett.110.244801}%
  \BibitemOpen
  \bibfield  {author} {\bibinfo {author} {\bibnamefont {Dunning}, \bibfnamefont
  {M.}}, \bibinfo {author} {\bibfnamefont {E.}~\bibnamefont {Hemsing}},
  \bibinfo {author} {\bibfnamefont {C.}~\bibnamefont {Hast}}, \bibinfo {author}
  {\bibfnamefont {T.~O.}\ \bibnamefont {Raubenheimer}}, \bibinfo {author}
  {\bibfnamefont {S.}~\bibnamefont {Weathersby}}, \bibinfo {author}
  {\bibfnamefont {D.}~\bibnamefont {Xiang}}, \ and\ \bibinfo {author}
  {\bibfnamefont {F.}~\bibnamefont {Fu}}} (\bibinfo {year}
  {2013}{\natexlab{b}}),\ \href
  {http://link.aps.org/doi/10.1103/PhysRevLett.110.244801} {\bibfield
  {journal} {\bibinfo  {journal} {Phys. Rev. Lett.}\ }\textbf {\bibinfo
  {volume} {110}},\ \bibinfo {pages} {244801}}\BibitemShut {NoStop}%
\bibitem [{\citenamefont {Duris}\ \emph
  {et~al.}(2012{\natexlab{a}})\citenamefont {Duris}, \citenamefont {Li},
  \citenamefont {Musumeci}, \citenamefont {Sakai}, \citenamefont {Threlkeld},
  \citenamefont {Williams}, \citenamefont {Fedurin}, \citenamefont {Kusche},
  \citenamefont {Pogorelsky}, \citenamefont {Polyanskiy},\ and\ \citenamefont
  {Yakimenko}}]{duris:440}%
  \BibitemOpen
  \bibfield  {author} {\bibinfo {author} {\bibnamefont {Duris}, \bibfnamefont
  {J.}}, \bibinfo {author} {\bibfnamefont {R.~K.}\ \bibnamefont {Li}}, \bibinfo
  {author} {\bibfnamefont {P.}~\bibnamefont {Musumeci}}, \bibinfo {author}
  {\bibfnamefont {Y.}~\bibnamefont {Sakai}}, \bibinfo {author} {\bibfnamefont
  {E.}~\bibnamefont {Threlkeld}}, \bibinfo {author} {\bibfnamefont
  {O.}~\bibnamefont {Williams}}, \bibinfo {author} {\bibfnamefont
  {M.}~\bibnamefont {Fedurin}}, \bibinfo {author} {\bibfnamefont
  {K.}~\bibnamefont {Kusche}}, \bibinfo {author} {\bibfnamefont
  {I.}~\bibnamefont {Pogorelsky}}, \bibinfo {author} {\bibfnamefont
  {M.}~\bibnamefont {Polyanskiy}}, \ and\ \bibinfo {author} {\bibfnamefont
  {V.}~\bibnamefont {Yakimenko}}} (\bibinfo {year} {2012}{\natexlab{a}}),\
  \href {http://link.aip.org/link/?APC/1507/440/1} {\bibfield  {journal}
  {\bibinfo  {journal} {AIP Conference Proceedings}\ }\textbf {\bibinfo
  {volume} {1507}}~(\bibinfo {number} {1}),\ \bibinfo {pages}
  {440}}\BibitemShut {NoStop}%
\bibitem [{\citenamefont {Duris}\ \emph
  {et~al.}(2012{\natexlab{b}})\citenamefont {Duris}, \citenamefont {Musumeci},\
  and\ \citenamefont {Li}}]{PhysRevSTAB.15.061301}%
  \BibitemOpen
  \bibfield  {author} {\bibinfo {author} {\bibnamefont {Duris}, \bibfnamefont
  {J.~P.}}, \bibinfo {author} {\bibfnamefont {P.}~\bibnamefont {Musumeci}}, \
  and\ \bibinfo {author} {\bibfnamefont {R.~K.}\ \bibnamefont {Li}}} (\bibinfo
  {year} {2012}{\natexlab{b}}),\ \href
  {http://link.aps.org/doi/10.1103/PhysRevSTAB.15.061301} {\bibfield  {journal}
  {\bibinfo  {journal} {Phys. Rev. ST Accel. Beams}\ }\textbf {\bibinfo
  {volume} {15}},\ \bibinfo {pages} {061301}}\BibitemShut {NoStop}%
\bibitem [{\citenamefont {Einfeld}\ \emph {et~al.}(1995)\citenamefont
  {Einfeld}, \citenamefont {Schaper},\ and\ \citenamefont {Plesko}}]{Einfeld}%
  \BibitemOpen
  \bibfield  {author} {\bibinfo {author} {\bibnamefont {Einfeld}, \bibfnamefont
  {D.}}, \bibinfo {author} {\bibfnamefont {J.}~\bibnamefont {Schaper}}, \ and\
  \bibinfo {author} {\bibfnamefont {M.}~\bibnamefont {Plesko}}} (\bibinfo
  {year} {1995}),\ \href@noop {} {\bibinfo  {journal} {Proc. Part. Acc. Conf.,
  PAC'1995}\ ,\ \bibinfo {pages} {177}}\BibitemShut {NoStop}%
\bibitem [{\citenamefont {Elder}\ \emph {et~al.}(1947)\citenamefont {Elder},
  \citenamefont {Gurewitsch}, \citenamefont {Langmuir},\ and\ \citenamefont
  {Pollock}}]{SR}%
  \BibitemOpen
\bibfield  {journal} {  }\bibfield  {author} {\bibinfo {author} {\bibnamefont
  {Elder}, \bibfnamefont {F.~R.}}, \bibinfo {author} {\bibfnamefont {A.~M.}\
  \bibnamefont {Gurewitsch}}, \bibinfo {author} {\bibfnamefont {R.~V.}\
  \bibnamefont {Langmuir}}, \ and\ \bibinfo {author} {\bibfnamefont {H.~C.}\
  \bibnamefont {Pollock}}} (\bibinfo {year} {1947}),\ \href@noop {} {\bibfield
  {journal} {\bibinfo  {journal} {Phys. Rev.}\ }\textbf {\bibinfo {volume}
  {71}},\ \bibinfo {pages} {829}}\BibitemShut {NoStop}%
\bibitem [{\citenamefont {Elias}\ \emph {et~al.}(1976)\citenamefont {Elias},
  \citenamefont {Fairbank}, \citenamefont {Madey}, \citenamefont {Schwettman},\
  and\ \citenamefont {Smith}}]{Madey2}%
  \BibitemOpen
  \bibfield  {author} {\bibinfo {author} {\bibnamefont {Elias}, \bibfnamefont
  {L.~R.}}, \bibinfo {author} {\bibfnamefont {W.~M.}\ \bibnamefont {Fairbank}},
  \bibinfo {author} {\bibfnamefont {J.~M.~J.}\ \bibnamefont {Madey}}, \bibinfo
  {author} {\bibfnamefont {H.~A.}\ \bibnamefont {Schwettman}}, \ and\ \bibinfo
  {author} {\bibfnamefont {T.~I.}\ \bibnamefont {Smith}}} (\bibinfo {year}
  {1976}),\ \href {http://link.aps.org/doi/10.1103/PhysRevLett.36.717}
  {\bibfield  {journal} {\bibinfo  {journal} {Phys. Rev. Lett.}\ }\textbf
  {\bibinfo {volume} {36}},\ \bibinfo {pages} {717}}\BibitemShut {NoStop}%
\bibitem [{\citenamefont {Emery}\ and\ \citenamefont {Borland}(1999)}]{Emery}%
  \BibitemOpen
  \bibfield  {author} {\bibinfo {author} {\bibnamefont {Emery}, \bibfnamefont
  {L.}}, \ and\ \bibinfo {author} {\bibfnamefont {M.}~\bibnamefont {Borland}}}
  (\bibinfo {year} {1999}),\ in\ \href@noop {} {\emph {\bibinfo {booktitle}
  {Proc. Part. Acc. Conf., PAC'1999, New York}}},\ p.\ \bibinfo {pages}
  {200}\BibitemShut {NoStop}%
\bibitem [{\citenamefont {Emma}\ \emph {et~al.}(2010)\citenamefont {Emma},
  \citenamefont {Akre}, \citenamefont {Arthur}, \citenamefont {Bionta},
  \citenamefont {Bostedt}, \citenamefont {Bozek}, \citenamefont {Brachmann},
  \citenamefont {Bucksbaum}, \citenamefont {Coffee}, \citenamefont {Decker},
  \citenamefont {Ding}, \citenamefont {Dowell}, \citenamefont {Edstrom},
  \citenamefont {Fisher}, \citenamefont {Frisch}, \citenamefont {Gilevich},
  \citenamefont {Hastings}, \citenamefont {Hays}, \citenamefont {Hering},
  \citenamefont {Huang}, \citenamefont {Iverson}, \citenamefont {Loos},
  \citenamefont {Messerschmidt}, \citenamefont {Miahnahri}, \citenamefont
  {Moeller}, \citenamefont {Nuhn}, \citenamefont {Pile}, \citenamefont
  {Ratner}, \citenamefont {Rzepiela}, \citenamefont {Schultz}, \citenamefont
  {Smith}, \citenamefont {Stefan}, \citenamefont {Tompkins}, \citenamefont
  {Turner}, \citenamefont {Welch}, \citenamefont {White}, \citenamefont {Wu},
  \citenamefont {Yocky},\ and\ \citenamefont {Galayda}}]{LCLS_2010}%
  \BibitemOpen
  \bibfield  {author} {\bibinfo {author} {\bibnamefont {Emma}, \bibfnamefont
  {P.}}, \bibinfo {author} {\bibfnamefont {A.}~\bibnamefont {Akre}}, \bibinfo
  {author} {\bibfnamefont {J.}~\bibnamefont {Arthur}}, \bibinfo {author}
  {\bibfnamefont {R.}~\bibnamefont {Bionta}}, \bibinfo {author} {\bibfnamefont
  {C.}~\bibnamefont {Bostedt}}, \bibinfo {author} {\bibfnamefont
  {J.}~\bibnamefont {Bozek}}, \bibinfo {author} {\bibfnamefont
  {A.}~\bibnamefont {Brachmann}}, \bibinfo {author} {\bibfnamefont
  {P.}~\bibnamefont {Bucksbaum}}, \bibinfo {author} {\bibfnamefont
  {R.}~\bibnamefont {Coffee}}, \bibinfo {author} {\bibfnamefont {F.-J.}\
  \bibnamefont {Decker}}, \bibinfo {author} {\bibfnamefont {Y.}~\bibnamefont
  {Ding}}, \bibinfo {author} {\bibfnamefont {D.}~\bibnamefont {Dowell}},
  \bibinfo {author} {\bibfnamefont {S.}~\bibnamefont {Edstrom}}, \bibinfo
  {author} {\bibfnamefont {A.}~\bibnamefont {Fisher}}, \bibinfo {author}
  {\bibfnamefont {J.}~\bibnamefont {Frisch}}, \bibinfo {author} {\bibfnamefont
  {S.}~\bibnamefont {Gilevich}}, \bibinfo {author} {\bibfnamefont
  {J.}~\bibnamefont {Hastings}}, \bibinfo {author} {\bibfnamefont
  {G.}~\bibnamefont {Hays}}, \bibinfo {author} {\bibfnamefont {P.}~\bibnamefont
  {Hering}}, \bibinfo {author} {\bibfnamefont {Z.}~\bibnamefont {Huang}},
  \bibinfo {author} {\bibfnamefont {R.}~\bibnamefont {Iverson}}, \bibinfo
  {author} {\bibfnamefont {H.}~\bibnamefont {Loos}}, \bibinfo {author}
  {\bibfnamefont {M.}~\bibnamefont {Messerschmidt}}, \bibinfo {author}
  {\bibfnamefont {A.}~\bibnamefont {Miahnahri}}, \bibinfo {author}
  {\bibfnamefont {S.}~\bibnamefont {Moeller}}, \bibinfo {author} {\bibfnamefont
  {H.-D.}\ \bibnamefont {Nuhn}}, \bibinfo {author} {\bibfnamefont
  {G.}~\bibnamefont {Pile}}, \bibinfo {author} {\bibfnamefont {D.}~\bibnamefont
  {Ratner}}, \bibinfo {author} {\bibfnamefont {J.}~\bibnamefont {Rzepiela}},
  \bibinfo {author} {\bibfnamefont {D.}~\bibnamefont {Schultz}}, \bibinfo
  {author} {\bibfnamefont {T.}~\bibnamefont {Smith}}, \bibinfo {author}
  {\bibfnamefont {P.}~\bibnamefont {Stefan}}, \bibinfo {author} {\bibfnamefont
  {H.}~\bibnamefont {Tompkins}}, \bibinfo {author} {\bibfnamefont
  {J.}~\bibnamefont {Turner}}, \bibinfo {author} {\bibfnamefont
  {J.}~\bibnamefont {Welch}}, \bibinfo {author} {\bibfnamefont
  {W.}~\bibnamefont {White}}, \bibinfo {author} {\bibfnamefont
  {J.}~\bibnamefont {Wu}}, \bibinfo {author} {\bibfnamefont {G.}~\bibnamefont
  {Yocky}}, \ and\ \bibinfo {author} {\bibfnamefont {J.}~\bibnamefont
  {Galayda}}} (\bibinfo {year} {2010}),\ \href@noop {} {\bibfield  {journal}
  {\bibinfo  {journal} {Nature Photon.}\ }\textbf {\bibinfo {volume} {4}},\
  \bibinfo {pages} {641}}\BibitemShut {NoStop}%
\bibitem [{\citenamefont {Emma}\ \emph {et~al.}(2004)\citenamefont {Emma},
  \citenamefont {Bane}, \citenamefont {Cornacchia}, \citenamefont {Huang},
  \citenamefont {Schlarb}, \citenamefont {Stupakov},\ and\ \citenamefont
  {Walz}}]{PE2004}%
  \BibitemOpen
  \bibfield  {author} {\bibinfo {author} {\bibnamefont {Emma}, \bibfnamefont
  {P.}}, \bibinfo {author} {\bibfnamefont {K.}~\bibnamefont {Bane}}, \bibinfo
  {author} {\bibfnamefont {M.}~\bibnamefont {Cornacchia}}, \bibinfo {author}
  {\bibfnamefont {Z.}~\bibnamefont {Huang}}, \bibinfo {author} {\bibfnamefont
  {H.}~\bibnamefont {Schlarb}}, \bibinfo {author} {\bibfnamefont
  {G.}~\bibnamefont {Stupakov}}, \ and\ \bibinfo {author} {\bibfnamefont
  {D.}~\bibnamefont {Walz}}} (\bibinfo {year} {2004}),\ \href
  {http://link.aps.org/doi/10.1103/PhysRevLett.92.074801} {\bibfield  {journal}
  {\bibinfo  {journal} {Phys. Rev. Lett.}\ }\textbf {\bibinfo {volume} {92}},\
  \bibinfo {pages} {074801}}\BibitemShut {NoStop}%
\bibitem [{\citenamefont {Emma}\ \emph {et~al.}(2006)\citenamefont {Emma},
  \citenamefont {Huang}, \citenamefont {Kim},\ and\ \citenamefont
  {Piot}}]{EEX2}%
  \BibitemOpen
  \bibfield  {author} {\bibinfo {author} {\bibnamefont {Emma}, \bibfnamefont
  {P.}}, \bibinfo {author} {\bibfnamefont {Z.}~\bibnamefont {Huang}}, \bibinfo
  {author} {\bibfnamefont {K.-J.}\ \bibnamefont {Kim}}, \ and\ \bibinfo
  {author} {\bibfnamefont {P.}~\bibnamefont {Piot}}} (\bibinfo {year} {2006}),\
  \href {http://link.aps.org/doi/10.1103/PhysRevSTAB.9.100702} {\bibfield
  {journal} {\bibinfo  {journal} {Phys. Rev. ST Accel. Beams}\ }\textbf
  {\bibinfo {volume} {9}},\ \bibinfo {pages} {100702}}\BibitemShut {NoStop}%
\bibitem [{\citenamefont {Emma}\ \emph {et~al.}(2012)\citenamefont {Emma},
  \citenamefont {Marcus}, \citenamefont {Penn}, \citenamefont {Prosnitz},\ and\
  \citenamefont {Reinsch}}]{NGLSTN}%
  \BibitemOpen
  \bibfield  {author} {\bibinfo {author} {\bibnamefont {Emma}, \bibfnamefont
  {P.}}, \bibinfo {author} {\bibfnamefont {G.}~\bibnamefont {Marcus}}, \bibinfo
  {author} {\bibfnamefont {G.}~\bibnamefont {Penn}}, \bibinfo {author}
  {\bibfnamefont {D.}~\bibnamefont {Prosnitz}}, \ and\ \bibinfo {author}
  {\bibfnamefont {M.}~\bibnamefont {Reinsch}}} (\bibinfo {year} {2012}),\
  \href@noop {} {\emph {\bibinfo {title} {Possible FEL Designs for the
  NGLS}}},\ \bibinfo {type} {Preprint}\ \bibinfo {number}
  {NGLS-Technical-Note-33}\BibitemShut {NoStop}%
\bibitem [{\citenamefont {Emma}\ and\ \citenamefont
  {Stupakov}(2003)}]{emma03s}%
  \BibitemOpen
  \bibfield  {author} {\bibinfo {author} {\bibnamefont {Emma}, \bibfnamefont
  {P.}}, \ and\ \bibinfo {author} {\bibfnamefont {G.~V.}\ \bibnamefont
  {Stupakov}}} (\bibinfo {year} {2003}),\ \href@noop {} {\bibfield  {journal}
  {\bibinfo  {journal} {Phys. Rev. ST Accel. Beams}\ }\textbf {\bibinfo
  {volume} {6}},\ \bibinfo {pages} {030701}}\BibitemShut {NoStop}%
\bibitem [{\citenamefont {Evain}\ \emph {et~al.}(2012)\citenamefont {Evain},
  \citenamefont {Loulergue}, \citenamefont {Nadji}, \citenamefont {Filhol},
  \citenamefont {Couprie},\ and\ \citenamefont {Zholents}}]{EEHGSOLEIL}%
  \BibitemOpen
  \bibfield  {author} {\bibinfo {author} {\bibnamefont {Evain}, \bibfnamefont
  {C.}}, \bibinfo {author} {\bibfnamefont {A.}~\bibnamefont {Loulergue}},
  \bibinfo {author} {\bibfnamefont {A.}~\bibnamefont {Nadji}}, \bibinfo
  {author} {\bibfnamefont {J.}~\bibnamefont {Filhol}}, \bibinfo {author}
  {\bibfnamefont {M.}~\bibnamefont {Couprie}}, \ and\ \bibinfo {author}
  {\bibfnamefont {A.}~\bibnamefont {Zholents}}} (\bibinfo {year} {2012}),\
  \href@noop {} {\bibfield  {journal} {\bibinfo  {journal} {New J. Phys.}\
  }\textbf {\bibinfo {volume} {14}},\ \bibinfo {pages} {023003}}\BibitemShut
  {NoStop}%
\bibitem [{\citenamefont {Fawley}(2008)}]{Fawley2008}%
  \BibitemOpen
  \bibfield  {author} {\bibinfo {author} {\bibnamefont {Fawley}, \bibfnamefont
  {W.}}} (\bibinfo {year} {2008}),\ \href
  {http://dx.doi.org/10.1016/j.nima.2008.04.051} {\bibfield  {journal}
  {\bibinfo  {journal} {Nucl. Instr. and Methods in Phys. Res. A}\ }\textbf
  {\bibinfo {volume} {593}}~(\bibinfo {number} {1€-2}),\ \bibinfo {pages}
  {111}}\BibitemShut {NoStop}%
\bibitem [{\citenamefont {Feikes}\ \emph {et~al.}(2009)\citenamefont {Feikes},
  \citenamefont {von Hartrott}, \citenamefont {W\"{u}stefeld}, \citenamefont
  {Hoehl}, \citenamefont {Klein}, \citenamefont {Muller},\ and\ \citenamefont
  {Ulm}}]{Feik2}%
  \BibitemOpen
  \bibfield  {author} {\bibinfo {author} {\bibnamefont {Feikes}, \bibfnamefont
  {J.}}, \bibinfo {author} {\bibfnamefont {M.}~\bibnamefont {von Hartrott}},
  \bibinfo {author} {\bibfnamefont {G.}~\bibnamefont {W\"{u}stefeld}}, \bibinfo
  {author} {\bibfnamefont {A.}~\bibnamefont {Hoehl}}, \bibinfo {author}
  {\bibfnamefont {R.}~\bibnamefont {Klein}}, \bibinfo {author} {\bibfnamefont
  {R.}~\bibnamefont {Muller}}, \ and\ \bibinfo {author} {\bibfnamefont
  {G.}~\bibnamefont {Ulm}}} (\bibinfo {year} {2009}),\ in\ \href@noop {} {\emph
  {\bibinfo {booktitle} {Proc. of the 2009 Particle Accelerator Conference,
  Vancouver, BC, Canada}}}\BibitemShut {NoStop}%
\bibitem [{\citenamefont {Feikes}\ \emph {et~al.}(2004)\citenamefont {Feikes},
  \citenamefont {Holldack}, \citenamefont {Kuske},\ and\ \citenamefont
  {W\"{u}stefeld}}]{Feik1}%
  \BibitemOpen
  \bibfield  {author} {\bibinfo {author} {\bibnamefont {Feikes}, \bibfnamefont
  {J.}}, \bibinfo {author} {\bibfnamefont {K.}~\bibnamefont {Holldack}},
  \bibinfo {author} {\bibfnamefont {P.}~\bibnamefont {Kuske}}, \ and\ \bibinfo
  {author} {\bibfnamefont {G.}~\bibnamefont {W\"{u}stefeld}}} (\bibinfo {year}
  {2004}),\ in\ \href@noop {} {\emph {\bibinfo {booktitle} {Proc. of the 2004
  European Particle Accelerator Conference, Lucerne, Switzerland}}},\ p.\
  \bibinfo {pages} {2290}\BibitemShut {NoStop}%
\bibitem [{\citenamefont {Feldhaus}\ \emph {et~al.}(1997)\citenamefont
  {Feldhaus}, \citenamefont {Saldin}, \citenamefont {Schneider}, \citenamefont
  {Schneidmiller},\ and\ \citenamefont {Yurkov}}]{Feldhaus1997341}%
  \BibitemOpen
  \bibfield  {author} {\bibinfo {author} {\bibnamefont {Feldhaus},
  \bibfnamefont {J.}}, \bibinfo {author} {\bibfnamefont {E.}~\bibnamefont
  {Saldin}}, \bibinfo {author} {\bibfnamefont {J.}~\bibnamefont {Schneider}},
  \bibinfo {author} {\bibfnamefont {E.}~\bibnamefont {Schneidmiller}}, \ and\
  \bibinfo {author} {\bibfnamefont {M.}~\bibnamefont {Yurkov}}} (\bibinfo
  {year} {1997}),\ \href
  {http://www.sciencedirect.com/science/article/pii/S0030401897001636}
  {\bibfield  {journal} {\bibinfo  {journal} {Opt. Commun.}\ }\textbf {\bibinfo
  {volume} {140}}~(\bibinfo {number} {4-6}),\ \bibinfo {pages} {341
  }}\BibitemShut {NoStop}%
\bibitem [{\citenamefont {Feng}\ \emph {et~al.}(2012)\citenamefont {Feng},
  \citenamefont {Chen},\ and\ \citenamefont {Zhao}}]{ML4}%
  \BibitemOpen
  \bibfield  {author} {\bibinfo {author} {\bibnamefont {Feng}, \bibfnamefont
  {C.}}, \bibinfo {author} {\bibfnamefont {J.}~\bibnamefont {Chen}}, \ and\
  \bibinfo {author} {\bibfnamefont {Z.}~\bibnamefont {Zhao}}} (\bibinfo {year}
  {2012}),\ \href@noop {} {\bibfield  {journal} {\bibinfo  {journal} {Phys.
  Rev. ST Accel. Beams}\ }\textbf {\bibinfo {volume} {15}},\ \bibinfo {pages}
  {080703}}\BibitemShut {NoStop}%
\bibitem [{\citenamefont {Feng}\ \emph {et~al.}(2013)\citenamefont {Feng},
  \citenamefont {Shen}, \citenamefont {Zhang}, \citenamefont {Wang},
  \citenamefont {Zhao},\ and\ \citenamefont {Xiang}}]{CPACHG2}%
  \BibitemOpen
  \bibfield  {author} {\bibinfo {author} {\bibnamefont {Feng}, \bibfnamefont
  {C.}}, \bibinfo {author} {\bibfnamefont {L.}~\bibnamefont {Shen}}, \bibinfo
  {author} {\bibfnamefont {M.}~\bibnamefont {Zhang}}, \bibinfo {author}
  {\bibfnamefont {D.}~\bibnamefont {Wang}}, \bibinfo {author} {\bibfnamefont
  {Z.}~\bibnamefont {Zhao}}, \ and\ \bibinfo {author} {\bibfnamefont
  {D.}~\bibnamefont {Xiang}}} (\bibinfo {year} {2013}),\ \href@noop {}
  {\bibfield  {journal} {\bibinfo  {journal} {Nucl. Instrum. Methods Phys. Res.
  A}\ }\textbf {\bibinfo {volume} {712}},\ \bibinfo {pages} {113}}\BibitemShut
  {NoStop}%
\bibitem [{\citenamefont {Fiocco}\ and\ \citenamefont {Thompson}(1963)}]{1963}%
  \BibitemOpen
  \bibfield  {author} {\bibinfo {author} {\bibnamefont {Fiocco}, \bibfnamefont
  {G.}}, \ and\ \bibinfo {author} {\bibfnamefont {E.}~\bibnamefont {Thompson}}}
  (\bibinfo {year} {1963}),\ \href
  {http://link.aps.org/doi/10.1103/PhysRevLett.10.89} {\bibfield  {journal}
  {\bibinfo  {journal} {Phys. Rev. Lett.}\ }\textbf {\bibinfo {volume} {10}},\
  \bibinfo {pages} {89}}\BibitemShut {NoStop}%
\bibitem [{\citenamefont {Frassetto}\ \emph {et~al.}(2008)\citenamefont
  {Frassetto}, \citenamefont {Giannessi},\ and\ \citenamefont
  {Poletto}}]{GratingEFF}%
  \BibitemOpen
  \bibfield  {author} {\bibinfo {author} {\bibnamefont {Frassetto},
  \bibfnamefont {F.}}, \bibinfo {author} {\bibfnamefont {L.}~\bibnamefont
  {Giannessi}}, \ and\ \bibinfo {author} {\bibfnamefont {L.}~\bibnamefont
  {Poletto}}} (\bibinfo {year} {2008}),\ \href@noop {} {\bibfield  {journal}
  {\bibinfo  {journal} {Nucl. Instrum. Methods Phys. Res. A}\ }\textbf
  {\bibinfo {volume} {593}},\ \bibinfo {pages} {14}}\BibitemShut {NoStop}%
\bibitem [{\citenamefont {Freund}\ and\ \citenamefont
  {O'Shea}(2000)}]{FreundTC}%
  \BibitemOpen
  \bibfield  {author} {\bibinfo {author} {\bibnamefont {Freund}, \bibfnamefont
  {H.~P.}}, \ and\ \bibinfo {author} {\bibfnamefont {P.~G.}\ \bibnamefont
  {O'Shea}}} (\bibinfo {year} {2000}),\ \href@noop {} {\bibfield  {journal}
  {\bibinfo  {journal} {Phys. Rev. Lett.}\ }\textbf {\bibinfo {volume} {84}},\
  \bibinfo {pages} {2861}}\BibitemShut {NoStop}%
\bibitem [{\citenamefont {Gai}\ \emph {et~al.}(1988)\citenamefont {Gai},
  \citenamefont {Schoessow}, \citenamefont {Cole}, \citenamefont {Konecny},
  \citenamefont {Norem}, \citenamefont {Rosenzweig},\ and\ \citenamefont
  {Simpson}}]{Gai}%
  \BibitemOpen
  \bibfield  {author} {\bibinfo {author} {\bibnamefont {Gai}, \bibfnamefont
  {W.}}, \bibinfo {author} {\bibfnamefont {P.}~\bibnamefont {Schoessow}},
  \bibinfo {author} {\bibfnamefont {B.}~\bibnamefont {Cole}}, \bibinfo {author}
  {\bibfnamefont {R.}~\bibnamefont {Konecny}}, \bibinfo {author} {\bibfnamefont
  {J.}~\bibnamefont {Norem}}, \bibinfo {author} {\bibfnamefont
  {J.}~\bibnamefont {Rosenzweig}}, \ and\ \bibinfo {author} {\bibfnamefont
  {J.}~\bibnamefont {Simpson}}} (\bibinfo {year} {1988}),\ \href@noop {}
  {\bibfield  {journal} {\bibinfo  {journal} {Phys. Rev. Lett.}\ }\textbf
  {\bibinfo {volume} {61}},\ \bibinfo {pages} {2756}}\BibitemShut {NoStop}%
\bibitem [{\citenamefont {Gallardo}\ \emph {et~al.}(1988)\citenamefont
  {Gallardo}, \citenamefont {Fernow}, \citenamefont {Palmer},\ and\
  \citenamefont {Pellegrini}}]{Gallardo}%
  \BibitemOpen
  \bibfield  {author} {\bibinfo {author} {\bibnamefont {Gallardo},
  \bibfnamefont {J.}}, \bibinfo {author} {\bibfnamefont {R.}~\bibnamefont
  {Fernow}}, \bibinfo {author} {\bibfnamefont {R.}~\bibnamefont {Palmer}}, \
  and\ \bibinfo {author} {\bibfnamefont {C.}~\bibnamefont {Pellegrini}}}
  (\bibinfo {year} {1988}),\ \href {\doibase 10.1109/3.7085} {\bibfield
  {journal} {\bibinfo  {journal} {Quantum Electronics, IEEE Journal of}\
  }\textbf {\bibinfo {volume} {24}}~(\bibinfo {number} {8}),\ \bibinfo {pages}
  {1557}}\BibitemShut {NoStop}%
\bibitem [{\citenamefont {Gandhi}\ \emph {et~al.}(2013)\citenamefont {Gandhi},
  \citenamefont {Penn}, \citenamefont {Reinsch}, \citenamefont {Wurtele},\ and\
  \citenamefont {Fawley}}]{Gandhi}%
  \BibitemOpen
  \bibfield  {author} {\bibinfo {author} {\bibnamefont {Gandhi}, \bibfnamefont
  {P.}}, \bibinfo {author} {\bibfnamefont {G.}~\bibnamefont {Penn}}, \bibinfo
  {author} {\bibfnamefont {M.}~\bibnamefont {Reinsch}}, \bibinfo {author}
  {\bibfnamefont {J.~S.}\ \bibnamefont {Wurtele}}, \ and\ \bibinfo {author}
  {\bibfnamefont {W.~M.}\ \bibnamefont {Fawley}}} (\bibinfo {year} {2013}),\
  \href@noop {} {\bibfield  {journal} {\bibinfo  {journal} {Phys. Rev. ST
  Accel. Beams}\ }\textbf {\bibinfo {volume} {16}},\ \bibinfo {pages}
  {020703}}\BibitemShut {NoStop}%
\bibitem [{\citenamefont {Geloni}\ \emph {et~al.}(2010)\citenamefont {Geloni},
  \citenamefont {Kocharyan},\ and\ \citenamefont {Saldin}}]{GeloniTC}%
  \BibitemOpen
  \bibfield  {author} {\bibinfo {author} {\bibnamefont {Geloni}, \bibfnamefont
  {G.}}, \bibinfo {author} {\bibfnamefont {V.}~\bibnamefont {Kocharyan}}, \
  and\ \bibinfo {author} {\bibfnamefont {E.}~\bibnamefont {Saldin}}} (\bibinfo
  {year} {2010}),\ \href@noop {} {\enquote {\bibinfo {title} {{Generation of
  doublet spectral lines at self-seeded X-ray FELs}},}\ }\Eprint
  {http://arxiv.org/abs/1011.3910} {arXiv:1011.3910} \BibitemShut {NoStop}%
\bibitem [{\citenamefont {Geloni}\ \emph {et~al.}(2011)\citenamefont {Geloni},
  \citenamefont {Kocharyan},\ and\ \citenamefont
  {Saldin}}]{doi:10.1080/09500340.2011.586473}%
  \BibitemOpen
  \bibfield  {author} {\bibinfo {author} {\bibnamefont {Geloni}, \bibfnamefont
  {G.}}, \bibinfo {author} {\bibfnamefont {V.}~\bibnamefont {Kocharyan}}, \
  and\ \bibinfo {author} {\bibfnamefont {E.}~\bibnamefont {Saldin}}} (\bibinfo
  {year} {2011}),\ \href
  {http://www.tandfonline.com/doi/abs/10.1080/09500340.2011.586473} {\bibfield
  {journal} {\bibinfo  {journal} {Journal of Modern Optics}\ }\textbf {\bibinfo
  {volume} {58}}~(\bibinfo {number} {16}),\ \bibinfo {pages}
  {1391}}\BibitemShut {NoStop}%
\bibitem [{\citenamefont {Geloni}\ \emph {et~al.}(2007)\citenamefont {Geloni},
  \citenamefont {Saldin}, \citenamefont {Schneidmiller},\ and\ \citenamefont
  {Yurkov}}]{SCGeloni}%
  \BibitemOpen
  \bibfield  {author} {\bibinfo {author} {\bibnamefont {Geloni}, \bibfnamefont
  {G.}}, \bibinfo {author} {\bibfnamefont {E.}~\bibnamefont {Saldin}}, \bibinfo
  {author} {\bibfnamefont {E.}~\bibnamefont {Schneidmiller}}, \ and\ \bibinfo
  {author} {\bibfnamefont {M.}~\bibnamefont {Yurkov}}} (\bibinfo {year}
  {2007}),\ \href@noop {} {\bibfield  {journal} {\bibinfo  {journal} {Nucl.
  Instrum. Methods Phys. Res. A}\ }\textbf {\bibinfo {volume} {583}},\ \bibinfo
  {pages} {228}}\BibitemShut {NoStop}%
\bibitem [{\citenamefont {Giannessi}\ \emph {et~al.}(2011)\citenamefont
  {Giannessi}, \citenamefont {Bacci}, \citenamefont {Bellaveglia},
  \citenamefont {Briquez}, \citenamefont {Castellano}, \citenamefont
  {Chiadroni}, \citenamefont {Cianchi}, \citenamefont {Ciocci}, \citenamefont
  {Couprie}, \citenamefont {Cultrera}, \citenamefont {Dattoli}, \citenamefont
  {Filippetto}, \citenamefont {Del~Franco}, \citenamefont {Di~Pirro},
  \citenamefont {Ferrario}, \citenamefont {Ficcadenti}, \citenamefont
  {Frassetto}, \citenamefont {Gallo}, \citenamefont {Gatti}, \citenamefont
  {Labat}, \citenamefont {Marcus}, \citenamefont {Moreno}, \citenamefont
  {Mostacci}, \citenamefont {Pace}, \citenamefont {Petralia}, \citenamefont
  {Petrillo}, \citenamefont {Poletto}, \citenamefont {Quattromini},
  \citenamefont {Rau}, \citenamefont {Ronsivalle}, \citenamefont {Rosenzweig},
  \citenamefont {Rossi}, \citenamefont {Rossi~Albertini}, \citenamefont
  {Sabia}, \citenamefont {Serluca}, \citenamefont {Spampinati}, \citenamefont
  {Spassovsky}, \citenamefont {Spataro}, \citenamefont {Surrenti},
  \citenamefont {Vaccarezza},\ and\ \citenamefont
  {Vicario}}]{GiannessiChirpedSeeded}%
  \BibitemOpen
  \bibfield  {author} {\bibinfo {author} {\bibnamefont {Giannessi},
  \bibfnamefont {L.}}, \bibinfo {author} {\bibfnamefont {A.}~\bibnamefont
  {Bacci}}, \bibinfo {author} {\bibfnamefont {M.}~\bibnamefont {Bellaveglia}},
  \bibinfo {author} {\bibfnamefont {F.}~\bibnamefont {Briquez}}, \bibinfo
  {author} {\bibfnamefont {M.}~\bibnamefont {Castellano}}, \bibinfo {author}
  {\bibfnamefont {E.}~\bibnamefont {Chiadroni}}, \bibinfo {author}
  {\bibfnamefont {A.}~\bibnamefont {Cianchi}}, \bibinfo {author} {\bibfnamefont
  {F.}~\bibnamefont {Ciocci}}, \bibinfo {author} {\bibfnamefont {M.~E.}\
  \bibnamefont {Couprie}}, \bibinfo {author} {\bibfnamefont {L.}~\bibnamefont
  {Cultrera}}, \bibinfo {author} {\bibfnamefont {G.}~\bibnamefont {Dattoli}},
  \bibinfo {author} {\bibfnamefont {D.}~\bibnamefont {Filippetto}}, \bibinfo
  {author} {\bibfnamefont {M.}~\bibnamefont {Del~Franco}}, \bibinfo {author}
  {\bibfnamefont {G.}~\bibnamefont {Di~Pirro}}, \bibinfo {author}
  {\bibfnamefont {M.}~\bibnamefont {Ferrario}}, \bibinfo {author}
  {\bibfnamefont {L.}~\bibnamefont {Ficcadenti}}, \bibinfo {author}
  {\bibfnamefont {F.}~\bibnamefont {Frassetto}}, \bibinfo {author}
  {\bibfnamefont {A.}~\bibnamefont {Gallo}}, \bibinfo {author} {\bibfnamefont
  {G.}~\bibnamefont {Gatti}}, \bibinfo {author} {\bibfnamefont
  {M.}~\bibnamefont {Labat}}, \bibinfo {author} {\bibfnamefont
  {G.}~\bibnamefont {Marcus}}, \bibinfo {author} {\bibfnamefont
  {M.}~\bibnamefont {Moreno}}, \bibinfo {author} {\bibfnamefont
  {A.}~\bibnamefont {Mostacci}}, \bibinfo {author} {\bibfnamefont
  {E.}~\bibnamefont {Pace}}, \bibinfo {author} {\bibfnamefont {A.}~\bibnamefont
  {Petralia}}, \bibinfo {author} {\bibfnamefont {V.}~\bibnamefont {Petrillo}},
  \bibinfo {author} {\bibfnamefont {L.}~\bibnamefont {Poletto}}, \bibinfo
  {author} {\bibfnamefont {M.}~\bibnamefont {Quattromini}}, \bibinfo {author}
  {\bibfnamefont {J.~V.}\ \bibnamefont {Rau}}, \bibinfo {author} {\bibfnamefont
  {C.}~\bibnamefont {Ronsivalle}}, \bibinfo {author} {\bibfnamefont
  {J.}~\bibnamefont {Rosenzweig}}, \bibinfo {author} {\bibfnamefont {A.~R.}\
  \bibnamefont {Rossi}}, \bibinfo {author} {\bibfnamefont {V.}~\bibnamefont
  {Rossi~Albertini}}, \bibinfo {author} {\bibfnamefont {E.}~\bibnamefont
  {Sabia}}, \bibinfo {author} {\bibfnamefont {M.}~\bibnamefont {Serluca}},
  \bibinfo {author} {\bibfnamefont {S.}~\bibnamefont {Spampinati}}, \bibinfo
  {author} {\bibfnamefont {I.}~\bibnamefont {Spassovsky}}, \bibinfo {author}
  {\bibfnamefont {B.}~\bibnamefont {Spataro}}, \bibinfo {author} {\bibfnamefont
  {V.}~\bibnamefont {Surrenti}}, \bibinfo {author} {\bibfnamefont
  {C.}~\bibnamefont {Vaccarezza}}, \ and\ \bibinfo {author} {\bibfnamefont
  {C.}~\bibnamefont {Vicario}}} (\bibinfo {year} {2011}),\ \href
  {http://link.aps.org/doi/10.1103/PhysRevLett.106.144801} {\bibfield
  {journal} {\bibinfo  {journal} {Phys. Rev. Lett.}\ }\textbf {\bibinfo
  {volume} {106}},\ \bibinfo {pages} {144801}}\BibitemShut {NoStop}%
\bibitem [{\citenamefont {Girard}\ \emph {et~al.}(1984)\citenamefont {Girard},
  \citenamefont {Lapierre}, \citenamefont {Ortega}, \citenamefont {Bazin},
  \citenamefont {Billardon}, \citenamefont {Elleaume}, \citenamefont {Bergher},
  \citenamefont {Velghe},\ and\ \citenamefont {Petroff}}]{LURE}%
  \BibitemOpen
  \bibfield  {author} {\bibinfo {author} {\bibnamefont {Girard}, \bibfnamefont
  {B.}}, \bibinfo {author} {\bibfnamefont {Y.}~\bibnamefont {Lapierre}},
  \bibinfo {author} {\bibfnamefont {J.~M.}\ \bibnamefont {Ortega}}, \bibinfo
  {author} {\bibfnamefont {C.}~\bibnamefont {Bazin}}, \bibinfo {author}
  {\bibfnamefont {M.}~\bibnamefont {Billardon}}, \bibinfo {author}
  {\bibfnamefont {P.}~\bibnamefont {Elleaume}}, \bibinfo {author}
  {\bibfnamefont {M.}~\bibnamefont {Bergher}}, \bibinfo {author} {\bibfnamefont
  {M.}~\bibnamefont {Velghe}}, \ and\ \bibinfo {author} {\bibfnamefont
  {Y.}~\bibnamefont {Petroff}}} (\bibinfo {year} {1984}),\ \href@noop {}
  {\bibfield  {journal} {\bibinfo  {journal} {Phys. Rev. Lett.}\ }\textbf
  {\bibinfo {volume} {53}},\ \bibinfo {pages} {2405}}\BibitemShut {NoStop}%
\bibitem [{\citenamefont {Gordon}\ \emph {et~al.}(2001)\citenamefont {Gordon},
  \citenamefont {Sprangle}, \citenamefont {Hafizi},\ and\ \citenamefont
  {Roberson}}]{Gordon2001190}%
  \BibitemOpen
  \bibfield  {author} {\bibinfo {author} {\bibnamefont {Gordon}, \bibfnamefont
  {D.}}, \bibinfo {author} {\bibfnamefont {P.}~\bibnamefont {Sprangle}},
  \bibinfo {author} {\bibfnamefont {B.}~\bibnamefont {Hafizi}}, \ and\ \bibinfo
  {author} {\bibfnamefont {C.}~\bibnamefont {Roberson}}} (\bibinfo {year}
  {2001}),\ \href
  {http://www.sciencedirect.com/science/article/pii/S0168900201016941}
  {\bibfield  {journal} {\bibinfo  {journal} {Nuclear Instruments and Methods
  in Physics Research Section A: Accelerators, Spectrometers, Detectors and
  Associated Equipment}\ }\textbf {\bibinfo {volume} {475}}~(\bibinfo {number}
  {1-€"3}),\ \bibinfo {pages} {190 }}\BibitemShut {NoStop}%
\bibitem [{\citenamefont {Gover}\ and\ \citenamefont
  {Dyunin}(2009)}]{gover09d}%
  \BibitemOpen
  \bibfield  {author} {\bibinfo {author} {\bibnamefont {Gover}, \bibfnamefont
  {A.}}, \ and\ \bibinfo {author} {\bibfnamefont {E.}~\bibnamefont {Dyunin}}}
  (\bibinfo {year} {2009}),\ \href@noop {} {\bibfield  {journal} {\bibinfo
  {journal} {Phys. Rev. Lett.}\ }\textbf {\bibinfo {volume} {102}},\ \bibinfo
  {pages} {154801}}\BibitemShut {NoStop}%
\bibitem [{\citenamefont {Gover}\ \emph {et~al.}(2012)\citenamefont {Gover},
  \citenamefont {Nause}, \citenamefont {Dyunin},\ and\ \citenamefont
  {Fedurin}}]{gover_nature}%
  \BibitemOpen
  \bibfield  {author} {\bibinfo {author} {\bibnamefont {Gover}, \bibfnamefont
  {A.}}, \bibinfo {author} {\bibfnamefont {A.}~\bibnamefont {Nause}}, \bibinfo
  {author} {\bibfnamefont {E.}~\bibnamefont {Dyunin}}, \ and\ \bibinfo {author}
  {\bibfnamefont {M.}~\bibnamefont {Fedurin}}} (\bibinfo {year} {2012}),\
  \href@noop {} {\bibfield  {journal} {\bibinfo  {journal} {Nature Phys.}\
  }\textbf {\bibinfo {volume} {8}}~(\bibinfo {number} {12}),\ \bibinfo {pages}
  {877}}\BibitemShut {NoStop}%
\bibitem [{\citenamefont {Han}\ \emph {et~al.}(2012)\citenamefont {Han},
  \citenamefont {Kang},\ and\ \citenamefont {Ko}}]{laser_heater_PAL-FEL}%
  \BibitemOpen
  \bibfield  {author} {\bibinfo {author} {\bibnamefont {Han}, \bibfnamefont
  {J.-H.}}, \bibinfo {author} {\bibfnamefont {H.-S.}\ \bibnamefont {Kang}}, \
  and\ \bibinfo {author} {\bibfnamefont {I.~S.}\ \bibnamefont {Ko}}} (\bibinfo
  {year} {2012}),\ in\ \href@noop {} {\emph {\bibinfo {booktitle} {2012
  International Particle Accelerator Conference}}},\ p.\ \bibinfo {pages}
  {1735}\BibitemShut {NoStop}%
\bibitem [{\citenamefont {Heifets}\ \emph {et~al.}(2002)\citenamefont
  {Heifets}, \citenamefont {Stupakov},\ and\ \citenamefont
  {Krinsky}}]{heifets02sk}%
  \BibitemOpen
  \bibfield  {author} {\bibinfo {author} {\bibnamefont {Heifets}, \bibfnamefont
  {S.}}, \bibinfo {author} {\bibfnamefont {G.}~\bibnamefont {Stupakov}}, \ and\
  \bibinfo {author} {\bibfnamefont {S.}~\bibnamefont {Krinsky}}} (\bibinfo
  {year} {2002}),\ \href@noop {} {\bibfield  {journal} {\bibinfo  {journal}
  {Phys. Rev. ST Accel. Beams}\ }\textbf {\bibinfo {volume} {5}},\ \bibinfo
  {pages} {064401}}\BibitemShut {NoStop}%
\bibitem [{\citenamefont {Hemsing}\ \emph {et~al.}(2014)\citenamefont
  {Hemsing}, \citenamefont {Dunning}, \citenamefont {Hast}, \citenamefont
  {Raubenheimer}, \citenamefont {Weathersby},\ and\ \citenamefont
  {Xiang}}]{echo14}%
  \BibitemOpen
  \bibfield  {author} {\bibinfo {author} {\bibnamefont {Hemsing}, \bibfnamefont
  {E.}}, \bibinfo {author} {\bibfnamefont {M.}~\bibnamefont {Dunning}},
  \bibinfo {author} {\bibfnamefont {C.}~\bibnamefont {Hast}}, \bibinfo {author}
  {\bibfnamefont {T.}~\bibnamefont {Raubenheimer}}, \bibinfo {author}
  {\bibfnamefont {S.}~\bibnamefont {Weathersby}}, \ and\ \bibinfo {author}
  {\bibfnamefont {D.}~\bibnamefont {Xiang}}} (\bibinfo {year} {2014}),\
  \href@noop {} {\enquote {\bibinfo {title} {{Demonstration of highly coherent
  vacuum ultraviolet radiation from an echo-enabled harmonic generation
  free-electron laser}},}\ }\bibinfo {note} {To be published}\BibitemShut
  {NoStop}%
\bibitem [{\citenamefont {Hemsing}\ \emph {et~al.}(2013)\citenamefont
  {Hemsing}, \citenamefont {Knyazik}, \citenamefont {Dunning}, \citenamefont
  {Xiang}, \citenamefont {Marinelli}, \citenamefont {Hast},\ and\ \citenamefont
  {Rosenzweig}}]{hemsing2013coherent}%
  \BibitemOpen
  \bibfield  {author} {\bibinfo {author} {\bibnamefont {Hemsing}, \bibfnamefont
  {E.}}, \bibinfo {author} {\bibfnamefont {A.}~\bibnamefont {Knyazik}},
  \bibinfo {author} {\bibfnamefont {M.}~\bibnamefont {Dunning}}, \bibinfo
  {author} {\bibfnamefont {D.}~\bibnamefont {Xiang}}, \bibinfo {author}
  {\bibfnamefont {A.}~\bibnamefont {Marinelli}}, \bibinfo {author}
  {\bibfnamefont {C.}~\bibnamefont {Hast}}, \ and\ \bibinfo {author}
  {\bibfnamefont {J.~B.}\ \bibnamefont {Rosenzweig}}} (\bibinfo {year}
  {2013}),\ \href@noop {} {\bibfield  {journal} {\bibinfo  {journal} {Nature
  Phys.}\ }\textbf {\bibinfo {volume} {9}}~(\bibinfo {number} {9}),\ \bibinfo
  {pages} {549}}\BibitemShut {NoStop}%
\bibitem [{\citenamefont {Hemsing}\ \emph {et~al.}(2012)\citenamefont
  {Hemsing}, \citenamefont {Knyazik}, \citenamefont {O'Shea}, \citenamefont
  {Marinelli}, \citenamefont {Musumeci}, \citenamefont {Williams},
  \citenamefont {Tochitsky},\ and\ \citenamefont
  {Rosenzweig}}]{hemsing:091110}%
  \BibitemOpen
  \bibfield  {author} {\bibinfo {author} {\bibnamefont {Hemsing}, \bibfnamefont
  {E.}}, \bibinfo {author} {\bibfnamefont {A.}~\bibnamefont {Knyazik}},
  \bibinfo {author} {\bibfnamefont {F.}~\bibnamefont {O'Shea}}, \bibinfo
  {author} {\bibfnamefont {A.}~\bibnamefont {Marinelli}}, \bibinfo {author}
  {\bibfnamefont {P.}~\bibnamefont {Musumeci}}, \bibinfo {author}
  {\bibfnamefont {O.}~\bibnamefont {Williams}}, \bibinfo {author}
  {\bibfnamefont {S.}~\bibnamefont {Tochitsky}}, \ and\ \bibinfo {author}
  {\bibfnamefont {J.~B.}\ \bibnamefont {Rosenzweig}}} (\bibinfo {year}
  {2012}),\ \href {http://link.aip.org/link/?APL/100/091110/1} {\bibfield
  {journal} {\bibinfo  {journal} {Applied Physics Letters}\ }\textbf {\bibinfo
  {volume} {100}}~(\bibinfo {number} {9}),\ \bibinfo {eid}
  {091110}}\BibitemShut {NoStop}%
\bibitem [{\citenamefont {Hemsing}\ and\ \citenamefont
  {Marinelli}(2012)}]{HemsingECHOOAM}%
  \BibitemOpen
  \bibfield  {author} {\bibinfo {author} {\bibnamefont {Hemsing}, \bibfnamefont
  {E.}}, \ and\ \bibinfo {author} {\bibfnamefont {A.}~\bibnamefont
  {Marinelli}}} (\bibinfo {year} {2012}),\ \href@noop {} {\bibfield  {journal}
  {\bibinfo  {journal} {Phys. Rev. Lett.}\ }\textbf {\bibinfo {volume} {109}},\
  \bibinfo {pages} {224801}}\BibitemShut {NoStop}%
\bibitem [{\citenamefont {Hemsing}\ \emph {et~al.}(2011)\citenamefont
  {Hemsing}, \citenamefont {Marinelli},\ and\ \citenamefont
  {Rosenzweig}}]{HemsingOAMFEL}%
  \BibitemOpen
  \bibfield  {author} {\bibinfo {author} {\bibnamefont {Hemsing}, \bibfnamefont
  {E.}}, \bibinfo {author} {\bibfnamefont {A.}~\bibnamefont {Marinelli}}, \
  and\ \bibinfo {author} {\bibfnamefont {J.~B.}\ \bibnamefont {Rosenzweig}}}
  (\bibinfo {year} {2011}),\ \href@noop {} {\bibfield  {journal} {\bibinfo
  {journal} {Phys. Rev. Lett.}\ }\textbf {\bibinfo {volume} {106}}~(\bibinfo
  {number} {16}),\ \bibinfo {pages} {164803}}\BibitemShut {NoStop}%
\bibitem [{\citenamefont {Hemsing}\ \emph
  {et~al.}(2009{\natexlab{a}})\citenamefont {Hemsing}, \citenamefont
  {Musumeci}, \citenamefont {Marinelli},\ and\ \citenamefont
  {Rosenzweig}}]{HemsingFEL09}%
  \BibitemOpen
  \bibfield  {author} {\bibinfo {author} {\bibnamefont {Hemsing}, \bibfnamefont
  {E.}}, \bibinfo {author} {\bibfnamefont {P.}~\bibnamefont {Musumeci}},
  \bibinfo {author} {\bibfnamefont {A.}~\bibnamefont {Marinelli}}, \ and\
  \bibinfo {author} {\bibfnamefont {J.~B.}\ \bibnamefont {Rosenzweig}}}
  (\bibinfo {year} {2009}{\natexlab{a}}),\ in\ \href@noop {} {\emph {\bibinfo
  {booktitle} {Proceedings of the 2009 FEL Conference}}}\ (\bibinfo {address}
  {Liverpool, UK})\ p.\ \bibinfo {pages} {188}\BibitemShut {NoStop}%
\bibitem [{\citenamefont {Hemsing}\ \emph
  {et~al.}(2009{\natexlab{b}})\citenamefont {Hemsing}, \citenamefont
  {Musumeci}, \citenamefont {Reiche}, \citenamefont {Tikhoplav}, \citenamefont
  {Marinelli}, \citenamefont {Rosenzweig},\ and\ \citenamefont {Gover}}]{OAM1}%
  \BibitemOpen
  \bibfield  {author} {\bibinfo {author} {\bibnamefont {Hemsing}, \bibfnamefont
  {E.}}, \bibinfo {author} {\bibfnamefont {P.}~\bibnamefont {Musumeci}},
  \bibinfo {author} {\bibfnamefont {S.}~\bibnamefont {Reiche}}, \bibinfo
  {author} {\bibfnamefont {R.}~\bibnamefont {Tikhoplav}}, \bibinfo {author}
  {\bibfnamefont {A.}~\bibnamefont {Marinelli}}, \bibinfo {author}
  {\bibfnamefont {J.~B.}\ \bibnamefont {Rosenzweig}}, \ and\ \bibinfo {author}
  {\bibfnamefont {A.}~\bibnamefont {Gover}}} (\bibinfo {year}
  {2009}{\natexlab{b}}),\ \href@noop {} {\bibfield  {journal} {\bibinfo
  {journal} {Phys. Rev. Lett.}\ }\textbf {\bibinfo {volume} {102}},\ \bibinfo
  {pages} {174801}}\BibitemShut {NoStop}%
\bibitem [{\citenamefont {Hemsing}\ and\ \citenamefont
  {Xiang}(2013)}]{PhysRevSTAB.16.010706}%
  \BibitemOpen
  \bibfield  {author} {\bibinfo {author} {\bibnamefont {Hemsing}, \bibfnamefont
  {E.}}, \ and\ \bibinfo {author} {\bibfnamefont {D.}~\bibnamefont {Xiang}}}
  (\bibinfo {year} {2013}),\ \href
  {http://link.aps.org/doi/10.1103/PhysRevSTAB.16.010706} {\bibfield  {journal}
  {\bibinfo  {journal} {Phys. Rev. ST Accel. Beams}\ }\textbf {\bibinfo
  {volume} {16}},\ \bibinfo {pages} {010706}}\BibitemShut {NoStop}%
\bibitem [{\citenamefont {Henderson}\ and\ \citenamefont
  {McNeil}(2012)}]{EEHGML}%
  \BibitemOpen
  \bibfield  {author} {\bibinfo {author} {\bibnamefont {Henderson},
  \bibfnamefont {J.}}, \ and\ \bibinfo {author} {\bibfnamefont
  {B.}~\bibnamefont {McNeil}}} (\bibinfo {year} {2012}),\ \href@noop {}
  {\bibfield  {journal} {\bibinfo  {journal} {Europhysics Letters}\ }\textbf
  {\bibinfo {volume} {100}},\ \bibinfo {pages} {64001}}\BibitemShut {NoStop}%
\bibitem [{\citenamefont {Holldack}\ \emph {et~al.}(2006)\citenamefont
  {Holldack}, \citenamefont {Khan}, \citenamefont {Mitzner},\ and\
  \citenamefont {Quas}}]{Holl}%
  \BibitemOpen
  \bibfield  {author} {\bibinfo {author} {\bibnamefont {Holldack},
  \bibfnamefont {K.}}, \bibinfo {author} {\bibfnamefont {S.}~\bibnamefont
  {Khan}}, \bibinfo {author} {\bibfnamefont {R.}~\bibnamefont {Mitzner}}, \
  and\ \bibinfo {author} {\bibfnamefont {T.}~\bibnamefont {Quas}}} (\bibinfo
  {year} {2006}),\ \href@noop {} {\bibfield  {journal} {\bibinfo  {journal}
  {Phys. Rev. Lett}\ }\textbf {\bibinfo {volume} {96}},\ \bibinfo {pages}
  {054801}}\BibitemShut {NoStop}%
\bibitem [{\citenamefont {Huang}\ \emph
  {et~al.}(2004{\natexlab{a}})\citenamefont {Huang}, \citenamefont {Borland},
  \citenamefont {Emma}, \citenamefont {Wu}, \citenamefont {Limborg},
  \citenamefont {Stupakov},\ and\ \citenamefont {Welch}}]{huang_laser_heater}%
  \BibitemOpen
  \bibfield  {author} {\bibinfo {author} {\bibnamefont {Huang}, \bibfnamefont
  {Z.}}, \bibinfo {author} {\bibfnamefont {M.}~\bibnamefont {Borland}},
  \bibinfo {author} {\bibfnamefont {P.}~\bibnamefont {Emma}}, \bibinfo {author}
  {\bibfnamefont {J.}~\bibnamefont {Wu}}, \bibinfo {author} {\bibfnamefont
  {C.}~\bibnamefont {Limborg}}, \bibinfo {author} {\bibfnamefont
  {G.}~\bibnamefont {Stupakov}}, \ and\ \bibinfo {author} {\bibfnamefont
  {J.}~\bibnamefont {Welch}}} (\bibinfo {year} {2004}{\natexlab{a}}),\
  \href@noop {} {\bibfield  {journal} {\bibinfo  {journal} {Phys. Rev. ST
  Accel. Beams}\ }\textbf {\bibinfo {volume} {7}},\ \bibinfo {pages}
  {074401}}\BibitemShut {NoStop}%
\bibitem [{\citenamefont {Huang}\ \emph {et~al.}(2010)\citenamefont {Huang},
  \citenamefont {Brachmann}, \citenamefont {Decker}, \citenamefont {Ding},
  \citenamefont {Dowell}, \citenamefont {Emma}, \citenamefont {Frisch},
  \citenamefont {Gilevich}, \citenamefont {Hays}, \citenamefont {Hering},
  \citenamefont {Iverson}, \citenamefont {Loos}, \citenamefont {Miahnahri},
  \citenamefont {Nuhn}, \citenamefont {Ratner}, \citenamefont {Stupakov},
  \citenamefont {Turner}, \citenamefont {Welch}, \citenamefont {White},
  \citenamefont {Wu},\ and\ \citenamefont {Xiang}}]{lcls_heater}%
  \BibitemOpen
  \bibfield  {author} {\bibinfo {author} {\bibnamefont {Huang}, \bibfnamefont
  {Z.}}, \bibinfo {author} {\bibfnamefont {A.}~\bibnamefont {Brachmann}},
  \bibinfo {author} {\bibfnamefont {F.-J.}\ \bibnamefont {Decker}}, \bibinfo
  {author} {\bibfnamefont {Y.}~\bibnamefont {Ding}}, \bibinfo {author}
  {\bibfnamefont {D.}~\bibnamefont {Dowell}}, \bibinfo {author} {\bibfnamefont
  {P.}~\bibnamefont {Emma}}, \bibinfo {author} {\bibfnamefont {J.}~\bibnamefont
  {Frisch}}, \bibinfo {author} {\bibfnamefont {S.}~\bibnamefont {Gilevich}},
  \bibinfo {author} {\bibfnamefont {G.}~\bibnamefont {Hays}}, \bibinfo {author}
  {\bibfnamefont {P.}~\bibnamefont {Hering}}, \bibinfo {author} {\bibfnamefont
  {R.}~\bibnamefont {Iverson}}, \bibinfo {author} {\bibfnamefont
  {H.}~\bibnamefont {Loos}}, \bibinfo {author} {\bibfnamefont {A.}~\bibnamefont
  {Miahnahri}}, \bibinfo {author} {\bibfnamefont {H.-D.}\ \bibnamefont {Nuhn}},
  \bibinfo {author} {\bibfnamefont {D.}~\bibnamefont {Ratner}}, \bibinfo
  {author} {\bibfnamefont {G.}~\bibnamefont {Stupakov}}, \bibinfo {author}
  {\bibfnamefont {J.}~\bibnamefont {Turner}}, \bibinfo {author} {\bibfnamefont
  {J.}~\bibnamefont {Welch}}, \bibinfo {author} {\bibfnamefont
  {W.}~\bibnamefont {White}}, \bibinfo {author} {\bibfnamefont
  {J.}~\bibnamefont {Wu}}, \ and\ \bibinfo {author} {\bibfnamefont
  {D.}~\bibnamefont {Xiang}}} (\bibinfo {year} {2010}),\ \href@noop {}
  {\bibfield  {journal} {\bibinfo  {journal} {Phys. Rev. ST Accel. Beams}\
  }\textbf {\bibinfo {volume} {13}},\ \bibinfo {pages} {020703}}\BibitemShut
  {NoStop}%
\bibitem [{\citenamefont {Huang}\ and\ \citenamefont {Kim}(2002)}]{huang02k}%
  \BibitemOpen
  \bibfield  {author} {\bibinfo {author} {\bibnamefont {Huang}, \bibfnamefont
  {Z.}}, \ and\ \bibinfo {author} {\bibfnamefont {K.-J.}\ \bibnamefont {Kim}}}
  (\bibinfo {year} {2002}),\ \href@noop {} {\bibfield  {journal} {\bibinfo
  {journal} {Phys. Rev. ST Accel. Beams}\ }\textbf {\bibinfo {volume} {5}},\
  \bibinfo {pages} {074401}}\BibitemShut {NoStop}%
\bibitem [{\citenamefont {Huang}\ and\ \citenamefont
  {Kim}(2007)}]{PhysRevSTAB.10.034801}%
  \BibitemOpen
  \bibfield  {author} {\bibinfo {author} {\bibnamefont {Huang}, \bibfnamefont
  {Z.}}, \ and\ \bibinfo {author} {\bibfnamefont {K.-J.}\ \bibnamefont {Kim}}}
  (\bibinfo {year} {2007}),\ \href
  {http://link.aps.org/doi/10.1103/PhysRevSTAB.10.034801} {\bibfield  {journal}
  {\bibinfo  {journal} {Phys. Rev. ST Accel. Beams}\ }\textbf {\bibinfo
  {volume} {10}},\ \bibinfo {pages} {034801}}\BibitemShut {NoStop}%
\bibitem [{\citenamefont {Huang}\ and\ \citenamefont {Ruth}(2006)}]{XrayRGA}%
  \BibitemOpen
  \bibfield  {author} {\bibinfo {author} {\bibnamefont {Huang}, \bibfnamefont
  {Z.}}, \ and\ \bibinfo {author} {\bibfnamefont {R.~D.}\ \bibnamefont {Ruth}}}
  (\bibinfo {year} {2006}),\ \href
  {http://link.aps.org/doi/10.1103/PhysRevLett.96.144801} {\bibfield  {journal}
  {\bibinfo  {journal} {Phys. Rev. Lett.}\ }\textbf {\bibinfo {volume} {96}},\
  \bibinfo {pages} {144801}}\BibitemShut {NoStop}%
\bibitem [{\citenamefont {Huang}\ \emph
  {et~al.}(2004{\natexlab{b}})\citenamefont {Huang}, \citenamefont {Stupakov},\
  and\ \citenamefont {Zolotorev}}]{huang04sz}%
  \BibitemOpen
  \bibfield  {author} {\bibinfo {author} {\bibnamefont {Huang}, \bibfnamefont
  {Z.}}, \bibinfo {author} {\bibfnamefont {G.}~\bibnamefont {Stupakov}}, \ and\
  \bibinfo {author} {\bibfnamefont {M.}~\bibnamefont {Zolotorev}}} (\bibinfo
  {year} {2004}{\natexlab{b}}),\ \href@noop {} {\bibfield  {journal} {\bibinfo
  {journal} {Phys. Rev. ST Accel. Beams}\ }\textbf {\bibinfo {volume} {7}},\
  \bibinfo {pages} {011302}}\BibitemShut {NoStop}%
\bibitem [{\citenamefont {Ishikawa}\ \emph {et~al.}(2012)\citenamefont
  {Ishikawa}, \citenamefont {Aoyagi}, \citenamefont {Asaka}, \citenamefont
  {Asano}, \citenamefont {Azumi}, \citenamefont {Bizen}, \citenamefont {Ego},
  \citenamefont {andT. Fukui}, \citenamefont {Furukawa}, \citenamefont {Goto},
  \citenamefont {Hanaki}, \citenamefont {Hara}, \citenamefont {Hasegawa},
  \citenamefont {Hatsui}, \citenamefont {Higashiya}, \citenamefont {Hirono},
  \citenamefont {Hosoda}, \citenamefont {Ishii}, \citenamefont {Inagaki},
  \citenamefont {Inubushi}, \citenamefont {Itoga}, \citenamefont {Joti},
  \citenamefont {Kago}, \citenamefont {Kameshima}, \citenamefont {Kimura},
  \citenamefont {Kirihara}, \citenamefont {Kiyomichi}, \citenamefont
  {Kobayashi}, \citenamefont {Kondo}, \citenamefont {Kudo}, \citenamefont
  {Maesaka}, \citenamefont {Marecha}, \citenamefont {Masuda}, \citenamefont
  {Matsubara}, \citenamefont {Matsumoto}, \citenamefont {Matsushita},
  \citenamefont {Matsui}, \citenamefont {Nagasono}, \citenamefont {Nariyama},
  \citenamefont {Ohashi}, \citenamefont {Ohata}, \citenamefont {Ohshima},
  \citenamefont {Ono}, \citenamefont {Otake}, \citenamefont {Saji},
  \citenamefont {Sakurai}, \citenamefont {Sato}, \citenamefont {Sawada},
  \citenamefont {Seike}, \citenamefont {Shirasawa}, \citenamefont {Sugimoto},
  \citenamefont {Suzuki}, \citenamefont {Takahashi}, \citenamefont {Takebe},
  \citenamefont {Takeshita}, \citenamefont {Tamasaku}, \citenamefont {Tanaka},
  \citenamefont {Tanaka}, \citenamefont {Tanaka}, \citenamefont {Togashi},
  \citenamefont {Togawa}, \citenamefont {Tokuhisa}, \citenamefont {Tomizawa},
  \citenamefont {Tono}, \citenamefont {Wu}, \citenamefont {Yabashi},
  \citenamefont {Yamaga}, \citenamefont {Yamashita}, \citenamefont {Yanagida},
  \citenamefont {Zhang}, \citenamefont {Shintake}, \citenamefont {Kitamura},\
  and\ \citenamefont {Kumagai}}]{SACLA}%
  \BibitemOpen
  \bibfield  {author} {\bibinfo {author} {\bibnamefont {Ishikawa},
  \bibfnamefont {T.}}, \bibinfo {author} {\bibfnamefont {H.}~\bibnamefont
  {Aoyagi}}, \bibinfo {author} {\bibfnamefont {T.}~\bibnamefont {Asaka}},
  \bibinfo {author} {\bibfnamefont {Y.}~\bibnamefont {Asano}}, \bibinfo
  {author} {\bibfnamefont {N.}~\bibnamefont {Azumi}}, \bibinfo {author}
  {\bibfnamefont {T.}~\bibnamefont {Bizen}}, \bibinfo {author} {\bibfnamefont
  {H.}~\bibnamefont {Ego}}, \bibinfo {author} {\bibfnamefont {K.~F.}\
  \bibnamefont {andT. Fukui}}, \bibinfo {author} {\bibfnamefont
  {Y.}~\bibnamefont {Furukawa}}, \bibinfo {author} {\bibfnamefont
  {S.}~\bibnamefont {Goto}}, \bibinfo {author} {\bibfnamefont {H.}~\bibnamefont
  {Hanaki}}, \bibinfo {author} {\bibfnamefont {T.}~\bibnamefont {Hara}},
  \bibinfo {author} {\bibfnamefont {T.}~\bibnamefont {Hasegawa}}, \bibinfo
  {author} {\bibfnamefont {T.}~\bibnamefont {Hatsui}}, \bibinfo {author}
  {\bibfnamefont {A.}~\bibnamefont {Higashiya}}, \bibinfo {author}
  {\bibfnamefont {T.}~\bibnamefont {Hirono}}, \bibinfo {author} {\bibfnamefont
  {N.}~\bibnamefont {Hosoda}}, \bibinfo {author} {\bibfnamefont
  {M.}~\bibnamefont {Ishii}}, \bibinfo {author} {\bibfnamefont
  {T.}~\bibnamefont {Inagaki}}, \bibinfo {author} {\bibfnamefont
  {Y.}~\bibnamefont {Inubushi}}, \bibinfo {author} {\bibfnamefont
  {T.}~\bibnamefont {Itoga}}, \bibinfo {author} {\bibfnamefont
  {Y.}~\bibnamefont {Joti}}, \bibinfo {author} {\bibfnamefont {M.}~\bibnamefont
  {Kago}}, \bibinfo {author} {\bibfnamefont {T.}~\bibnamefont {Kameshima}},
  \bibinfo {author} {\bibfnamefont {H.}~\bibnamefont {Kimura}}, \bibinfo
  {author} {\bibfnamefont {Y.}~\bibnamefont {Kirihara}}, \bibinfo {author}
  {\bibfnamefont {A.}~\bibnamefont {Kiyomichi}}, \bibinfo {author}
  {\bibfnamefont {T.}~\bibnamefont {Kobayashi}}, \bibinfo {author}
  {\bibfnamefont {C.}~\bibnamefont {Kondo}}, \bibinfo {author} {\bibfnamefont
  {T.}~\bibnamefont {Kudo}}, \bibinfo {author} {\bibfnamefont {H.}~\bibnamefont
  {Maesaka}}, \bibinfo {author} {\bibfnamefont {X.}~\bibnamefont {Marecha}},
  \bibinfo {author} {\bibfnamefont {T.}~\bibnamefont {Masuda}}, \bibinfo
  {author} {\bibfnamefont {S.}~\bibnamefont {Matsubara}}, \bibinfo {author}
  {\bibfnamefont {T.}~\bibnamefont {Matsumoto}}, \bibinfo {author}
  {\bibfnamefont {T.}~\bibnamefont {Matsushita}}, \bibinfo {author}
  {\bibfnamefont {S.}~\bibnamefont {Matsui}}, \bibinfo {author} {\bibfnamefont
  {M.}~\bibnamefont {Nagasono}}, \bibinfo {author} {\bibfnamefont
  {N.}~\bibnamefont {Nariyama}}, \bibinfo {author} {\bibfnamefont
  {H.}~\bibnamefont {Ohashi}}, \bibinfo {author} {\bibfnamefont
  {T.}~\bibnamefont {Ohata}}, \bibinfo {author} {\bibfnamefont
  {T.}~\bibnamefont {Ohshima}}, \bibinfo {author} {\bibfnamefont
  {S.}~\bibnamefont {Ono}}, \bibinfo {author} {\bibfnamefont {Y.}~\bibnamefont
  {Otake}}, \bibinfo {author} {\bibfnamefont {C.}~\bibnamefont {Saji}},
  \bibinfo {author} {\bibfnamefont {T.}~\bibnamefont {Sakurai}}, \bibinfo
  {author} {\bibfnamefont {T.}~\bibnamefont {Sato}}, \bibinfo {author}
  {\bibfnamefont {K.}~\bibnamefont {Sawada}}, \bibinfo {author} {\bibfnamefont
  {T.}~\bibnamefont {Seike}}, \bibinfo {author} {\bibfnamefont
  {K.}~\bibnamefont {Shirasawa}}, \bibinfo {author} {\bibfnamefont
  {T.}~\bibnamefont {Sugimoto}}, \bibinfo {author} {\bibfnamefont
  {S.}~\bibnamefont {Suzuki}}, \bibinfo {author} {\bibfnamefont
  {S.}~\bibnamefont {Takahashi}}, \bibinfo {author} {\bibfnamefont
  {H.}~\bibnamefont {Takebe}}, \bibinfo {author} {\bibfnamefont
  {K.}~\bibnamefont {Takeshita}}, \bibinfo {author} {\bibfnamefont
  {K.}~\bibnamefont {Tamasaku}}, \bibinfo {author} {\bibfnamefont
  {H.}~\bibnamefont {Tanaka}}, \bibinfo {author} {\bibfnamefont
  {R.}~\bibnamefont {Tanaka}}, \bibinfo {author} {\bibfnamefont
  {T.}~\bibnamefont {Tanaka}}, \bibinfo {author} {\bibfnamefont
  {T.}~\bibnamefont {Togashi}}, \bibinfo {author} {\bibfnamefont
  {K.}~\bibnamefont {Togawa}}, \bibinfo {author} {\bibfnamefont
  {A.}~\bibnamefont {Tokuhisa}}, \bibinfo {author} {\bibfnamefont
  {H.}~\bibnamefont {Tomizawa}}, \bibinfo {author} {\bibfnamefont
  {K.}~\bibnamefont {Tono}}, \bibinfo {author} {\bibfnamefont {S.}~\bibnamefont
  {Wu}}, \bibinfo {author} {\bibfnamefont {M.}~\bibnamefont {Yabashi}},
  \bibinfo {author} {\bibfnamefont {M.}~\bibnamefont {Yamaga}}, \bibinfo
  {author} {\bibfnamefont {A.}~\bibnamefont {Yamashita}}, \bibinfo {author}
  {\bibfnamefont {K.}~\bibnamefont {Yanagida}}, \bibinfo {author}
  {\bibfnamefont {C.}~\bibnamefont {Zhang}}, \bibinfo {author} {\bibfnamefont
  {T.}~\bibnamefont {Shintake}}, \bibinfo {author} {\bibfnamefont
  {H.}~\bibnamefont {Kitamura}}, \ and\ \bibinfo {author} {\bibfnamefont
  {N.}~\bibnamefont {Kumagai}}} (\bibinfo {year} {2012}),\ \href@noop {}
  {\bibfield  {journal} {\bibinfo  {journal} {Nature Photon.}\ }\textbf
  {\bibinfo {volume} {6}},\ \bibinfo {pages} {540}}\BibitemShut {NoStop}%
\bibitem [{\citenamefont {Jackson}(1999)}]{jackson}%
  \BibitemOpen
  \bibfield  {author} {\bibinfo {author} {\bibnamefont {Jackson}, \bibfnamefont
  {J.~D.}}} (\bibinfo {year} {1999}),\ \href@noop {} {\emph {\bibinfo {title}
  {Classical Electrodynamics}}},\ \bibinfo {edition} {3rd}\ ed.\ (\bibinfo
  {publisher} {Wiley},\ \bibinfo {address} {New~York})\BibitemShut {NoStop}%
\bibitem [{\citenamefont {Jia}(2008)}]{HGHGS3}%
  \BibitemOpen
  \bibfield  {author} {\bibinfo {author} {\bibnamefont {Jia}, \bibfnamefont
  {Q.}}} (\bibinfo {year} {2008}),\ \href@noop {} {\bibfield  {journal}
  {\bibinfo  {journal} {Appl. Phys. Lett.}\ }\textbf {\bibinfo {volume} {93}},\
  \bibinfo {pages} {141102}}\BibitemShut {NoStop}%
\bibitem [{\citenamefont {Jiang}\ \emph {et~al.}(2011)\citenamefont {Jiang},
  \citenamefont {Power}, \citenamefont {Lindberg}, \citenamefont {Liu},\ and\
  \citenamefont {Gai}}]{EEX6}%
  \BibitemOpen
  \bibfield  {author} {\bibinfo {author} {\bibnamefont {Jiang}, \bibfnamefont
  {B.}}, \bibinfo {author} {\bibfnamefont {J.~G.}\ \bibnamefont {Power}},
  \bibinfo {author} {\bibfnamefont {R.}~\bibnamefont {Lindberg}}, \bibinfo
  {author} {\bibfnamefont {W.}~\bibnamefont {Liu}}, \ and\ \bibinfo {author}
  {\bibfnamefont {W.}~\bibnamefont {Gai}}} (\bibinfo {year} {2011}),\ \href
  {http://link.aps.org/doi/10.1103/PhysRevLett.106.114801} {\bibfield
  {journal} {\bibinfo  {journal} {Phys. Rev. Lett.}\ }\textbf {\bibinfo
  {volume} {106}},\ \bibinfo {pages} {114801}}\BibitemShut {NoStop}%
\bibitem [{\citenamefont {Jiao}\ \emph {et~al.}(2011)\citenamefont {Jiao},
  \citenamefont {Ratner},\ and\ \citenamefont {Chao}}]{Jiao2011}%
  \BibitemOpen
  \bibfield  {author} {\bibinfo {author} {\bibnamefont {Jiao}, \bibfnamefont
  {Y.}}, \bibinfo {author} {\bibfnamefont {D.~F.}\ \bibnamefont {Ratner}}, \
  and\ \bibinfo {author} {\bibfnamefont {A.~W.}\ \bibnamefont {Chao}}}
  (\bibinfo {year} {2011}),\ \href
  {http://link.aps.org/doi/10.1103/PhysRevSTAB.14.110702} {\bibfield  {journal}
  {\bibinfo  {journal} {Phys. Rev. ST Accel. Beams}\ }\textbf {\bibinfo
  {volume} {14}},\ \bibinfo {pages} {110702}}\BibitemShut {NoStop}%
\bibitem [{\citenamefont {Johnson}\ \emph {et~al.}(2008)\citenamefont
  {Johnson}, \citenamefont {Beaud}, \citenamefont {Milne}, \citenamefont
  {Krasniqi}, \citenamefont {Zijlstra}, \citenamefont {Garcia}, \citenamefont
  {Kaiser}, \citenamefont {Grolimund}, \citenamefont {Abela},\ and\
  \citenamefont {Ingold}}]{LSapplication3}%
  \BibitemOpen
  \bibfield  {author} {\bibinfo {author} {\bibnamefont {Johnson}, \bibfnamefont
  {S.~L.}}, \bibinfo {author} {\bibfnamefont {P.}~\bibnamefont {Beaud}},
  \bibinfo {author} {\bibfnamefont {C.~J.}\ \bibnamefont {Milne}}, \bibinfo
  {author} {\bibfnamefont {F.~S.}\ \bibnamefont {Krasniqi}}, \bibinfo {author}
  {\bibfnamefont {E.~S.}\ \bibnamefont {Zijlstra}}, \bibinfo {author}
  {\bibfnamefont {M.~E.}\ \bibnamefont {Garcia}}, \bibinfo {author}
  {\bibfnamefont {M.}~\bibnamefont {Kaiser}}, \bibinfo {author} {\bibfnamefont
  {D.}~\bibnamefont {Grolimund}}, \bibinfo {author} {\bibfnamefont
  {R.}~\bibnamefont {Abela}}, \ and\ \bibinfo {author} {\bibfnamefont
  {G.}~\bibnamefont {Ingold}}} (\bibinfo {year} {2008}),\ \href@noop {}
  {\bibfield  {journal} {\bibinfo  {journal} {Phys. Rev. Lett.}\ }\textbf
  {\bibinfo {volume} {100}},\ \bibinfo {pages} {155501}}\BibitemShut {NoStop}%
\bibitem [{\citenamefont {Jones}\ \emph {et~al.}(2000)\citenamefont {Jones},
  \citenamefont {Diddams}, \citenamefont {Ranka}, \citenamefont {Stentz},
  \citenamefont {Windeler}, \citenamefont {Hall},\ and\ \citenamefont
  {Cundiff}}]{Jones}%
  \BibitemOpen
  \bibfield  {author} {\bibinfo {author} {\bibnamefont {Jones}, \bibfnamefont
  {D.}}, \bibinfo {author} {\bibfnamefont {S.~A.}\ \bibnamefont {Diddams}},
  \bibinfo {author} {\bibfnamefont {J.~K.}\ \bibnamefont {Ranka}}, \bibinfo
  {author} {\bibfnamefont {A.}~\bibnamefont {Stentz}}, \bibinfo {author}
  {\bibfnamefont {R.~S.}\ \bibnamefont {Windeler}}, \bibinfo {author}
  {\bibfnamefont {J.~L.}\ \bibnamefont {Hall}}, \ and\ \bibinfo {author}
  {\bibfnamefont {S.~T.}\ \bibnamefont {Cundiff}}} (\bibinfo {year} {2000}),\
  \href@noop {} {\bibfield  {journal} {\bibinfo  {journal} {Science}\ }\textbf
  {\bibinfo {volume} {288}},\ \bibinfo {pages} {635}}\BibitemShut {NoStop}%
\bibitem [{\citenamefont {Joshi}\ \emph {et~al.}(1984)\citenamefont {Joshi},
  \citenamefont {Mori}, \citenamefont {Katsouleas}, \citenamefont {Dawson},
  \citenamefont {Kindel},\ and\ \citenamefont
  {Forslund}}]{plasma_acceleration}%
  \BibitemOpen
  \bibfield  {author} {\bibinfo {author} {\bibnamefont {Joshi}, \bibfnamefont
  {C.}}, \bibinfo {author} {\bibfnamefont {W.~B.}\ \bibnamefont {Mori}},
  \bibinfo {author} {\bibfnamefont {T.}~\bibnamefont {Katsouleas}}, \bibinfo
  {author} {\bibfnamefont {J.~M.}\ \bibnamefont {Dawson}}, \bibinfo {author}
  {\bibfnamefont {J.~M.}\ \bibnamefont {Kindel}}, \ and\ \bibinfo {author}
  {\bibfnamefont {D.~W.}\ \bibnamefont {Forslund}}} (\bibinfo {year} {1984}),\
  \href@noop {} {\bibfield  {journal} {\bibinfo  {journal} {Nature}\ }\textbf
  {\bibinfo {volume} {311}},\ \bibinfo {pages} {525}}\BibitemShut {NoStop}%
\bibitem [{\citenamefont {Kezerashvili}\ and\ \citenamefont
  {Skrinsky}(1991)}]{Kezerashvili}%
  \BibitemOpen
  \bibfield  {author} {\bibinfo {author} {\bibnamefont {Kezerashvili},
  \bibfnamefont {G.~Y.}}, \ and\ \bibinfo {author} {\bibfnamefont {A.~N.}\
  \bibnamefont {Skrinsky}}} (\bibinfo {year} {1991}),\ \href@noop {} {\bibfield
   {journal} {\bibinfo  {journal} {Nucl. Instrum. Methods Phys. Res. A}\
  }\textbf {\bibinfo {volume} {307}},\ \bibinfo {pages} {179}}\BibitemShut
  {NoStop}%
\bibitem [{\citenamefont {Khan}\ \emph {et~al.}(2006)\citenamefont {Khan},
  \citenamefont {Holldack}, \citenamefont {Kachel}, \citenamefont {Mitzner},\
  and\ \citenamefont {Quast}}]{laser-slicingBESSY}%
  \BibitemOpen
  \bibfield  {author} {\bibinfo {author} {\bibnamefont {Khan}, \bibfnamefont
  {S.}}, \bibinfo {author} {\bibfnamefont {K.}~\bibnamefont {Holldack}},
  \bibinfo {author} {\bibfnamefont {T.}~\bibnamefont {Kachel}}, \bibinfo
  {author} {\bibfnamefont {R.}~\bibnamefont {Mitzner}}, \ and\ \bibinfo
  {author} {\bibfnamefont {T.}~\bibnamefont {Quast}}} (\bibinfo {year}
  {2006}),\ \href@noop {} {\bibfield  {journal} {\bibinfo  {journal} {Phys.
  Rev. Lett.}\ }\textbf {\bibinfo {volume} {97}},\ \bibinfo {pages}
  {074801}}\BibitemShut {NoStop}%
\bibitem [{\citenamefont {Kim}(1986{\natexlab{a}})}]{KJKim}%
  \BibitemOpen
  \bibfield  {author} {\bibinfo {author} {\bibnamefont {Kim}, \bibfnamefont
  {K.-J.}}} (\bibinfo {year} {1986}{\natexlab{a}}),\ \href
  {http://dx.doi.org/10.1016/0168-9002(86)90047-1} {\bibfield  {journal}
  {\bibinfo  {journal} {Nucl. Instr. Meth. Phys. Res. A}\ }\textbf {\bibinfo
  {volume} {246}}~(\bibinfo {number} {1-€"3}),\ \bibinfo {pages}
  {67}}\BibitemShut {NoStop}%
\bibitem [{\citenamefont {Kim}(1986{\natexlab{b}})}]{PhysRevLett.57.1871}%
  \BibitemOpen
  \bibfield  {author} {\bibinfo {author} {\bibnamefont {Kim}, \bibfnamefont
  {K.-J.}}} (\bibinfo {year} {1986}{\natexlab{b}}),\ \href
  {http://link.aps.org/doi/10.1103/PhysRevLett.57.1871} {\bibfield  {journal}
  {\bibinfo  {journal} {Phys. Rev. Lett.}\ }\textbf {\bibinfo {volume} {57}},\
  \bibinfo {pages} {1871}}\BibitemShut {NoStop}%
\bibitem [{\citenamefont {Kim}\ and\ \citenamefont
  {Lindberg}(2011)}]{FEL11_kim_lindberg}%
  \BibitemOpen
  \bibfield  {author} {\bibinfo {author} {\bibnamefont {Kim}, \bibfnamefont
  {K.-J.}}, \ and\ \bibinfo {author} {\bibfnamefont {R.~R.}\ \bibnamefont
  {Lindberg}}} (\bibinfo {year} {2011}),\ in\ \href@noop {} {\emph {\bibinfo
  {booktitle} {Proceedings of the 2011 FEL Conference}}}\ (\bibinfo {address}
  {Shanghai, China})\ p.\ \bibinfo {pages} {156}\BibitemShut {NoStop}%
\bibitem [{\citenamefont {Kim}\ \emph {et~al.}(2008)\citenamefont {Kim},
  \citenamefont {Shvyd'ko},\ and\ \citenamefont {Reiche}}]{XrayOCL}%
  \BibitemOpen
  \bibfield  {author} {\bibinfo {author} {\bibnamefont {Kim}, \bibfnamefont
  {K.-J.}}, \bibinfo {author} {\bibfnamefont {Y.}~\bibnamefont {Shvyd'ko}}, \
  and\ \bibinfo {author} {\bibfnamefont {S.}~\bibnamefont {Reiche}}} (\bibinfo
  {year} {2008}),\ \href
  {http://link.aps.org/doi/10.1103/PhysRevLett.100.244802} {\bibfield
  {journal} {\bibinfo  {journal} {Phys. Rev. Lett.}\ }\textbf {\bibinfo
  {volume} {100}},\ \bibinfo {pages} {244802}}\BibitemShut {NoStop}%
\bibitem [{\citenamefont {Kimura}\ \emph {et~al.}(2004)\citenamefont {Kimura},
  \citenamefont {Babzien}, \citenamefont {Ben-Zvi}, \citenamefont {Campbell},
  \citenamefont {Cline}, \citenamefont {Dilley}, \citenamefont {Gallardo},
  \citenamefont {Gottschalk}, \citenamefont {Kusche}, \citenamefont {Pantell},
  \citenamefont {Pogorelsky}, \citenamefont {Quimby}, \citenamefont {Skaritka},
  \citenamefont {Steinhauer}, \citenamefont {Yakimenko},\ and\ \citenamefont
  {Zhou}}]{KimuraSTELLAPRL2004}%
  \BibitemOpen
  \bibfield  {author} {\bibinfo {author} {\bibnamefont {Kimura}, \bibfnamefont
  {W.~D.}}, \bibinfo {author} {\bibfnamefont {M.}~\bibnamefont {Babzien}},
  \bibinfo {author} {\bibfnamefont {I.}~\bibnamefont {Ben-Zvi}}, \bibinfo
  {author} {\bibfnamefont {L.~P.}\ \bibnamefont {Campbell}}, \bibinfo {author}
  {\bibfnamefont {D.~B.}\ \bibnamefont {Cline}}, \bibinfo {author}
  {\bibfnamefont {C.~E.}\ \bibnamefont {Dilley}}, \bibinfo {author}
  {\bibfnamefont {J.~C.}\ \bibnamefont {Gallardo}}, \bibinfo {author}
  {\bibfnamefont {S.~C.}\ \bibnamefont {Gottschalk}}, \bibinfo {author}
  {\bibfnamefont {K.~P.}\ \bibnamefont {Kusche}}, \bibinfo {author}
  {\bibfnamefont {R.~H.}\ \bibnamefont {Pantell}}, \bibinfo {author}
  {\bibfnamefont {I.~V.}\ \bibnamefont {Pogorelsky}}, \bibinfo {author}
  {\bibfnamefont {D.~C.}\ \bibnamefont {Quimby}}, \bibinfo {author}
  {\bibfnamefont {J.}~\bibnamefont {Skaritka}}, \bibinfo {author}
  {\bibfnamefont {L.~C.}\ \bibnamefont {Steinhauer}}, \bibinfo {author}
  {\bibfnamefont {V.}~\bibnamefont {Yakimenko}}, \ and\ \bibinfo {author}
  {\bibfnamefont {F.}~\bibnamefont {Zhou}}} (\bibinfo {year} {2004}),\ \href
  {http://link.aps.org/doi/10.1103/PhysRevLett.92.054801} {\bibfield  {journal}
  {\bibinfo  {journal} {Phys. Rev. Lett.}\ }\textbf {\bibinfo {volume} {92}},\
  \bibinfo {pages} {054801}}\BibitemShut {NoStop}%
\bibitem [{\citenamefont {Kimura}\ \emph {et~al.}(2001)\citenamefont {Kimura},
  \citenamefont {van Steenbergen}, \citenamefont {Babzien}, \citenamefont
  {Ben-Zvi}, \citenamefont {Campbell}, \citenamefont {Cline}, \citenamefont
  {Dilley}, \citenamefont {Gallardo}, \citenamefont {Gottschalk}, \citenamefont
  {He}, \citenamefont {Kusche}, \citenamefont {Liu}, \citenamefont {Pantell},
  \citenamefont {Pogorelsky}, \citenamefont {Quimby}, \citenamefont {Skaritka},
  \citenamefont {Steinhauer},\ and\ \citenamefont
  {Yakimenko}}]{KimuraSTELLAPRL2001}%
  \BibitemOpen
  \bibfield  {author} {\bibinfo {author} {\bibnamefont {Kimura}, \bibfnamefont
  {W.~D.}}, \bibinfo {author} {\bibfnamefont {A.}~\bibnamefont {van
  Steenbergen}}, \bibinfo {author} {\bibfnamefont {M.}~\bibnamefont {Babzien}},
  \bibinfo {author} {\bibfnamefont {I.}~\bibnamefont {Ben-Zvi}}, \bibinfo
  {author} {\bibfnamefont {L.~P.}\ \bibnamefont {Campbell}}, \bibinfo {author}
  {\bibfnamefont {D.~B.}\ \bibnamefont {Cline}}, \bibinfo {author}
  {\bibfnamefont {C.~E.}\ \bibnamefont {Dilley}}, \bibinfo {author}
  {\bibfnamefont {J.~C.}\ \bibnamefont {Gallardo}}, \bibinfo {author}
  {\bibfnamefont {S.~C.}\ \bibnamefont {Gottschalk}}, \bibinfo {author}
  {\bibfnamefont {P.}~\bibnamefont {He}}, \bibinfo {author} {\bibfnamefont
  {K.~P.}\ \bibnamefont {Kusche}}, \bibinfo {author} {\bibfnamefont
  {Y.}~\bibnamefont {Liu}}, \bibinfo {author} {\bibfnamefont {R.~H.}\
  \bibnamefont {Pantell}}, \bibinfo {author} {\bibfnamefont {I.~V.}\
  \bibnamefont {Pogorelsky}}, \bibinfo {author} {\bibfnamefont {D.~C.}\
  \bibnamefont {Quimby}}, \bibinfo {author} {\bibfnamefont {J.}~\bibnamefont
  {Skaritka}}, \bibinfo {author} {\bibfnamefont {L.~C.}\ \bibnamefont
  {Steinhauer}}, \ and\ \bibinfo {author} {\bibfnamefont {V.}~\bibnamefont
  {Yakimenko}}} (\bibinfo {year} {2001}),\ \href {\doibase
  10.1103/PhysRevLett.86.4041} {\bibfield  {journal} {\bibinfo  {journal}
  {Phys. Rev. Lett.}\ }\textbf {\bibinfo {volume} {86}},\ \bibinfo {pages}
  {4041}}\BibitemShut {NoStop}%
\bibitem [{\citenamefont {Kincaid}\ \emph {et~al.}(1984)\citenamefont
  {Kincaid}, \citenamefont {Freeman}, \citenamefont {Pellegrini}, \citenamefont
  {Luccio}, \citenamefont {Krinsky}, \citenamefont {van Steenbergen},\ and\
  \citenamefont {DeMartini}}]{Kinkaid}%
  \BibitemOpen
  \bibfield  {author} {\bibinfo {author} {\bibnamefont {Kincaid}, \bibfnamefont
  {B.}}, \bibinfo {author} {\bibfnamefont {R.}~\bibnamefont {Freeman}},
  \bibinfo {author} {\bibfnamefont {C.}~\bibnamefont {Pellegrini}}, \bibinfo
  {author} {\bibfnamefont {A.}~\bibnamefont {Luccio}}, \bibinfo {author}
  {\bibfnamefont {S.}~\bibnamefont {Krinsky}}, \bibinfo {author} {\bibfnamefont
  {A.}~\bibnamefont {van Steenbergen}}, \ and\ \bibinfo {author} {\bibfnamefont
  {F.}~\bibnamefont {DeMartini}}} (\bibinfo {year} {1984}),\ in\ \href@noop {}
  {\emph {\bibinfo {booktitle} {AIP Conf. Proc.}}},\ Vol.\ \bibinfo {volume}
  {118},\ p.\ \bibinfo {pages} {110}\BibitemShut {NoStop}%
\bibitem [{\citenamefont {Kowalski}\ and\ \citenamefont
  {Enge}(1972)}]{beamrotator}%
  \BibitemOpen
  \bibfield  {author} {\bibinfo {author} {\bibnamefont {Kowalski},
  \bibfnamefont {S.}}, \ and\ \bibinfo {author} {\bibfnamefont
  {H.}~\bibnamefont {Enge}}} (\bibinfo {year} {1972}),\ in\ \href@noop {}
  {\emph {\bibinfo {booktitle} {Proceedings of the International Conference on
  Magnet Technology}}},\ p.\ \bibinfo {pages} {182}\BibitemShut {NoStop}%
\bibitem [{\citenamefont {Krafft}\ and\ \citenamefont
  {Priebe}(2010)}]{compton_scattering}%
  \BibitemOpen
  \bibfield  {author} {\bibinfo {author} {\bibnamefont {Krafft}, \bibfnamefont
  {G.~A.}}, \ and\ \bibinfo {author} {\bibfnamefont {G.}~\bibnamefont
  {Priebe}}} (\bibinfo {year} {2010}),\ \href {\doibase
  10.1142/S1793626810000440} {\bibfield  {journal} {\bibinfo  {journal}
  {Reviews of Accelerator Science and Technology}\ }\textbf {\bibinfo {volume}
  {03}}~(\bibinfo {number} {01}),\ \bibinfo {pages} {147}}\BibitemShut
  {NoStop}%
\bibitem [{\citenamefont {Krausz}\ and\ \citenamefont
  {Ivanov}(2009)}]{KraussIvanov:2009}%
  \BibitemOpen
  \bibfield  {author} {\bibinfo {author} {\bibnamefont {Krausz}, \bibfnamefont
  {F.}}, \ and\ \bibinfo {author} {\bibfnamefont {M.}~\bibnamefont {Ivanov}}}
  (\bibinfo {year} {2009}),\ \href@noop {} {\bibfield  {journal} {\bibinfo
  {journal} {Rev. Mod. Phys.}\ }\textbf {\bibinfo {volume} {88}},\ \bibinfo
  {pages} {163}}\BibitemShut {NoStop}%
\bibitem [{\citenamefont {Krinsky}(2002)}]{krinsky02}%
  \BibitemOpen
  \bibfield  {author} {\bibinfo {author} {\bibnamefont {Krinsky}, \bibfnamefont
  {S.}}} (\bibinfo {year} {2002}),\ \href
  {http://link.aip.org/link/?APC/648/23/1} {\bibfield  {journal} {\bibinfo
  {journal} {AIP Conference Proceedings}\ }\textbf {\bibinfo {volume}
  {648}}~(\bibinfo {number} {1}),\ \bibinfo {pages} {23}}\BibitemShut {NoStop}%
\bibitem [{\citenamefont {Kumar}\ \emph {et~al.}(2010)\citenamefont {Kumar},
  \citenamefont {Pukhov},\ and\ \citenamefont {Lotov}}]{LHC}%
  \BibitemOpen
  \bibfield  {author} {\bibinfo {author} {\bibnamefont {Kumar}, \bibfnamefont
  {N.}}, \bibinfo {author} {\bibfnamefont {A.}~\bibnamefont {Pukhov}}, \ and\
  \bibinfo {author} {\bibfnamefont {K.}~\bibnamefont {Lotov}}} (\bibinfo {year}
  {2010}),\ \href@noop {} {\bibfield  {journal} {\bibinfo  {journal} {Phys.
  Rev. Lett.}\ }\textbf {\bibinfo {volume} {104}},\ \bibinfo {pages}
  {255003}}\BibitemShut {NoStop}%
\bibitem [{\citenamefont {Kur}\ \emph {et~al.}(2011)\citenamefont {Kur},
  \citenamefont {Dunning}, \citenamefont {McNeil}, \citenamefont {Wurtele},\
  and\ \citenamefont {Zholents}}]{Kur}%
  \BibitemOpen
  \bibfield  {author} {\bibinfo {author} {\bibnamefont {Kur}, \bibfnamefont
  {E.}}, \bibinfo {author} {\bibfnamefont {D.~J.}\ \bibnamefont {Dunning}},
  \bibinfo {author} {\bibfnamefont {B.~W.~J.}\ \bibnamefont {McNeil}}, \bibinfo
  {author} {\bibfnamefont {J.}~\bibnamefont {Wurtele}}, \ and\ \bibinfo
  {author} {\bibfnamefont {A.~A.}\ \bibnamefont {Zholents}}} (\bibinfo {year}
  {2011}),\ \href@noop {} {\bibfield  {journal} {\bibinfo  {journal} {New J. of
  Phys.}\ }\textbf {\bibinfo {volume} {13}},\ \bibinfo {pages}
  {063012}}\BibitemShut {NoStop}%
\bibitem [{\citenamefont {Labat}\ \emph {et~al.}(2011)\citenamefont {Labat},
  \citenamefont {Bellaveglia}, \citenamefont {Bougeard}, \citenamefont
  {Carr\'e}, \citenamefont {Ciocci}, \citenamefont {Chiadroni}, \citenamefont
  {Cianchi}, \citenamefont {Couprie}, \citenamefont {Cultrera}, \citenamefont
  {Del~Franco}, \citenamefont {Di~Pirro}, \citenamefont {Drago}, \citenamefont
  {Ferrario}, \citenamefont {Filippetto}, \citenamefont {Frassetto},
  \citenamefont {Gallo}, \citenamefont {Garzella}, \citenamefont {Gatti},
  \citenamefont {Giannessi}, \citenamefont {Lambert}, \citenamefont {Mostacci},
  \citenamefont {Petralia}, \citenamefont {Petrillo}, \citenamefont {Poletto},
  \citenamefont {Quattromini}, \citenamefont {Rau}, \citenamefont {Ronsivalle},
  \citenamefont {Sabia}, \citenamefont {Serluca}, \citenamefont {Spassovsky},
  \citenamefont {Surrenti}, \citenamefont {Vaccarezza},\ and\ \citenamefont
  {Vicario}}]{SPARCHHG}%
  \BibitemOpen
  \bibfield  {author} {\bibinfo {author} {\bibnamefont {Labat}, \bibfnamefont
  {M.}}, \bibinfo {author} {\bibfnamefont {M.}~\bibnamefont {Bellaveglia}},
  \bibinfo {author} {\bibfnamefont {M.}~\bibnamefont {Bougeard}}, \bibinfo
  {author} {\bibfnamefont {B.}~\bibnamefont {Carr\'e}}, \bibinfo {author}
  {\bibfnamefont {F.}~\bibnamefont {Ciocci}}, \bibinfo {author} {\bibfnamefont
  {E.}~\bibnamefont {Chiadroni}}, \bibinfo {author} {\bibfnamefont
  {A.}~\bibnamefont {Cianchi}}, \bibinfo {author} {\bibfnamefont {M.~E.}\
  \bibnamefont {Couprie}}, \bibinfo {author} {\bibfnamefont {L.}~\bibnamefont
  {Cultrera}}, \bibinfo {author} {\bibfnamefont {M.}~\bibnamefont
  {Del~Franco}}, \bibinfo {author} {\bibfnamefont {G.}~\bibnamefont
  {Di~Pirro}}, \bibinfo {author} {\bibfnamefont {A.}~\bibnamefont {Drago}},
  \bibinfo {author} {\bibfnamefont {M.}~\bibnamefont {Ferrario}}, \bibinfo
  {author} {\bibfnamefont {D.}~\bibnamefont {Filippetto}}, \bibinfo {author}
  {\bibfnamefont {F.}~\bibnamefont {Frassetto}}, \bibinfo {author}
  {\bibfnamefont {A.}~\bibnamefont {Gallo}}, \bibinfo {author} {\bibfnamefont
  {D.}~\bibnamefont {Garzella}}, \bibinfo {author} {\bibfnamefont
  {G.}~\bibnamefont {Gatti}}, \bibinfo {author} {\bibfnamefont
  {L.}~\bibnamefont {Giannessi}}, \bibinfo {author} {\bibfnamefont
  {G.}~\bibnamefont {Lambert}}, \bibinfo {author} {\bibfnamefont
  {A.}~\bibnamefont {Mostacci}}, \bibinfo {author} {\bibfnamefont
  {A.}~\bibnamefont {Petralia}}, \bibinfo {author} {\bibfnamefont
  {V.}~\bibnamefont {Petrillo}}, \bibinfo {author} {\bibfnamefont
  {L.}~\bibnamefont {Poletto}}, \bibinfo {author} {\bibfnamefont
  {M.}~\bibnamefont {Quattromini}}, \bibinfo {author} {\bibfnamefont {J.~V.}\
  \bibnamefont {Rau}}, \bibinfo {author} {\bibfnamefont {C.}~\bibnamefont
  {Ronsivalle}}, \bibinfo {author} {\bibfnamefont {E.}~\bibnamefont {Sabia}},
  \bibinfo {author} {\bibfnamefont {M.}~\bibnamefont {Serluca}}, \bibinfo
  {author} {\bibfnamefont {I.}~\bibnamefont {Spassovsky}}, \bibinfo {author}
  {\bibfnamefont {V.}~\bibnamefont {Surrenti}}, \bibinfo {author}
  {\bibfnamefont {C.}~\bibnamefont {Vaccarezza}}, \ and\ \bibinfo {author}
  {\bibfnamefont {C.}~\bibnamefont {Vicario}}} (\bibinfo {year} {2011}),\ \href
  {http://link.aps.org/doi/10.1103/PhysRevLett.107.224801} {\bibfield
  {journal} {\bibinfo  {journal} {Phys. Rev. Lett.}\ }\textbf {\bibinfo
  {volume} {107}},\ \bibinfo {pages} {224801}}\BibitemShut {NoStop}%
\bibitem [{\citenamefont {Labat}\ \emph {et~al.}(2007)\citenamefont {Labat},
  \citenamefont {Hosaka}, \citenamefont {Mochihashi}, \citenamefont {Shimada},
  \citenamefont {Katoh}, \citenamefont {Lambert}, \citenamefont {Hara},
  \citenamefont {Takashima},\ and\ \citenamefont {Couprie}}]{UVSOR1}%
  \BibitemOpen
  \bibfield  {author} {\bibinfo {author} {\bibnamefont {Labat}, \bibfnamefont
  {M.}}, \bibinfo {author} {\bibfnamefont {M.}~\bibnamefont {Hosaka}}, \bibinfo
  {author} {\bibfnamefont {A.}~\bibnamefont {Mochihashi}}, \bibinfo {author}
  {\bibfnamefont {M.}~\bibnamefont {Shimada}}, \bibinfo {author} {\bibfnamefont
  {M.}~\bibnamefont {Katoh}}, \bibinfo {author} {\bibfnamefont
  {G.}~\bibnamefont {Lambert}}, \bibinfo {author} {\bibfnamefont
  {T.}~\bibnamefont {Hara}}, \bibinfo {author} {\bibfnamefont {Y.}~\bibnamefont
  {Takashima}}, \ and\ \bibinfo {author} {\bibfnamefont {M.}~\bibnamefont
  {Couprie}}} (\bibinfo {year} {2007}),\ \href@noop {} {\bibfield  {journal}
  {\bibinfo  {journal} {Eur. Phys. J. D}\ }\textbf {\bibinfo {volume} {44}},\
  \bibinfo {pages} {187}}\BibitemShut {NoStop}%
\bibitem [{\citenamefont {Lambert}\ \emph {et~al.}(2008)\citenamefont
  {Lambert}, \citenamefont {Hara}, \citenamefont {Garzella}, \citenamefont
  {Tanikawa}, \citenamefont {Labat}, \citenamefont {Carre}, \citenamefont
  {Kitamura}, \citenamefont {Shintake}, \citenamefont {Bougeard}, \citenamefont
  {Inoue}, \citenamefont {Tanaka}, \citenamefont {Salieres}, \citenamefont
  {Merdji}, \citenamefont {Chubar}, \citenamefont {Gobert}, \citenamefont
  {Tahara},\ and\ \citenamefont {Couprie}}]{HHG160nm}%
  \BibitemOpen
  \bibfield  {author} {\bibinfo {author} {\bibnamefont {Lambert}, \bibfnamefont
  {G.}}, \bibinfo {author} {\bibfnamefont {T.}~\bibnamefont {Hara}}, \bibinfo
  {author} {\bibfnamefont {D.}~\bibnamefont {Garzella}}, \bibinfo {author}
  {\bibfnamefont {T.}~\bibnamefont {Tanikawa}}, \bibinfo {author}
  {\bibfnamefont {M.}~\bibnamefont {Labat}}, \bibinfo {author} {\bibfnamefont
  {B.}~\bibnamefont {Carre}}, \bibinfo {author} {\bibfnamefont
  {H.}~\bibnamefont {Kitamura}}, \bibinfo {author} {\bibfnamefont
  {T.}~\bibnamefont {Shintake}}, \bibinfo {author} {\bibfnamefont
  {M.}~\bibnamefont {Bougeard}}, \bibinfo {author} {\bibfnamefont
  {S.}~\bibnamefont {Inoue}}, \bibinfo {author} {\bibfnamefont
  {Y.}~\bibnamefont {Tanaka}}, \bibinfo {author} {\bibfnamefont
  {P.}~\bibnamefont {Salieres}}, \bibinfo {author} {\bibfnamefont
  {H.}~\bibnamefont {Merdji}}, \bibinfo {author} {\bibfnamefont
  {O.}~\bibnamefont {Chubar}}, \bibinfo {author} {\bibfnamefont
  {O.}~\bibnamefont {Gobert}}, \bibinfo {author} {\bibfnamefont
  {K.}~\bibnamefont {Tahara}}, \ and\ \bibinfo {author} {\bibfnamefont {M.~E.}\
  \bibnamefont {Couprie}}} (\bibinfo {year} {2008}),\ \href
  {http://dx.doi.org/10.1038/nphys889} {\bibfield  {journal} {\bibinfo
  {journal} {Nature Phys.}\ }\textbf {\bibinfo {volume} {4}}~(\bibinfo {number}
  {4}),\ \bibinfo {pages} {296}}\BibitemShut {NoStop}%
\bibitem [{\citenamefont {Landau}\ and\ \citenamefont
  {Lifshitz}(1979)}]{landau_lifshitz_ctf}%
  \BibitemOpen
  \bibfield  {author} {\bibinfo {author} {\bibnamefont {Landau}, \bibfnamefont
  {L.~D.}}, \ and\ \bibinfo {author} {\bibfnamefont {E.~M.}\ \bibnamefont
  {Lifshitz}}} (\bibinfo {year} {1979}),\ \href@noop {} {\emph {\bibinfo
  {title} {The Classical Theory of Fields}}},\ \bibinfo {edition} {4th}\ ed.,\
  \bibinfo {series} {Course of Theoretical Physics}, Vol.~\bibinfo {volume}
  {2}\ (\bibinfo  {publisher} {Pergamon},\ \bibinfo {address} {London})\
  \bibinfo {note} {(Translated from the Russian)}\BibitemShut {NoStop}%
\bibitem [{\citenamefont {Lawler}\ \emph {et~al.}(2013)\citenamefont {Lawler},
  \citenamefont {Bisognano}, \citenamefont {Bosch}, \citenamefont {Chiang},
  \citenamefont {Green}, \citenamefont {Jacobs}, \citenamefont {Miller},
  \citenamefont {Wehlitz}, \citenamefont {Yavuz},\ and\ \citenamefont
  {York}}]{Bisognano}%
  \BibitemOpen
  \bibfield  {author} {\bibinfo {author} {\bibnamefont {Lawler}, \bibfnamefont
  {J.~E.}}, \bibinfo {author} {\bibfnamefont {J.}~\bibnamefont {Bisognano}},
  \bibinfo {author} {\bibfnamefont {R.~A.}\ \bibnamefont {Bosch}}, \bibinfo
  {author} {\bibfnamefont {T.~C.}\ \bibnamefont {Chiang}}, \bibinfo {author}
  {\bibfnamefont {M.~A.}\ \bibnamefont {Green}}, \bibinfo {author}
  {\bibfnamefont {K.}~\bibnamefont {Jacobs}}, \bibinfo {author} {\bibfnamefont
  {T.}~\bibnamefont {Miller}}, \bibinfo {author} {\bibfnamefont
  {R.}~\bibnamefont {Wehlitz}}, \bibinfo {author} {\bibfnamefont
  {D.}~\bibnamefont {Yavuz}}, \ and\ \bibinfo {author} {\bibfnamefont {R.~C.}\
  \bibnamefont {York}}} (\bibinfo {year} {2013}),\ \href
  {http://stacks.iop.org/0022-3727/46/i=32/a=325501} {\bibfield  {journal}
  {\bibinfo  {journal} {Journal of Physics D: Applied Physics}\ }\textbf
  {\bibinfo {volume} {46}}~(\bibinfo {number} {32}),\ \bibinfo {pages}
  {325501}}\BibitemShut {NoStop}%
\bibitem [{\citenamefont {Lee}(1999)}]{SYLee}%
  \BibitemOpen
  \bibfield  {author} {\bibinfo {author} {\bibnamefont {Lee}, \bibfnamefont
  {S.~Y.}}} (\bibinfo {year} {1999}),\ \href@noop {} {\emph {\bibinfo {title}
  {Accelerator Physics}}}\ (\bibinfo  {publisher} {World
  Scientific})\BibitemShut {NoStop}%
\bibitem [{\citenamefont {Leemann}\ \emph {et~al.}(2009)\citenamefont
  {Leemann}, \citenamefont {Andersson}, \citenamefont {Eriksson}, \citenamefont
  {Lindgren}, \citenamefont {Wall\'en}, \citenamefont {Bengtsson},\ and\
  \citenamefont {Streun}}]{Leemann}%
  \BibitemOpen
  \bibfield  {author} {\bibinfo {author} {\bibnamefont {Leemann}, \bibfnamefont
  {S.~C.}}, \bibinfo {author} {\bibfnamefont {A.}~\bibnamefont {Andersson}},
  \bibinfo {author} {\bibfnamefont {M.}~\bibnamefont {Eriksson}}, \bibinfo
  {author} {\bibfnamefont {L.-J.}\ \bibnamefont {Lindgren}}, \bibinfo {author}
  {\bibfnamefont {E.}~\bibnamefont {Wall\'en}}, \bibinfo {author}
  {\bibfnamefont {J.}~\bibnamefont {Bengtsson}}, \ and\ \bibinfo {author}
  {\bibfnamefont {A.}~\bibnamefont {Streun}}} (\bibinfo {year} {2009}),\ \href
  {http://link.aps.org/doi/10.1103/PhysRevSTAB.12.120701} {\bibfield  {journal}
  {\bibinfo  {journal} {Phys. Rev. ST Accel. Beams}\ }\textbf {\bibinfo
  {volume} {12}},\ \bibinfo {pages} {120701}}\BibitemShut {NoStop}%
\bibitem [{\citenamefont {Leemans}\ \emph {et~al.}(1996)\citenamefont
  {Leemans}, \citenamefont {Schoenlein}, \citenamefont {Volfbeyn},
  \citenamefont {Chin}, \citenamefont {Glover}, \citenamefont {Balling},
  \citenamefont {Zolotorev}, \citenamefont {Kim}, \citenamefont
  {Chattopadhyay},\ and\ \citenamefont {Shank}}]{Wim1996}%
  \BibitemOpen
  \bibfield  {author} {\bibinfo {author} {\bibnamefont {Leemans}, \bibfnamefont
  {W.~P.}}, \bibinfo {author} {\bibfnamefont {R.~W.}\ \bibnamefont
  {Schoenlein}}, \bibinfo {author} {\bibfnamefont {P.}~\bibnamefont
  {Volfbeyn}}, \bibinfo {author} {\bibfnamefont {A.~H.}\ \bibnamefont {Chin}},
  \bibinfo {author} {\bibfnamefont {T.~E.}\ \bibnamefont {Glover}}, \bibinfo
  {author} {\bibfnamefont {P.}~\bibnamefont {Balling}}, \bibinfo {author}
  {\bibfnamefont {M.}~\bibnamefont {Zolotorev}}, \bibinfo {author}
  {\bibfnamefont {K.~J.}\ \bibnamefont {Kim}}, \bibinfo {author} {\bibfnamefont
  {S.}~\bibnamefont {Chattopadhyay}}, \ and\ \bibinfo {author} {\bibfnamefont
  {C.~V.}\ \bibnamefont {Shank}}} (\bibinfo {year} {1996}),\ \href
  {http://link.aps.org/doi/10.1103/PhysRevLett.77.4182} {\bibfield  {journal}
  {\bibinfo  {journal} {Phys. Rev. Lett.}\ }\textbf {\bibinfo {volume} {77}},\
  \bibinfo {pages} {4182}}\BibitemShut {NoStop}%
\bibitem [{\citenamefont {Limborg}(1998)}]{Limb}%
  \BibitemOpen
  \bibfield  {author} {\bibinfo {author} {\bibnamefont {Limborg}, \bibfnamefont
  {C.}}} (\bibinfo {year} {1998}),\ in\ \href@noop {} {\emph {\bibinfo
  {booktitle} {Proc. SPIE}}},\ Vol.\ \bibinfo {volume} {3451},\ p.~\bibinfo
  {pages} {72}\BibitemShut {NoStop}%
\bibitem [{\citenamefont {Liu}\ \emph {et~al.}(2013)\citenamefont {Liu},
  \citenamefont {Li}, \citenamefont {Chen}, \citenamefont {Chen}, \citenamefont
  {Deng}, \citenamefont {Ding}, \citenamefont {Fan}, \citenamefont {Fang},
  \citenamefont {Feng}, \citenamefont {Feng}, \citenamefont {Gu}, \citenamefont
  {Gu}, \citenamefont {Guo}, \citenamefont {Huang}, \citenamefont {Huang},
  \citenamefont {Huang}, \citenamefont {Jia}, \citenamefont {Lan},
  \citenamefont {Leng}, \citenamefont {Li}, \citenamefont {Li}, \citenamefont
  {Li}, \citenamefont {Lin}, \citenamefont {Shang}, \citenamefont {Shen},
  \citenamefont {Tang}, \citenamefont {Wang}, \citenamefont {Wang},
  \citenamefont {Wang}, \citenamefont {Wang}, \citenamefont {Wang},
  \citenamefont {Wang}, \citenamefont {Yao}, \citenamefont {Ye}, \citenamefont
  {Yin}, \citenamefont {Yu}, \citenamefont {Zhang}, \citenamefont {Zhang},
  \citenamefont {Zhang}, \citenamefont {Zhang}, \citenamefont {Zhang},
  \citenamefont {Zhong}, \citenamefont {Zhou}, \citenamefont {Wang},\ and\
  \citenamefont {Zhao}}]{HGHGSINAP}%
  \BibitemOpen
  \bibfield  {author} {\bibinfo {author} {\bibnamefont {Liu}, \bibfnamefont
  {B.}}, \bibinfo {author} {\bibfnamefont {W.~B.}\ \bibnamefont {Li}}, \bibinfo
  {author} {\bibfnamefont {J.~H.}\ \bibnamefont {Chen}}, \bibinfo {author}
  {\bibfnamefont {Z.~H.}\ \bibnamefont {Chen}}, \bibinfo {author}
  {\bibfnamefont {H.~X.}\ \bibnamefont {Deng}}, \bibinfo {author}
  {\bibfnamefont {J.~G.}\ \bibnamefont {Ding}}, \bibinfo {author}
  {\bibfnamefont {Y.}~\bibnamefont {Fan}}, \bibinfo {author} {\bibfnamefont
  {G.~P.}\ \bibnamefont {Fang}}, \bibinfo {author} {\bibfnamefont
  {C.}~\bibnamefont {Feng}}, \bibinfo {author} {\bibfnamefont {L.}~\bibnamefont
  {Feng}}, \bibinfo {author} {\bibfnamefont {Q.}~\bibnamefont {Gu}}, \bibinfo
  {author} {\bibfnamefont {M.}~\bibnamefont {Gu}}, \bibinfo {author}
  {\bibfnamefont {C.}~\bibnamefont {Guo}}, \bibinfo {author} {\bibfnamefont
  {D.~Z.}\ \bibnamefont {Huang}}, \bibinfo {author} {\bibfnamefont {M.~M.}\
  \bibnamefont {Huang}}, \bibinfo {author} {\bibfnamefont {W.~H.}\ \bibnamefont
  {Huang}}, \bibinfo {author} {\bibfnamefont {Q.~K.}\ \bibnamefont {Jia}},
  \bibinfo {author} {\bibfnamefont {T.~H.}\ \bibnamefont {Lan}}, \bibinfo
  {author} {\bibfnamefont {Y.~B.}\ \bibnamefont {Leng}}, \bibinfo {author}
  {\bibfnamefont {D.~G.}\ \bibnamefont {Li}}, \bibinfo {author} {\bibfnamefont
  {W.~M.}\ \bibnamefont {Li}}, \bibinfo {author} {\bibfnamefont
  {X.}~\bibnamefont {Li}}, \bibinfo {author} {\bibfnamefont {G.~Q.}\
  \bibnamefont {Lin}}, \bibinfo {author} {\bibfnamefont {L.}~\bibnamefont
  {Shang}}, \bibinfo {author} {\bibfnamefont {L.}~\bibnamefont {Shen}},
  \bibinfo {author} {\bibfnamefont {C.~X.}\ \bibnamefont {Tang}}, \bibinfo
  {author} {\bibfnamefont {G.~L.}\ \bibnamefont {Wang}}, \bibinfo {author}
  {\bibfnamefont {L.}~\bibnamefont {Wang}}, \bibinfo {author} {\bibfnamefont
  {R.}~\bibnamefont {Wang}}, \bibinfo {author} {\bibfnamefont {X.~T.}\
  \bibnamefont {Wang}}, \bibinfo {author} {\bibfnamefont {Z.~S.}\ \bibnamefont
  {Wang}}, \bibinfo {author} {\bibfnamefont {Z.~S.}\ \bibnamefont {Wang}},
  \bibinfo {author} {\bibfnamefont {H.~F.}\ \bibnamefont {Yao}}, \bibinfo
  {author} {\bibfnamefont {K.~R.}\ \bibnamefont {Ye}}, \bibinfo {author}
  {\bibfnamefont {L.~X.}\ \bibnamefont {Yin}}, \bibinfo {author} {\bibfnamefont
  {L.~Y.}\ \bibnamefont {Yu}}, \bibinfo {author} {\bibfnamefont {J.~Q.}\
  \bibnamefont {Zhang}}, \bibinfo {author} {\bibfnamefont {M.}~\bibnamefont
  {Zhang}}, \bibinfo {author} {\bibfnamefont {M.}~\bibnamefont {Zhang}},
  \bibinfo {author} {\bibfnamefont {T.}~\bibnamefont {Zhang}}, \bibinfo
  {author} {\bibfnamefont {W.~Y.}\ \bibnamefont {Zhang}}, \bibinfo {author}
  {\bibfnamefont {S.~P.}\ \bibnamefont {Zhong}}, \bibinfo {author}
  {\bibfnamefont {Q.~G.}\ \bibnamefont {Zhou}}, \bibinfo {author}
  {\bibfnamefont {D.}~\bibnamefont {Wang}}, \ and\ \bibinfo {author}
  {\bibfnamefont {Z.~T.}\ \bibnamefont {Zhao}}} (\bibinfo {year} {2013}),\
  \href@noop {} {\bibfield  {journal} {\bibinfo  {journal} {Phys. Rev. ST
  Accel. Beams}\ }\textbf {\bibinfo {volume} {16}},\ \bibinfo {pages}
  {020704}}\BibitemShut {NoStop}%
\bibitem [{\citenamefont {Liu}\ and\ \citenamefont
  {Neil}(1993)}]{beam_cond_neil}%
  \BibitemOpen
  \bibfield  {author} {\bibinfo {author} {\bibnamefont {Liu}, \bibfnamefont
  {H.}}, \ and\ \bibinfo {author} {\bibfnamefont {G.}~\bibnamefont {Neil}}}
  (\bibinfo {year} {1993}),\ in\ \href@noop {} {\emph {\bibinfo {booktitle}
  {Proc. {IEEE} Particle Accelerator Conference, Washington, 1993}}}\ (\bibinfo
   {publisher} {IEEE},\ \bibinfo {address} {Piscataway, NJ})\ p.\ \bibinfo
  {pages} {279}\BibitemShut {NoStop}%
\bibitem [{\citenamefont {Lutman}\ \emph {et~al.}(2013)\citenamefont {Lutman},
  \citenamefont {Coffee}, \citenamefont {Ding}, \citenamefont {Huang},
  \citenamefont {Krzywinski}, \citenamefont {Maxwell}, \citenamefont
  {Messerschmidt},\ and\ \citenamefont {Nuhn}}]{Lutman}%
  \BibitemOpen
  \bibfield  {author} {\bibinfo {author} {\bibnamefont {Lutman}, \bibfnamefont
  {A.~A.}}, \bibinfo {author} {\bibfnamefont {R.}~\bibnamefont {Coffee}},
  \bibinfo {author} {\bibfnamefont {Y.}~\bibnamefont {Ding}}, \bibinfo {author}
  {\bibfnamefont {Z.}~\bibnamefont {Huang}}, \bibinfo {author} {\bibfnamefont
  {J.}~\bibnamefont {Krzywinski}}, \bibinfo {author} {\bibfnamefont
  {T.}~\bibnamefont {Maxwell}}, \bibinfo {author} {\bibfnamefont
  {M.}~\bibnamefont {Messerschmidt}}, \ and\ \bibinfo {author} {\bibfnamefont
  {H.-D.}\ \bibnamefont {Nuhn}}} (\bibinfo {year} {2013}),\ \href@noop {}
  {\bibfield  {journal} {\bibinfo  {journal} {Phys. Rev. Lett.}\ }\textbf
  {\bibinfo {volume} {110}},\ \bibinfo {pages} {134801}}\BibitemShut {NoStop}%
\bibitem [{\citenamefont {Madey}(1971)}]{Madey}%
  \BibitemOpen
  \bibfield  {author} {\bibinfo {author} {\bibnamefont {Madey}, \bibfnamefont
  {J.}}} (\bibinfo {year} {1971}),\ \href@noop {} {\bibfield  {journal}
  {\bibinfo  {journal} {J. Appl. Phys.}\ }\textbf {\bibinfo {volume} {42}},\
  \bibinfo {pages} {1906}}\BibitemShut {NoStop}%
\bibitem [{\citenamefont {Madey}(2010)}]{Madey1}%
  \BibitemOpen
  \bibfield  {author} {\bibinfo {author} {\bibnamefont {Madey}, \bibfnamefont
  {J.~M.~J.}}} (\bibinfo {year} {2010}),\ \href@noop {} {\bibfield  {journal}
  {\bibinfo  {journal} {Reviews of Accelerator Science and Technology}\
  }\textbf {\bibinfo {volume} {3}},\ \bibinfo {pages} {1}}\BibitemShut
  {NoStop}%
\bibitem [{\citenamefont {Marinelli}\ \emph {et~al.}(2013)\citenamefont
  {Marinelli}, \citenamefont {Lutman}, \citenamefont {Wu}, \citenamefont
  {Ding}, \citenamefont {Krzywinski}, \citenamefont {Nuhn}, \citenamefont
  {Feng}, \citenamefont {Coffee},\ and\ \citenamefont
  {Pellegrini}}]{PhysRevLett.111.134801}%
  \BibitemOpen
  \bibfield  {author} {\bibinfo {author} {\bibnamefont {Marinelli},
  \bibfnamefont {A.}}, \bibinfo {author} {\bibfnamefont {A.~A.}\ \bibnamefont
  {Lutman}}, \bibinfo {author} {\bibfnamefont {J.}~\bibnamefont {Wu}}, \bibinfo
  {author} {\bibfnamefont {Y.}~\bibnamefont {Ding}}, \bibinfo {author}
  {\bibfnamefont {J.}~\bibnamefont {Krzywinski}}, \bibinfo {author}
  {\bibfnamefont {H.-D.}\ \bibnamefont {Nuhn}}, \bibinfo {author}
  {\bibfnamefont {Y.}~\bibnamefont {Feng}}, \bibinfo {author} {\bibfnamefont
  {R.~N.}\ \bibnamefont {Coffee}}, \ and\ \bibinfo {author} {\bibfnamefont
  {C.}~\bibnamefont {Pellegrini}}} (\bibinfo {year} {2013}),\ \href
  {http://link.aps.org/doi/10.1103/PhysRevLett.111.134801} {\bibfield
  {journal} {\bibinfo  {journal} {Phys. Rev. Lett.}\ }\textbf {\bibinfo
  {volume} {111}},\ \bibinfo {pages} {134801}}\BibitemShut {NoStop}%
\bibitem [{\citenamefont {Martin}\ and\ \citenamefont
  {Bartolini}(2011)}]{Martin}%
  \BibitemOpen
  \bibfield  {author} {\bibinfo {author} {\bibnamefont {Martin}, \bibfnamefont
  {I.~P.~S.}}, \ and\ \bibinfo {author} {\bibfnamefont {R.}~\bibnamefont
  {Bartolini}}} (\bibinfo {year} {2011}),\ \href@noop {} {\bibfield  {journal}
  {\bibinfo  {journal} {Phys. Rev. Spec. Topics Accel. and Beams}\ }\textbf
  {\bibinfo {volume} {14}},\ \bibinfo {pages} {030702}}\BibitemShut {NoStop}%
\bibitem [{\citenamefont {McNeil}\ \emph {et~al.}(2005)\citenamefont {McNeil},
  \citenamefont {Robb},\ and\ \citenamefont {Poole}}]{HGHGS1}%
  \BibitemOpen
  \bibfield  {author} {\bibinfo {author} {\bibnamefont {McNeil}, \bibfnamefont
  {B.}}, \bibinfo {author} {\bibfnamefont {G.}~\bibnamefont {Robb}}, \ and\
  \bibinfo {author} {\bibfnamefont {M.}~\bibnamefont {Poole}}} (\bibinfo {year}
  {2005}),\ in\ \href@noop {} {\emph {\bibinfo {booktitle} {Proceedings of 2005
  Particle Accelerator Conference}}},\ p.\ \bibinfo {pages} {1718}\BibitemShut
  {NoStop}%
\bibitem [{\citenamefont {McNeil}\ and\ \citenamefont
  {Thompson}(2010)}]{McNeilNature}%
  \BibitemOpen
  \bibfield  {author} {\bibinfo {author} {\bibnamefont {McNeil}, \bibfnamefont
  {B.}}, \ and\ \bibinfo {author} {\bibnamefont {Thompson}}} (\bibinfo {year}
  {2010}),\ \href@noop {} {\bibfield  {journal} {\bibinfo  {journal} {Nature
  Photonics}\ }\textbf {\bibinfo {volume} {4}},\ \bibinfo {pages}
  {814}}\BibitemShut {NoStop}%
\bibitem [{\citenamefont {McNeil}\ \emph {et~al.}(2011)\citenamefont {McNeil},
  \citenamefont {Thompson}, \citenamefont {Dunning},\ and\ \citenamefont
  {Sheehy}}]{HHGML}%
  \BibitemOpen
  \bibfield  {author} {\bibinfo {author} {\bibnamefont {McNeil}, \bibfnamefont
  {B.}}, \bibinfo {author} {\bibfnamefont {N.}~\bibnamefont {Thompson}},
  \bibinfo {author} {\bibfnamefont {D.}~\bibnamefont {Dunning}}, \ and\
  \bibinfo {author} {\bibfnamefont {B.}~\bibnamefont {Sheehy}}} (\bibinfo
  {year} {2011}),\ \href@noop {} {\bibfield  {journal} {\bibinfo  {journal} {J.
  Phys. B: At. Mol. Opt. Phys.}\ }\textbf {\bibinfo {volume} {44}},\ \bibinfo
  {pages} {065404}}\BibitemShut {NoStop}%
\bibitem [{\citenamefont {McNeil}\ \emph {et~al.}(2006)\citenamefont {McNeil},
  \citenamefont {Robb}, \citenamefont {Poole},\ and\ \citenamefont
  {Thompson}}]{PhysRevLett.96.084801}%
  \BibitemOpen
  \bibfield  {author} {\bibinfo {author} {\bibnamefont {McNeil}, \bibfnamefont
  {B.~W.~J.}}, \bibinfo {author} {\bibfnamefont {G.~R.~M.}\ \bibnamefont
  {Robb}}, \bibinfo {author} {\bibfnamefont {M.~W.}\ \bibnamefont {Poole}}, \
  and\ \bibinfo {author} {\bibfnamefont {N.~R.}\ \bibnamefont {Thompson}}}
  (\bibinfo {year} {2006}),\ \href {\doibase 10.1103/PhysRevLett.96.084801}
  {\bibfield  {journal} {\bibinfo  {journal} {Phys. Rev. Lett.}\ }\textbf
  {\bibinfo {volume} {96}},\ \bibinfo {pages} {084801}}\BibitemShut {NoStop}%
\bibitem [{\citenamefont {McNeil}\ \emph {et~al.}(2013)\citenamefont {McNeil},
  \citenamefont {Thompson},\ and\ \citenamefont {Dunning}}]{HBSASE}%
  \BibitemOpen
  \bibfield  {author} {\bibinfo {author} {\bibnamefont {McNeil}, \bibfnamefont
  {B.~W.~J.}}, \bibinfo {author} {\bibfnamefont {N.~R.}\ \bibnamefont
  {Thompson}}, \ and\ \bibinfo {author} {\bibfnamefont {D.~J.}\ \bibnamefont
  {Dunning}}} (\bibinfo {year} {2013}),\ \href
  {http://link.aps.org/doi/10.1103/PhysRevLett.110.134802} {\bibfield
  {journal} {\bibinfo  {journal} {Phys. Rev. Lett.}\ }\textbf {\bibinfo
  {volume} {110}},\ \bibinfo {pages} {134802}}\BibitemShut {NoStop}%
\bibitem [{\citenamefont {Molina-Terriza}\ \emph {et~al.}(2007)\citenamefont
  {Molina-Terriza}, \citenamefont {Torres},\ and\ \citenamefont
  {Torner}}]{MolinaNature}%
  \BibitemOpen
  \bibfield  {author} {\bibinfo {author} {\bibnamefont {Molina-Terriza},
  \bibfnamefont {G.}}, \bibinfo {author} {\bibfnamefont {J.~P.}\ \bibnamefont
  {Torres}}, \ and\ \bibinfo {author} {\bibfnamefont {L.}~\bibnamefont
  {Torner}}} (\bibinfo {year} {2007}),\ \href
  {http://dx.doi.org/10.1038/nphys607} {\bibfield  {journal} {\bibinfo
  {journal} {Nature Phys.}\ }\textbf {\bibinfo {volume} {3}}~(\bibinfo {number}
  {5}),\ \bibinfo {pages} {305}}\BibitemShut {NoStop}%
\bibitem [{\citenamefont {Molo}\ \emph {et~al.}(2011)\citenamefont {Molo},
  \citenamefont {Bakr}, \citenamefont {Honer}, \citenamefont {Huck},
  \citenamefont {Khan}, \citenamefont {Nowaczyk}, \citenamefont {Schick},
  \citenamefont {Ungelenk},\ and\ \citenamefont {Zeinalzadeh}}]{DELTA}%
  \BibitemOpen
  \bibfield  {author} {\bibinfo {author} {\bibnamefont {Molo}, \bibfnamefont
  {R.}}, \bibinfo {author} {\bibfnamefont {M.}~\bibnamefont {Bakr}}, \bibinfo
  {author} {\bibfnamefont {M.}~\bibnamefont {Honer}}, \bibinfo {author}
  {\bibfnamefont {H.}~\bibnamefont {Huck}}, \bibinfo {author} {\bibfnamefont
  {S.}~\bibnamefont {Khan}}, \bibinfo {author} {\bibfnamefont {A.}~\bibnamefont
  {Nowaczyk}}, \bibinfo {author} {\bibfnamefont {A.}~\bibnamefont {Schick}},
  \bibinfo {author} {\bibfnamefont {P.}~\bibnamefont {Ungelenk}}, \ and\
  \bibinfo {author} {\bibfnamefont {M.}~\bibnamefont {Zeinalzadeh}}} (\bibinfo
  {year} {2011}),\ in\ \href@noop {} {\emph {\bibinfo {booktitle} {2011
  Free-electron laser Conference}}},\ p.\ \bibinfo {pages} {219}\BibitemShut
  {NoStop}%
\bibitem [{\citenamefont {Muggli}\ \emph {et~al.}(2008)\citenamefont {Muggli},
  \citenamefont {Yakimenko}, \citenamefont {Babzien}, \citenamefont {Kallos},\
  and\ \citenamefont {Kusche}}]{Muggli}%
  \BibitemOpen
  \bibfield  {author} {\bibinfo {author} {\bibnamefont {Muggli}, \bibfnamefont
  {P.}}, \bibinfo {author} {\bibfnamefont {V.}~\bibnamefont {Yakimenko}},
  \bibinfo {author} {\bibfnamefont {M.}~\bibnamefont {Babzien}}, \bibinfo
  {author} {\bibfnamefont {E.}~\bibnamefont {Kallos}}, \ and\ \bibinfo {author}
  {\bibfnamefont {K.}~\bibnamefont {Kusche}}} (\bibinfo {year} {2008}),\
  \href@noop {} {\bibfield  {journal} {\bibinfo  {journal} {Phys. Rev. Lett.}\
  }\textbf {\bibinfo {volume} {101}},\ \bibinfo {pages} {054801}}\BibitemShut
  {NoStop}%
\bibitem [{\citenamefont {Mukamel}(1995)}]{Mukamel}%
  \BibitemOpen
  \bibfield  {author} {\bibinfo {author} {\bibnamefont {Mukamel}, \bibfnamefont
  {S.}}} (\bibinfo {year} {1995}),\ \href@noop {} {\emph {\bibinfo {title}
  {Principles of nonlinear optical spectroscopy}}}\ (\bibinfo  {publisher}
  {Oxford Univeristy},\ \bibinfo {address} {New York})\BibitemShut {NoStop}%
\bibitem [{\citenamefont {Murphy}\ and\ \citenamefont
  {Pellegrini}(1990)}]{LaserHandbook}%
  \BibitemOpen
  \bibfield  {author} {\bibinfo {author} {\bibnamefont {Murphy}, \bibfnamefont
  {J.~B.}}, \ and\ \bibinfo {author} {\bibfnamefont {C.}~\bibnamefont
  {Pellegrini}}} (\bibinfo {year} {1990}),\ in\ \href@noop {} {\emph {\bibinfo
  {booktitle} {Laser Handbook}}},\ Vol.~\bibinfo {volume} {6},\ \bibinfo
  {editor} {edited by\ \bibinfo {editor} {\bibfnamefont {W.}~\bibnamefont
  {Colson}}, \bibinfo {editor} {\bibfnamefont {C.}~\bibnamefont {Pellegrini}},
  \ and\ \bibinfo {editor} {\bibfnamefont {A.}~\bibnamefont {Renieri}}},\
  Chap.~\bibinfo {chapter} {5}\ (\bibinfo  {publisher} {North Holland,
  Amsterdam})\BibitemShut {NoStop}%
\bibitem [{\citenamefont {Musumeci}\ \emph
  {et~al.}(2005{\natexlab{a}})\citenamefont {Musumeci}, \citenamefont
  {Pellegrini},\ and\ \citenamefont {Rosenzweig}}]{PhysRevE.72.016501}%
  \BibitemOpen
  \bibfield  {author} {\bibinfo {author} {\bibnamefont {Musumeci},
  \bibfnamefont {P.}}, \bibinfo {author} {\bibfnamefont {C.}~\bibnamefont
  {Pellegrini}}, \ and\ \bibinfo {author} {\bibfnamefont {J.~B.}\ \bibnamefont
  {Rosenzweig}}} (\bibinfo {year} {2005}{\natexlab{a}}),\ \href
  {http://link.aps.org/doi/10.1103/PhysRevE.72.016501} {\bibfield  {journal}
  {\bibinfo  {journal} {Phys. Rev. E}\ }\textbf {\bibinfo {volume} {72}},\
  \bibinfo {pages} {016501}}\BibitemShut {NoStop}%
\bibitem [{\citenamefont {Musumeci}\ \emph
  {et~al.}(2005{\natexlab{b}})\citenamefont {Musumeci}, \citenamefont
  {Tochitsky}, \citenamefont {Boucher}, \citenamefont {Clayton}, \citenamefont
  {Doyuran}, \citenamefont {England}, \citenamefont {Joshi}, \citenamefont
  {Pellegrini}, \citenamefont {Ralph}, \citenamefont {Rosenzweig},
  \citenamefont {Sung}, \citenamefont {Tolmachev}, \citenamefont {Travish},
  \citenamefont {Varfolomeev}, \citenamefont {Varfolomeev}, \citenamefont
  {Yarovoi},\ and\ \citenamefont {Yoder}}]{PhysRevLett.94.154801}%
  \BibitemOpen
  \bibfield  {author} {\bibinfo {author} {\bibnamefont {Musumeci},
  \bibfnamefont {P.}}, \bibinfo {author} {\bibfnamefont {S.~Y.}\ \bibnamefont
  {Tochitsky}}, \bibinfo {author} {\bibfnamefont {S.}~\bibnamefont {Boucher}},
  \bibinfo {author} {\bibfnamefont {C.~E.}\ \bibnamefont {Clayton}}, \bibinfo
  {author} {\bibfnamefont {A.}~\bibnamefont {Doyuran}}, \bibinfo {author}
  {\bibfnamefont {R.~J.}\ \bibnamefont {England}}, \bibinfo {author}
  {\bibfnamefont {C.}~\bibnamefont {Joshi}}, \bibinfo {author} {\bibfnamefont
  {C.}~\bibnamefont {Pellegrini}}, \bibinfo {author} {\bibfnamefont {J.~E.}\
  \bibnamefont {Ralph}}, \bibinfo {author} {\bibfnamefont {J.~B.}\ \bibnamefont
  {Rosenzweig}}, \bibinfo {author} {\bibfnamefont {C.}~\bibnamefont {Sung}},
  \bibinfo {author} {\bibfnamefont {S.}~\bibnamefont {Tolmachev}}, \bibinfo
  {author} {\bibfnamefont {G.}~\bibnamefont {Travish}}, \bibinfo {author}
  {\bibfnamefont {A.~A.}\ \bibnamefont {Varfolomeev}}, \bibinfo {author}
  {\bibfnamefont {A.~A.}\ \bibnamefont {Varfolomeev}}, \bibinfo {author}
  {\bibfnamefont {T.}~\bibnamefont {Yarovoi}}, \ and\ \bibinfo {author}
  {\bibfnamefont {R.~B.}\ \bibnamefont {Yoder}}} (\bibinfo {year}
  {2005}{\natexlab{b}}),\ \href
  {http://link.aps.org/doi/10.1103/PhysRevLett.94.154801} {\bibfield  {journal}
  {\bibinfo  {journal} {Phys. Rev. Lett.}\ }\textbf {\bibinfo {volume} {94}},\
  \bibinfo {pages} {154801}}\BibitemShut {NoStop}%
\bibitem [{\citenamefont {Neutze}\ \emph {et~al.}(2000)\citenamefont {Neutze},
  \citenamefont {Wouts}, \citenamefont {Van~der Spoel}, \citenamefont
  {Weckert},\ and\ \citenamefont {Hajdu}}]{DBD}%
  \BibitemOpen
  \bibfield  {author} {\bibinfo {author} {\bibnamefont {Neutze}, \bibfnamefont
  {R.}}, \bibinfo {author} {\bibfnamefont {R.}~\bibnamefont {Wouts}}, \bibinfo
  {author} {\bibfnamefont {D.}~\bibnamefont {Van~der Spoel}}, \bibinfo {author}
  {\bibfnamefont {E.}~\bibnamefont {Weckert}}, \ and\ \bibinfo {author}
  {\bibfnamefont {J.}~\bibnamefont {Hajdu}}} (\bibinfo {year} {2000}),\
  \href@noop {} {\bibfield  {journal} {\bibinfo  {journal} {Nature}\ }\textbf
  {\bibinfo {volume} {406}},\ \bibinfo {pages} {752}}\BibitemShut {NoStop}%
\bibitem [{\citenamefont {Padgett}\ and\ \citenamefont
  {Bowman}(2011)}]{PadgettNature}%
  \BibitemOpen
  \bibfield  {author} {\bibinfo {author} {\bibnamefont {Padgett}, \bibfnamefont
  {M.}}, \ and\ \bibinfo {author} {\bibfnamefont {R.}~\bibnamefont {Bowman}}}
  (\bibinfo {year} {2011}),\ \href {http://dx.doi.org/10.1038/nphoton.2011.81}
  {\bibfield  {journal} {\bibinfo  {journal} {Nature Photon.}\ }\textbf
  {\bibinfo {volume} {5}}~(\bibinfo {number} {6}),\ \bibinfo {pages}
  {343}}\BibitemShut {NoStop}%
\bibitem [{\citenamefont {Palmer}(1988)}]{Palmer:1987hi}%
  \BibitemOpen
  \bibfield  {author} {\bibinfo {author} {\bibnamefont {Palmer}, \bibfnamefont
  {R.}}} (\bibinfo {year} {1988}),\ in\ \href@noop {} {\emph {\bibinfo
  {booktitle} {Lect. Notes Phys.}}},\ Vol.\ \bibinfo {volume} {296},\ pp.\
  \bibinfo {pages} {607--635}\BibitemShut {NoStop}%
\bibitem [{\citenamefont {Palmer}(1995)}]{Palmer:1994tj}%
  \BibitemOpen
  \bibfield  {author} {\bibinfo {author} {\bibnamefont {Palmer}, \bibfnamefont
  {R.}}} (\bibinfo {year} {1995}),\ in\ \href@noop {} {\emph {\bibinfo
  {booktitle} {AIP Conf. Proc.}}},\ Vol.\ \bibinfo {volume} {335},\ pp.\
  \bibinfo {pages} {90--100}\BibitemShut {NoStop}%
\bibitem [{\citenamefont {Palmer}(1972)}]{palmer:3014}%
  \BibitemOpen
  \bibfield  {author} {\bibinfo {author} {\bibnamefont {Palmer}, \bibfnamefont
  {R.~B.}}} (\bibinfo {year} {1972}),\ \href
  {http://link.aip.org/link/?JAP/43/3014/1} {\bibfield  {journal} {\bibinfo
  {journal} {Journal of Applied Physics}\ }\textbf {\bibinfo {volume}
  {43}}~(\bibinfo {number} {7}),\ \bibinfo {pages} {3014}}\BibitemShut
  {NoStop}%
\bibitem [{\citenamefont {Panofsky}\ and\ \citenamefont
  {Wenzel}(1956)}]{panofsky56w}%
  \BibitemOpen
  \bibfield  {author} {\bibinfo {author} {\bibnamefont {Panofsky},
  \bibfnamefont {W.~K.~H.}}, \ and\ \bibinfo {author} {\bibfnamefont
  {W.}~\bibnamefont {Wenzel}}} (\bibinfo {year} {1956}),\ \href@noop {}
  {\bibfield  {journal} {\bibinfo  {journal} {Rev. Sci. Instr.}\ }\textbf
  {\bibinfo {volume} {27}},\ \bibinfo {pages} {967}}\BibitemShut {NoStop}%
\bibitem [{\citenamefont {Papadichev}(1995)}]{Papadichev1995ABS79}%
  \BibitemOpen
  \bibfield  {author} {\bibinfo {author} {\bibnamefont {Papadichev},
  \bibfnamefont {V.}}} (\bibinfo {year} {1995}),\ \href
  {http://www.sciencedirect.com/science/article/pii/0168900294015856}
  {\bibfield  {journal} {\bibinfo  {journal} {Nucl. Instrum. Methods Phys. Res.
  A}\ }\textbf {\bibinfo {volume} {358}}~(\bibinfo {number} {1-3}),\ \bibinfo
  {pages} {ABS79}}\BibitemShut {NoStop}%
\bibitem [{\citenamefont {Patterson}(2010)}]{PattersonLCLS}%
  \BibitemOpen
  \bibfield  {author} {\bibinfo {author} {\bibnamefont {Patterson},
  \bibfnamefont {B.}}} (\bibinfo {year} {2010}),\ \href@noop {} {\emph
  {\bibinfo {title} {Resource letter on stimulated inelastic x-ray scattering
  at an XFEL}}},\ \bibinfo {type} {Preprint}\ \bibinfo {number}
  {SLAC-TN-10-026}\BibitemShut {NoStop}%
\bibitem [{\citenamefont {Pellegrini}\ and\ \citenamefont
  {Sessler}(1971)}]{Pell}%
  \BibitemOpen
  \bibfield  {author} {\bibinfo {author} {\bibnamefont {Pellegrini},
  \bibfnamefont {C.}}, \ and\ \bibinfo {author} {\bibfnamefont
  {A.}~\bibnamefont {Sessler}}} (\bibinfo {year} {1971}),\ \href@noop {}
  {\bibfield  {journal} {\bibinfo  {journal} {Nuovo Cimento}\ }\textbf
  {\bibinfo {volume} {3}},\ \bibinfo {pages} {116}}\BibitemShut {NoStop}%
\bibitem [{\citenamefont {Peralta}\ \emph {et~al.}(2013)\citenamefont
  {Peralta}, \citenamefont {Soong}, \citenamefont {England}, \citenamefont
  {Colby}, \citenamefont {Wu}, \citenamefont {Montazeri}, \citenamefont
  {McGuinness}, \citenamefont {McNeur}, \citenamefont {Leedle}, \citenamefont
  {Walz}, \citenamefont {Sozer}, \citenamefont {Cowan}, \citenamefont
  {Schwartz}, \citenamefont {G.},\ and\ \citenamefont
  {Byer}}]{Peralta:2013vpa}%
  \BibitemOpen
  \bibfield  {author} {\bibinfo {author} {\bibnamefont {Peralta}, \bibfnamefont
  {E.}}, \bibinfo {author} {\bibfnamefont {K.}~\bibnamefont {Soong}}, \bibinfo
  {author} {\bibfnamefont {R.}~\bibnamefont {England}}, \bibinfo {author}
  {\bibfnamefont {E.}~\bibnamefont {Colby}}, \bibinfo {author} {\bibfnamefont
  {Z.}~\bibnamefont {Wu}}, \bibinfo {author} {\bibfnamefont {B.}~\bibnamefont
  {Montazeri}}, \bibinfo {author} {\bibfnamefont {C.}~\bibnamefont
  {McGuinness}}, \bibinfo {author} {\bibfnamefont {J.}~\bibnamefont {McNeur}},
  \bibinfo {author} {\bibfnamefont {K.}~\bibnamefont {Leedle}}, \bibinfo
  {author} {\bibfnamefont {D.}~\bibnamefont {Walz}}, \bibinfo {author}
  {\bibfnamefont {E.}~\bibnamefont {Sozer}}, \bibinfo {author} {\bibfnamefont
  {B.}~\bibnamefont {Cowan}}, \bibinfo {author} {\bibfnamefont
  {B.}~\bibnamefont {Schwartz}}, \bibinfo {author} {\bibfnamefont
  {T.}~\bibnamefont {G.}}, \ and\ \bibinfo {author} {\bibfnamefont
  {R.}~\bibnamefont {Byer}}} (\bibinfo {year} {2013}),\ \href@noop {} {\bibinfo
   {journal} {Nature}\ }\BibitemShut {NoStop}%
\bibitem [{\citenamefont {Petrillo}\ \emph {et~al.}(2013)\citenamefont
  {Petrillo}, \citenamefont {Anania}, \citenamefont {Artioli}, \citenamefont
  {Bacci}, \citenamefont {Bellaveglia}, \citenamefont {Chiadroni},
  \citenamefont {Cianchi}, \citenamefont {Ciocci}, \citenamefont {Dattoli},
  \citenamefont {Di~Giovenale}, \citenamefont {Di~Pirro}, \citenamefont
  {Ferrario}, \citenamefont {Gatti}, \citenamefont {Giannessi}, \citenamefont
  {Mostacci}, \citenamefont {Musumeci}, \citenamefont {Petralia}, \citenamefont
  {Pompili}, \citenamefont {Quattromini}, \citenamefont {Rau}, \citenamefont
  {Ronsivalle}, \citenamefont {Rossi}, \citenamefont {Sabia}, \citenamefont
  {Vaccarezza},\ and\ \citenamefont {Villa}}]{PetrilloTwoBeam}%
  \BibitemOpen
\bibfield  {journal} {  }\bibfield  {author} {\bibinfo {author} {\bibnamefont
  {Petrillo}, \bibfnamefont {V.}}, \bibinfo {author} {\bibfnamefont {M.~P.}\
  \bibnamefont {Anania}}, \bibinfo {author} {\bibfnamefont {M.}~\bibnamefont
  {Artioli}}, \bibinfo {author} {\bibfnamefont {A.}~\bibnamefont {Bacci}},
  \bibinfo {author} {\bibfnamefont {M.}~\bibnamefont {Bellaveglia}}, \bibinfo
  {author} {\bibfnamefont {E.}~\bibnamefont {Chiadroni}}, \bibinfo {author}
  {\bibfnamefont {A.}~\bibnamefont {Cianchi}}, \bibinfo {author} {\bibfnamefont
  {F.}~\bibnamefont {Ciocci}}, \bibinfo {author} {\bibfnamefont
  {G.}~\bibnamefont {Dattoli}}, \bibinfo {author} {\bibfnamefont
  {D.}~\bibnamefont {Di~Giovenale}}, \bibinfo {author} {\bibfnamefont
  {G.}~\bibnamefont {Di~Pirro}}, \bibinfo {author} {\bibfnamefont
  {M.}~\bibnamefont {Ferrario}}, \bibinfo {author} {\bibfnamefont
  {G.}~\bibnamefont {Gatti}}, \bibinfo {author} {\bibfnamefont
  {L.}~\bibnamefont {Giannessi}}, \bibinfo {author} {\bibfnamefont
  {A.}~\bibnamefont {Mostacci}}, \bibinfo {author} {\bibfnamefont
  {P.}~\bibnamefont {Musumeci}}, \bibinfo {author} {\bibfnamefont
  {A.}~\bibnamefont {Petralia}}, \bibinfo {author} {\bibfnamefont
  {R.}~\bibnamefont {Pompili}}, \bibinfo {author} {\bibfnamefont
  {M.}~\bibnamefont {Quattromini}}, \bibinfo {author} {\bibfnamefont {J.~V.}\
  \bibnamefont {Rau}}, \bibinfo {author} {\bibfnamefont {C.}~\bibnamefont
  {Ronsivalle}}, \bibinfo {author} {\bibfnamefont {A.~R.}\ \bibnamefont
  {Rossi}}, \bibinfo {author} {\bibfnamefont {E.}~\bibnamefont {Sabia}},
  \bibinfo {author} {\bibfnamefont {C.}~\bibnamefont {Vaccarezza}}, \ and\
  \bibinfo {author} {\bibfnamefont {F.}~\bibnamefont {Villa}}} (\bibinfo {year}
  {2013}),\ \href {http://link.aps.org/doi/10.1103/PhysRevLett.111.114802}
  {\bibfield  {journal} {\bibinfo  {journal} {Phys. Rev. Lett.}\ }\textbf
  {\bibinfo {volume} {111}},\ \bibinfo {pages} {114802}}\BibitemShut {NoStop}%
\bibitem [{\citenamefont {Piot}\ \emph {et~al.}(2011)\citenamefont {Piot},
  \citenamefont {Sun}, \citenamefont {Power},\ and\ \citenamefont
  {Rihaoui}}]{EEX5}%
  \BibitemOpen
  \bibfield  {author} {\bibinfo {author} {\bibnamefont {Piot}, \bibfnamefont
  {P.}}, \bibinfo {author} {\bibfnamefont {Y.-E.}\ \bibnamefont {Sun}},
  \bibinfo {author} {\bibfnamefont {J.~G.}\ \bibnamefont {Power}}, \ and\
  \bibinfo {author} {\bibfnamefont {M.}~\bibnamefont {Rihaoui}}} (\bibinfo
  {year} {2011}),\ \href
  {http://link.aps.org/doi/10.1103/PhysRevSTAB.14.022801} {\bibfield  {journal}
  {\bibinfo  {journal} {Phys. Rev. ST Accel. Beams}\ }\textbf {\bibinfo
  {volume} {14}},\ \bibinfo {pages} {022801}}\BibitemShut {NoStop}%
\bibitem [{\citenamefont {Plettner}\ \emph {et~al.}(2005)\citenamefont
  {Plettner}, \citenamefont {Byer}, \citenamefont {Colby}, \citenamefont
  {Cowan}, \citenamefont {Sears}, \citenamefont {Spencer},\ and\ \citenamefont
  {Siemann}}]{PhysRevLett.95.134801}%
  \BibitemOpen
  \bibfield  {author} {\bibinfo {author} {\bibnamefont {Plettner},
  \bibfnamefont {T.}}, \bibinfo {author} {\bibfnamefont {R.~L.}\ \bibnamefont
  {Byer}}, \bibinfo {author} {\bibfnamefont {E.}~\bibnamefont {Colby}},
  \bibinfo {author} {\bibfnamefont {B.}~\bibnamefont {Cowan}}, \bibinfo
  {author} {\bibfnamefont {C.~M.~S.}\ \bibnamefont {Sears}}, \bibinfo {author}
  {\bibfnamefont {J.~E.}\ \bibnamefont {Spencer}}, \ and\ \bibinfo {author}
  {\bibfnamefont {R.~H.}\ \bibnamefont {Siemann}}} (\bibinfo {year} {2005}),\
  \href@noop {} {\bibfield  {journal} {\bibinfo  {journal} {Phys. Rev. Lett.}\
  }\textbf {\bibinfo {volume} {95}},\ \bibinfo {pages} {134801}}\BibitemShut
  {NoStop}%
\bibitem [{\citenamefont {Popmintchev}\ \emph {et~al.}(2010)\citenamefont
  {Popmintchev}, \citenamefont {Chen}, \citenamefont {Arpin}, \citenamefont
  {Murnane},\ and\ \citenamefont {Kapteyn}}]{HHGreview}%
  \BibitemOpen
  \bibfield  {author} {\bibinfo {author} {\bibnamefont {Popmintchev},
  \bibfnamefont {T.}}, \bibinfo {author} {\bibfnamefont {M.}~\bibnamefont
  {Chen}}, \bibinfo {author} {\bibfnamefont {P.}~\bibnamefont {Arpin}},
  \bibinfo {author} {\bibfnamefont {M.}~\bibnamefont {Murnane}}, \ and\
  \bibinfo {author} {\bibfnamefont {H.}~\bibnamefont {Kapteyn}}} (\bibinfo
  {year} {2010}),\ \href@noop {} {\bibfield  {journal} {\bibinfo  {journal}
  {Nature Photon.}\ }\textbf {\bibinfo {volume} {4}},\ \bibinfo {pages}
  {822}}\BibitemShut {NoStop}%
\bibitem [{\citenamefont {{Pottorf}}\ and\ \citenamefont
  {{Wang}}(2002)}]{WangHermIFEL2002}%
  \BibitemOpen
  \bibfield  {author} {\bibinfo {author} {\bibnamefont {{Pottorf}},
  \bibfnamefont {S.}}, \ and\ \bibinfo {author} {\bibfnamefont {X.~J.}\
  \bibnamefont {{Wang}}}} (\bibinfo {year} {2002}),\ in\ \href@noop {} {\emph
  {\bibinfo {booktitle} {Quantum Aspects of Beam Physics}}},\ \bibinfo {editor}
  {edited by\ \bibinfo {editor} {\bibfnamefont {P.}~\bibnamefont {{Chen}}}},\
  pp.\ \bibinfo {pages} {232--241}\BibitemShut {NoStop}%
\bibitem [{\citenamefont {Qiang}\ and\ \citenamefont
  {Wu}(2011{\natexlab{a}})}]{MC4}%
  \BibitemOpen
  \bibfield  {author} {\bibinfo {author} {\bibnamefont {Qiang}, \bibfnamefont
  {J.}}, \ and\ \bibinfo {author} {\bibfnamefont {J.}~\bibnamefont {Wu}}}
  (\bibinfo {year} {2011}{\natexlab{a}}),\ \href@noop {} {\bibfield  {journal}
  {\bibinfo  {journal} {Nucl. Instrum. Methods Phys. Res. A}\ }\textbf
  {\bibinfo {volume} {640}},\ \bibinfo {pages} {228}}\BibitemShut {NoStop}%
\bibitem [{\citenamefont {Qiang}\ and\ \citenamefont
  {Wu}(2011{\natexlab{b}})}]{CHGQiang1}%
  \BibitemOpen
  \bibfield  {author} {\bibinfo {author} {\bibnamefont {Qiang}, \bibfnamefont
  {J.}}, \ and\ \bibinfo {author} {\bibfnamefont {J.}~\bibnamefont {Wu}}}
  (\bibinfo {year} {2011}{\natexlab{b}}),\ \href@noop {} {\bibfield  {journal}
  {\bibinfo  {journal} {Appl. Phys. Lett.}\ }\textbf {\bibinfo {volume} {99}},\
  \bibinfo {pages} {081101}}\BibitemShut {NoStop}%
\bibitem [{\citenamefont {Ratner}\ and\ \citenamefont
  {Chao}(2011)}]{FEL11ratner_chao}%
  \BibitemOpen
  \bibfield  {author} {\bibinfo {author} {\bibnamefont {Ratner}, \bibfnamefont
  {D.}}, \ and\ \bibinfo {author} {\bibfnamefont {A.}~\bibnamefont {Chao}}}
  (\bibinfo {year} {2011}),\ in\ \href@noop {} {\emph {\bibinfo {booktitle}
  {Proceedings of the 2011 FEL Conference}}}\ (\bibinfo {address} {Shanghai,
  China})\ p.~\bibinfo {pages} {53}\BibitemShut {NoStop}%
\bibitem [{\citenamefont {Ratner}\ \emph
  {et~al.}(2011{\natexlab{a}})\citenamefont {Ratner}, \citenamefont {Chao},\
  and\ \citenamefont {Huang}}]{MC3}%
  \BibitemOpen
  \bibfield  {author} {\bibinfo {author} {\bibnamefont {Ratner}, \bibfnamefont
  {D.}}, \bibinfo {author} {\bibfnamefont {A.}~\bibnamefont {Chao}}, \ and\
  \bibinfo {author} {\bibfnamefont {Z.}~\bibnamefont {Huang}}} (\bibinfo {year}
  {2011}{\natexlab{a}}),\ \href
  {http://link.aps.org/doi/10.1103/PhysRevSTAB.14.020701} {\bibfield  {journal}
  {\bibinfo  {journal} {Phys. Rev. ST Accel. Beams}\ }\textbf {\bibinfo
  {volume} {14}},\ \bibinfo {pages} {020701}}\BibitemShut {NoStop}%
\bibitem [{\citenamefont {Ratner}\ \emph {et~al.}(2012)\citenamefont {Ratner},
  \citenamefont {Fry}, \citenamefont {Stupakov},\ and\ \citenamefont
  {White}}]{ratner_et_al_12}%
  \BibitemOpen
  \bibfield  {author} {\bibinfo {author} {\bibnamefont {Ratner}, \bibfnamefont
  {D.}}, \bibinfo {author} {\bibfnamefont {A.}~\bibnamefont {Fry}}, \bibinfo
  {author} {\bibfnamefont {G.}~\bibnamefont {Stupakov}}, \ and\ \bibinfo
  {author} {\bibfnamefont {W.}~\bibnamefont {White}}} (\bibinfo {year}
  {2012}),\ \href@noop {} {\bibfield  {journal} {\bibinfo  {journal} {Phys.
  Rev. ST Accel. Beams}\ }\textbf {\bibinfo {volume} {15}},\ \bibinfo {pages}
  {030702}}\BibitemShut {NoStop}%
\bibitem [{\citenamefont {Ratner}\ \emph
  {et~al.}(2011{\natexlab{b}})\citenamefont {Ratner}, \citenamefont {Huang},\
  and\ \citenamefont {Stupakov}}]{ratner11hs}%
  \BibitemOpen
  \bibfield  {author} {\bibinfo {author} {\bibnamefont {Ratner}, \bibfnamefont
  {D.}}, \bibinfo {author} {\bibfnamefont {Z.}~\bibnamefont {Huang}}, \ and\
  \bibinfo {author} {\bibfnamefont {G.}~\bibnamefont {Stupakov}}} (\bibinfo
  {year} {2011}{\natexlab{b}}),\ \href@noop {} {\bibfield  {journal} {\bibinfo
  {journal} {Phys. Rev. ST Accel. Beams}\ }\textbf {\bibinfo {volume} {14}},\
  \bibinfo {pages} {060710}}\BibitemShut {NoStop}%
\bibitem [{\citenamefont {Ratner}\ and\ \citenamefont
  {Stupakov}(2012)}]{ratner_stupakov_12}%
  \BibitemOpen
  \bibfield  {author} {\bibinfo {author} {\bibnamefont {Ratner}, \bibfnamefont
  {D.}}, \ and\ \bibinfo {author} {\bibfnamefont {G.}~\bibnamefont {Stupakov}}}
  (\bibinfo {year} {2012}),\ \href@noop {} {\bibfield  {journal} {\bibinfo
  {journal} {Phys. Rev. Lett.}\ }\textbf {\bibinfo {volume} {109}},\ \bibinfo
  {pages} {034801}}\BibitemShut {NoStop}%
\bibitem [{\citenamefont {Ratner}\ and\ \citenamefont
  {Chao}(2010)}]{PhysRevLett.105.154801}%
  \BibitemOpen
  \bibfield  {author} {\bibinfo {author} {\bibnamefont {Ratner}, \bibfnamefont
  {D.~F.}}, \ and\ \bibinfo {author} {\bibfnamefont {A.~W.}\ \bibnamefont
  {Chao}}} (\bibinfo {year} {2010}),\ \href
  {http://link.aps.org/doi/10.1103/PhysRevLett.105.154801} {\bibfield
  {journal} {\bibinfo  {journal} {Phys. Rev. Lett.}\ }\textbf {\bibinfo
  {volume} {105}},\ \bibinfo {pages} {154801}}\BibitemShut {NoStop}%
\bibitem [{\citenamefont {Reiche}\ \emph {et~al.}(2008)\citenamefont {Reiche},
  \citenamefont {Musumeci}, \citenamefont {Pellegrini},\ and\ \citenamefont
  {Rosenzweig}}]{Reiche200845}%
  \BibitemOpen
  \bibfield  {author} {\bibinfo {author} {\bibnamefont {Reiche}, \bibfnamefont
  {S.}}, \bibinfo {author} {\bibfnamefont {P.}~\bibnamefont {Musumeci}},
  \bibinfo {author} {\bibfnamefont {C.}~\bibnamefont {Pellegrini}}, \ and\
  \bibinfo {author} {\bibfnamefont {J.}~\bibnamefont {Rosenzweig}}} (\bibinfo
  {year} {2008}),\ \href
  {http://www.sciencedirect.com/science/article/pii/S0168900208006207}
  {\bibfield  {journal} {\bibinfo  {journal} {Nucl. Instrum. Methods Phys. Res.
  A}\ }\textbf {\bibinfo {volume} {593}}~(\bibinfo {number} {1-2}),\ \bibinfo
  {pages} {45}}\BibitemShut {NoStop}%
\bibitem [{\citenamefont {Rosenzweig}\ \emph {et~al.}(2008)\citenamefont
  {Rosenzweig}, \citenamefont {Alesini}, \citenamefont {Andonian},
  \citenamefont {Boscolo}, \citenamefont {Dunning}, \citenamefont {Faillace},
  \citenamefont {Ferrario}, \citenamefont {Fukusawa}, \citenamefont
  {Giannessi}, \citenamefont {Hemsing}, \citenamefont {Marcus}, \citenamefont
  {Marinelli}, \citenamefont {Musumeci}, \citenamefont {O'Shea}, \citenamefont
  {Palumbo}, \citenamefont {Pellegrini}, \citenamefont {Petrillo},
  \citenamefont {Reiche}, \citenamefont {Ronsivalle}, \citenamefont {Spataro},\
  and\ \citenamefont {Vaccarezza}}]{Rosenzweig200839}%
  \BibitemOpen
  \bibfield  {author} {\bibinfo {author} {\bibnamefont {Rosenzweig},
  \bibfnamefont {J.}}, \bibinfo {author} {\bibfnamefont {D.}~\bibnamefont
  {Alesini}}, \bibinfo {author} {\bibfnamefont {G.}~\bibnamefont {Andonian}},
  \bibinfo {author} {\bibfnamefont {M.}~\bibnamefont {Boscolo}}, \bibinfo
  {author} {\bibfnamefont {M.}~\bibnamefont {Dunning}}, \bibinfo {author}
  {\bibfnamefont {L.}~\bibnamefont {Faillace}}, \bibinfo {author}
  {\bibfnamefont {M.}~\bibnamefont {Ferrario}}, \bibinfo {author}
  {\bibfnamefont {A.}~\bibnamefont {Fukusawa}}, \bibinfo {author}
  {\bibfnamefont {L.}~\bibnamefont {Giannessi}}, \bibinfo {author}
  {\bibfnamefont {E.}~\bibnamefont {Hemsing}}, \bibinfo {author} {\bibfnamefont
  {G.}~\bibnamefont {Marcus}}, \bibinfo {author} {\bibfnamefont
  {A.}~\bibnamefont {Marinelli}}, \bibinfo {author} {\bibfnamefont
  {P.}~\bibnamefont {Musumeci}}, \bibinfo {author} {\bibfnamefont
  {B.}~\bibnamefont {O'Shea}}, \bibinfo {author} {\bibfnamefont
  {L.}~\bibnamefont {Palumbo}}, \bibinfo {author} {\bibfnamefont
  {C.}~\bibnamefont {Pellegrini}}, \bibinfo {author} {\bibfnamefont
  {V.}~\bibnamefont {Petrillo}}, \bibinfo {author} {\bibfnamefont
  {S.}~\bibnamefont {Reiche}}, \bibinfo {author} {\bibfnamefont
  {C.}~\bibnamefont {Ronsivalle}}, \bibinfo {author} {\bibfnamefont
  {B.}~\bibnamefont {Spataro}}, \ and\ \bibinfo {author} {\bibfnamefont
  {C.}~\bibnamefont {Vaccarezza}}} (\bibinfo {year} {2008}),\ \href
  {http://www.sciencedirect.com/science/article/pii/S0168900208006190}
  {\bibfield  {journal} {\bibinfo  {journal} {Nucl. Instrum. Methods Phys. Res.
  A}\ }\textbf {\bibinfo {volume} {593}}~(\bibinfo {number} {1-2}),\ \bibinfo
  {pages} {39}}\BibitemShut {NoStop}%
\bibitem [{\citenamefont {Rousse}\ \emph {et~al.}(2001)\citenamefont {Rousse},
  \citenamefont {Rischel},\ and\ \citenamefont {Gauthier}}]{LSapplication1}%
  \BibitemOpen
  \bibfield  {author} {\bibinfo {author} {\bibnamefont {Rousse}, \bibfnamefont
  {A.}}, \bibinfo {author} {\bibfnamefont {C.}~\bibnamefont {Rischel}}, \ and\
  \bibinfo {author} {\bibfnamefont {J.-C.}\ \bibnamefont {Gauthier}}} (\bibinfo
  {year} {2001}),\ \href@noop {} {\bibfield  {journal} {\bibinfo  {journal}
  {Rev. Mod. Phys.}\ }\textbf {\bibinfo {volume} {73}},\ \bibinfo {pages}
  {17}}\BibitemShut {NoStop}%
\bibitem [{\citenamefont {Saldin}\ \emph
  {et~al.}(2006{\natexlab{a}})\citenamefont {Saldin}, \citenamefont
  {E.A.Schneidmiller},\ and\ \citenamefont {M.V.Yurkov}}]{Sald3}%
  \BibitemOpen
  \bibfield  {author} {\bibinfo {author} {\bibnamefont {Saldin}, \bibfnamefont
  {E.}}, \bibinfo {author} {\bibnamefont {E.A.Schneidmiller}}, \ and\ \bibinfo
  {author} {\bibnamefont {M.V.Yurkov}}} (\bibinfo {year}
  {2006}{\natexlab{a}}),\ \href@noop {} {\bibfield  {journal} {\bibinfo
  {journal} {Phys. Rev. Spec. Topics Accel. and Beams}\ }\textbf {\bibinfo
  {volume} {9}},\ \bibinfo {pages} {050702}}\BibitemShut {NoStop}%
\bibitem [{\citenamefont {Saldin}\ \emph {et~al.}(1998)\citenamefont {Saldin},
  \citenamefont {Schneidmiller},\ and\ \citenamefont
  {Yurkov}}]{Saldinstatistical}%
  \BibitemOpen
  \bibfield  {author} {\bibinfo {author} {\bibnamefont {Saldin}, \bibfnamefont
  {E.}}, \bibinfo {author} {\bibfnamefont {E.}~\bibnamefont {Schneidmiller}}, \
  and\ \bibinfo {author} {\bibfnamefont {M.}~\bibnamefont {Yurkov}}} (\bibinfo
  {year} {1998}),\ \href@noop {} {\bibfield  {journal} {\bibinfo  {journal}
  {Optics Communications}\ }\textbf {\bibinfo {volume} {148}},\ \bibinfo
  {pages} {383}}\BibitemShut {NoStop}%
\bibitem [{\citenamefont {Saldin}\ \emph {et~al.}(2004)\citenamefont {Saldin},
  \citenamefont {Schneidmiller},\ and\ \citenamefont
  {Yurkov}}]{Saldin-laser-heater}%
  \BibitemOpen
  \bibfield  {author} {\bibinfo {author} {\bibnamefont {Saldin}, \bibfnamefont
  {E.}}, \bibinfo {author} {\bibfnamefont {E.}~\bibnamefont {Schneidmiller}}, \
  and\ \bibinfo {author} {\bibfnamefont {M.}~\bibnamefont {Yurkov}}} (\bibinfo
  {year} {2004}),\ \href@noop {} {\bibfield  {journal} {\bibinfo  {journal}
  {Nucl. Instrum. Methods Phys. Res. A}\ }\textbf {\bibinfo {volume}
  {528}}~(\bibinfo {number} {1-2}),\ \bibinfo {pages} {355 }}\BibitemShut
  {NoStop}%
\bibitem [{\citenamefont {Saldin}\ \emph {et~al.}(2005)\citenamefont {Saldin},
  \citenamefont {Schneidmiller},\ and\ \citenamefont {Yurkov}}]{ORS}%
  \BibitemOpen
  \bibfield  {author} {\bibinfo {author} {\bibnamefont {Saldin}, \bibfnamefont
  {E.}}, \bibinfo {author} {\bibfnamefont {E.}~\bibnamefont {Schneidmiller}}, \
  and\ \bibinfo {author} {\bibfnamefont {M.}~\bibnamefont {Yurkov}}} (\bibinfo
  {year} {2005}),\ \href@noop {} {\bibfield  {journal} {\bibinfo  {journal}
  {Nucl. Instrum. Methods Phys. Res. A}\ }\textbf {\bibinfo {volume} {539}},\
  \bibinfo {pages} {499}}\BibitemShut {NoStop}%
\bibitem [{\citenamefont {Saldin}\ \emph {et~al.}(2000)\citenamefont {Saldin},
  \citenamefont {Schneidmiller},\ and\ \citenamefont {Yurkov}}]{SaldinBook}%
  \BibitemOpen
  \bibfield  {author} {\bibinfo {author} {\bibnamefont {Saldin}, \bibfnamefont
  {E.~L.}}, \bibinfo {author} {\bibfnamefont {E.~A.}\ \bibnamefont
  {Schneidmiller}}, \ and\ \bibinfo {author} {\bibfnamefont {M.~V.}\
  \bibnamefont {Yurkov}}} (\bibinfo {year} {2000}),\ \href@noop {} {\emph
  {\bibinfo {title} {The physics of free electron lasers}}}\ (\bibinfo
  {publisher} {Springer},\ \bibinfo {address} {Berlin})\BibitemShut {NoStop}%
\bibitem [{\citenamefont {Saldin}\ \emph
  {et~al.}(2002{\natexlab{a}})\citenamefont {Saldin}, \citenamefont
  {Schneidmiller},\ and\ \citenamefont {Yurkov}}]{saldin02-1sy}%
  \BibitemOpen
  \bibfield  {author} {\bibinfo {author} {\bibnamefont {Saldin}, \bibfnamefont
  {E.~L.}}, \bibinfo {author} {\bibfnamefont {E.~A.}\ \bibnamefont
  {Schneidmiller}}, \ and\ \bibinfo {author} {\bibfnamefont {M.~V.}\
  \bibnamefont {Yurkov}}} (\bibinfo {year} {2002}{\natexlab{a}}),\ \href@noop
  {} {\bibfield  {journal} {\bibinfo  {journal} {Nucl. Instrum. Methods Phys.
  Res. A}\ }\textbf {\bibinfo {volume} {490}},\ \bibinfo {pages}
  {1}}\BibitemShut {NoStop}%
\bibitem [{\citenamefont {Saldin}\ \emph
  {et~al.}(2002{\natexlab{b}})\citenamefont {Saldin}, \citenamefont
  {Schneidmiller},\ and\ \citenamefont {Yurkov}}]{saldin02sy}%
  \BibitemOpen
  \bibfield  {author} {\bibinfo {author} {\bibnamefont {Saldin}, \bibfnamefont
  {E.~L.}}, \bibinfo {author} {\bibfnamefont {E.~A.}\ \bibnamefont
  {Schneidmiller}}, \ and\ \bibinfo {author} {\bibfnamefont {M.~V.}\
  \bibnamefont {Yurkov}}} (\bibinfo {year} {2002}{\natexlab{b}}),\ \href@noop
  {} {\bibfield  {journal} {\bibinfo  {journal} {Opt. Commun.}\ }\textbf
  {\bibinfo {volume} {202}},\ \bibinfo {pages} {169}}\BibitemShut {NoStop}%
\bibitem [{\citenamefont {Saldin}\ \emph
  {et~al.}(2006{\natexlab{b}})\citenamefont {Saldin}, \citenamefont
  {Schneidmiller},\ and\ \citenamefont {Yurkov}}]{attosecond2}%
  \BibitemOpen
  \bibfield  {author} {\bibinfo {author} {\bibnamefont {Saldin}, \bibfnamefont
  {E.~L.}}, \bibinfo {author} {\bibfnamefont {E.~A.}\ \bibnamefont
  {Schneidmiller}}, \ and\ \bibinfo {author} {\bibfnamefont {M.~V.}\
  \bibnamefont {Yurkov}}} (\bibinfo {year} {2006}{\natexlab{b}}),\ \href@noop
  {} {\bibfield  {journal} {\bibinfo  {journal} {Phys. Rev. ST Accel. Beams}\
  }\textbf {\bibinfo {volume} {9}},\ \bibinfo {pages} {050702}}\BibitemShut
  {NoStop}%
\bibitem [{\citenamefont {Saldin}\ \emph {et~al.}(2010)\citenamefont {Saldin},
  \citenamefont {Schneidmiller},\ and\ \citenamefont
  {Yurkov}}]{1367-2630-12-3-035010}%
  \BibitemOpen
  \bibfield  {author} {\bibinfo {author} {\bibnamefont {Saldin}, \bibfnamefont
  {E.~L.}}, \bibinfo {author} {\bibfnamefont {E.~A.}\ \bibnamefont
  {Schneidmiller}}, \ and\ \bibinfo {author} {\bibfnamefont {M.~V.}\
  \bibnamefont {Yurkov}}} (\bibinfo {year} {2010}),\ \href
  {http://stacks.iop.org/1367-2630/12/i=3/a=035010} {\bibfield  {journal}
  {\bibinfo  {journal} {New Journal of Physics}\ }\textbf {\bibinfo {volume}
  {12}}~(\bibinfo {number} {3}),\ \bibinfo {pages} {035010}}\BibitemShut
  {NoStop}%
\bibitem [{\citenamefont {Sal\'{e}n}\ \emph {et~al.}(2011)\citenamefont
  {Sal\'{e}n}, \citenamefont {Hamberg}, \citenamefont {Van~der Meulen},
  \citenamefont {Larsson}, \citenamefont {Angelova-Hamberg}, \citenamefont
  {Ziemann}, \citenamefont {Schlarb}, \citenamefont {L\"{o}hl}, \citenamefont
  {Saldin}, \citenamefont {Schneidmiller}, \citenamefont {Yurkov},
  \citenamefont {B\"{o}dewadt}, \citenamefont {Winter}, \citenamefont {Khan},\
  and\ \citenamefont {Meseck}}]{ORS1}%
  \BibitemOpen
  \bibfield  {author} {\bibinfo {author} {\bibnamefont {Sal\'{e}n},
  \bibfnamefont {P.}}, \bibinfo {author} {\bibfnamefont {M.}~\bibnamefont
  {Hamberg}}, \bibinfo {author} {\bibfnamefont {P.}~\bibnamefont {Van~der
  Meulen}}, \bibinfo {author} {\bibfnamefont {M.}~\bibnamefont {Larsson}},
  \bibinfo {author} {\bibfnamefont {G.}~\bibnamefont {Angelova-Hamberg}},
  \bibinfo {author} {\bibfnamefont {V.}~\bibnamefont {Ziemann}}, \bibinfo
  {author} {\bibfnamefont {H.}~\bibnamefont {Schlarb}}, \bibinfo {author}
  {\bibfnamefont {F.}~\bibnamefont {L\"{o}hl}}, \bibinfo {author}
  {\bibfnamefont {E.}~\bibnamefont {Saldin}}, \bibinfo {author} {\bibfnamefont
  {E.}~\bibnamefont {Schneidmiller}}, \bibinfo {author} {\bibfnamefont
  {M.}~\bibnamefont {Yurkov}}, \bibinfo {author} {\bibfnamefont
  {J.}~\bibnamefont {B\"{o}dewadt}}, \bibinfo {author} {\bibfnamefont
  {A.}~\bibnamefont {Winter}}, \bibinfo {author} {\bibfnamefont
  {S.}~\bibnamefont {Khan}}, \ and\ \bibinfo {author} {\bibfnamefont
  {A.}~\bibnamefont {Meseck}}} (\bibinfo {year} {2011}),\ in\ \href@noop {}
  {\emph {\bibinfo {booktitle} {2011 Free-electron laser Conference}}},\ p.\
  \bibinfo {pages} {366}\BibitemShut {NoStop}%
\bibitem [{\citenamefont {Sands}(1970)}]{Sands}%
  \BibitemOpen
  \bibfield  {author} {\bibinfo {author} {\bibnamefont {Sands}, \bibfnamefont
  {M.}}} (\bibinfo {year} {1970}),\ \href@noop {} {\bibinfo  {journal} {SLAC-
  121, UC-28}\ }\BibitemShut {NoStop}%
\bibitem [{\citenamefont {Sasaki}\ and\ \citenamefont
  {McNulty}(2008)}]{Sasaki}%
  \BibitemOpen
\bibfield  {journal} {  }\bibfield  {author} {\bibinfo {author} {\bibnamefont
  {Sasaki}, \bibfnamefont {S.}}, \ and\ \bibinfo {author} {\bibfnamefont
  {I.}~\bibnamefont {McNulty}}} (\bibinfo {year} {2008}),\ \href@noop {}
  {\bibfield  {journal} {\bibinfo  {journal} {Phys. Rev. Lett.}\ }\textbf
  {\bibinfo {volume} {100}}~(\bibinfo {number} {12}),\ \bibinfo {eid}
  {124801}}\BibitemShut {NoStop}%
\bibitem [{\citenamefont {Sch\"achter}(1995)}]{PASERtheoryPRA1995}%
  \BibitemOpen
  \bibfield  {author} {\bibinfo {author} {\bibnamefont {Sch\"achter},
  \bibfnamefont {L.}}} (\bibinfo {year} {1995}),\ \href
  {http://www.sciencedirect.com/science/article/pii/037596019500619E}
  {\bibfield  {journal} {\bibinfo  {journal} {Physics Letters A}\ }\textbf
  {\bibinfo {volume} {205}}~(\bibinfo {number} {5 - 6}),\ \bibinfo {pages} {355
  }}\BibitemShut {NoStop}%
\bibitem [{\citenamefont {Schlott}\ \emph {et~al.}(2006)\citenamefont
  {Schlott}, \citenamefont {Abramsohn}, \citenamefont {Ingold}, \citenamefont
  {Lerch},\ and\ \citenamefont {Beaud}}]{Schl}%
  \BibitemOpen
  \bibfield  {author} {\bibinfo {author} {\bibnamefont {Schlott}, \bibfnamefont
  {V.}}, \bibinfo {author} {\bibfnamefont {D.}~\bibnamefont {Abramsohn}},
  \bibinfo {author} {\bibfnamefont {G.}~\bibnamefont {Ingold}}, \bibinfo
  {author} {\bibfnamefont {P.}~\bibnamefont {Lerch}}, \ and\ \bibinfo {author}
  {\bibfnamefont {P.}~\bibnamefont {Beaud}}} (\bibinfo {year} {2006}),\ in\
  \href@noop {} {\emph {\bibinfo {booktitle} {Proc. of the 2004 European
  Particle Accelerator Conference, Edinburg, Scotland}}},\ p.\ \bibinfo {pages}
  {1229}\BibitemShut {NoStop}%
\bibitem [{\citenamefont {Schm\"{u}ser}\ \emph {et~al.}(2008)\citenamefont
  {Schm\"{u}ser}, \citenamefont {Dohlus},\ and\ \citenamefont
  {Rossbach}}]{Schmuser}%
  \BibitemOpen
  \bibfield  {author} {\bibinfo {author} {\bibnamefont {Schm\"{u}ser},
  \bibfnamefont {P.}}, \bibinfo {author} {\bibfnamefont {M.}~\bibnamefont
  {Dohlus}}, \ and\ \bibinfo {author} {\bibfnamefont {J.}~\bibnamefont
  {Rossbach}}} (\bibinfo {year} {2008}),\ \href@noop {} {\emph {\bibinfo
  {title} {Ultraviolet and soft x-ray free-electron lasers}}}\ (\bibinfo
  {publisher} {Springer})\BibitemShut {NoStop}%
\bibitem [{\citenamefont {Schneider}(2010)}]{Schneider}%
  \BibitemOpen
  \bibfield  {author} {\bibinfo {author} {\bibnamefont {Schneider},
  \bibfnamefont {J.~R.}}} (\bibinfo {year} {2010}),\ \href@noop {} {\bibfield
  {journal} {\bibinfo  {journal} {Reviews of Accelerator Science and
  Technology}\ }\textbf {\bibinfo {volume} {3}},\ \bibinfo {pages}
  {13}}\BibitemShut {NoStop}%
\bibitem [{\citenamefont {Schneidmiller}\ and\ \citenamefont
  {Yurkov}(2012)}]{HLSchneidmiller}%
  \BibitemOpen
  \bibfield  {author} {\bibinfo {author} {\bibnamefont {Schneidmiller},
  \bibfnamefont {E.~A.}}, \ and\ \bibinfo {author} {\bibfnamefont {M.~V.}\
  \bibnamefont {Yurkov}}} (\bibinfo {year} {2012}),\ \href
  {http://link.aps.org/doi/10.1103/PhysRevSTAB.15.080702} {\bibfield  {journal}
  {\bibinfo  {journal} {Phys. Rev. ST Accel. Beams}\ }\textbf {\bibinfo
  {volume} {15}},\ \bibinfo {pages} {080702}}\BibitemShut {NoStop}%
\bibitem [{\citenamefont {Schoenlein}\ \emph
  {et~al.}(2000{\natexlab{a}})\citenamefont {Schoenlein}, \citenamefont
  {Chattopadhyay}, \citenamefont {Chong}, \citenamefont {Glover}, \citenamefont
  {Heimann}, \citenamefont {C.V.~Shank},\ and\ \citenamefont
  {Zolotorev}}]{Scho1}%
  \BibitemOpen
  \bibfield  {author} {\bibinfo {author} {\bibnamefont {Schoenlein},
  \bibfnamefont {R.}}, \bibinfo {author} {\bibfnamefont {S.}~\bibnamefont
  {Chattopadhyay}}, \bibinfo {author} {\bibfnamefont {H.~W.}\ \bibnamefont
  {Chong}}, \bibinfo {author} {\bibfnamefont {T.}~\bibnamefont {Glover}},
  \bibinfo {author} {\bibfnamefont {P.}~\bibnamefont {Heimann}}, \bibinfo
  {author} {\bibfnamefont {A.~Z.}\ \bibnamefont {C.V.~Shank}}, \ and\ \bibinfo
  {author} {\bibfnamefont {M.}~\bibnamefont {Zolotorev}}} (\bibinfo {year}
  {2000}{\natexlab{a}}),\ \href@noop {} {\bibfield  {journal} {\bibinfo
  {journal} {Appl. Phys. B}\ }\textbf {\bibinfo {volume} {71}},\ \bibinfo
  {pages} {1}}\BibitemShut {NoStop}%
\bibitem [{\citenamefont {Schoenlein}\ \emph
  {et~al.}(2000{\natexlab{b}})\citenamefont {Schoenlein}, \citenamefont
  {Chattopadhyay}, \citenamefont {Chong}, \citenamefont {Glover}, \citenamefont
  {Heimann}, \citenamefont {Shank}, \citenamefont {Zholents},\ and\
  \citenamefont {Zolotorev}}]{laser-slicingLBNL}%
  \BibitemOpen
  \bibfield  {author} {\bibinfo {author} {\bibnamefont {Schoenlein},
  \bibfnamefont {R.~W.}}, \bibinfo {author} {\bibfnamefont {S.}~\bibnamefont
  {Chattopadhyay}}, \bibinfo {author} {\bibfnamefont {H.~H.~W.}\ \bibnamefont
  {Chong}}, \bibinfo {author} {\bibfnamefont {T.~E.}\ \bibnamefont {Glover}},
  \bibinfo {author} {\bibfnamefont {P.~A.}\ \bibnamefont {Heimann}}, \bibinfo
  {author} {\bibfnamefont {C.~V.}\ \bibnamefont {Shank}}, \bibinfo {author}
  {\bibfnamefont {A.~A.}\ \bibnamefont {Zholents}}, \ and\ \bibinfo {author}
  {\bibfnamefont {M.~S.}\ \bibnamefont {Zolotorev}}} (\bibinfo {year}
  {2000}{\natexlab{b}}),\ \href@noop {} {\bibfield  {journal} {\bibinfo
  {journal} {Science}\ }\textbf {\bibinfo {volume} {287}},\ \bibinfo {pages}
  {2237}}\BibitemShut {NoStop}%
\bibitem [{\citenamefont {Schroeder}\ \emph {et~al.}(2004)\citenamefont
  {Schroeder}, \citenamefont {Esarey},\ and\ \citenamefont
  {Leemans}}]{schroeder_conditioner}%
  \BibitemOpen
  \bibfield  {author} {\bibinfo {author} {\bibnamefont {Schroeder},
  \bibfnamefont {C.~B.}}, \bibinfo {author} {\bibfnamefont {E.}~\bibnamefont
  {Esarey}}, \ and\ \bibinfo {author} {\bibfnamefont {W.~P.}\ \bibnamefont
  {Leemans}}} (\bibinfo {year} {2004}),\ \href
  {http://link.aps.org/doi/10.1103/PhysRevLett.93.194801} {\bibfield  {journal}
  {\bibinfo  {journal} {Phys. Rev. Lett.}\ }\textbf {\bibinfo {volume} {93}},\
  \bibinfo {pages} {194801}}\BibitemShut {NoStop}%
\bibitem [{\citenamefont {Schweigert}\ and\ \citenamefont
  {Mukamel}(2007)}]{Schweigert}%
  \BibitemOpen
  \bibfield  {author} {\bibinfo {author} {\bibnamefont {Schweigert},
  \bibfnamefont {I.}}, \ and\ \bibinfo {author} {\bibfnamefont
  {S.}~\bibnamefont {Mukamel}}} (\bibinfo {year} {2007}),\ \href@noop {}
  {\bibfield  {journal} {\bibinfo  {journal} {Phys. Rev. A}\ }\textbf {\bibinfo
  {volume} {76}},\ \bibinfo {pages} {012504}}\BibitemShut {NoStop}%
\bibitem [{\citenamefont {Sears}\ \emph {et~al.}(2008)\citenamefont {Sears},
  \citenamefont {Colby}, \citenamefont {England}, \citenamefont {Ischebeck},
  \citenamefont {McGuinness}, \citenamefont {Nelson}, \citenamefont {Noble},
  \citenamefont {Siemann}, \citenamefont {Spencer}, \citenamefont {Walz},
  \citenamefont {Plettner},\ and\ \citenamefont
  {Byer}}]{PhysRevSTAB.11.101301}%
  \BibitemOpen
  \bibfield  {author} {\bibinfo {author} {\bibnamefont {Sears}, \bibfnamefont
  {C.~M.~S.}}, \bibinfo {author} {\bibfnamefont {E.}~\bibnamefont {Colby}},
  \bibinfo {author} {\bibfnamefont {R.~J.}\ \bibnamefont {England}}, \bibinfo
  {author} {\bibfnamefont {R.}~\bibnamefont {Ischebeck}}, \bibinfo {author}
  {\bibfnamefont {C.}~\bibnamefont {McGuinness}}, \bibinfo {author}
  {\bibfnamefont {J.}~\bibnamefont {Nelson}}, \bibinfo {author} {\bibfnamefont
  {R.}~\bibnamefont {Noble}}, \bibinfo {author} {\bibfnamefont {R.~H.}\
  \bibnamefont {Siemann}}, \bibinfo {author} {\bibfnamefont {J.}~\bibnamefont
  {Spencer}}, \bibinfo {author} {\bibfnamefont {D.}~\bibnamefont {Walz}},
  \bibinfo {author} {\bibfnamefont {T.}~\bibnamefont {Plettner}}, \ and\
  \bibinfo {author} {\bibfnamefont {R.~L.}\ \bibnamefont {Byer}}} (\bibinfo
  {year} {2008}),\ \href
  {http://link.aps.org/doi/10.1103/PhysRevSTAB.11.101301} {\bibfield  {journal}
  {\bibinfo  {journal} {Phys. Rev. ST Accel. Beams}\ }\textbf {\bibinfo
  {volume} {11}},\ \bibinfo {pages} {101301}}\BibitemShut {NoStop}%
\bibitem [{\citenamefont {Sessler}\ \emph {et~al.}(1992)\citenamefont
  {Sessler}, \citenamefont {Whittum},\ and\ \citenamefont {Yu}}]{sessler92wy}%
  \BibitemOpen
  \bibfield  {author} {\bibinfo {author} {\bibnamefont {Sessler}, \bibfnamefont
  {A.~M.}}, \bibinfo {author} {\bibfnamefont {D.~H.}\ \bibnamefont {Whittum}},
  \ and\ \bibinfo {author} {\bibfnamefont {L.-H.}\ \bibnamefont {Yu}}}
  (\bibinfo {year} {1992}),\ \href@noop {} {\bibfield  {journal} {\bibinfo
  {journal} {Phys. Rev. Lett.}\ }\textbf {\bibinfo {volume} {68}},\ \bibinfo
  {pages} {309}}\BibitemShut {NoStop}%
\bibitem [{\citenamefont {Shen}\ \emph {et~al.}(2011)\citenamefont {Shen},
  \citenamefont {Yang}, \citenamefont {Carr}, \citenamefont {Hidaka},
  \citenamefont {Murphy},\ and\ \citenamefont {Wang}}]{Shen}%
  \BibitemOpen
  \bibfield  {author} {\bibinfo {author} {\bibnamefont {Shen}, \bibfnamefont
  {Y.}}, \bibinfo {author} {\bibfnamefont {X.}~\bibnamefont {Yang}}, \bibinfo
  {author} {\bibfnamefont {G.}~\bibnamefont {Carr}}, \bibinfo {author}
  {\bibfnamefont {Y.}~\bibnamefont {Hidaka}}, \bibinfo {author} {\bibfnamefont
  {J.}~\bibnamefont {Murphy}}, \ and\ \bibinfo {author} {\bibfnamefont
  {X.}~\bibnamefont {Wang}}} (\bibinfo {year} {2011}),\ \href@noop {}
  {\bibfield  {journal} {\bibinfo  {journal} {Phys. Rev. Lett.}\ }\textbf
  {\bibinfo {volume} {107}},\ \bibinfo {pages} {204801}}\BibitemShut {NoStop}%
\bibitem [{\citenamefont {Shersby-Harvie}(1948)}]{Shersby}%
  \BibitemOpen
  \bibfield  {author} {\bibinfo {author} {\bibnamefont {Shersby-Harvie},
  \bibfnamefont {R.~B.~R.}}} (\bibinfo {year} {1948}),\ \href@noop {}
  {\bibfield  {journal} {\bibinfo  {journal} {Nature}\ }\textbf {\bibinfo
  {volume} {162}},\ \bibinfo {pages} {890}}\BibitemShut {NoStop}%
\bibitem [{\citenamefont {Shintake}(1992)}]{Shintake}%
  \BibitemOpen
  \bibfield  {author} {\bibinfo {author} {\bibnamefont {Shintake},
  \bibfnamefont {T.}}} (\bibinfo {year} {1992}),\ \href@noop {} {\bibfield
  {journal} {\bibinfo  {journal} {Nucl. Instrum. Methods Phys. Res. A}\
  }\textbf {\bibinfo {volume} {311}},\ \bibinfo {pages} {453}}\BibitemShut
  {NoStop}%
\bibitem [{\citenamefont {Shintake}(2007)}]{MC2}%
  \BibitemOpen
  \bibfield  {author} {\bibinfo {author} {\bibnamefont {Shintake},
  \bibfnamefont {T.}}} (\bibinfo {year} {2007}),\ in\ \href@noop {} {\emph
  {\bibinfo {booktitle} {Proceedings of the 2007 FEL Conference}}}\ (\bibinfo
  {address} {Novosibirsk, Russia})\ p.\ \bibinfo {pages} {378}\BibitemShut
  {NoStop}%
\bibitem [{\citenamefont {Siegman}(1986)}]{siegman}%
  \BibitemOpen
  \bibfield  {author} {\bibinfo {author} {\bibnamefont {Siegman}, \bibfnamefont
  {A.~E.}}} (\bibinfo {year} {1986}),\ \href@noop {} {\emph {\bibinfo {title}
  {Lasers}}}\ (\bibinfo  {publisher} {University Science Books})\BibitemShut
  {NoStop}%
\bibitem [{\citenamefont {Spampinati}\ \emph {et~al.}(2012)\citenamefont
  {Spampinati}, \citenamefont {Allaria}, \citenamefont {Badano}, \citenamefont
  {Bassanese}, \citenamefont {Castronovo}, \citenamefont {Danailov},
  \citenamefont {Demidovich}, \citenamefont {Mitri}, \citenamefont {Diviacco},
  \citenamefont {Fawley}, \citenamefont {Froelish}, \citenamefont {Penco},
  \citenamefont {Spezzani}, \citenamefont {Tr\'{o}vo}, \citenamefont {DeNinno},
  \citenamefont {Ferrari},\ and\ \citenamefont
  {Giannessi}}]{laser_heater_FERMI}%
  \BibitemOpen
  \bibfield  {author} {\bibinfo {author} {\bibnamefont {Spampinati},
  \bibfnamefont {S.}}, \bibinfo {author} {\bibfnamefont {E.}~\bibnamefont
  {Allaria}}, \bibinfo {author} {\bibfnamefont {L.}~\bibnamefont {Badano}},
  \bibinfo {author} {\bibfnamefont {S.}~\bibnamefont {Bassanese}}, \bibinfo
  {author} {\bibfnamefont {D.}~\bibnamefont {Castronovo}}, \bibinfo {author}
  {\bibfnamefont {M.~B.}\ \bibnamefont {Danailov}}, \bibinfo {author}
  {\bibfnamefont {A.}~\bibnamefont {Demidovich}}, \bibinfo {author}
  {\bibfnamefont {S.~D.}\ \bibnamefont {Mitri}}, \bibinfo {author}
  {\bibfnamefont {B.}~\bibnamefont {Diviacco}}, \bibinfo {author}
  {\bibfnamefont {W.~M.}\ \bibnamefont {Fawley}}, \bibinfo {author}
  {\bibfnamefont {L.}~\bibnamefont {Froelish}}, \bibinfo {author}
  {\bibfnamefont {G.}~\bibnamefont {Penco}}, \bibinfo {author} {\bibfnamefont
  {C.}~\bibnamefont {Spezzani}}, \bibinfo {author} {\bibfnamefont
  {M.}~\bibnamefont {Tr\'{o}vo}}, \bibinfo {author} {\bibfnamefont
  {G.}~\bibnamefont {DeNinno}}, \bibinfo {author} {\bibfnamefont
  {E.}~\bibnamefont {Ferrari}}, \ and\ \bibinfo {author} {\bibfnamefont
  {L.}~\bibnamefont {Giannessi}}} (\bibinfo {year} {2012}),\ in\ \href@noop {}
  {\emph {\bibinfo {booktitle} {Proceeding of the FEL Conference 2012}}},\ p.\
  \bibinfo {pages} {177}\BibitemShut {NoStop}%
\bibitem [{\citenamefont {Sprangle}\ \emph {et~al.}(1993)\citenamefont
  {Sprangle}, \citenamefont {Hafizi}, \citenamefont {Joyce},\ and\
  \citenamefont {Serafim}}]{Sprangle19936}%
  \BibitemOpen
  \bibfield  {author} {\bibinfo {author} {\bibnamefont {Sprangle},
  \bibfnamefont {P.}}, \bibinfo {author} {\bibfnamefont {B.}~\bibnamefont
  {Hafizi}}, \bibinfo {author} {\bibfnamefont {G.}~\bibnamefont {Joyce}}, \
  and\ \bibinfo {author} {\bibfnamefont {P.}~\bibnamefont {Serafim}}} (\bibinfo
  {year} {1993}),\ \href
  {http://www.sciencedirect.com/science/article/pii/0168900293900053}
  {\bibfield  {journal} {\bibinfo  {journal} {Nucl. Instrum. Methods Phys. Res.
  A}\ }\textbf {\bibinfo {volume} {331}}~(\bibinfo {number} {1-3}),\ \bibinfo
  {pages} {6}}\BibitemShut {NoStop}%
\bibitem [{\citenamefont {Sprangle}\ \emph {et~al.}(2009)\citenamefont
  {Sprangle}, \citenamefont {Hafizi},\ and\ \citenamefont
  {Pe\~nano}}]{Sprangle}%
  \BibitemOpen
  \bibfield  {author} {\bibinfo {author} {\bibnamefont {Sprangle},
  \bibfnamefont {P.}}, \bibinfo {author} {\bibfnamefont {B.}~\bibnamefont
  {Hafizi}}, \ and\ \bibinfo {author} {\bibfnamefont {J.~R.}\ \bibnamefont
  {Pe\~nano}}} (\bibinfo {year} {2009}),\ \href {\doibase
  10.1103/PhysRevSTAB.12.050702} {\bibfield  {journal} {\bibinfo  {journal}
  {Phys. Rev. ST Accel. Beams}\ }\textbf {\bibinfo {volume} {12}},\ \bibinfo
  {pages} {050702}}\BibitemShut {NoStop}%
\bibitem [{\citenamefont {Stamm}\ \emph {et~al.}(2007)\citenamefont {Stamm},
  \citenamefont {Kachel}, \citenamefont {Pontius}, \citenamefont {Mitzner},
  \citenamefont {Quast}, \citenamefont {Holldack}, \citenamefont {Khan},
  \citenamefont {Lupulescu}, \citenamefont {Aziz}, \citenamefont {Wietstruk},
  \citenamefont {Durr},\ and\ \citenamefont {Eberhardt}}]{LSapplication5}%
  \BibitemOpen
  \bibfield  {author} {\bibinfo {author} {\bibnamefont {Stamm}, \bibfnamefont
  {C.}}, \bibinfo {author} {\bibfnamefont {T.}~\bibnamefont {Kachel}}, \bibinfo
  {author} {\bibfnamefont {N.}~\bibnamefont {Pontius}}, \bibinfo {author}
  {\bibfnamefont {R.}~\bibnamefont {Mitzner}}, \bibinfo {author} {\bibfnamefont
  {T.}~\bibnamefont {Quast}}, \bibinfo {author} {\bibfnamefont
  {K.}~\bibnamefont {Holldack}}, \bibinfo {author} {\bibfnamefont
  {S.}~\bibnamefont {Khan}}, \bibinfo {author} {\bibfnamefont {C.}~\bibnamefont
  {Lupulescu}}, \bibinfo {author} {\bibfnamefont {E.}~\bibnamefont {Aziz}},
  \bibinfo {author} {\bibfnamefont {M.}~\bibnamefont {Wietstruk}}, \bibinfo
  {author} {\bibfnamefont {H.}~\bibnamefont {Durr}}, \ and\ \bibinfo {author}
  {\bibfnamefont {W.}~\bibnamefont {Eberhardt}}} (\bibinfo {year} {2007}),\
  \href@noop {} {\bibfield  {journal} {\bibinfo  {journal} {Nat. Materials}\
  }\textbf {\bibinfo {volume} {6}},\ \bibinfo {pages} {740}}\BibitemShut
  {NoStop}%
\bibitem [{\citenamefont {van Steenbergen}\ \emph {et~al.}(1996)\citenamefont
  {van Steenbergen}, \citenamefont {Gallardo}, \citenamefont {Sandweiss},\ and\
  \citenamefont {Fang}}]{PhysRevLett.77.2690}%
  \BibitemOpen
  \bibfield  {author} {\bibinfo {author} {\bibnamefont {van Steenbergen},
  \bibfnamefont {A.}}, \bibinfo {author} {\bibfnamefont {J.}~\bibnamefont
  {Gallardo}}, \bibinfo {author} {\bibfnamefont {J.}~\bibnamefont {Sandweiss}},
  \ and\ \bibinfo {author} {\bibfnamefont {J.-M.}\ \bibnamefont {Fang}}}
  (\bibinfo {year} {1996}),\ \href
  {http://link.aps.org/doi/10.1103/PhysRevLett.77.2690} {\bibfield  {journal}
  {\bibinfo  {journal} {Phys. Rev. Lett.}\ }\textbf {\bibinfo {volume} {77}},\
  \bibinfo {pages} {2690}}\BibitemShut {NoStop}%
\bibitem [{\citenamefont {Steier}\ \emph {et~al.}(2005)\citenamefont {Steier},
  \citenamefont {Robin}, \citenamefont {Sannibale}, \citenamefont {Schoenlein},
  \citenamefont {Wan}, \citenamefont {Wittmer},\ and\ \citenamefont
  {Zholents}}]{Stei}%
  \BibitemOpen
  \bibfield  {author} {\bibinfo {author} {\bibnamefont {Steier}, \bibfnamefont
  {C.}}, \bibinfo {author} {\bibfnamefont {D.}~\bibnamefont {Robin}}, \bibinfo
  {author} {\bibfnamefont {F.}~\bibnamefont {Sannibale}}, \bibinfo {author}
  {\bibfnamefont {R.}~\bibnamefont {Schoenlein}}, \bibinfo {author}
  {\bibfnamefont {W.}~\bibnamefont {Wan}}, \bibinfo {author} {\bibfnamefont
  {W.}~\bibnamefont {Wittmer}}, \ and\ \bibinfo {author} {\bibfnamefont
  {A.}~\bibnamefont {Zholents}}} (\bibinfo {year} {2005}),\ in\ \href@noop {}
  {\emph {\bibinfo {booktitle} {Proc. of the 2005 Particle Accelerator
  Conference, Knoxville, Tennessee}}},\ p.\ \bibinfo {pages} {4096}\BibitemShut
  {NoStop}%
\bibitem [{\citenamefont {Stenger}\ \emph {et~al.}(2002)\citenamefont
  {Stenger}, \citenamefont {Schnatz}, \citenamefont {Tamm},\ and\ \citenamefont
  {Telle}}]{Stenger_et_al:2002}%
  \BibitemOpen
  \bibfield  {author} {\bibinfo {author} {\bibnamefont {Stenger}, \bibfnamefont
  {J.}}, \bibinfo {author} {\bibfnamefont {H.}~\bibnamefont {Schnatz}},
  \bibinfo {author} {\bibfnamefont {C.}~\bibnamefont {Tamm}}, \ and\ \bibinfo
  {author} {\bibfnamefont {H.~R.}\ \bibnamefont {Telle}}} (\bibinfo {year}
  {2002}),\ \href@noop {} {\bibfield  {journal} {\bibinfo  {journal} {Phys.
  Rev. Lett.}\ }\textbf {\bibinfo {volume} {88}},\ \bibinfo {pages}
  {073601}}\BibitemShut {NoStop}%
\bibitem [{\citenamefont {Strickl}\ and\ \citenamefont {Mourou}(1985)}]{CPA}%
  \BibitemOpen
  \bibfield  {author} {\bibinfo {author} {\bibnamefont {Strickl}, \bibfnamefont
  {D.}}, \ and\ \bibinfo {author} {\bibfnamefont {G.}~\bibnamefont {Mourou}}}
  (\bibinfo {year} {1985}),\ \href@noop {} {\bibfield  {journal} {\bibinfo
  {journal} {Opt. Commun.}\ }\textbf {\bibinfo {volume} {56}},\ \bibinfo
  {pages} {219}}\BibitemShut {NoStop}%
\bibitem [{\citenamefont {Stupakov}(2009)}]{EEHG1}%
  \BibitemOpen
  \bibfield  {author} {\bibinfo {author} {\bibnamefont {Stupakov},
  \bibfnamefont {G.}}} (\bibinfo {year} {2009}),\ \href@noop {} {\bibfield
  {journal} {\bibinfo  {journal} {Phys. Rev. Lett.}\ }\textbf {\bibinfo
  {volume} {102}},\ \bibinfo {pages} {074801}}\BibitemShut {NoStop}%
\bibitem [{\citenamefont {Stupakov}(2011)}]{FEL11stupakov_2}%
  \BibitemOpen
  \bibfield  {author} {\bibinfo {author} {\bibnamefont {Stupakov},
  \bibfnamefont {G.}}} (\bibinfo {year} {2011}),\ in\ \href@noop {} {\emph
  {\bibinfo {booktitle} {Proceedings of the 2011 FEL Conference}}}\ (\bibinfo
  {address} {Shanghai, China})\ p.~\bibinfo {pages} {49}\BibitemShut {NoStop}%
\bibitem [{\citenamefont {Stupakov}(2013)}]{FEL13stupakov_1}%
  \BibitemOpen
  \bibfield  {author} {\bibinfo {author} {\bibnamefont {Stupakov},
  \bibfnamefont {G.}}} (\bibinfo {year} {2013}),\ in\ \href@noop {} {\emph
  {\bibinfo {booktitle} {Proceedings of the 2013 FEL Conference}}}\ (\bibinfo
  {address} {New York, USA})\BibitemShut {NoStop}%
\bibitem [{\citenamefont {Stupakov}\ \emph {et~al.}(2010)\citenamefont
  {Stupakov}, \citenamefont {Huang},\ and\ \citenamefont
  {Ratner}}]{FEL10stupakov_4}%
  \BibitemOpen
  \bibfield  {author} {\bibinfo {author} {\bibnamefont {Stupakov},
  \bibfnamefont {G.}}, \bibinfo {author} {\bibfnamefont {Z.}~\bibnamefont
  {Huang}}, \ and\ \bibinfo {author} {\bibfnamefont {D.}~\bibnamefont
  {Ratner}}} (\bibinfo {year} {2010}),\ in\ \href@noop {} {\emph {\bibinfo
  {booktitle} {Proceedings of the 2010 FEL Conference}}}\ (\bibinfo {address}
  {Malm\"{o} City, Sweden})\ p.\ \bibinfo {pages} {278}\BibitemShut {NoStop}%
\bibitem [{\citenamefont {Stupakov}\ \emph {et~al.}(2013)\citenamefont
  {Stupakov}, \citenamefont {Sessler},\ and\ \citenamefont
  {Zolotorev}}]{COOL13}%
  \BibitemOpen
  \bibfield  {author} {\bibinfo {author} {\bibnamefont {Stupakov},
  \bibfnamefont {G.}}, \bibinfo {author} {\bibfnamefont {A.}~\bibnamefont
  {Sessler}}, \ and\ \bibinfo {author} {\bibfnamefont {M.~S.}\ \bibnamefont
  {Zolotorev}}} (\bibinfo {year} {2013}),\ in\ \href@noop {} {\emph {\bibinfo
  {booktitle} {Proceedings of COOL13 Conference}}}\ (\bibinfo {address}
  {M\"{u}rren, Switzerland})\BibitemShut {NoStop}%
\bibitem [{\citenamefont {Stupakov}\ and\ \citenamefont
  {Zolotorev}(2011)}]{FEL11stupakov_1}%
  \BibitemOpen
  \bibfield  {author} {\bibinfo {author} {\bibnamefont {Stupakov},
  \bibfnamefont {G.}}, \ and\ \bibinfo {author} {\bibfnamefont
  {M.}~\bibnamefont {Zolotorev}}} (\bibinfo {year} {2011}),\ in\ \href@noop {}
  {\emph {\bibinfo {booktitle} {Proceedings of the 2011 FEL Conference}}}\
  (\bibinfo {address} {Shanghai, China})\ p.~\bibinfo {pages} {45}\BibitemShut
  {NoStop}%
\bibitem [{\citenamefont {Sun}\ \emph {et~al.}(2010)\citenamefont {Sun},
  \citenamefont {Piot}, \citenamefont {Johnson}, \citenamefont {Lumpkin},
  \citenamefont {Maxwell}, \citenamefont {Ruan},\ and\ \citenamefont
  {Thurman-Keup}}]{EEX4}%
  \BibitemOpen
  \bibfield  {author} {\bibinfo {author} {\bibnamefont {Sun}, \bibfnamefont
  {Y.-E.}}, \bibinfo {author} {\bibfnamefont {P.}~\bibnamefont {Piot}},
  \bibinfo {author} {\bibfnamefont {A.}~\bibnamefont {Johnson}}, \bibinfo
  {author} {\bibfnamefont {A.~H.}\ \bibnamefont {Lumpkin}}, \bibinfo {author}
  {\bibfnamefont {T.~J.}\ \bibnamefont {Maxwell}}, \bibinfo {author}
  {\bibfnamefont {J.}~\bibnamefont {Ruan}}, \ and\ \bibinfo {author}
  {\bibfnamefont {R.}~\bibnamefont {Thurman-Keup}}} (\bibinfo {year} {2010}),\
  \href {http://link.aps.org/doi/10.1103/PhysRevLett.105.234801} {\bibfield
  {journal} {\bibinfo  {journal} {Phys. Rev. Lett.}\ }\textbf {\bibinfo
  {volume} {105}},\ \bibinfo {pages} {234801}}\BibitemShut {NoStop}%
\bibitem [{\citenamefont {Talman}(1995)}]{Mobius}%
  \BibitemOpen
  \bibfield  {author} {\bibinfo {author} {\bibnamefont {Talman}, \bibfnamefont
  {R.}}} (\bibinfo {year} {1995}),\ \href
  {http://link.aps.org/doi/10.1103/PhysRevLett.74.1590} {\bibfield  {journal}
  {\bibinfo  {journal} {Phys. Rev. Lett.}\ }\textbf {\bibinfo {volume} {74}},\
  \bibinfo {pages} {1590}}\BibitemShut {NoStop}%
\bibitem [{\citenamefont {Tanaka}(2013)}]{5TW}%
  \BibitemOpen
  \bibfield  {author} {\bibinfo {author} {\bibnamefont {Tanaka}, \bibfnamefont
  {T.}}} (\bibinfo {year} {2013}),\ \href@noop {} {\bibfield  {journal}
  {\bibinfo  {journal} {Phys. Rev. Lett.}\ }\textbf {\bibinfo {volume} {110}},\
  \bibinfo {pages} {084801}}\BibitemShut {NoStop}%
\bibitem [{\citenamefont {Tanikawa}\ \emph {et~al.}(2010)\citenamefont
  {Tanikawa}, \citenamefont {Adachi}, \citenamefont {Katoh}, \citenamefont
  {Yamazaki}, \citenamefont {Zen}, \citenamefont {Hosaka}, \citenamefont
  {Taira},\ and\ \citenamefont {Yamamoto}}]{UVSOR2}%
  \BibitemOpen
  \bibfield  {author} {\bibinfo {author} {\bibnamefont {Tanikawa},
  \bibfnamefont {T.}}, \bibinfo {author} {\bibfnamefont {M.}~\bibnamefont
  {Adachi}}, \bibinfo {author} {\bibfnamefont {M.}~\bibnamefont {Katoh}},
  \bibinfo {author} {\bibfnamefont {J.}~\bibnamefont {Yamazaki}}, \bibinfo
  {author} {\bibfnamefont {H.}~\bibnamefont {Zen}}, \bibinfo {author}
  {\bibfnamefont {M.}~\bibnamefont {Hosaka}}, \bibinfo {author} {\bibfnamefont
  {Y.}~\bibnamefont {Taira}}, \ and\ \bibinfo {author} {\bibfnamefont
  {N.}~\bibnamefont {Yamamoto}}} (\bibinfo {year} {2010}),\ in\ \href@noop {}
  {\emph {\bibinfo {booktitle} {2010 International Particle Accelerator
  Conference}}},\ p.\ \bibinfo {pages} {2206}\BibitemShut {NoStop}%
\bibitem [{\citenamefont {Thompson}\ and\ \citenamefont
  {McNeil}(2008)}]{PhysRevLett.100.203901}%
  \BibitemOpen
  \bibfield  {author} {\bibinfo {author} {\bibnamefont {Thompson},
  \bibfnamefont {N.~R.}}, \ and\ \bibinfo {author} {\bibfnamefont {B.~W.~J.}\
  \bibnamefont {McNeil}}} (\bibinfo {year} {2008}),\ \href
  {http://link.aps.org/doi/10.1103/PhysRevLett.100.203901} {\bibfield
  {journal} {\bibinfo  {journal} {Phys. Rev. Lett.}\ }\textbf {\bibinfo
  {volume} {100}},\ \bibinfo {pages} {203901}}\BibitemShut {NoStop}%
\bibitem [{\citenamefont {Thompson}\ and\ \citenamefont
  {McNeil}(2013)}]{ThompsonMcNeilModeZepta:2013}%
  \BibitemOpen
  \bibfield  {author} {\bibinfo {author} {\bibnamefont {Thompson},
  \bibfnamefont {N.~R.}}, \ and\ \bibinfo {author} {\bibfnamefont {B.~W.~J.}\
  \bibnamefont {McNeil}}} (\bibinfo {year} {2013}),\ \href@noop {} {\bibfield
  {journal} {\bibinfo  {journal} {Phys. Rev. Lett.}\ }\textbf {\bibinfo
  {volume} {110}},\ \bibinfo {pages} {104801}}\BibitemShut {NoStop}%
\bibitem [{\citenamefont {Togashi}\ \emph {et~al.}(2011)\citenamefont
  {Togashi}, \citenamefont {Takahashi}, \citenamefont {Midorikawa},
  \citenamefont {Aoyama}, \citenamefont {Yamakawa}, \citenamefont {Sato},
  \citenamefont {Iwasaki}, \citenamefont {Owada}, \citenamefont {Okino},
  \citenamefont {Yamanouchi}, \citenamefont {Kannari}, \citenamefont
  {Yagishita}, \citenamefont {Nakano}, \citenamefont {Couprie}, \citenamefont
  {Fukami}, \citenamefont {Hatsui}, \citenamefont {Hara}, \citenamefont
  {Kameshima}, \citenamefont {Kitamura}, \citenamefont {Kumagai}, \citenamefont
  {Matsubara}, \citenamefont {Nagasono}, \citenamefont {Ohashi}, \citenamefont
  {Ohshima}, \citenamefont {Otake}, \citenamefont {Shintake}, \citenamefont
  {Tamasaku}, \citenamefont {Tanaka}, \citenamefont {Tanaka}, \citenamefont
  {Togawa}, \citenamefont {Tomizawa}, \citenamefont {Watanabe}, \citenamefont
  {Yabashi},\ and\ \citenamefont {Ishikawa}}]{Togashi:11}%
  \BibitemOpen
  \bibfield  {author} {\bibinfo {author} {\bibnamefont {Togashi}, \bibfnamefont
  {T.}}, \bibinfo {author} {\bibfnamefont {E.~J.}\ \bibnamefont {Takahashi}},
  \bibinfo {author} {\bibfnamefont {K.}~\bibnamefont {Midorikawa}}, \bibinfo
  {author} {\bibfnamefont {M.}~\bibnamefont {Aoyama}}, \bibinfo {author}
  {\bibfnamefont {K.}~\bibnamefont {Yamakawa}}, \bibinfo {author}
  {\bibfnamefont {T.}~\bibnamefont {Sato}}, \bibinfo {author} {\bibfnamefont
  {A.}~\bibnamefont {Iwasaki}}, \bibinfo {author} {\bibfnamefont
  {S.}~\bibnamefont {Owada}}, \bibinfo {author} {\bibfnamefont
  {T.}~\bibnamefont {Okino}}, \bibinfo {author} {\bibfnamefont
  {K.}~\bibnamefont {Yamanouchi}}, \bibinfo {author} {\bibfnamefont
  {F.}~\bibnamefont {Kannari}}, \bibinfo {author} {\bibfnamefont
  {A.}~\bibnamefont {Yagishita}}, \bibinfo {author} {\bibfnamefont
  {H.}~\bibnamefont {Nakano}}, \bibinfo {author} {\bibfnamefont {M.~E.}\
  \bibnamefont {Couprie}}, \bibinfo {author} {\bibfnamefont {K.}~\bibnamefont
  {Fukami}}, \bibinfo {author} {\bibfnamefont {T.}~\bibnamefont {Hatsui}},
  \bibinfo {author} {\bibfnamefont {T.}~\bibnamefont {Hara}}, \bibinfo {author}
  {\bibfnamefont {T.}~\bibnamefont {Kameshima}}, \bibinfo {author}
  {\bibfnamefont {H.}~\bibnamefont {Kitamura}}, \bibinfo {author}
  {\bibfnamefont {N.}~\bibnamefont {Kumagai}}, \bibinfo {author} {\bibfnamefont
  {S.}~\bibnamefont {Matsubara}}, \bibinfo {author} {\bibfnamefont
  {M.}~\bibnamefont {Nagasono}}, \bibinfo {author} {\bibfnamefont
  {H.}~\bibnamefont {Ohashi}}, \bibinfo {author} {\bibfnamefont
  {T.}~\bibnamefont {Ohshima}}, \bibinfo {author} {\bibfnamefont
  {Y.}~\bibnamefont {Otake}}, \bibinfo {author} {\bibfnamefont
  {T.}~\bibnamefont {Shintake}}, \bibinfo {author} {\bibfnamefont
  {K.}~\bibnamefont {Tamasaku}}, \bibinfo {author} {\bibfnamefont
  {H.}~\bibnamefont {Tanaka}}, \bibinfo {author} {\bibfnamefont
  {T.}~\bibnamefont {Tanaka}}, \bibinfo {author} {\bibfnamefont
  {K.}~\bibnamefont {Togawa}}, \bibinfo {author} {\bibfnamefont
  {H.}~\bibnamefont {Tomizawa}}, \bibinfo {author} {\bibfnamefont
  {T.}~\bibnamefont {Watanabe}}, \bibinfo {author} {\bibfnamefont
  {M.}~\bibnamefont {Yabashi}}, \ and\ \bibinfo {author} {\bibfnamefont
  {T.}~\bibnamefont {Ishikawa}}} (\bibinfo {year} {2011}),\ \href@noop {}
  {\bibfield  {journal} {\bibinfo  {journal} {Opt. Express}\ }\textbf {\bibinfo
  {volume} {19}}~(\bibinfo {number} {1}),\ \bibinfo {pages} {317}}\BibitemShut
  {NoStop}%
\bibitem [{\citenamefont {Trebino}\ \emph {et~al.}(1997)\citenamefont
  {Trebino}, \citenamefont {DeLong}, \citenamefont {Fittinghoff}, \citenamefont
  {Sweetser}, \citenamefont {Krumb\"{u}gel}, \citenamefont {Richman},\ and\
  \citenamefont {Kane}}]{FROG}%
  \BibitemOpen
  \bibfield  {author} {\bibinfo {author} {\bibnamefont {Trebino}, \bibfnamefont
  {R.}}, \bibinfo {author} {\bibfnamefont {K.}~\bibnamefont {DeLong}}, \bibinfo
  {author} {\bibfnamefont {D.}~\bibnamefont {Fittinghoff}}, \bibinfo {author}
  {\bibfnamefont {J.}~\bibnamefont {Sweetser}}, \bibinfo {author}
  {\bibfnamefont {M.}~\bibnamefont {Krumb\"{u}gel}}, \bibinfo {author}
  {\bibfnamefont {B.}~\bibnamefont {Richman}}, \ and\ \bibinfo {author}
  {\bibfnamefont {D.}~\bibnamefont {Kane}}} (\bibinfo {year} {1997}),\
  \href@noop {} {\bibfield  {journal} {\bibinfo  {journal} {Rev. Sci.
  Instrum.}\ }\textbf {\bibinfo {volume} {68}},\ \bibinfo {pages}
  {3277}}\BibitemShut {NoStop}%
\bibitem [{\citenamefont {Vinokurov}(1996)}]{vinokurov96}%
  \BibitemOpen
  \bibfield  {author} {\bibinfo {author} {\bibnamefont {Vinokurov},
  \bibfnamefont {N.~A.}}} (\bibinfo {year} {1996}),\ \href@noop {} {\bibfield
  {journal} {\bibinfo  {journal} {Nucl. Instrum. Methods Phys. Res. A}\
  }\textbf {\bibinfo {volume} {375}},\ \bibinfo {pages} {264}}\BibitemShut
  {NoStop}%
\bibitem [{\citenamefont {Wang}(1999)}]{XJWang1999}%
  \BibitemOpen
  \bibfield  {author} {\bibinfo {author} {\bibnamefont {Wang}, \bibfnamefont
  {X.}}} (\bibinfo {year} {1999}),\ in\ \href@noop {} {\emph {\bibinfo
  {booktitle} {Proceedings of the 1999 Particle Accelerator Conference}}}\
  (\bibinfo {address} {New York, USA})\ p.\ \bibinfo {pages} {229}\BibitemShut
  {NoStop}%
\bibitem [{\citenamefont {Wang}\ and\ \citenamefont
  {Chang}(2003)}]{XJWang2003}%
  \BibitemOpen
  \bibfield  {author} {\bibinfo {author} {\bibnamefont {Wang}, \bibfnamefont
  {X.}}, \ and\ \bibinfo {author} {\bibfnamefont {X.}~\bibnamefont {Chang}}}
  (\bibinfo {year} {2003}),\ \href@noop {} {\bibfield  {journal} {\bibinfo
  {journal} {Nucl. Instrum. Methods Phys. Res. A}\ }\textbf {\bibinfo {volume}
  {507}},\ \bibinfo {pages} {310}}\BibitemShut {NoStop}%
\bibitem [{\citenamefont {Wang}\ \emph {et~al.}(2006)\citenamefont {Wang},
  \citenamefont {Shen}, \citenamefont {Watanabe}, \citenamefont {Murphy},
  \citenamefont {Rose},\ and\ \citenamefont {Tsang}}]{HGHG4th}%
  \BibitemOpen
  \bibfield  {author} {\bibinfo {author} {\bibnamefont {Wang}, \bibfnamefont
  {X.}}, \bibinfo {author} {\bibfnamefont {Y.}~\bibnamefont {Shen}}, \bibinfo
  {author} {\bibfnamefont {T.}~\bibnamefont {Watanabe}}, \bibinfo {author}
  {\bibfnamefont {J.}~\bibnamefont {Murphy}}, \bibinfo {author} {\bibfnamefont
  {J.}~\bibnamefont {Rose}}, \ and\ \bibinfo {author} {\bibfnamefont
  {T.}~\bibnamefont {Tsang}}} (\bibinfo {year} {2006}),\ in\ \href@noop {}
  {\emph {\bibinfo {booktitle} {Proceeding of the FEL Conference 2006}}},\
  p.~\bibinfo {pages} {18}\BibitemShut {NoStop}%
\bibitem [{\citenamefont {Weling}\ and\ \citenamefont {Auston}(1996)}]{Weling}%
  \BibitemOpen
  \bibfield  {author} {\bibinfo {author} {\bibnamefont {Weling}, \bibfnamefont
  {A.~S.}}, \ and\ \bibinfo {author} {\bibfnamefont {D.~H.}\ \bibnamefont
  {Auston}}} (\bibinfo {year} {1996}),\ \href@noop {} {\bibfield  {journal}
  {\bibinfo  {journal} {J. Opt. Soc. Am. B}\ }\textbf {\bibinfo {volume}
  {13}},\ \bibinfo {pages} {2783}}\BibitemShut {NoStop}%
\bibitem [{\citenamefont {Wernick}\ and\ \citenamefont
  {Marshall}(1992)}]{PhysRevA.46.3566}%
  \BibitemOpen
  \bibfield  {author} {\bibinfo {author} {\bibnamefont {Wernick}, \bibfnamefont
  {I.}}, \ and\ \bibinfo {author} {\bibfnamefont {T.~C.}\ \bibnamefont
  {Marshall}}} (\bibinfo {year} {1992}),\ \href
  {http://link.aps.org/doi/10.1103/PhysRevA.46.3566} {\bibfield  {journal}
  {\bibinfo  {journal} {Phys. Rev. A}\ }\textbf {\bibinfo {volume} {46}},\
  \bibinfo {pages} {3566}}\BibitemShut {NoStop}%
\bibitem [{\citenamefont {Wiedemann}(2003)}]{Wiedemann}%
  \BibitemOpen
  \bibfield  {author} {\bibinfo {author} {\bibnamefont {Wiedemann},
  \bibfnamefont {H.}}} (\bibinfo {year} {2003}),\ \href@noop {} {\emph
  {\bibinfo {title} {Synchrotron Radiation}}}\ (\bibinfo  {publisher}
  {Springer-Verlag})\BibitemShut {NoStop}%
\bibitem [{\citenamefont {Winick}(1995)}]{Winick}%
  \BibitemOpen
  \bibfield  {author} {\bibinfo {author} {\bibnamefont {Winick}, \bibfnamefont
  {H.}}} (\bibinfo {year} {1995}),\ \href@noop {} {\emph {\bibinfo {title} {The
  Synchrotron Radiation Sources: A Primer}}}\ (\bibinfo  {publisher} {World
  Scientific})\BibitemShut {NoStop}%
\bibitem [{\citenamefont {Wolski}\ \emph {et~al.}(2004)\citenamefont {Wolski},
  \citenamefont {Penn}, \citenamefont {Sessler},\ and\ \citenamefont
  {Wurtele}}]{wolsky_conditioner}%
  \BibitemOpen
  \bibfield  {author} {\bibinfo {author} {\bibnamefont {Wolski}, \bibfnamefont
  {A.}}, \bibinfo {author} {\bibfnamefont {G.}~\bibnamefont {Penn}}, \bibinfo
  {author} {\bibfnamefont {A.}~\bibnamefont {Sessler}}, \ and\ \bibinfo
  {author} {\bibfnamefont {J.}~\bibnamefont {Wurtele}}} (\bibinfo {year}
  {2004}),\ \href {http://link.aps.org/doi/10.1103/PhysRevSTAB.7.080701}
  {\bibfield  {journal} {\bibinfo  {journal} {Phys. Rev. ST Accel. Beams}\
  }\textbf {\bibinfo {volume} {7}},\ \bibinfo {pages} {080701}}\BibitemShut
  {NoStop}%
\bibitem [{\citenamefont {Wu}\ \emph {et~al.}(2012)\citenamefont {Wu},
  \citenamefont {Marinelli},\ and\ \citenamefont {Pellegrini}}]{iSASEFEL12}%
  \BibitemOpen
  \bibfield  {author} {\bibinfo {author} {\bibnamefont {Wu}, \bibfnamefont
  {J.}}, \bibinfo {author} {\bibfnamefont {A.}~\bibnamefont {Marinelli}}, \
  and\ \bibinfo {author} {\bibfnamefont {C.}~\bibnamefont {Pellegrini}}}
  (\bibinfo {year} {2012}),\ in\ \href@noop {} {\emph {\bibinfo {booktitle}
  {Proceedings of the 2012 FEL Conference}}}\ (\bibinfo {address} {Nara,
  Japan})\ p.\ \bibinfo {pages} {237}\BibitemShut {NoStop}%
\bibitem [{\citenamefont {Wu}\ \emph {et~al.}(2013)\citenamefont {Wu},
  \citenamefont {Pellegrini},\ and\ \citenamefont {Marinelli}}]{iSASE}%
  \BibitemOpen
  \bibfield  {author} {\bibinfo {author} {\bibnamefont {Wu}, \bibfnamefont
  {J.}}, \bibinfo {author} {\bibfnamefont {C.}~\bibnamefont {Pellegrini}}, \
  and\ \bibinfo {author} {\bibfnamefont {A.}~\bibnamefont {Marinelli}}}
  (\bibinfo {year} {2013}),\ \href@noop {} {\enquote {\bibinfo {title}
  {{iSASE}},}\ }\bibinfo {note} {To be published}\BibitemShut {NoStop}%
\bibitem [{\citenamefont {W\"{u}stefeld}(2008)}]{Wust}%
  \BibitemOpen
  \bibfield  {author} {\bibinfo {author} {\bibnamefont {W\"{u}stefeld},
  \bibfnamefont {G.}}} (\bibinfo {year} {2008}),\ in\ \href@noop {} {\emph
  {\bibinfo {booktitle} {Proc. of the 2008 European Particle Accelerator
  Conference, Genoa, Italy}}},\ p.~\bibinfo {pages} {26}\BibitemShut {NoStop}%
\bibitem [{\citenamefont {Xiang}(2010)}]{LAEE}%
  \BibitemOpen
  \bibfield  {author} {\bibinfo {author} {\bibnamefont {Xiang}, \bibfnamefont
  {D.}}} (\bibinfo {year} {2010}),\ \href
  {http://link.aps.org/doi/10.1103/PhysRevSTAB.13.010701} {\bibfield  {journal}
  {\bibinfo  {journal} {Phys. Rev. ST Accel. Beams}\ }\textbf {\bibinfo
  {volume} {13}},\ \bibinfo {pages} {010701}}\BibitemShut {NoStop}%
\bibitem [{\citenamefont {Xiang}(2012)}]{ICFA-Xiang}%
  \BibitemOpen
  \bibfield  {author} {\bibinfo {author} {\bibnamefont {Xiang}, \bibfnamefont
  {D.}}} (\bibinfo {year} {2012}),\ \href@noop {} {\bibfield  {journal}
  {\bibinfo  {journal} {ICFA Beam Dynamics Newsletter}\ }\textbf {\bibinfo
  {volume} {59}},\ \bibinfo {pages} {39}}\BibitemShut {NoStop}%
\bibitem [{\citenamefont {Xiang}\ and\ \citenamefont {Chao}(2011)}]{EEX3}%
  \BibitemOpen
  \bibfield  {author} {\bibinfo {author} {\bibnamefont {Xiang}, \bibfnamefont
  {D.}}, \ and\ \bibinfo {author} {\bibfnamefont {A.}~\bibnamefont {Chao}}}
  (\bibinfo {year} {2011}),\ \href
  {http://link.aps.org/doi/10.1103/PhysRevSTAB.14.114001} {\bibfield  {journal}
  {\bibinfo  {journal} {Phys. Rev. ST Accel. Beams}\ }\textbf {\bibinfo
  {volume} {14}},\ \bibinfo {pages} {114001}}\BibitemShut {NoStop}%
\bibitem [{\citenamefont {Xiang}\ \emph {et~al.}(2011)\citenamefont {Xiang},
  \citenamefont {Colby}, \citenamefont {Dunning}, \citenamefont {Gilevich},
  \citenamefont {Hast}, \citenamefont {Jobe}, \citenamefont {McCormick},
  \citenamefont {Nelson}, \citenamefont {Raubenheimer}, \citenamefont {Soong},
  \citenamefont {Stupakov}, \citenamefont {Szalata}, \citenamefont {Walz},
  \citenamefont {Weathersby},\ and\ \citenamefont {Woodley}}]{EMmeasurement}%
  \BibitemOpen
  \bibfield  {author} {\bibinfo {author} {\bibnamefont {Xiang}, \bibfnamefont
  {D.}}, \bibinfo {author} {\bibfnamefont {E.}~\bibnamefont {Colby}}, \bibinfo
  {author} {\bibfnamefont {M.}~\bibnamefont {Dunning}}, \bibinfo {author}
  {\bibfnamefont {S.}~\bibnamefont {Gilevich}}, \bibinfo {author}
  {\bibfnamefont {C.}~\bibnamefont {Hast}}, \bibinfo {author} {\bibfnamefont
  {K.}~\bibnamefont {Jobe}}, \bibinfo {author} {\bibfnamefont {D.}~\bibnamefont
  {McCormick}}, \bibinfo {author} {\bibfnamefont {J.}~\bibnamefont {Nelson}},
  \bibinfo {author} {\bibfnamefont {T.~O.}\ \bibnamefont {Raubenheimer}},
  \bibinfo {author} {\bibfnamefont {K.}~\bibnamefont {Soong}}, \bibinfo
  {author} {\bibfnamefont {G.}~\bibnamefont {Stupakov}}, \bibinfo {author}
  {\bibfnamefont {Z.}~\bibnamefont {Szalata}}, \bibinfo {author} {\bibfnamefont
  {D.}~\bibnamefont {Walz}}, \bibinfo {author} {\bibfnamefont {S.}~\bibnamefont
  {Weathersby}}, \ and\ \bibinfo {author} {\bibfnamefont {M.}~\bibnamefont
  {Woodley}}} (\bibinfo {year} {2011}),\ \href@noop {} {\bibfield  {journal}
  {\bibinfo  {journal} {Phys. Rev. ST Accel. Beams}\ }\textbf {\bibinfo
  {volume} {14}},\ \bibinfo {pages} {112801}}\BibitemShut {NoStop}%
\bibitem [{\citenamefont {Xiang}\ \emph
  {et~al.}(2012{\natexlab{a}})\citenamefont {Xiang}, \citenamefont {Colby},
  \citenamefont {Dunning}, \citenamefont {Gilevich}, \citenamefont {Hast},
  \citenamefont {Jobe}, \citenamefont {McCormick}, \citenamefont {Nelson},
  \citenamefont {Raubenheimer}, \citenamefont {Soong}, \citenamefont
  {Stupakov}, \citenamefont {Szalata}, \citenamefont {Walz}, \citenamefont
  {Weathersby},\ and\ \citenamefont {Woodley}}]{Echo7}%
  \BibitemOpen
  \bibfield  {author} {\bibinfo {author} {\bibnamefont {Xiang}, \bibfnamefont
  {D.}}, \bibinfo {author} {\bibfnamefont {E.}~\bibnamefont {Colby}}, \bibinfo
  {author} {\bibfnamefont {M.}~\bibnamefont {Dunning}}, \bibinfo {author}
  {\bibfnamefont {S.}~\bibnamefont {Gilevich}}, \bibinfo {author}
  {\bibfnamefont {C.}~\bibnamefont {Hast}}, \bibinfo {author} {\bibfnamefont
  {K.}~\bibnamefont {Jobe}}, \bibinfo {author} {\bibfnamefont {D.}~\bibnamefont
  {McCormick}}, \bibinfo {author} {\bibfnamefont {J.}~\bibnamefont {Nelson}},
  \bibinfo {author} {\bibfnamefont {T.~O.}\ \bibnamefont {Raubenheimer}},
  \bibinfo {author} {\bibfnamefont {K.}~\bibnamefont {Soong}}, \bibinfo
  {author} {\bibfnamefont {G.}~\bibnamefont {Stupakov}}, \bibinfo {author}
  {\bibfnamefont {Z.}~\bibnamefont {Szalata}}, \bibinfo {author} {\bibfnamefont
  {D.}~\bibnamefont {Walz}}, \bibinfo {author} {\bibfnamefont {S.}~\bibnamefont
  {Weathersby}}, \ and\ \bibinfo {author} {\bibfnamefont {M.}~\bibnamefont
  {Woodley}}} (\bibinfo {year} {2012}{\natexlab{a}}),\ \href
  {http://link.aps.org/doi/10.1103/PhysRevLett.108.024802} {\bibfield
  {journal} {\bibinfo  {journal} {Phys. Rev. Lett.}\ }\textbf {\bibinfo
  {volume} {108}},\ \bibinfo {pages} {024802}}\BibitemShut {NoStop}%
\bibitem [{\citenamefont {Xiang}\ \emph {et~al.}(2010)\citenamefont {Xiang},
  \citenamefont {Colby}, \citenamefont {Dunning}, \citenamefont {Gilevich},
  \citenamefont {Hast}, \citenamefont {Jobe}, \citenamefont {McCormick},
  \citenamefont {Nelson}, \citenamefont {Raubenheimer}, \citenamefont {Soong},
  \citenamefont {Stupakov}, \citenamefont {Szalata}, \citenamefont {Walz},
  \citenamefont {Weathersby}, \citenamefont {Woodley},\ and\ \citenamefont
  {Pernet}}]{Echo3}%
  \BibitemOpen
  \bibfield  {author} {\bibinfo {author} {\bibnamefont {Xiang}, \bibfnamefont
  {D.}}, \bibinfo {author} {\bibfnamefont {E.}~\bibnamefont {Colby}}, \bibinfo
  {author} {\bibfnamefont {M.}~\bibnamefont {Dunning}}, \bibinfo {author}
  {\bibfnamefont {S.}~\bibnamefont {Gilevich}}, \bibinfo {author}
  {\bibfnamefont {C.}~\bibnamefont {Hast}}, \bibinfo {author} {\bibfnamefont
  {K.}~\bibnamefont {Jobe}}, \bibinfo {author} {\bibfnamefont {D.}~\bibnamefont
  {McCormick}}, \bibinfo {author} {\bibfnamefont {J.}~\bibnamefont {Nelson}},
  \bibinfo {author} {\bibfnamefont {T.~O.}\ \bibnamefont {Raubenheimer}},
  \bibinfo {author} {\bibfnamefont {K.}~\bibnamefont {Soong}}, \bibinfo
  {author} {\bibfnamefont {G.}~\bibnamefont {Stupakov}}, \bibinfo {author}
  {\bibfnamefont {Z.}~\bibnamefont {Szalata}}, \bibinfo {author} {\bibfnamefont
  {D.}~\bibnamefont {Walz}}, \bibinfo {author} {\bibfnamefont {S.}~\bibnamefont
  {Weathersby}}, \bibinfo {author} {\bibfnamefont {M.}~\bibnamefont {Woodley}},
  \ and\ \bibinfo {author} {\bibfnamefont {P.-L.}\ \bibnamefont {Pernet}}}
  (\bibinfo {year} {2010}),\ \href
  {http://link.aps.org/doi/10.1103/PhysRevLett.105.114801} {\bibfield
  {journal} {\bibinfo  {journal} {Phys. Rev. Lett.}\ }\textbf {\bibinfo
  {volume} {105}},\ \bibinfo {pages} {114801}}\BibitemShut {NoStop}%
\bibitem [{\citenamefont {Xiang}\ \emph {et~al.}(2013)\citenamefont {Xiang},
  \citenamefont {Ding}, \citenamefont {Huang},\ and\ \citenamefont
  {Deng}}]{pSASE}%
  \BibitemOpen
  \bibfield  {author} {\bibinfo {author} {\bibnamefont {Xiang}, \bibfnamefont
  {D.}}, \bibinfo {author} {\bibfnamefont {Y.}~\bibnamefont {Ding}}, \bibinfo
  {author} {\bibfnamefont {Z.}~\bibnamefont {Huang}}, \ and\ \bibinfo {author}
  {\bibfnamefont {H.}~\bibnamefont {Deng}}} (\bibinfo {year} {2013}),\ \href
  {http://link.aps.org/doi/10.1103/PhysRevSTAB.16.010703} {\bibfield  {journal}
  {\bibinfo  {journal} {Phys. Rev. ST Accel. Beams}\ }\textbf {\bibinfo
  {volume} {16}},\ \bibinfo {pages} {010703}}\BibitemShut {NoStop}%
\bibitem [{\citenamefont {Xiang}\ \emph
  {et~al.}(2012{\natexlab{b}})\citenamefont {Xiang}, \citenamefont {Ding},
  \citenamefont {Raubenheimer},\ and\ \citenamefont {Wu}}]{ML3}%
  \BibitemOpen
  \bibfield  {author} {\bibinfo {author} {\bibnamefont {Xiang}, \bibfnamefont
  {D.}}, \bibinfo {author} {\bibfnamefont {Y.}~\bibnamefont {Ding}}, \bibinfo
  {author} {\bibfnamefont {T.}~\bibnamefont {Raubenheimer}}, \ and\ \bibinfo
  {author} {\bibfnamefont {J.}~\bibnamefont {Wu}}} (\bibinfo {year}
  {2012}{\natexlab{b}}),\ \href@noop {} {\bibfield  {journal} {\bibinfo
  {journal} {Phys. Rev. ST Accel. Beams}\ }\textbf {\bibinfo {volume} {15}},\
  \bibinfo {pages} {050707}}\BibitemShut {NoStop}%
\bibitem [{\citenamefont {Xiang}\ \emph {et~al.}(2009)\citenamefont {Xiang},
  \citenamefont {Huang},\ and\ \citenamefont {Stupakov}}]{EEHG-as}%
  \BibitemOpen
  \bibfield  {author} {\bibinfo {author} {\bibnamefont {Xiang}, \bibfnamefont
  {D.}}, \bibinfo {author} {\bibfnamefont {Z.}~\bibnamefont {Huang}}, \ and\
  \bibinfo {author} {\bibfnamefont {G.}~\bibnamefont {Stupakov}}} (\bibinfo
  {year} {2009}),\ \href@noop {} {\bibfield  {journal} {\bibinfo  {journal}
  {Phys. Rev. ST Accel. Beams}\ }\textbf {\bibinfo {volume} {12}},\ \bibinfo
  {pages} {060701}}\BibitemShut {NoStop}%
\bibitem [{\citenamefont {Xiang}\ and\ \citenamefont
  {Stupakov}(2009)}]{XiangStupakov2009}%
  \BibitemOpen
  \bibfield  {author} {\bibinfo {author} {\bibnamefont {Xiang}, \bibfnamefont
  {D.}}, \ and\ \bibinfo {author} {\bibfnamefont {G.}~\bibnamefont {Stupakov}}}
  (\bibinfo {year} {2009}),\ \href@noop {} {\bibfield  {journal} {\bibinfo
  {journal} {Phys. Rev. ST Accel. Beams}\ }\textbf {\bibinfo {volume} {12}},\
  \bibinfo {pages} {080701}}\BibitemShut {NoStop}%
\bibitem [{\citenamefont {Xiang}\ and\ \citenamefont
  {Wan}(2010)}]{CHGdivergence}%
  \BibitemOpen
  \bibfield  {author} {\bibinfo {author} {\bibnamefont {Xiang}, \bibfnamefont
  {D.}}, \ and\ \bibinfo {author} {\bibfnamefont {W.}~\bibnamefont {Wan}}}
  (\bibinfo {year} {2010}),\ \href@noop {} {\bibfield  {journal} {\bibinfo
  {journal} {Phys. Rev. Lett.}\ }\textbf {\bibinfo {volume} {104}},\ \bibinfo
  {pages} {084803}}\BibitemShut {NoStop}%
\bibitem [{\citenamefont {Xie}(2003)}]{xie03}%
  \BibitemOpen
  \bibfield  {author} {\bibinfo {author} {\bibnamefont {Xie}, \bibfnamefont
  {M.}}} (\bibinfo {year} {2003}),\ in\ \href@noop {} {\emph {\bibinfo
  {booktitle} {2003 Particle Accelerator Conference}}},\ p.\ \bibinfo {pages}
  {1843}\BibitemShut {NoStop}%
\bibitem [{\citenamefont {Yoder}\ \emph {et~al.}(2001)\citenamefont {Yoder},
  \citenamefont {Marshall},\ and\ \citenamefont
  {Hirshfield}}]{PhysRevLett.86.1765}%
  \BibitemOpen
  \bibfield  {author} {\bibinfo {author} {\bibnamefont {Yoder}, \bibfnamefont
  {R.~B.}}, \bibinfo {author} {\bibfnamefont {T.~C.}\ \bibnamefont {Marshall}},
  \ and\ \bibinfo {author} {\bibfnamefont {J.~L.}\ \bibnamefont {Hirshfield}}}
  (\bibinfo {year} {2001}),\ \href
  {http://link.aps.org/doi/10.1103/PhysRevLett.86.1765} {\bibfield  {journal}
  {\bibinfo  {journal} {Phys. Rev. Lett.}\ }\textbf {\bibinfo {volume} {86}},\
  \bibinfo {pages} {1765}}\BibitemShut {NoStop}%
\bibitem [{\citenamefont {Yu}(1991)}]{HGHG1}%
  \BibitemOpen
  \bibfield  {author} {\bibinfo {author} {\bibnamefont {Yu}, \bibfnamefont
  {L.~H.}}} (\bibinfo {year} {1991}),\ \href@noop {} {\bibfield  {journal}
  {\bibinfo  {journal} {Phys. Rev. A}\ }\textbf {\bibinfo {volume} {44}},\
  \bibinfo {pages} {5178}}\BibitemShut {NoStop}%
\bibitem [{\citenamefont {Yu}\ \emph {et~al.}(2000)\citenamefont {Yu},
  \citenamefont {Babzien}, \citenamefont {Ben-Zvi}, \citenamefont {DiMauro},
  \citenamefont {Doyuran}, \citenamefont {Graves}, \citenamefont {Johnson},
  \citenamefont {Krinsky}, \citenamefont {Malone}, \citenamefont {Pogorelsky},
  \citenamefont {Skaritka}, \citenamefont {Rakowsky}, \citenamefont {Solomon},
  \citenamefont {Wang}, \citenamefont {Woodle}, \citenamefont {Yakimenko},
  \citenamefont {Biedron}, \citenamefont {Galayda}, \citenamefont {Gluskin},
  \citenamefont {Jagger}, \citenamefont {Sajaev},\ and\ \citenamefont
  {Vasserman}}]{HGHGScience}%
  \BibitemOpen
  \bibfield  {author} {\bibinfo {author} {\bibnamefont {Yu}, \bibfnamefont
  {L.-H.}}, \bibinfo {author} {\bibfnamefont {M.}~\bibnamefont {Babzien}},
  \bibinfo {author} {\bibfnamefont {I.}~\bibnamefont {Ben-Zvi}}, \bibinfo
  {author} {\bibfnamefont {L.~F.}\ \bibnamefont {DiMauro}}, \bibinfo {author}
  {\bibfnamefont {A.}~\bibnamefont {Doyuran}}, \bibinfo {author} {\bibfnamefont
  {W.}~\bibnamefont {Graves}}, \bibinfo {author} {\bibfnamefont
  {E.}~\bibnamefont {Johnson}}, \bibinfo {author} {\bibfnamefont
  {S.}~\bibnamefont {Krinsky}}, \bibinfo {author} {\bibfnamefont
  {R.}~\bibnamefont {Malone}}, \bibinfo {author} {\bibfnamefont
  {I.}~\bibnamefont {Pogorelsky}}, \bibinfo {author} {\bibfnamefont
  {J.}~\bibnamefont {Skaritka}}, \bibinfo {author} {\bibfnamefont
  {G.}~\bibnamefont {Rakowsky}}, \bibinfo {author} {\bibfnamefont
  {L.}~\bibnamefont {Solomon}}, \bibinfo {author} {\bibfnamefont {X.~J.}\
  \bibnamefont {Wang}}, \bibinfo {author} {\bibfnamefont {M.}~\bibnamefont
  {Woodle}}, \bibinfo {author} {\bibfnamefont {V.}~\bibnamefont {Yakimenko}},
  \bibinfo {author} {\bibfnamefont {S.~G.}\ \bibnamefont {Biedron}}, \bibinfo
  {author} {\bibfnamefont {J.~N.}\ \bibnamefont {Galayda}}, \bibinfo {author}
  {\bibfnamefont {E.}~\bibnamefont {Gluskin}}, \bibinfo {author} {\bibfnamefont
  {J.}~\bibnamefont {Jagger}}, \bibinfo {author} {\bibfnamefont
  {V.}~\bibnamefont {Sajaev}}, \ and\ \bibinfo {author} {\bibfnamefont
  {I.}~\bibnamefont {Vasserman}}} (\bibinfo {year} {2000}),\ \href@noop {}
  {\bibfield  {journal} {\bibinfo  {journal} {Science}\ }\textbf {\bibinfo
  {volume} {289}},\ \bibinfo {pages} {932}}\BibitemShut {NoStop}%
\bibitem [{\citenamefont {Yu}\ \emph {et~al.}(2003)\citenamefont {Yu},
  \citenamefont {DiMauro}, \citenamefont {Doyuran}, \citenamefont {Graves},
  \citenamefont {Johnson}, \citenamefont {Heese}, \citenamefont {Krinsky},
  \citenamefont {Loos}, \citenamefont {Murphy}, \citenamefont {Rakowsky},
  \citenamefont {Rose}, \citenamefont {Shaftan}, \citenamefont {Sheehy},
  \citenamefont {Skaritka}, \citenamefont {Wang},\ and\ \citenamefont
  {Wu}}]{HGHG3rd}%
  \BibitemOpen
  \bibfield  {author} {\bibinfo {author} {\bibnamefont {Yu}, \bibfnamefont
  {L.~H.}}, \bibinfo {author} {\bibfnamefont {L.}~\bibnamefont {DiMauro}},
  \bibinfo {author} {\bibfnamefont {A.}~\bibnamefont {Doyuran}}, \bibinfo
  {author} {\bibfnamefont {W.~S.}\ \bibnamefont {Graves}}, \bibinfo {author}
  {\bibfnamefont {E.~D.}\ \bibnamefont {Johnson}}, \bibinfo {author}
  {\bibfnamefont {R.}~\bibnamefont {Heese}}, \bibinfo {author} {\bibfnamefont
  {S.}~\bibnamefont {Krinsky}}, \bibinfo {author} {\bibfnamefont
  {H.}~\bibnamefont {Loos}}, \bibinfo {author} {\bibfnamefont {J.~B.}\
  \bibnamefont {Murphy}}, \bibinfo {author} {\bibfnamefont {G.}~\bibnamefont
  {Rakowsky}}, \bibinfo {author} {\bibfnamefont {J.}~\bibnamefont {Rose}},
  \bibinfo {author} {\bibfnamefont {T.}~\bibnamefont {Shaftan}}, \bibinfo
  {author} {\bibfnamefont {B.}~\bibnamefont {Sheehy}}, \bibinfo {author}
  {\bibfnamefont {J.}~\bibnamefont {Skaritka}}, \bibinfo {author}
  {\bibfnamefont {X.~J.}\ \bibnamefont {Wang}}, \ and\ \bibinfo {author}
  {\bibfnamefont {Z.}~\bibnamefont {Wu}}} (\bibinfo {year} {2003}),\ \href
  {http://link.aps.org/doi/10.1103/PhysRevLett.91.074801} {\bibfield  {journal}
  {\bibinfo  {journal} {Phys. Rev. Lett.}\ }\textbf {\bibinfo {volume} {91}},\
  \bibinfo {pages} {074801}}\BibitemShut {NoStop}%
\bibitem [{\citenamefont {Zen}\ \emph {et~al.}(2011)\citenamefont {Zen},
  \citenamefont {Adachi}, \citenamefont {Katoh}, \citenamefont {Tanikawa},
  \citenamefont {Yamamoto},\ and\ \citenamefont {Hosaka}}]{CPACHG1}%
  \BibitemOpen
  \bibfield  {author} {\bibinfo {author} {\bibnamefont {Zen}, \bibfnamefont
  {H.}}, \bibinfo {author} {\bibfnamefont {M.}~\bibnamefont {Adachi}}, \bibinfo
  {author} {\bibfnamefont {M.}~\bibnamefont {Katoh}}, \bibinfo {author}
  {\bibfnamefont {T.}~\bibnamefont {Tanikawa}}, \bibinfo {author}
  {\bibfnamefont {N.}~\bibnamefont {Yamamoto}}, \ and\ \bibinfo {author}
  {\bibfnamefont {M.}~\bibnamefont {Hosaka}}} (\bibinfo {year} {2011}),\ in\
  \href@noop {} {\emph {\bibinfo {booktitle} {2011 Free-electron laser
  Conference}}},\ p.\ \bibinfo {pages} {366}\BibitemShut {NoStop}%
\bibitem [{\citenamefont {Zhao}(2010)}]{Zhao:RAST}%
  \BibitemOpen
  \bibfield  {author} {\bibinfo {author} {\bibnamefont {Zhao}, \bibfnamefont
  {Z.~T.}}} (\bibinfo {year} {2010}),\ \href@noop {} {\bibfield  {journal}
  {\bibinfo  {journal} {Reviews of Accelerator Science and Technology}\
  }\textbf {\bibinfo {volume} {3}},\ \bibinfo {pages} {57}}\BibitemShut
  {NoStop}%
\bibitem [{\citenamefont {Zhao}\ \emph {et~al.}(2012)\citenamefont {Zhao},
  \citenamefont {Wang}, \citenamefont {Chen}, \citenamefont {Chen},
  \citenamefont {Deng}, \citenamefont {Ding}, \citenamefont {Feng},
  \citenamefont {Gu}, \citenamefont {Huang}, \citenamefont {Lan}, \citenamefont
  {Leng}, \citenamefont {Li}, \citenamefont {Lin}, \citenamefont {Liu},
  \citenamefont {Prat}, \citenamefont {Wang}, \citenamefont {Wang},
  \citenamefont {Ye}, \citenamefont {Yu}, \citenamefont {Zhang}, \citenamefont
  {Zhang}, \citenamefont {Zhang}, \citenamefont {Zhang}, \citenamefont {Zhang},
  \citenamefont {Zhong},\ and\ \citenamefont {Zhou}}]{ZhaoZ12:first}%
  \BibitemOpen
  \bibfield  {author} {\bibinfo {author} {\bibnamefont {Zhao}, \bibfnamefont
  {Z.~T.}}, \bibinfo {author} {\bibfnamefont {D.}~\bibnamefont {Wang}},
  \bibinfo {author} {\bibfnamefont {J.~H.}\ \bibnamefont {Chen}}, \bibinfo
  {author} {\bibfnamefont {Z.~H.}\ \bibnamefont {Chen}}, \bibinfo {author}
  {\bibfnamefont {H.~X.}\ \bibnamefont {Deng}}, \bibinfo {author}
  {\bibfnamefont {J.~G.}\ \bibnamefont {Ding}}, \bibinfo {author}
  {\bibfnamefont {C.}~\bibnamefont {Feng}}, \bibinfo {author} {\bibfnamefont
  {Q.}~\bibnamefont {Gu}}, \bibinfo {author} {\bibfnamefont {M.~M.}\
  \bibnamefont {Huang}}, \bibinfo {author} {\bibfnamefont {T.~H.}\ \bibnamefont
  {Lan}}, \bibinfo {author} {\bibfnamefont {Y.~B.}\ \bibnamefont {Leng}},
  \bibinfo {author} {\bibfnamefont {D.~G.}\ \bibnamefont {Li}}, \bibinfo
  {author} {\bibfnamefont {G.~Q.}\ \bibnamefont {Lin}}, \bibinfo {author}
  {\bibfnamefont {B.}~\bibnamefont {Liu}}, \bibinfo {author} {\bibfnamefont
  {E.}~\bibnamefont {Prat}}, \bibinfo {author} {\bibfnamefont {X.~T.}\
  \bibnamefont {Wang}}, \bibinfo {author} {\bibfnamefont {Z.~S.}\ \bibnamefont
  {Wang}}, \bibinfo {author} {\bibfnamefont {K.~R.}\ \bibnamefont {Ye}},
  \bibinfo {author} {\bibfnamefont {L.~Y.}\ \bibnamefont {Yu}}, \bibinfo
  {author} {\bibfnamefont {H.~O.}\ \bibnamefont {Zhang}}, \bibinfo {author}
  {\bibfnamefont {J.~Q.}\ \bibnamefont {Zhang}}, \bibinfo {author}
  {\bibfnamefont {M.}~\bibnamefont {Zhang}}, \bibinfo {author} {\bibfnamefont
  {M.}~\bibnamefont {Zhang}}, \bibinfo {author} {\bibfnamefont
  {T.}~\bibnamefont {Zhang}}, \bibinfo {author} {\bibfnamefont {S.~P.}\
  \bibnamefont {Zhong}}, \ and\ \bibinfo {author} {\bibfnamefont {Q.~G.}\
  \bibnamefont {Zhou}}} (\bibinfo {year} {2012}),\ \href@noop {} {\bibfield
  {journal} {\bibinfo  {journal} {Nature Photon.}\ }\textbf {\bibinfo {volume}
  {6}},\ \bibinfo {pages} {360}}\BibitemShut {NoStop}%
\bibitem [{\citenamefont {Zholents}\ and\ \citenamefont
  {Decking}(2000)}]{Zhol2}%
  \BibitemOpen
  \bibfield  {author} {\bibinfo {author} {\bibnamefont {Zholents},
  \bibfnamefont {A.}}, \ and\ \bibinfo {author} {\bibfnamefont
  {W.}~\bibnamefont {Decking}}} (\bibinfo {year} {2000}),\ \href@noop {}
  {\bibinfo  {journal} {Proc. of the 2000 European Particle Accelerator
  Conference,Vienna, Austria}\ }\BibitemShut {NoStop}%
\bibitem [{\citenamefont {Zholents}\ and\ \citenamefont
  {Fawley}(2004)}]{Zhol6}%
  \BibitemOpen
\bibfield  {journal} {  }\bibfield  {author} {\bibinfo {author} {\bibnamefont
  {Zholents}, \bibfnamefont {A.}}, \ and\ \bibinfo {author} {\bibfnamefont
  {W.}~\bibnamefont {Fawley}}} (\bibinfo {year} {2004}),\ \href@noop {}
  {\bibfield  {journal} {\bibinfo  {journal} {Phys. Rev. Lett.}\ }\textbf
  {\bibinfo {volume} {92}},\ \bibinfo {pages} {224801}}\BibitemShut {NoStop}%
\bibitem [{\citenamefont {Zholents}\ \emph {et~al.}(2004)\citenamefont
  {Zholents}, \citenamefont {Fawley}, \citenamefont {Emma}, \citenamefont
  {Huang}, \citenamefont {Stupakov},\ and\ \citenamefont {Reiche}}]{Zhol5}%
  \BibitemOpen
  \bibfield  {author} {\bibinfo {author} {\bibnamefont {Zholents},
  \bibfnamefont {A.}}, \bibinfo {author} {\bibfnamefont {W.}~\bibnamefont
  {Fawley}}, \bibinfo {author} {\bibfnamefont {P.}~\bibnamefont {Emma}},
  \bibinfo {author} {\bibfnamefont {Z.}~\bibnamefont {Huang}}, \bibinfo
  {author} {\bibfnamefont {G.}~\bibnamefont {Stupakov}}, \ and\ \bibinfo
  {author} {\bibfnamefont {S.}~\bibnamefont {Reiche}}} (\bibinfo {year}
  {2004}),\ \href@noop {} {\bibinfo  {journal} {Proc. of the 2004 Free Electron
  Laser Conference, Trieste, Italy}\ ,\ \bibinfo {pages} {582}}\BibitemShut
  {NoStop}%
\bibitem [{\citenamefont {Zholents}\ \emph {et~al.}(1999)\citenamefont
  {Zholents}, \citenamefont {Heimann}, \citenamefont {Zolotorev},\ and\
  \citenamefont {Byrd}}]{Sasha2TCAV}%
  \BibitemOpen
\bibfield  {journal} {  }\bibfield  {author} {\bibinfo {author} {\bibnamefont
  {Zholents}, \bibfnamefont {A.}}, \bibinfo {author} {\bibfnamefont
  {P.}~\bibnamefont {Heimann}}, \bibinfo {author} {\bibfnamefont
  {M.}~\bibnamefont {Zolotorev}}, \ and\ \bibinfo {author} {\bibfnamefont
  {J.}~\bibnamefont {Byrd}}} (\bibinfo {year} {1999}),\ \href@noop {}
  {\bibfield  {journal} {\bibinfo  {journal} {Nucl. Instrum. Methods Phys. Res.
  A}\ }\textbf {\bibinfo {volume} {425}},\ \bibinfo {pages} {385}}\BibitemShut
  {NoStop}%
\bibitem [{\citenamefont {Zholents}\ and\ \citenamefont
  {Holldack}(2006)}]{Zhol3}%
  \BibitemOpen
  \bibfield  {author} {\bibinfo {author} {\bibnamefont {Zholents},
  \bibfnamefont {A.}}, \ and\ \bibinfo {author} {\bibfnamefont
  {K.}~\bibnamefont {Holldack}}} (\bibinfo {year} {2006}),\ \href@noop {}
  {\bibinfo  {journal} {Proc. of the 2006 Free Electron Laser Conference,
  Berlin, Germany}\ ,\ \bibinfo {pages} {725}}\BibitemShut {NoStop}%
\bibitem [{\citenamefont {Zholents}\ and\ \citenamefont {Penn}(2010)}]{ZholTC}%
  \BibitemOpen
\bibfield  {journal} {  }\bibfield  {author} {\bibinfo {author} {\bibnamefont
  {Zholents}, \bibfnamefont {A.}}, \ and\ \bibinfo {author} {\bibfnamefont
  {G.}~\bibnamefont {Penn}}} (\bibinfo {year} {2010}),\ \href@noop {}
  {\bibfield  {journal} {\bibinfo  {journal} {Nucl. Instrum. Methods Phys. Res.
  A}\ }\textbf {\bibinfo {volume} {612}},\ \bibinfo {pages} {254}}\BibitemShut
  {NoStop}%
\bibitem [{\citenamefont {Zholents}\ and\ \citenamefont
  {Zolotorev}(2011)}]{EEX7}%
  \BibitemOpen
  \bibfield  {author} {\bibinfo {author} {\bibnamefont {Zholents},
  \bibfnamefont {A.}}, \ and\ \bibinfo {author} {\bibfnamefont
  {M.}~\bibnamefont {Zolotorev}}} (\bibinfo {year} {2011}),\ \href@noop {}
  {\emph {\bibinfo {title} {A New Type of Bunch Compressor and Seeding of a
  Short Wave Length Coherent Radiation}}},\ \bibinfo {type} {Preprint}\
  \bibinfo {number} {ANL/APS/LS-327}\BibitemShut {NoStop}%
\bibitem [{\citenamefont
  {Zholents}(2005{\natexlab{a}})}]{zholents_conditioner}%
  \BibitemOpen
  \bibfield  {author} {\bibinfo {author} {\bibnamefont {Zholents},
  \bibfnamefont {A.~A.}}} (\bibinfo {year} {2005}{\natexlab{a}}),\ \href
  {http://link.aps.org/doi/10.1103/PhysRevSTAB.8.050701} {\bibfield  {journal}
  {\bibinfo  {journal} {Phys. Rev. ST Accel. Beams}\ }\textbf {\bibinfo
  {volume} {8}},\ \bibinfo {pages} {050701}}\BibitemShut {NoStop}%
\bibitem [{\citenamefont {Zholents}(2005{\natexlab{b}})}]{Zhol4}%
  \BibitemOpen
  \bibfield  {author} {\bibinfo {author} {\bibnamefont {Zholents},
  \bibfnamefont {A.~A.}}} (\bibinfo {year} {2005}{\natexlab{b}}),\ \href
  {http://link.aps.org/doi/10.1103/PhysRevSTAB.8.040701} {\bibfield  {journal}
  {\bibinfo  {journal} {Phys. Rev. ST Accel. Beams}\ }\textbf {\bibinfo
  {volume} {8}},\ \bibinfo {pages} {040701}}\BibitemShut {NoStop}%
\bibitem [{\citenamefont {Zholents}\ and\ \citenamefont
  {Penn}(2005)}]{attosecond1}%
  \BibitemOpen
  \bibfield  {author} {\bibinfo {author} {\bibnamefont {Zholents},
  \bibfnamefont {A.~A.}}, \ and\ \bibinfo {author} {\bibfnamefont
  {G.}~\bibnamefont {Penn}}} (\bibinfo {year} {2005}),\ \href@noop {}
  {\bibfield  {journal} {\bibinfo  {journal} {Phys. Rev. ST Accel. Beams}\
  }\textbf {\bibinfo {volume} {8}},\ \bibinfo {pages} {050704}}\BibitemShut
  {NoStop}%
\bibitem [{\citenamefont {Zholents}\ and\ \citenamefont
  {Zolotorev}(2008)}]{zholents2008}%
  \BibitemOpen
  \bibfield  {author} {\bibinfo {author} {\bibnamefont {Zholents},
  \bibfnamefont {A.~A.}}, \ and\ \bibinfo {author} {\bibfnamefont
  {M.}~\bibnamefont {Zolotorev}}} (\bibinfo {year} {2008}),\ \href@noop {}
  {\bibfield  {journal} {\bibinfo  {journal} {New J. Physics}\ }\textbf
  {\bibinfo {volume} {10}},\ \bibinfo {pages} {025005}}\BibitemShut {NoStop}%
\bibitem [{\citenamefont {Zholents}\ and\ \citenamefont
  {Zolotorev}(1996)}]{laser-slicing96}%
  \BibitemOpen
  \bibfield  {author} {\bibinfo {author} {\bibnamefont {Zholents},
  \bibfnamefont {A.~A.}}, \ and\ \bibinfo {author} {\bibfnamefont {M.~S.}\
  \bibnamefont {Zolotorev}}} (\bibinfo {year} {1996}),\ \href@noop {}
  {\bibfield  {journal} {\bibinfo  {journal} {Phys. Rev. Lett.}\ }\textbf
  {\bibinfo {volume} {76}},\ \bibinfo {pages} {912}}\BibitemShut {NoStop}%
\bibitem [{\citenamefont {Zolotorev}(2002)}]{Zolotorev2002445}%
  \BibitemOpen
  \bibfield  {author} {\bibinfo {author} {\bibnamefont {Zolotorev},
  \bibfnamefont {M.}}} (\bibinfo {year} {2002}),\ \href
  {http://www.sciencedirect.com/science/article/pii/S0168900202003595}
  {\bibfield  {journal} {\bibinfo  {journal} {Nuclear Instruments and Methods
  in Physics Research Section A: Accelerators, Spectrometers, Detectors and
  Associated Equipment}\ }\textbf {\bibinfo {volume} {483}}~(\bibinfo {number}
  {1-2}),\ \bibinfo {pages} {445 }}\BibitemShut {NoStop}%
\end{thebibliography}%

\end{document}